\documentclass[phd,tocprelim]{cornell}
\usepackage{graphicx,pstricks}
\usepackage{graphics}
\usepackage{moreverb}
\usepackage{subfigure}
\usepackage{epsfig}
\usepackage{subfigure}
\usepackage{hangcaption}
\usepackage{txfonts}
\usepackage{palatino}
\graphicspath{{figs/}}
\renewcommand{\caption}[1]{\singlespacing\hangcaption{#1}\normalspacing}

\newcommand{\etal}{\textit{et al.}}

\title {Production and Decay of Charmed Mesons in the Upsilon Energy Region}
\author {David Russell Perticone}
\conferraldate {May}{1988}
\degreefield {Ph.D.}
\copyrightholder{David Russell Perticone}
\copyrightyear{1988}

\begin{document}


\def\particleone{\rm}
\def\particletwo{\sl}

\def\tiny{\vrule width 0pt}

\def\conventionone{

 \def\PM{\ifmmode{\pm}\else{$\pm$}\fi}

 \def\decays{\ifmmode{\rightarrow}\else{$\rightarrow$}\fi\tiny}

 \def\EPEM{\ifmmode{e^+e^-}\else{$e^+e^-$}\fi}
 \def\epem{\ifmmode{e^+e^-}\else{$e^+e^-$}\fi}


 \def\G{\ifmmode{\gamma}\else{$\gamma$}\fi}


 \def\W{\ifmmode{{\particleone W}}\else{{\particleone W}}\fi}
 \def\WP{\ifmmode{{\particleone W}^+}\else{{\particleone W}$^+$}\fi}
 \def\WM{\ifmmode{{\particleone W}^-}\else{{\particleone W}$^-$}\fi}
 \def\WPM{\ifmmode{{\particleone W}^\pm}\else{{\particleone W}$^\pm$}\fi}
 \def\WMP{\ifmmode{{\particleone W}^\mp}\else{{\particleone W}$^\mp$}\fi}


 \def\Z{\ifmmode{{\particleone Z}}\else{{\particleone Z}}\fi}
 \def\ZZ{\ifmmode{{\particleone Z}^0}\else{{\particleone Z}$^0$}\fi}


 \def\N{\ifmmode{\nu}\else{$\nu$}\fi}
 \def\NB{\ifmmode{\overline{\nu}}
	\else{$\overline{\nu}$}\fi}

 \def\NE{\ifmmode{\nu_e}\else{$\nu_e$}\fi}
 \def\NEB{\ifmmode{\overline{\nu}\tiny_e}
	\else{$\overline{\nu}\tiny_e$}\fi}


 \def\E{\ifmmode{e}\else{$e$}\fi}
 \def\EP{\ifmmode{e^+}\else{$e^+$}\fi}
 \def\EM{\ifmmode{e^-}\else{$e^-$}\fi}
 \def\EPM{\ifmmode{e^\pm}\else{$e^\pm$}\fi}
 \def\EMP{\ifmmode{e^\mp}\else{$e^\mp$}\fi}


 \def\NM{\ifmmode{\nu_\mu}\else{$\nu_\mu$}\fi}
 \def\NMB{\ifmmode{\overline{\nu}\tiny_\mu}
	\else{$\overline{\nu}\tiny_\mu$}\fi}


 \def\M{\ifmmode{\mu}\else{$\mu$}\fi}
 \def\MP{\ifmmode{\mu^+}\else{$\mu^+$}\fi}
 \def\MM{\ifmmode{\mu^-}\else{$\mu^-$}\fi}
 \def\MPM{\ifmmode{\mu^\pm}\else{$\mu^\pm$}\fi}
 \def\MMP{\ifmmode{\mu^\mp}\else{$\mu^\mp$}\fi}


 \def\NT{\ifmmode{\nu_\tau}\else{$\nu_\tau$}\fi}
 \def\NTB{\ifmmode{\overline{\nu}\tiny_\tau}
	\else{$\overline{\nu}\tiny_\tau$}\fi}


 \def\T{\ifmmode{\tau}\else{$\tau$}\fi}
 \def\TP{\ifmmode{\tau^+}\else{$\tau^+$}\fi}
 \def\TM{\ifmmode{\tau^-}\else{$\tau^-$}\fi}
 \def\TPM{\ifmmode{\tau^\pm}\else{$\tau^\pm$}\fi}
 \def\TMP{\ifmmode{\tau^\mp}\else{$\tau^\mp$}\fi}


 \def\NL{\ifmmode{\nu_\ell}\else{$\nu_\ell$}\fi}
 \def\NLB{\ifmmode{\overline{\nu}\tiny_\ell}
	\else{$\overline{\nu}\tiny_\ell$}\fi}


 \def\L{\ifmmode{\ell}\else{$\ell$}\fi}
 \def\LP{\ifmmode{\ell^+}\else{$\ell^+$}\fi}
 \def\LM{\ifmmode{\ell^-}\else{$\ell^-$}\fi}
 \def\LPM{\ifmmode{\ell^\pm}\else{$\ell^\pm$}\fi}
 \def\LMP{\ifmmode{\ell^\mp}\else{$\ell^\mp$}\fi}


 \def\PI{\ifmmode{\pi}\else{$\pi$}\fi}
 \def\PIP{\ifmmode{\pi^+}\else{$\pi^+$}\fi}
 \def\PIZ{\ifmmode{\pi^0}\else{$\pi^0$}\fi}
 \def\PIM{\ifmmode{\pi^-}\else{$\pi^-$}\fi}
 \def\PIPM{\ifmmode{\pi^\pm}\else{$\pi^\pm$}\fi}
 \def\PIMP{\ifmmode{\pi^\mp}\else{$\pi^\mp$}\fi}
 \def\PIPMZ{\ifmmode{\pi^{\pm,0}}\else{$\pi^{\pm,0}$}\fi}


 \def\ET{\ifmmode{\eta}\else{$\eta$}\fi}
 \def\ETZ{\ifmmode{\eta^0}\else{$\eta^0$}\fi}


 \def\K{\ifmmode{{\particleone K}}\else{{\particleone K}}\fi}
 \def\KB{\ifmmode{\overline{{\particleone K}}}
	\else{$\overline{{\particleone K}}$}\fi}

 \def\KZ{\ifmmode{{\particleone K}^0}\else{{\particleone K}$^0$}\fi}
 \def\KSH{\ifmmode{{\particleone K}^0_S}\else{{\particleone K}$^0_S$}\fi}
 \def\KLO{\ifmmode{{\particleone K}^0_L}\else{{\particleone K}$^0_L$}\fi}
 \def\KZB{\ifmmode{\overline{{\particleone K}}\tiny^0}
	\else{$\overline{{\particleone K}}\tiny^0$}\fi}

 \def\KP{\ifmmode{{\particleone K}^+}\else{{\particleone K}$^+$}\fi}
 \def\KM{\ifmmode{{\particleone K}^-}\else{{\particleone K}$^-$}\fi}
 \def\KPM{\ifmmode{{\particleone K}^\pm}\else{{\particleone K}$^\pm$}\fi}
 \def\KMP{\ifmmode{{\particleone K}^\mp}\else{{\particleone K}$^\mp$}\fi}


 \def\D{\ifmmode{{\particleone D}}\else{{\particleone D}}\fi\tiny}
 \def\DB{\ifmmode{\overline{{\particleone D}}}
	\else{$\overline{{\particleone D}}$}\fi}
 \def\DF{\ifmmode{{\particleone D}^+_s}\else{{\particleone D}$^+_s$}\fi}
 \def\DZ{\ifmmode{{\particleone D}^0}\else{{\particleone D}$^0$}\fi}
 \def\DZB{\ifmmode{\overline{{\particleone D}}\tiny^0}
	\else{$\overline{{\particleone D}}\tiny^0$}\fi}

 \def\DP{\ifmmode{{\particleone D}^+}\else{{\particleone D}$^+$}\fi}
 \def\DM{\ifmmode{{\particleone D}^-}\else{{\particleone D}$^-$}\fi}
 \def\DPM{\ifmmode{{\particleone D}^\pm}\else{{\particleone D}$^\pm$}\fi}
 \def\DMP{\ifmmode{{\particleone D}^\mp}\else{{\particleone D}$^\mp$}\fi}


 \def\F{\ifmmode{{\particleone F}}\else{{\particleone F}}\fi}
 \def\FP{\ifmmode{{\particleone F}^+}\else{{\particleone F}$^+$}\fi}
 \def\FM{\ifmmode{{\particleone F}^-}\else{{\particleone F}$^-$}\fi}
 \def\FPM{\ifmmode{{\particleone F}^\pm}\else{{\particleone F}$^\pm$}\fi}
 \def\FMP{\ifmmode{{\particleone F}^\mp}\else{{\particleone F}$^\mp$}\fi}


 \def\B{\ifmmode{{\particleone B}}\else{{\particleone B}}\fi\tiny}
 \def\BB{\ifmmode{\overline{{\particleone B}}}
	\else{$\overline{{\particleone B}}$}\fi}

 \def\BZ{\ifmmode{{\particleone B}^0}\else{{\particleone B}$^0$}\fi}
 \def\BZB{\ifmmode{\overline{{\particleone B}}\tiny^0}
	\else{$\overline{{\particleone B}}\tiny^0$}\fi}

 \def\BP{\ifmmode{{\particleone B}^+}\else{{\particleone B}$^+$}\fi}
 \def\BM{\ifmmode{{\particleone B}^-}\else{{\particleone B}$^-$}\fi}
 \def\BPM{\ifmmode{{\particleone B}^\pm}\else{{\particleone B}$^\pm$}\fi}
 \def\BMP{\ifmmode{{\particleone B}^\mp}\else{{\particleone B}$^\mp$}\fi}


 \def\PR{\ifmmode{{\particleone p}}\else{{\particleone p}}\fi}
 \def\PB{\ifmmode{\overline{{\particleone p}}}
	\else{$\overline{{\particleone p}}$}\fi}


 \def\NR{\ifmmode{{\particleone n}}\else{{\particleone n}}\fi}
 \def\NB{\ifmmode{\overline{{\particleone n}}}
	\else{$\overline{{\particleone n}}$}\fi}


 \def\LA{\ifmmode{\Lambda}\else{$\Lambda$}\fi}
 \def\LAB{\ifmmode{\overline{\Lambda}}
	\else{$\overline{\Lambda}$}\fi}
 \def\LAZ{\ifmmode{\Lambda^0}\else{$\Lambda^0$}\fi}
 \def\LAZB{\ifmmode{\overline{\Lambda}\tiny^0}
	\else{$\overline{\Lambda}\tiny^0$}\fi}


 \def\SI{\ifmmode{\Sigma}\else{$\Sigma$}\fi}
 \def\SIB{\ifmmode{\overline{\Sigma}}
	\else{$\overline{\Sigma}$}\fi}
 \def\SIZ{\ifmmode{\Sigma^0}\else{$\Sigma^0$}\fi}
 \def\SIZB{\ifmmode{\overline{\Sigma}\tiny^0}
	\else{$\overline{\Sigma}\tiny^0$}\fi}
 \def\SIP{\ifmmode{\Sigma^+}\else{$\Sigma^+$}\fi}
 \def\SIM{\ifmmode{\Sigma^-}\else{$\Sigma^-$}\fi}
 \def\SIPM{\ifmmode{\Sigma^\pm}\else{$\Sigma^\pm$}\fi}
 \def\SIMP{\ifmmode{\Sigma^\mp}\else{$\Sigma^\mp$}\fi}
 \def\SIPB{\ifmmode{\overline{\Sigma}\tiny^+}
	\else{$\overline{\Sigma}\tiny^+$}\fi}
 \def\SIMB{\ifmmode{\overline{\Sigma}\tiny^-}
	\else{$\overline{\Sigma}\tiny^-$}\fi}
 \def\SIPMB{\ifmmode{\overline{\Sigma}\tiny^\pm}
	\else{$\overline{\Sigma}\tiny^\pm$}\fi}
 \def\SIMPB{\ifmmode{\overline{\Sigma}\tiny^\mp}
	\else{$\overline{\Sigma}\tiny^\mp$}\fi}


 \def\XI{\ifmmode{\Xi}\else{$\Xi$}\fi}
 \def\XIB{\ifmmode{\overline{\Xi}}
	\else{$\overline{\Xi}$}\fi}
 \def\XIZ{\ifmmode{\Xi^0}\else{$\Xi^0$}\fi}
 \def\XIZB{\ifmmode{\overline{\Xi}\tiny^0}
	\else{$\overline{\Xi}\tiny^0$}\fi}
 \def\XIM{\ifmmode{\Xi^-}\else{$\Xi^-$}\fi}
 \def\XIPB{\ifmmode{\overline{\Xi}\tiny^+}
	\else{$\overline{\Xi}\tiny^+$}\fi}


 \def\OMM{\ifmmode{\Omega^-}\else{$\Omega^-$}\fi}
 \def\OMPB{\ifmmode{\overline{\Omega}\tiny^+}
	\else{$\overline{\Omega}\tiny^+$}\fi}


 \def\LC{\ifmmode{\Lambda_c}\else{$\Lambda_c$}\fi}
 \def\LCB{\ifmmode{\overline{\Lambda_c}}
	\else{$\overline{\Lambda}\tiny_c$}\fi}
 \def\LCP{\ifmmode{\Lambda^+_c}\else{$\Lambda^+_c$}\fi}
 \def\LCMB{\ifmmode{\overline{\Lambda}\tiny^-_c}
	\else{$\overline{\Lambda}\tiny^-_c$}\fi}


 \def\RH{\ifmmode{\rho}\else{$\rho$}\fi}
 \def\RHP{\ifmmode{\rho^+}\else{$\rho^+$}\fi}
 \def\RHZ{\ifmmode{\rho^0}\else{$\rho^0$}\fi}
 \def\RHM{\ifmmode{\rho^-}\else{$\rho^-$}\fi}
 \def\RHPM{\ifmmode{\rho^\pm}\else{$\rho^\pm$}\fi}
 \def\RHMP{\ifmmode{\rho^\mp}\else{$\rho^\mp$}\fi}
 \def\RHPMZ{\ifmmode{\rho^{\pm,0}}\else{$\rho^{\pm,0}$}\fi}


 \def\oM{\ifmmode{\omega}\else{$\omega$}\fi}
 \def\oMZ{\ifmmode{\omega^0}\else{$\omega^0$}\fi}


 \def\ETP{\ifmmode{\eta'}\else{$\eta'$}\fi}
 \def\ETPZ{\ifmmode{\eta'\tiny^0}\else{$\eta'\tiny^0$}\fi}


 \def\PH{\ifmmode{\phi}\else{$\phi$}\fi}


 \def\PS{\ifmmode{\psi}\else{$\psi$}\fi}
 \def\PSP{\ifmmode{\psi'}\else{$\psi'$}\fi}
 \def\PSPP{\ifmmode{\psi''}\else{$\psi''$}\fi}
 \def\PSPPP{\ifmmode{\psi'''}\else{$\psi'''$}\fi}
 \def\PSPPPP{\ifmmode{\psi''''}\else{$\psi''''$}\fi}


 \def\US{\ifmmode{\Upsilon{\particleone (1S)}}
	\else{$\Upsilon{\particleone (1S)}$}\fi}
 \def\USS{\ifmmode{\Upsilon{\particleone (2S)}}
	\else{$\Upsilon{\particleone (2S)}$}\fi}
 \def\USSS{\ifmmode{\Upsilon{\particleone (3S)}}
	\else{$\Upsilon{\particleone (3S)}$}\fi}
 \def\USSSS{\ifmmode{\Upsilon{\particleone (4S)}}
	\else{$\Upsilon{\particleone (4S)}$}\fi}
 \def\USSSSS{\ifmmode{\Upsilon{\particleone (5S)}}
	\else{$\Upsilon{\particleone (5S)}$}\fi}
 \def\USSSSSS{\ifmmode{\Upsilon{\particleone (6S)}}
	\else{$\Upsilon{\particleone (6S)}$}\fi}


 \def\KS{\ifmmode{{\particleone K}^*}\else{{\particleone K}$^*$}\fi}
 \def\KSB{\ifmmode{\overline{{\particleone K}}\tiny^*}
	\else{$\overline{{\particleone K}}\tiny^*$}\fi}

 \def\KSZ{\ifmmode{{\particleone K}^{*0}}\else{{\particleone K}$^{*0}$}\fi}
 \def\KSZB{\ifmmode{\overline{{\particleone K}}\tiny^{*0}}
	\else{$\overline{{\particleone K}}\tiny^{*0}$}\fi}

 \def\KSP{\ifmmode{{\particleone K}^{*+}}\else{{\particleone K}$^{*+}$}\fi}
 \def\KSM{\ifmmode{{\particleone K}^{*-}}\else{{\particleone K}$^{*-}$}\fi}
 \def\KSPM{\ifmmode{{\particleone K}^{*\pm}}\else{{\particleone K}$^{*\pm}$}\fi}
 \def\KSMP{\ifmmode{{\particleone K}^{*\mp}}\else{{\particleone K}$^{*\mp}$}\fi}


 \def\DS{\ifmmode{{\particleone D}^*}\else{{\particleone D}$^*$}\fi}
 \def\DSB{\ifmmode{\overline{{\particleone D}}\tiny^*}
	\else{$\overline{{\particleone D}}\tiny^*$}\fi}

 \def\DSZ{\ifmmode{{\particleone D}^{*0}}\else{{\particleone D}$^{*0}$}\fi}
 \def\DSZB{\ifmmode{\overline{{\particleone D}}\tiny^{*0}}
	\else{$\overline{{\particleone D}}\tiny^{*0}$}\fi}

 \def\DSP{\ifmmode{{\particleone D}^{*+}}\else{{\particleone D}$^{*+}$}\fi}
 \def\DSM{\ifmmode{{\particleone D}^{*-}}\else{{\particleone D}$^{*-}$}\fi}
 \def\DSPM{\ifmmode{{\particleone D}^{*\pm}}\else{{\particleone D}$^{*\pm}$}\fi}
 \def\DSMP{\ifmmode{{\particleone D}^{*\mp}}\else{{\particleone D}$^{*\mp}$}\fi}


 \def\DDS{\ifmmode{{\particleone D}^{**}}\else{{\particleone D}$^{**}$}\fi}
 \def\DDSB{\ifmmode{\overline{{\particleone D}}\tiny^{**}}
	\else{$\overline{{\particleone D}}\tiny^{**}$}\fi}

 \def\DDSZ{\ifmmode{{\particleone D}^{**0}}\else{{\particleone D}$^{**0}$}\fi}
 \def\DDSZB{\ifmmode{\overline{{\particleone D}}\tiny^{**0}}
	\else{$\overline{{\particleone D}}\tiny^{**0}$}\fi}

 \def\DDSP{\ifmmode{{\particleone D}^{**+}}\else{{\particleone D}$^{**+}$}\fi}
 \def\DDSM{\ifmmode{{\particleone D}^{**-}}\else{{\particleone D}$^{**-}$}\fi}
 \def\DDSPM{\ifmmode{{\particleone D}^{**\pm}}
	\else{{\particleone D}$^{**\pm}$}\fi}
 \def\DDSMP{\ifmmode{{\particleone D}^{**\mp}}
	\else{{\particleone D}$^{**\mp}$}\fi}

}

\def\conventiontwo{

 \def\epem{\ifmmode{e^+e^-}\else{$e^+e^-$}\fi}

 \def\decays{\ifmmode{\rightarrow}\else{$\rightarrow$}\fi\tiny}


\def\gamma{\ifmmode{\mathchar"10D}\else{$\mathchar"10D$}\fi}


 \def\W{\ifmmode{{\particletwo W}}\else{{\particletwo W}}\fi}
 \def\Wplus{\ifmmode{{\particletwo W}^+}\else{{\particletwo W}$^+$}\fi}
 \def\Wminus{\ifmmode{{\particletwo W}^-}\else{{\particletwo W}$^-$}\fi}
 \def\Wpm{\ifmmode{{\particletwo W}^\pm}\else{{\particletwo W}$^\pm$}\fi}
 \def\Wmp{\ifmmode{{\particletwo W}^\mp}\else{{\particletwo W}$^\mp$}\fi}


 \def\Z{\ifmmode{{\particletwo Z}}\else{{\particletwo Z}}\fi}
 \def\Zzero{\ifmmode{{\particletwo Z}^0}\else{{\particletwo Z}$^0$}\fi}


 \def\nu{\ifmmode{\mathchar"117}\else{$\mathchar"117$}\fi}
 \def\nubar{\ifmmode{\overline{\nu}\tiny_e}
	\else{$\overline{\nu}\tiny_e$}\fi}

 \def\nue{\ifmmode{\nu_e}\else{$\nu_e$}\fi}
 \def\nuebar{\ifmmode{\overline{\nu}\tiny_e}
	\else{$\overline{\nu}\tiny_e$}\fi}


 \def\e{\ifmmode{e}\else{$e$}\fi}
 \def\eplus{\ifmmode{e^+}\else{$e^+$}\fi}
 \def\eminus{\ifmmode{e^-}\else{$e^-$}\fi}
 \def\epm{\ifmmode{e^\pm}\else{$e^\pm$}\fi}
 \def\emp{\ifmmode{e^\mp}\else{$e^\mp$}\fi}


 \def\numu{\ifmmode{\nu_\mu}\else{$\nu_\mu$}\fi}
 \def\numubar{\ifmmode{\overline{\nu}\tiny_\mu}
	\else{$\overline{\nu}\tiny_\mu$}\fi}


 \def\mu{\ifmmode{\mathchar"116}\else{$\mathchar"116$}\fi}
 \def\muplus{\ifmmode{\mu^+}\else{$\mu^+$}\fi}
 \def\muminus{\ifmmode{\mu^-}\else{$\mu^-$}\fi}
 \def\mupm{\ifmmode{\mu^\pm}\else{$\mu^\pm$}\fi}
 \def\mump{\ifmmode{\mu^\mp}\else{$\mu^\mp$}\fi}


 \def\nutau{\ifmmode{\nu_\tau}\else{$\nu_\tau$}\fi}
 \def\nutaubar{\ifmmode{\overline{\nu}\tiny_\tau}
	\else{$\overline{\nu}\tiny_\tau$}\fi}


 \def\tau{\ifmmode{\mathchar"11C}\else{$\mathchar"11C$}\fi}
 \def\tauplus{\ifmmode{\tau^+}\else{$\tau^+$}\fi}
 \def\tauminus{\ifmmode{\tau^-}\else{$\tau^-$}\fi}
 \def\taupm{\ifmmode{\tau^\pm}\else{$\tau^\pm$}\fi}
 \def\taump{\ifmmode{\tau^\mp}\else{$\tau^\mp$}\fi}


 \def\nulep{\ifmmode{\nu_\ell}\else{$\nu_\ell$}\fi}
 \def\nulepbar{\ifmmode{\overline{\nu}\tiny_\ell}
	\else{$\overline{\nu}\tiny_\ell$}\fi}


 \def\lep{\ifmmode{\ell}\else{$\ell$}\fi}
 \def\lepplus{\ifmmode{\ell^+}\else{$\ell^+$}\fi}
 \def\lepminus{\ifmmode{\ell^-}\else{$\ell^-$}\fi}
 \def\leppm{\ifmmode{\ell^\pm}\else{$\ell^\pm$}\fi}
 \def\lepmp{\ifmmode{\ell^\mp}\else{$\ell^\mp$}\fi}


 \def\pi{\ifmmode{\mathchar"119}\else{$\mathchar"119$}\fi}
 \def\piplus{\ifmmode{\pi^+}\else{$\pi^+$}\fi}
 \def\pizero{\ifmmode{\pi^0}\else{$\pi^0$}\fi}
 \def\piminus{\ifmmode{\pi^-}\else{$\pi^-$}\fi}
 \def\pipm{\ifmmode{\pi^\pm}\else{$\pi^\pm$}\fi}
 \def\pimp{\ifmmode{\pi^\mp}\else{$\pi^\mp$}\fi}
 \def\pipmz{\ifmmode{\pi^{\pm,0}}\else{$\pi^{\pm,0}$}\fi}


 \def\eta{\ifmmode{\mathchar"111}\else{$\mathchar"111$}\fi}
 \def\etazero{\ifmmode{\eta^0}\else{$\eta^0$}\fi}


 \def\K{\ifmmode{{\particletwo K}}\else{{\particletwo K}}\fi}
 \def\Kbar{\ifmmode{\overline{{\particletwo K}}}
	\else{$\overline{{\particletwo K}}$}\fi}

 \def\Kzero{\ifmmode{{\particletwo K}^0}\else{{\particletwo K}$^0$}\fi}
 \def\Kshort{\ifmmode{{\particletwo K}^0_S}\else{{\particletwo K}$^0_S$}\fi}
 \def\Klong{\ifmmode{{\particletwo K}^0_L}\else{{\particletwo K}$^0_L$}\fi}
 \def\Kzerobar{\ifmmode{\overline{{\particletwo K}}\tiny^0}
	\else{$\overline{{\particletwo K}}\tiny^0$}\fi}

 \def\Kplus{\ifmmode{{\particletwo K}^+}\else{{\particletwo K}$^+$}\fi}
 \def\Kminus{\ifmmode{{\particletwo K}^-}\else{{\particletwo K}$^-$}\fi}
 \def\Kpm{\ifmmode{{\particletwo K}^\pm}\else{{\particletwo K}$^\pm$}\fi}
 \def\Kmp{\ifmmode{{\particletwo K}^\mp}\else{{\particletwo K}$^\mp$}\fi}


 \def\D{\ifmmode{{\particletwo D}}\else{{\particletwo D}}\fi}
 \def\Dbar{\ifmmode{\overline{{\particletwo D}}}
	\else{$\overline{{\particletwo D}}$}\fi}

 \def\Dzero{\ifmmode{{\particletwo D}^0}\else{{\particletwo D}$^0$}\fi}
 \def\Dzerobar{\ifmmode{\overline{{\particletwo D}}\tiny^0}
	\else{$\overline{{\particletwo D}}\tiny^0$}\fi}

 \def\Dplus{\ifmmode{{\particletwo D}^+}\else{{\particletwo D}$^+$}\fi}
 \def\Dminus{\ifmmode{{\particletwo D}^-}\else{{\particletwo D}$^-$}\fi}
 \def\Dpm{\ifmmode{{\particletwo D}^\pm}\else{{\particletwo D}$^\pm$}\fi}
 \def\Dmp{\ifmmode{{\particletwo D}^\mp}\else{{\particletwo D}$^\mp$}\fi}


 \def\F{\ifmmode{{\particletwo F}}\else{{\particletwo F}}\fi}
 \def\Fplus{\ifmmode{{\particletwo F}^+}\else{{\particletwo F}$^+$}\fi}
 \def\Fminus{\ifmmode{{\particletwo F}^-}\else{{\particletwo F}$^-$}\fi}
 \def\Fpm{\ifmmode{{\particletwo F}^\pm}\else{{\particletwo F}$^\pm$}\fi}
 \def\Fmp{\ifmmode{{\particletwo F}^\mp}\else{{\particletwo F}$^\mp$}\fi}


 \def\B{\ifmmode{{\particletwo B}}\else{{\particletwo B}}\fi}
 \def\Bbar{\ifmmode{\overline{{\particletwo B}}}
	\else{$\overline{{\particletwo B}}$}\fi}

 \def\Bzero{\ifmmode{{\particletwo B}^0}\else{{\particletwo B}$^0$}\fi}
 \def\Bzerobar{\ifmmode{\overline{{\particletwo B}}\tiny^0}
	\else{$\overline{{\particletwo B}}\tiny^0$}\fi}

 \def\Bplus{\ifmmode{{\particletwo B}^+}\else{{\particletwo B}$^+$}\fi}
 \def\Bminus{\ifmmode{{\particletwo B}^-}\else{{\particletwo B}$^-$}\fi}
 \def\Bpm{\ifmmode{{\particletwo B}^\pm}\else{{\particletwo B}$^\pm$}\fi}
 \def\Bmp{\ifmmode{{\particletwo B}^\mp}\else{{\particletwo B}$^\mp$}\fi}


 \def\pro{\ifmmode{{\particletwo p}}\else{{\particletwo p}}\fi}
 \def\probar{\ifmmode{\overline{{\particletwo p}}}
	\else{$\overline{{\particletwo p}}$}\fi}


 \def\neu{\ifmmode{{\particletwo n}}\else{{\particletwo n}}\fi}
 \def\neubar{\ifmmode{\overline{{\particletwo n}}}
	\else{$\overline{{\particletwo n}}$}\fi}


 \def\Lambda{\ifmmode{\mathchar"7003}\else{$\mathchar"7003$}\fi}
 \def\Lambdabar{\ifmmode{\overline{\Lambda}}
	\else{$\overline{\Lambda}$}\fi}
 \def\Lambdazero{\ifmmode{\Lambda^0}\else{$\Lambda^0$}\fi}
 \def\Lambdazerobar{\ifmmode{\overline{\Lambda}\tiny^0}
	\else{$\overline{\Lambda}\tiny^0$}\fi}


 \def\Sigma{\ifmmode{\mathchar"7006}\else{$\mathchar"7006$}\fi}
 \def\Sigmabar{\ifmmode{\overline{\Sigma}}
	\else{$\overline{\Sigma}$}\fi}
 \def\Sigmazero{\ifmmode{\Sigma^0}\else{$\Sigma^0$}\fi}
 \def\Sigmazerobar{\ifmmode{\overline{\Sigma}\tiny^0}
	\else{$\overline{\Sigma}\tiny^0$}\fi}
 \def\Sigmaplus{\ifmmode{\Sigma^+}\else{$\Sigma^+$}\fi}
 \def\Sigmaminus{\ifmmode{\Sigma^-}\else{$\Sigma^-$}\fi}
 \def\Sigmapm{\ifmmode{\Sigma^\pm}\else{$\Sigma^\pm$}\fi}
 \def\Sigmamp{\ifmmode{\Sigma^\mp}\else{$\Sigma^\mp$}\fi}
 \def\Sigmaplusbar{\ifmmode{\overline{\Sigma}\tiny^+}
	\else{$\overline{\Sigma}\tiny^+$}\fi}
 \def\Sigmaminusbar{\ifmmode{\overline{\Sigma}\tiny^-}
	\else{$\overline{\Sigma}\tiny^-$}\fi}
 \def\Sigmapmbar{\ifmmode{\overline{\Sigma}\tiny^\pm}
	\else{$\overline{\Sigma}\tiny^\pm$}\fi}
 \def\Sigmampbar{\ifmmode{\overline{\Sigma}\tiny^\mp}
	\else{$\overline{\Sigma}\tiny^\mp$}\fi}


 \def\Xi{\ifmmode{\mathchar"7004}\else{$\mathchar"7004$}\fi}
 \def\Xibar{\ifmmode{\overline{\Xi}}
	\else{$\overline{\Xi}$}\fi}
 \def\Xizero{\ifmmode{\Xi^0}\else{$\Xi^0$}\fi}
 \def\Xizerobar{\ifmmode{\overline{\Xi}\tiny^0}
	\else{$\overline{\Xi}\tiny^0$}\fi}
 \def\Ximinus{\ifmmode{\Xi^-}\else{$\Xi^-$}\fi}
 \def\Xiplusbar{\ifmmode{\overline{\Xi}\tiny^+}
	\else{$\overline{\Xi}\tiny^+$}\fi}


 \def\Omegaminus{\ifmmode{\Omega^-}\else{$\Omega^-$}\fi}
 \def\Omegaplusbar{\ifmmode{\overline{\Omega}\tiny^+}
	\else{$\overline{\Omega}\tiny^+$}\fi}


 \def\Lambdac{\ifmmode{\Lambda_c}\else{$\Lambda_c$}\fi}
 \def\Lambdacbar{\ifmmode{\overline{\Lambda_c}}
	\else{$\overline{\Lambda}\tiny_c$}\fi}
 \def\Lambdacplus{\ifmmode{\Lambda^+_c}\else{$\Lambda^+_c$}\fi}
 \def\Lambdacminusbar{\ifmmode{\overline{\Lambda}\tiny^-_c}
	\else{$\overline{\Lambda}\tiny^-_c$}\fi}


 \def\rho{\ifmmode{\mathchar"11A}\else{$\mathchar"11A$}\fi}
 \def\rhoplus{\ifmmode{\rho^+}\else{$\rho^+$}\fi}
 \def\rhozero{\ifmmode{\rho^0}\else{$\rho^0$}\fi}
 \def\rhominus{\ifmmode{\rho^-}\else{$\rho^-$}\fi}
 \def\rhopm{\ifmmode{\rho^\pm}\else{$\rho^\pm$}\fi}
 \def\rhomp{\ifmmode{\rho^\mp}\else{$\rho^\mp$}\fi}
 \def\rhopmz{\ifmmode{\rho^{\pm,0}}\else{$\rho^{\pm,0}$}\fi}


 \def\omega{\ifmmode{\mathchar"121}\else{$\mathchar"121$}\fi}
 \def\omegazero{\ifmmode{\omega^0}\else{$\omega^0$}\fi}


 \def\etaprime{\ifmmode{\eta'}\else{$\eta'$}\fi}
 \def\etaprimezero{\ifmmode{\eta'\tiny^0}\else{$\eta'\tiny^0$}\fi}


 \def\phi{\ifmmode{\mathchar"11E}\else{$\mathchar"11E$}\fi}


 \def\psi{\ifmmode{\mathchar"120}\else{$\mathchar"120$}\fi}
 \def\psiprime{\ifmmode{\psi'}\else{$\psi'$}\fi}
 \def\psidoubleprime{\ifmmode{\psi''}\else{$\psi''$}\fi}
 \def\psitripleprime{\ifmmode{\psi'''}\else{$\psi'''$}\fi}
 \def\psifourprime{\ifmmode{\psi''''}\else{$\psi''''$}\fi}


 \def\Uones{\ifmmode{\Upsilon{\particletwo (1S)}}
	\else{$\Upsilon{\particletwo (1S)}$}\fi}
 \def\Utwos{\ifmmode{\Upsilon{\particletwo (2S)}}
	\else{$\Upsilon{\particletwo (2S)}$}\fi}
 \def\Uthrees{\ifmmode{\Upsilon{\particletwo (3S)}}
	\else{$\Upsilon{\particletwo (3S)}$}\fi}
 \def\Ufours{\ifmmode{\Upsilon{\particletwo (4S)}}
	\else{$\Upsilon{\particletwo (4S)}$}\fi}
 \def\Ufives{\ifmmode{\Upsilon{\particletwo (5S)}}
	\else{$\Upsilon{\particletwo (5S)}$}\fi}
 \def\Usixs{\ifmmode{\Upsilon{\particletwo (6S)}}
	\else{$\Upsilon{\particletwo (6S)}$}\fi}


 \def\Kstar{\ifmmode{{\particletwo K}^*}\else{{\particletwo K}$^*$}\fi}
 \def\Kstarbar{\ifmmode{\overline{{\particletwo K}}\tiny^*}
	\else{$\overline{{\particletwo K}}\tiny^*$}\fi}

 \def\Kstarzero{\ifmmode{{\particletwo K}^{*0}}\else{{\particletwo K}$^{*0}$}\fi}
 \def\Kstarzerobar{\ifmmode{\overline{{\particletwo K}}\tiny^{*0}}
	\else{$\overline{{\particletwo K}}\tiny^{*0}$}\fi}

 \def\Kstarplus{\ifmmode{{\particletwo K}^{*+}}\else{{\particletwo K}$^{*+}$}\fi}
 \def\Kstarminus{\ifmmode{{\particletwo K}^{*-}}\else{{\particletwo K}$^{*-}$}\fi}
 \def\Kstarpm{\ifmmode{{\particletwo K}^{*\pm}}\else{{\particletwo K}$^{*\pm}$}\fi}
 \def\Kstarmp{\ifmmode{{\particletwo K}^{*\mp}}\else{{\particletwo K}$^{*\mp}$}\fi}


 \def\Dstar{\ifmmode{{\particletwo D}^*}\else{{\particletwo D}$^*$}\fi}
 \def\Dstarbar{\ifmmode{\overline{{\particletwo D}}\tiny^*}
	\else{$\overline{{\particletwo D}}\tiny^*$}\fi}

 \def\Dstarzero{\ifmmode{{\particletwo D}^{*0}}\else{{\particletwo D}$^{*0}$}\fi}
 \def\Dstarzerobar{\ifmmode{\overline{{\particletwo D}}\tiny^{*0}}
	\else{$\overline{{\particletwo D}}\tiny^{*0}$}\fi}

 \def\Dstarplus{\ifmmode{{\particletwo D}^{*+}}\else{{\particletwo D}$^{*+}$}\fi}
 \def\Dstarminus{\ifmmode{{\particletwo D}^{*-}}\else{{\particletwo D}$^{*-}$}\fi}
 \def\Dstarpm{\ifmmode{{\particletwo D}^{*\pm}}\else{{\particletwo D}$^{*\pm}$}\fi}
 \def\Dstarmp{\ifmmode{{\particletwo D}^{*\mp}}\else{{\particletwo D}$^{*\mp}$}\fi}


 \def\Ddoublestar{\ifmmode{{\particletwo D}^{**}}\else{{\particletwo D}$^{**}$}\fi}
 \def\Ddoublestarbar{\ifmmode{\overline{{\particletwo D}}\tiny^{**}}
	\else{$\overline{{\particletwo D}}\tiny^{**}$}\fi}

 \def\Ddoublestarzero{\ifmmode{{\particletwo D}^{**0}}
	\else{{\particletwo D}$^{**0}$}\fi}
 \def\Ddoublestarzerobar{\ifmmode{\overline{{\particletwo D}}\tiny^{**0}}
	\else{$\overline{{\particletwo D}}\tiny^{**0}$}\fi}

 \def\Ddoublestarplus{\ifmmode{{\particletwo D}^{**+}}
	\else{{\particletwo D}$^{**+}$}\fi}
 \def\Ddoublestarminus{\ifmmode{{\particletwo D}^{**-}}
	\else{{\particletwo D}$^{**-}$}\fi}
 \def\Ddoublestarpm{\ifmmode{{\particletwo D}^{**\pm}}
	\else{{\particletwo D}$^{**\pm}$}\fi}
 \def\Ddoublestarmp{\ifmmode{{\particletwo D}^{**\mp}}
	\else{{\particletwo D}$^{**\mp}$}\fi}


 \def\Bstar{\ifmmode{{\particletwo B}^*}\else{{\particletwo B}$^*$}\fi}
 \def\Bstarbar{\ifmmode{\overline{{\particletwo B}}\tiny^*}
	\else{$\overline{{\particletwo B}}\tiny^*$}\fi}

 \def\Bstarzero{\ifmmode{{\particletwo B}^{*0}}\else{{\particletwo B}$^{*0}$}\fi}
 \def\Bstarzerobar{\ifmmode{\overline{{\particletwo B}}\tiny^{*0}}
	\else{$\overline{{\particletwo B}}\tiny^{*0}$}\fi}

 \def\Bstarplus{\ifmmode{{\particletwo B}^{*+}}\else{{\particletwo B}$^{*+}$}\fi}
 \def\Bstarminus{\ifmmode{{\particletwo B}^{*-}}\else{{\particletwo B}$^{*-}$}\fi}
 \def\Bstarpm{\ifmmode{{\particletwo B}^{*\pm}}\else{{\particletwo B}$^{*\pm}$}\fi}
 \def\Bstarmp{\ifmmode{{\particletwo B}^{*\mp}}\else{{\particletwo B}$^{*\mp}$}\fi}

 \def\Vud{\ifmmode{{\rm V}_{ud}}\else{{\rm V}$_{ud}$}\fi}
 \def\Vcd{\ifmmode{{\rm V}_{cd}}\else{{\rm V}$_{cd}$}\fi}
 \def\Vtd{\ifmmode{{\rm V}_{td}}\else{{\rm V}$_{td}$}\fi}
 \def\Vus{\ifmmode{{\rm V}_{us}}\else{{\rm V}$_{us}$}\fi}
 \def\Vcs{\ifmmode{{\rm V}_{cs}}\else{{\rm V}$_{cs}$}\fi}
 \def\Vts{\ifmmode{{\rm V}_{ts}}\else{{\rm V}$_{ts}$}\fi}
 \def\Vub{\ifmmode{{\rm V}_{ub}}\else{{\rm V}$_{ub}$}\fi}
 \def\Vcb{\ifmmode{{\rm V}_{cb}}\else{{\rm V}$_{cb}$}\fi}
 \def\Vtb{\ifmmode{{\rm V}_{tb}}\else{{\rm V}$_{tb}$}\fi}

}
%

\def\today{\ifcase\month\or
	January\or February\or March\or April\or May\or June\or
	July\or August\or September\or October\or November\or December\fi
	\space\number\day,~\number\year}

\def\cbxone#1#2{\bigskip\rightline{\vbox{\hbox{\today}
		  \hbox{CBX 87~---~#1}\hbox{#2}}}\bigskip}

\def\cbxtwo#1#2#3{\bigskip\rightline{\vbox{\hbox{\today}
		  \hbox{CBX 87~---~#1}\hbox{#2}\hbox{#3}}}\bigskip}

\def\cbxthree#1#2#3#4{\bigskip\rightline{\vbox{\hbox{\today}
		  \hbox{CBX 87~---~#1}\hbox{#2}\hbox{#3}\hbox{#4}}}\bigskip}

\def\cbxfour#1#2#3#4#5{\bigskip\rightline{\vbox{\hbox{\today}
	  \hbox{CBX 87~---~#1}\hbox{#2}\hbox{#3}\hbox{#4}\hbox{#5}}}\bigskip}

\def\csnone#1#2{\bigskip\rightline{\vbox{\hbox{\today}
		  \hbox{CSN~---~#1}\hbox{#2}}}\bigskip}

\def\csntwo#1#2#3{\bigskip\rightline{\vbox{\hbox{\today}
		  \hbox{CSN~---~#1}\hbox{#2}\hbox{#3}}}\bigskip}

\def\csnthree#1#2#3#4{\bigskip\rightline{\vbox{\hbox{\today}
		  \hbox{CSN~---~#1}\hbox{#2}\hbox{#3}\hbox{#4}}}\bigskip}

\def\csnfour#1#2#3#4#5{\bigskip\rightline{\vbox{\hbox{\today}
	  \hbox{CSN~---~#1}\hbox{#2}\hbox{#3}\hbox{#4}\hbox{#5}}}\bigskip}

\def\rightboxone#1{\bigskip\rightline{\vbox{\hbox{#1}}}\bigskip}

\def\rightboxtwo#1#2{\bigskip\rightline{\vbox{\hbox{#1}
	  \hbox{#2}}}\bigskip}

\def\rightboxthree#1#2#3{\bigskip\rightline{\vbox{\hbox{#1}
	  \hbox{#2}\hbox{#3}}}\bigskip}

\def\rightboxfour#1#2#3#4{\bigskip\rightline{\vbox{\hbox{#1}
	  \hbox{#2}\hbox{#3}\hbox{#4}}}\bigskip}

\def\rightboxfive#1#2#3#4#5{\bigskip\rightline{\vbox{\hbox{#1}
	  \hbox{#2}\hbox{#3}\hbox{#4}\hbox{#5}}}\bigskip}

\def\pb{$\rm pb^{-1}$} 
\def\pp{$\pm$}
\def\cv{$\chi^2_V$}
\def\et{ {\it et al.,} \rm}
\def\tpr{Phys. Rev.}
\conventionone
\maketitle
\makecopyright

\begin{abstract}
\par
We study the properties of charmed, nonstrange, \D\ mesons produced
in continuum \EPEM\ annihilations.  The \EPEM\ collisions were
generated in the energy region of the
\USSS\ and \USSSS\ resonances by the Cornell Electron Storage Ring,
and were analyzed by the CLEO detector.
We make extensive study of the decay topology \DZ, \DP\ \decays\
\KZB X, where the \KZB\ is observed in the final state \KSH\
\decays\ \PIP\PIM. The decay \DP\ \decays\ \KM\PIP\PIP\ is also
considered.
We analyze the hadronization process through which charmed quarks
become charmed hadrons. We measure the probability that a charmed
meson will be produced in a state of non zero angular momentum. A
comparison of the \DZ\ and \DP\ fragmentation distributions is
made to different fragmentation models, and to other charmed
hadrons.
A measurement of the relative production of several \DP\ decay
modes is made, and the total \DZ\ and \DP\ cross sections are
estimated. Techniques are presented for the analysis of satellite
mass peaks and $\DP \Leftrightarrow \DF$ reflections.
We search for nonspectator effects in the weak decays of charmed
mesons, and for evidence of hadronic final state interactions in
\DZ\ decay.

\end{abstract}

\begin{biosketch}
David Russell Perticone, son of Frank and Lena  B.
Perticone, was born and raised on America's North Coast in the city of Rochester, New York.
He  was educated in the West Irondequoit public school
system.  During the summer which followed his junior year at
Irondequoit high school, he spent six weeks traveling through
Europe and the Soviet Union as a student ambassador on the People
to People program. In his senior year, he was elected President of
the Robert Fulton Society. He also was a four year participant in
the interscholastic tennis program.
In the fall of 1979 he entered Rensselaer Polytechnic Institute as
a chemical engineering major, only to emerge four years later with
a degree in physics. During the summer of 1981 he was selected to
participate in the Summer Science Program of the Stanford Linear
Accelerator Center. It was there that he received his introduction
to experimental high energy physics and drift chamber technology.
He returned to SLAC the following summer.
In June of 1983, shortly after the completion of his B.S., 
he began graduate study at
Cornell University. At that time he joined the
Laboratory of Nuclear Studies. In his first year of graduate school 
he was supported  by a full graduate research assistant-ship, the only
Cornell physics graduate student in
recent history to hold such an appointment. He remained a G.R.A.
until completion of his degree, with a brief hiatus as a half time
teaching assistant during the spring of his third year.
An avid reader, his recreational diversions include  ice hockey,
croquet, and bicycling. Upon completion of his Ph.D. program he
has accepted a Research Associate position with the University of
Minnesota.
\end{biosketch}

\begin{acknowledgements}
It is in many ways ironic that an experimental high energy physics
paper bears the names of a hundred or so physicists, while a thesis
from the same experiment bears the name of only one. Many people
have contributed to this work, here I shall offer an incomplete list.
\par
Inspiration to achieve such a high level of education has come form
the many excellent instructors that I have had the good fortune to
be exposed to. Don Ross, who taught me trigonometry and calculus at
Irondequoit high school, instilled   early on   the virtues of
rigorous analysis. Seyffie Maleki, then completing his Ph.D. at
R.P.I., revived my interest in physics enough so that I made it my
major. Prof. Bill Atwood and Art Ogawa of SLAC group A made my
introduction to experimental high energy physics so enjoyable that
I decided to pursue it as a career.
\par
I have benefited greatly from being able to work with many
talented people at Cornell's Laboratory of Nuclear Studies.
Foremost thanks go to my special committee chairman  Prof. Don
Hartill,  for the kindness, support, and tremendous patience he has
shown me over the years. He allowed me the flexibility to
independently pursue my research interests, and the opportunity to
rely on his keen physical insight when things went awry. Prof. Gil
Gilchriese had the unique distinction of being the only committee member to
attend all my graduate
examinations, and has supplied me with several interesting projects.
Profs. Peter Lepage and T. M. Yan served as the theory members on
my committee. Together with Prof. Henry Tye, they provided an
amiable interface to the world of theoretical physics.
\par
Prof. Dave Cassel served as the chairman of my special committee
during the first 1.5 years of my   program. He has been a
perennial source of wisdom during my stay at Cornell. On several
occasions I worked with Prof. Rich Galik, which I enjoyed for among
other things his pleasant demeanor. Thanks are also due to Selden
Ball, Chris Bebek, Dan Coffman, John Dobbins, Curt Dunnam, Dave
Kreinick, Mike Ogg, and Mike $(\mu)$ Roman for assistance on a
variety of occasions. Also to the LNS CESR, computer, and technical
support groups for doing all the hard work.
\par
My work on the drift chamber was supervised by Prof. Paul Avery
while he was at Cornell. His lightning quick wit, which was only
exceeded by his typing speed, cheered up otherwise gloomy tasks.
In addition, he assisted in writing almost all of the software tools
vital to CLEO data analysis. My most heartfelt thanks go to J.
Kandaswamy, who always took time out of his busy schedule whenever
I needed help. He answered questions on hardware, software, data
analysis, theory, and just about anything else I could throw at him.
Outside of the lab, he has been a source of good conversation,
companionship at films and sporting events, and occasional Indian
cooking. His dedication to LNS  staff and facilities  has truly
been impressive. J.K., for all you do, this Bud's for you.
\par
Thanks also go to the many people and institutions that make up the
CLEO collaboration. Ian Brock, Mac Mestayer, and Gary Word have
provided valuable software assistance. My hat goes off to the
 members of the VD group, whose hard nosed efforts and classy
detector made this data something special. In particular, I would
like to express my gratitude to VD'er Steve Behrends, for his
friendship, his example as a physicist, and for making me do a much
better job with the drift chamber calibration than I ever would
have liked. Many of the projects in this thesis were inspired by
the recent CLEO analysis coordinators Prof. Gil Gilchriese, Prof.
Tom Ferguson, and Prof. Hassan Jawahery. Their ``encouragement'' to
write papers and give talks has led to a very important part of my
education. Thanks are also in order for the members of my review
committees.
\par
My long time  office mates Thierry (Skippy) Copi\'e, Jim Mueller, and
Dan Riley have provided a lively and stimulating environment to learn
and do physics. I am greatly indebted to them for the many lessons
I have learned, especially those on the croquet field. In
particular I would like to thank Skippy, who I have worked with in
close association since the day I arrived in Ithaca. He has
provided me with invaluable assistance all the way from problem
sets to thesis figures. Most importantly, he showed me that it was
possible to be a good physicist and enjoy life too. My regards to
the many  graduate student
comrades who made life in and out of the lab much more enjoyable.
Also to the woods, streams, waterfalls and ponds of Ithaca and the
music of Jane Siberry, in whose beauty I have found solace on many
occasions.
\par
Finally, an effort of this magnitude could not have been accomplished
without the support and encouragement of many people. My family and
friends have always managed to reward my long absences and
sometimes over zealous work habits with kindness and warmth. I
cannot thank them enough for this. Above all, the most important
contribution has come from my parents, Frank and Lena, without
whose love, dedication, and sacrifice this work would not have been
possible.

\end{acknowledgements}

\contentspage
\tablelistpage
\figurelistpage

\normalspacing \setcounter{page}{1} \pagenumbering{arabic}
\pagestyle{cornell} \addtolength{\parskip}{0.5\baselineskip}
\chapter{Introduction}
\begin{quote}
 After all a man cant only do what he has to do, 
 with what he has to do it with, with what he has 
 learned, to the best of his judgment. And I reckon
a hog is still a hog, no matter what it looks like. 
 So here goes, \\
 William Faulkner, \underline{Old Man}
\end{quote}
\par
\begin{figure}[t!]
\centering
\includegraphics[scale=0.75]{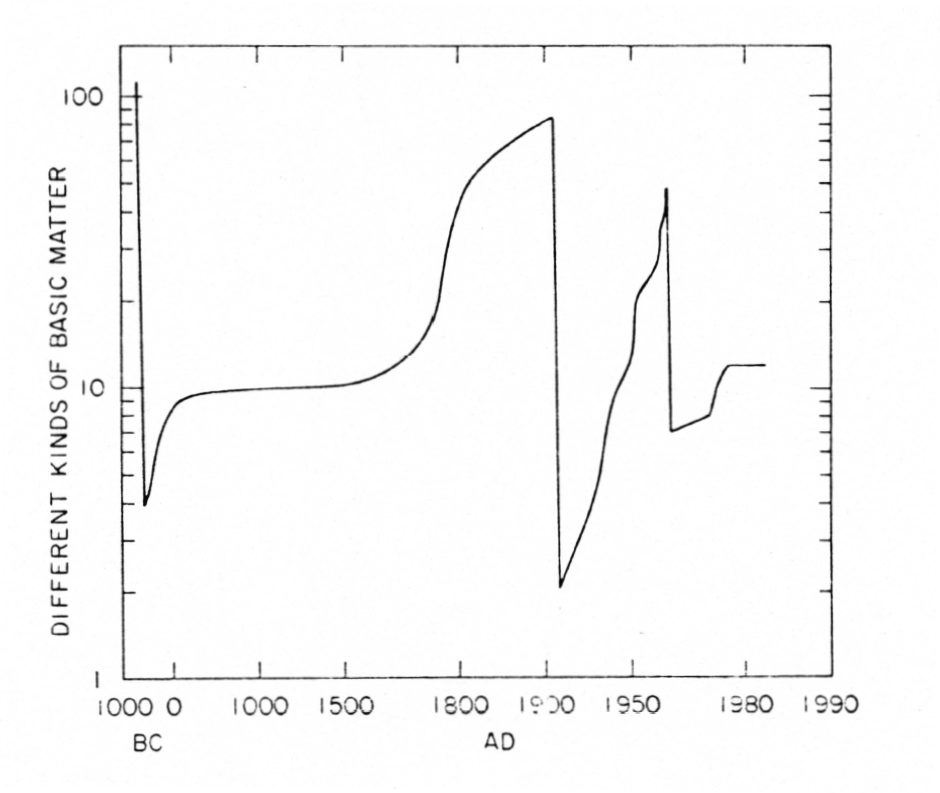}
\caption{Illustrative sketch of the approximate number of elementary
particles versus time.}
\label{fig:c1f1}
\end{figure}

Man's desire to understand the behavior of  nature at its most
fundamental level has existed for several millennia.   As
technology has advanced, ever finer distance scales have been
revealed. Accompanying these revelations have been revolutions in
what has been considered fundamental, as displayed in
Figure \ref{fig:c1f1}.
 Revolutions may be characterized by the
observation of a plethora of states considered to be fundamental,
which are ultimately explained as being composed of a newer set
of constituents. 
The theory currently in vogue is referred to as the
$SU(3)_c \times SU(2) \times SU(1)$ Standard Model of the Strong
and Electroweak interactions.  It describes the interactions
of twelve fundamental fermions (matter fields) via the
exchange of  gauge bosons belonging to three distinct forces
(strong, electromagnetic, weak).
It is an anomaly free, fully renormalizable theory, with an
impressive record of performance. Yet, in its minimal form it 
has a total of 18 parameters; three gauge couplings, two parameters
for the Higgs sector, nine masses for the fundamental fermions
(three are expected to be massless), and three angles plus a phase
which describe quark mixing in the weak interaction.
The centerpiece of the standard model is the Higgs boson, which
breaks the electroweak symmetry  and provides the mass of the
weak intermediate vector bosons and the fundamental fermions.
The Higgs remains unobserved experimentally, and many of the
parameters are not predicted by the theory. These must be 
determined experimentally
(the Higgs mass is among them). We briefly recount the salient
features of the fundamental fermions, and the three forces that
govern their dynamics. Detailed expositions on the standard model
can be found in several textbooks \cite{cpp,ql}.

\section{Matter Fields} 
Symmetry pervades our classification of the fundamental fermions.
The twelve are broken into two groups of 6, being the leptons and
the quarks. Of the six in each family there are two types of
particles which are distinguished by electric charge (-1,0 for
leptons, $+ {2\over 3}, -{1 \over 3}$ for quarks). It is then
convenient to group the $ -1 \choose 0$,  $ + {2\over 3}  \choose 
 -{1 \over 3} $ pairs into generations. For the left-handed states
each generation forms a natural isodoublet. This hierarchy is
summarized in Table~\ref{t:1p1}.
\begin{table}[t!]
\centering
\caption{Quark and Lepton Doublets}
\begin{tabular}{|c|c|c|}
\hline
Generation & Quarks & Leptons\\
1 & $ {u \choose d }_L$ & $ {\nu_e \choose e^- }_L$ \\ 
2 & $ {c \choose s}_L$ & $ {\nu_{\mu} \choose \mu^- }_L$ \\
3 & $ {t \choose b}_L$ & $ {\nu_{\tau} \choose \tau^- }_L$ \\
\hline
\end{tabular}
\label{t:1p1}
\end{table}
 The 
 electrically neutral leptons
are called neutrinos. Since neutrinos are thought to be massless, right handed
states are only allowed for anti-neutrinos. Thus the charged leptons
form right handed singlets, as do both types of electrically charged quarks.
 Despite the topological similarity, two rather profound differences
exist between the quarks and leptons:
\begin{enumerate}
\item The first great distinction
between the two families is that quarks carry color charge and are
thus allowed to participate in the strong interaction. Since there
are three types of color charge, there are in principle three times
as many  quarks as leptons. 
\item
The second is that each lepton generation
possess a distinctive identity, such that the total number of
leptons  $(L_i = 1)$  and anti-leptons $(L_i = -1)$ from each doublet
$(i = 1, 2, 3)$ are separately conserved for all known interactions.  Quarks
have a much weaker identity. The first generation forms an isospin
doublet, while each of the remaining four flavors each carry their own
quantum number. The quark  flavor quantum numbers are  only
preserved in the strong interaction.
\end{enumerate}
We shall now examine the ramifications of these differences.

\section{Gauge Fields}
\begin{table}[ht!]
\centering
\caption{Properties of Gauge Bosons}
\begin{tabular}{|c|c|c|c|c|c|}
\hline
Force & Symmetry  & Quanta & Mass & Coupling & Charge  \\ \hline
Strong & $SU(3)$ &   8 gluons (g)& No & $\alpha_s(Q^2)$ &
color\\ \hline
Electroweak & $U(1)$ & Photon ( \G )& No & $e$ & No \\ \hline
 Electroweak & $SU(2)$ & \Z\ & 92.6 GeV &$ g = $ & No\\
Electroweak & $SU(2)$ & 2 \WPM\ & 81.8 GeV & ${e / \sin\Theta_W}$ &
electric \\ \hline
\end{tabular}
\label{t:1p2}
\end{table}
 Identical fermions are prevented from occupying the
same spacetime point, integer  spin  (bosons) particles suffer no
such restriction and are thus naturally associated with fields.
Particles are subjected to a given force by exchanging the field
quanta of that particular force. Table~\ref{t:1p2}
presents the properties of the Gauge bosons, and 
\begin{figure}[htp!]
\centering
\includegraphics[scale=0.65]{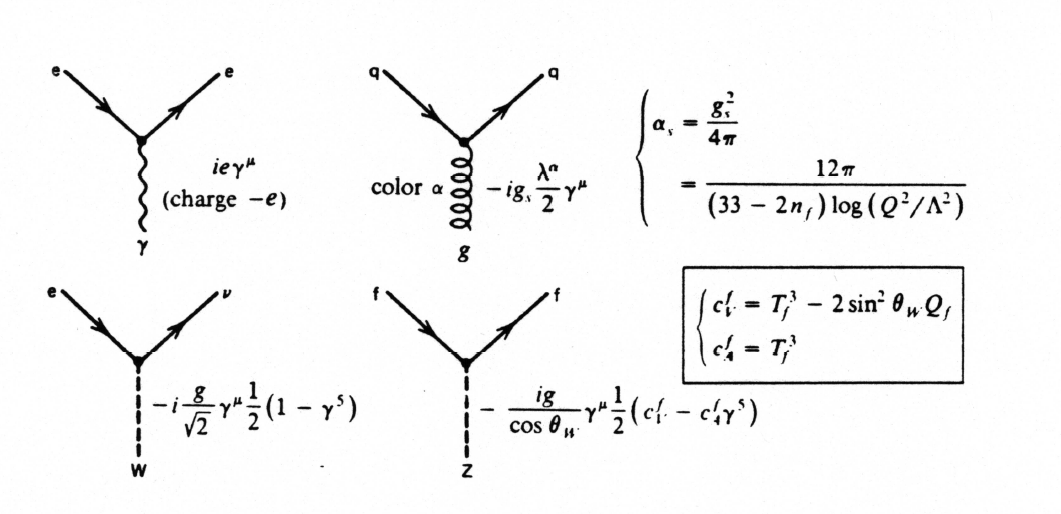}
\caption{Feynman
diagrams for the electric, strong, weak neutral, and weak charged
interactions, and associated parameters.}
\label{fig:c1f2}
\end{figure}
Figure~\ref{fig:c1f2} 
collects the Feynman
rules for these interactions. The coupling constants for these
forces are very different, obeying the approximate relationship
 $$ \rm Strong:EM:Weak \hskip .2in\ 1:{1\over 137}:10^{-5}$$
 \par The
dominant theme in the  Standard Model is local  gauge (or phase)
invariance. The Lagrangian for any interaction is the particle
physics analog  of DNA. It contains all the information and possible
reactions for a given interaction. All Lagrangians are required to
be invariant under the transformation $ e^{-i\alpha(x)}$. It is
precisely the terms which arise in the Lagrangian to preserve local
gauge invariance which insure renormalizability and give each force
its distinctive character.
 \subsection{Strong Interaction}
Gluons  themselves posses color charge, thus not only transmit the
strong force between colored objects, but are capable of
interacting with themselves. The strong interaction is Non-Abelian,
and part of this is manifested in that the coupling constant of the
strong force is not constant, but changes with
energy or 1/distance. To lowest order the coupling constant is:
$$ \alpha_s(Q^2) = {12\pi \over (33 - 2n_f)ln(Q^2/ \Lambda^2)} $$
where $n_f$ is the number of quark flavors and $\Lambda$ is the QCD
scale parameter.
This causes an effect referred to as asymptotic freedom. When quarks
are probed at very short  distance they appear almost free, however
as quarks are separated the force becomes so great that it is
energetically favorable to pop another q\=q pair from the vacuum.
Colored objects (i. e. quarks) have never been directly observed.
  They always combine in
two's (q\=q) and three's (qqq) to produce colorless final states.
 These hadrons are very complex structures, consisting of the
valence quarks, the gluons which bind them, and virtual (q\=q) pairs
(or sea  quarks). Deep inelastic scattering experiments, which
measure the momentum fractions of composite objects, found that
about
50\% of a proton's momentum is carried by neutral particles,
substantiating the role of gluons as the hadronic ``glue."
\begin{figure}[htp!]
\centering
\includegraphics[scale=0.65]{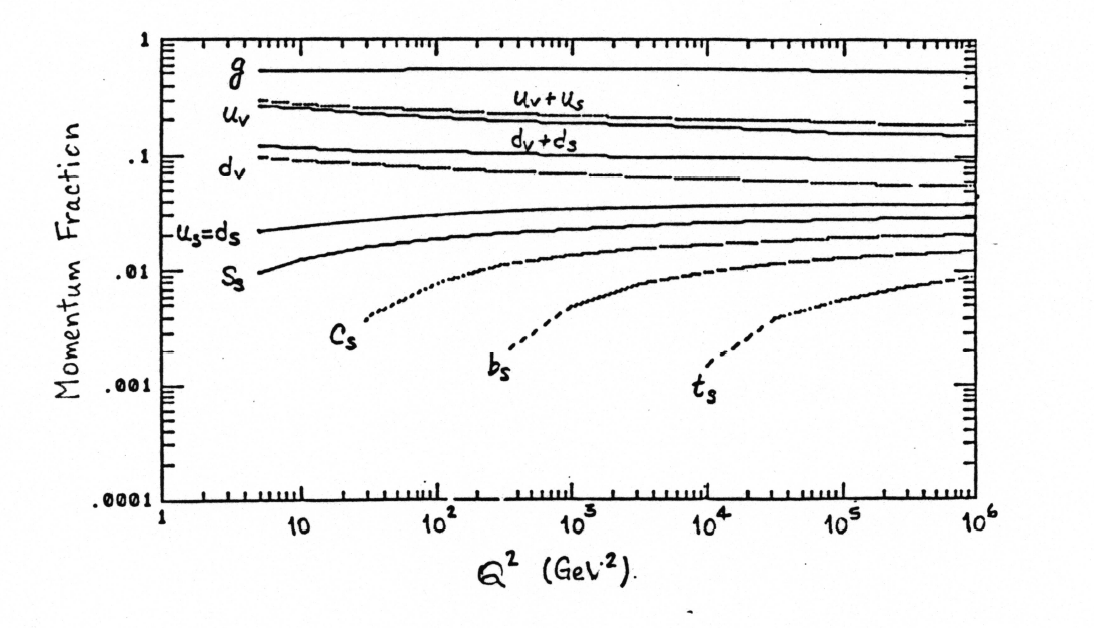}
\caption{Momentum fraction of   proton components  versus $Q^2$
of the probe.}
\label{fig:c1f3}
\end{figure}
Figure~\ref{fig:c1f3} shows a calculation of the momentum fraction of the components of
a proton versus $Q^2$ of the probe, done by Eitchen \cite{eicht}.
The theory of color interactions is known as quantum
chromodynamics (QCD). 
\subsection{Electroweak  Interaction}
In a master stroke, the weak and electromagnetic forces were unified
in the same fashion as the electric and magnetic forces were
unified by Maxwell. This is the so called Weinberg-Salam $SU(2)_L
\times U(1)_Y$ theory, which merited a Nobel prize. Here the four
electroweak vector bosons are split, where the three weak vector
bosons (two charged and one neutral) acquire a mass on order 100
GeV, and the photon remains massless. Because the weak bosons are
massive, the range over which the weak force can be transmitted is
greatly reduced.  \par
\begin{figure}[htp!]
\centering
\includegraphics[scale=0.60]{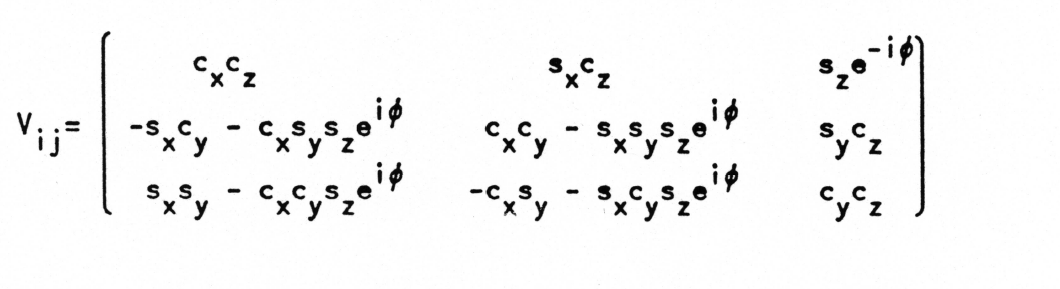}
\caption{Quark
mixing matrix in the parameterization of Chau and Keung. The three
angles are x, y, and z. $\rm s_x = \sin(x)$ and $\rm c_x = \cos(x)$.}
\label{fig:c1f4}
\end{figure}

The exchange of the charged \WPM\
bosons governs the transmutations of the quarks and leptons. The
most general transformation consists of a quark undergoing a flavor
and charge changing transition by emitting a virtual \W\ boson.
The \W\ then fragments into either a particle antiparticle pair
from a lepton doublet, or  quark pair. The pivotal factor is that
no constraint such as lepton number exists for weak transitions of
quarks. Thus a $ + {2 \over 3}$ quark could transform into any
lighter $ - {1 \over 3}$ quark. Similarly, the \W\ may also fragment
into quarks which do not belong to the same doublet.
   The quark ``mass" eigenstates 
are not the eigenstates of the weak interaction.  The
weak eigenstates are obtained by the rotation of the  $ - {1 \over
3}$ quarks $$ \left ( \matrix{   d' \cr   s' \cr  t' \cr } 
\right ) =  \left( \matrix{ V_{ud} & V_{us} & V_{ub} \cr  
 V_{cd} & V_{cs} & V_{cb} \cr
 V_{td} & V_{ts} & V_{tb} \cr} \right)
\left ( \matrix{   d  \cr   s  \cr  t  \cr } 
\right )
$$
  $V$ is a $ 3 \times 3$ unitary matrix often referred to as the
K-M matrix, after Kobayashi and Maskawa \cite{kmm}
who
first generalized the quark mixing matrix to three generations. It
contains three angles and a phase for three generations of quarks.
The parameterization of Chau and Keung \cite{chau}
 is shown in Figure~\ref{fig:c1f4}
 It has the approximate
value $$  \left( \matrix{ 1 &  s & s^3 \cr -s & 1 & s^2 \cr s^3
&-s^2 &1\cr} \right) \hskip 0.5in s \sim 0.23 $$ Off diagonal
transitions occur much less frequently, and are historically referred
to as Cabbibo suppressed. This fortunate feature of weak decays
lends a great deal of richness to the decays of heavy quarks. 
Flavor changing neutral currents, however, are excluded by the
unitarity of the K-M matrix.
\section{The Role of Charm}
\begin{figure}[htp!]
\centering
\includegraphics[scale=0.65]{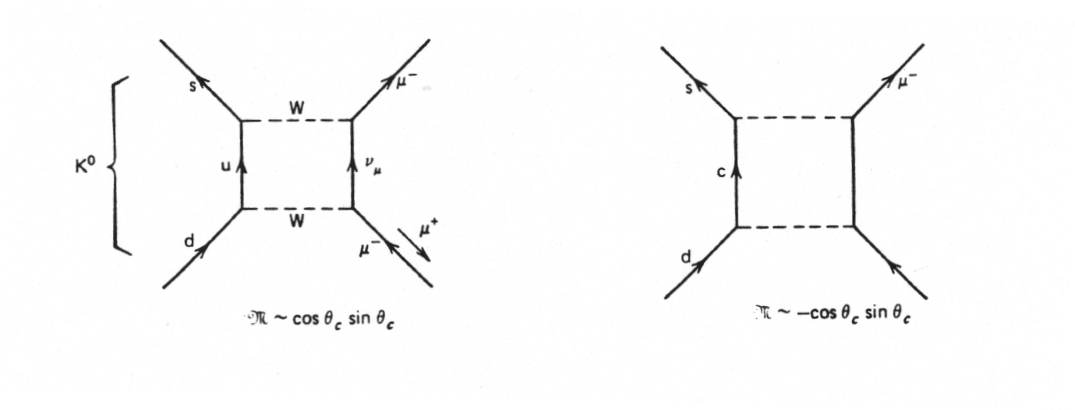}
\caption{Cancellation of the decay \KLO\ \decays\
\MP\MM\ via the GIM mechanism.}
\label{fig:c1f5}
\end{figure}

The prediction of charm and the four quark mixing (Cabbibo) matrix
were important progenitors to the three generation picture of the
standard model. Initially, three quarks $u,d,s$ formed an
approximate flavor symmetry, and were used to classify the known
states of the day. At that time the first two lepton generations
were known, and the lepton quark asymmetry was both technically and
aesthetically displeasing to theorists.
 It became apparent that certain kaon decays, such as \KLO\ \decays\
\MP\MM\
were notoriously absent. Unitarity of the quark mixing matrix was
first prognosticated by Glashow, Iliopolos, and Maiani \cite{gim}
who completed the Cabbibo mixing matrix, casting it in  the
form: $$ \pmatrix{ d' \cr s' \cr} = \pmatrix{ \cos\theta_c &
\sin\theta_c\cr -\sin\theta_c & \cos\theta_c } \pmatrix{ d  \cr s 
\cr}$$ This provided a clean mechanism to remove \KLO\
\decays\ \MP\MM\ (Figure~\ref{fig:c1f5}).
 This prediction was confirmed some
four years later by the observation of a c\=c bound state by groups
at SLAC and Brookhaven. This represented the first major step in
proliferation of the heavy quarks, which now totals two and a third
is expected. The observation of the charmed quark was an event of such
magnitude that the experimentalists who made the discovery were
awarded  a Nobel prize. This sparked off a flurry of activity in the
study of charmed particles, which still continues briskly today.


\chapter{ Theory of the Production and Decay of Charmed Mesons}

Here we review the progress made in understanding how charmed quarks 
hadronize and
eventually decay. These are two entirely different mechanisms, and this
is reflected in the paths taken in our theoretical understanding of
these processes. Quark hadronization is theoretically intractable, and
relies on extensive computer modeling and phenomenology. The theory of the
decays of
charmed mesons enjoys a (limited) quantitative stature, however the
predictive power of these theories has been poor. In both instances,
theoretical advancement has relied heavily on experimental analysis.

\section{Fragmentation Fundamentals}
Experimentally, we have not observed free quarks.
Fragmentation theory aims to describe the process through which
an uncombined quark 
evolves into the hadrons which we observe in our detectors.
 This  section aims to address the theoretical 
progress that has been made in understanding heavy quark
fragmentation.  It will emphasize the physics of charm quark
fragmentation to  charmed mesons in 
$e^+e^-$ collisions.
\par
One of the differences among the flavors  is the quark mass.
While this is in general an ill defined concept since quarks cannot
be observed in isolation, qualitative trends still exist. The
estimated  constituent (quark + surrounding gluons) quark masses
obey the  approximate proportions:
$$m_u:m_d:m_s:m_c:m_b \rightarrow 1:1:1.4:5.1:14.9$$
The first three flavors are referred to as `light' quarks or q, 
and the rest
as `heavy' quarks or Q.
The most important effect of mass occurs
during quark anti-quark production in the color field. 
When examined from a  ``thermodynamic" standpoint \cite{peterson}
quark pair production is governed by the expression  
$$ \rm rate \propto exp\left({-  2\mu c^2  \over  kT} \right)$$
where $\mu$ is the quark mass and $T$ is  the universal temperature,
corresponding to approximately to 160 MeV.
This leads to the hadron production ratios \cite{hadi}
$$ \PI  : \K  : \D : \B\
\simeq {1:   0.04 :10^{-6}:10^{-15} }$$ 
This suggests that 
  heavy Q\=Q pairs are almost never predicted in the
hadronization process.
    Our experience with the  electromagnetic production of quark
pairs points toward more of a threshold effect, where quark
electric charge and not mass is the dominant factor in the
fragmentation of a virtual photon.
 Gluons hadronize independently of quark flavor, and a sufficiently
hard gluon spectrum could be a prominent source of heavy quark
production. Charmed particle cross sections in \PR\PB\ collisions
have been much larger than originally anticipated. The inclusive
charmed particle cross section \cite{hadi}
$\sigma_{incl}$ ranges
from $\simeq
 5 \times 10^{-1}$ mb at $s = 2 \times 10^2\ \rm GeV^2$ to
$\simeq
  1 $ mb at $s = 1 \times 10^4\ \rm Ge{V^2}$\kern-.4em .
\ This is to be compared with a pion cross section that remains
essentially constant at $10^2$ mb. While this effect is not fully
understood, QCD  flavor-annihilation and flavor-excitation processes
are the most likely explanation. At CESR energies, it is highly
unlikely that radiative gluons are capable of producing c\=c pairs,
and it is a physically reasonable to assume that charmed quarks are
only produced electromagnetically.
\par
 The variation in mass will also introduce
kinematical differences between light and heavy quark fragmentation.
Attaching a light quark to a moving heavy quark will not decelerate
it very much. Thus hadrons formed from a primary heavy quark are
expected to be `hard', retaining 
a large fraction of the original quark momentum. This is in contrast
to light quark fragmentation, which is `soft.'

A particularly convenient way to observe hadronization is the reaction
$${e^+}{e^-} \rightarrow \gamma^\ast,Z^0 \rightarrow   f\bar f$$
where $f$ is a fermion. Electromagnetically, q\=q pairs are 
produced relative to muon pairs
by $3e^2_q$ where $e_q$ is the quark charge. Thus heavy quarks are
produced, above threshold,  as often as light quarks with the
same charge.

 If heavy quarks are not produced during
hadronization, a reconstructed heavy hadron can be directly compared
to the primary Q that may have caused it. One exception to this, of
course, would be heavy quarks produced in the  cascade 
decays of even heavier quarks.
The kinematics of these 
decays are well known and the contribution to the fragmentation data
of these hadrons can be sorted from those
of primary Q\=Q pairs.
Heavy quark production is not limited to $e^+e^-$ collisions.
Important early results in charm fragmentation were gleaned from
$\nu N \rightarrow \rm charm$ reactions and leptoproduction \cite{chirin}
but these will not be discussed
here. 
\par
Fragmentation of quarks into hadrons is  not well understood.
The principal reason for this is that the process is very violent,
and takes place where $\alpha_s(Q^2)$ is large.  We are therefore
unable  to apply perturbative QCD to do any calculations.  We have
circumvented this by proposing a probability function $D^H_q(z)$,
which describes the probability that a hadron H will be found in
a jet formed by  the fragmentation of q with a fraction z of the
original quarks energy-momentum. These are the so-called fragmentation
functions.  A number of different forms for these fragmentation
functions have been put forth, based on a variety of ideas.  This
section will attempt to illuminate the various approaches that have
used to describe fragmentation.  
Some of the fundamental tools needed to understand fragmentation
will also be described

There are three scenarios which have been put forward to describe
fragmentation. These are Independent Fragmentation (IF), String
Fragmentation  (SF), and, more recently, Cluster Fragmentation (CF).
IF was proposed by Field and Feynman \cite{fefe}
   as a
first guess model of jet formation. Its central theme is that all
partons fragment independently of each other. The SF model,
based on an idealization of the color field, was due initially
to Artu and Messier \cite{artu}
Their work was highly
embellished by  the Lund \cite{ander}
 group and provides the foundation
for many of the Monte Carlos
currently employed in high energy physics. A promising
development,  CF is based on leading log QCD.  Fragmentation takes
place as a QCD shower generated as an off shell parton comes on
shell.

\subsection{Independent Fragmentation}
\begin{figure}[htp!]
\centering
\includegraphics[scale=0.65]{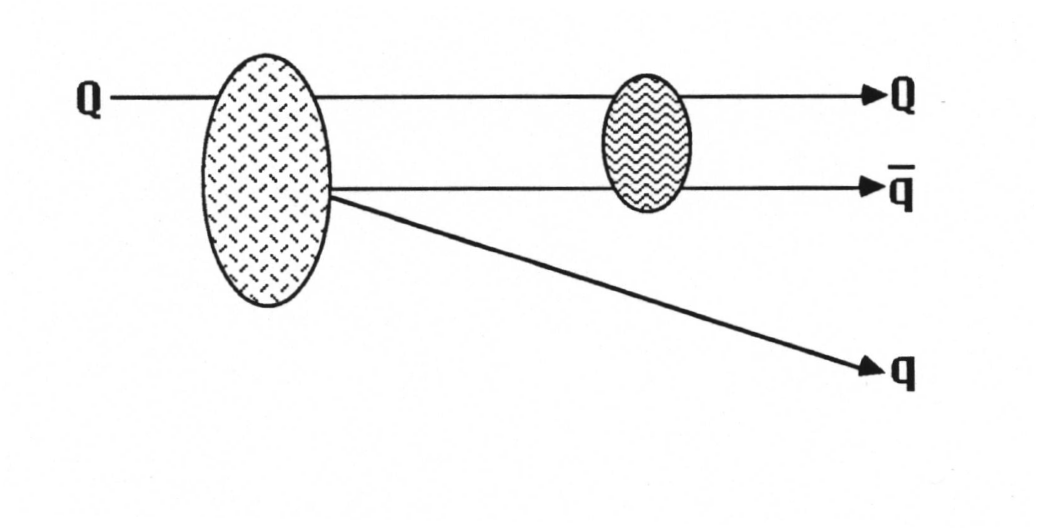}
\caption{Independent fragmentation.}
\label{fig:c2f1}
\end{figure}

IF is a relatively uncomplicated idea. It is described by the diagram
 shown in Figure~\ref{fig:c2f1}.
A $q_1\bar{q_1}$ pair is produced in the
color field of uncombined quark $q_0$ moving through spacetime. The initial
quark combines with the appropriate anti quark to form a
meson ($q_0\bar{q_1}$) while the remaining quark $q_1$ is left
uncombined to continue the fragmentation process. A  serious problem
with this concept is that there is eventually one quark left over,
which creates difficulties in terms of flavor and four momentum
conservation. This is dealt with in Monte Carlo at the end of event
generation by globally imposing energy and momentum conservation
throughout the entire event.  Despite its
counter intuitive nature, IF was unsophisticated, available
and became popular. A number of its deficiencies, particularly
the way it handles gluons, have made it less popular today in light
of the better models available.

\subsection{String Fragmentation}

Because the field quanta of QCD, the gluons, carry color charge,
the field lines of the color field collapse into a flux tube or
`string' between quarks.  String models are defined in 1$+$1 dimensions
\textit{i.e.} 1 time and 1 space or 1 energy and 1 momentum. The primary
parameter is the string tension constant $\kappa$, which has the
approximate value $\kappa \approx 0.2 \ {GeV}^2 \approx \ 1 \left( GeV
\over fermi \right)$. The key equations of SF are
$$ \left( dp \over dt \right) = \pm \kappa $$
$$ \left(dE \over dx \right) = \kappa$$
where the + -- refer to the string pulling the parton to the right
and left respectively. These two equations can be integrated and
combined with the  Lorentz invariant expression
$E^2 = p^2 + m^2$ to yield
$$ \left(x-x_0\right)^2  -\left(t-t_0\right)^2 = \left({\mu \over \kappa}
\right)^2 \eqno(1)$$
Thus, massive quarks move along hyperbolas in spacetime with light
cone asymptotes. `Massless' quarks would move along the light 
cone. There are several important questions that need to be addressed:
what determines how and when the string breaks, and is there
any relation between the color field and the mass of the hadron formed?
The first question is a pivotal issue with SF, and we will see later
how different approaches to this question lead to slightly different
fragmentation functions.
\subsection{Cluster Fragmentation}

A third and relatively new approach to fragmentation is the 
called Cluster Fragmentation.  It is based on leading log QCD 
and the Altareli-Parisi formalism. 
Each branching is considered to be an independent event, and the parton
comes more on shell at each branch. 

The classical branching probabilities are given by
$$P_{q\rightarrow qg} = {4\over 3} {\left(1 +x^2 \over 1-x\right)} $$
$$P_{g\rightarrow gg} = 3\left[ {1-x \over x} + {x \over 1-x}
+ x\left(1-x\right) \right] $$
$$P_{g\rightarrow qq} = \left[x^2 + (1-x^2)\right]$$
where $P_{a \rightarrow bc}$ represents the probability that $b$ will
retain a fraction x of $a$'s momentum.
Work has been done in this area by Field and Wolfram \cite{fiewo}
 Gottschalk \cite{gotts}
and the most popular by
Webber \cite{webbe}.
The shower is terminated at the point
which the parton reaches the appropriate quark mass or cutoff $Q_o$ for
gluons (to prevent infrared divergences). At this point all the gluons
are split into q\=q pairs and color flow is regulated to produce colorless
clusters which decay through phase space. 
The Webber model has  the advantage of requiring only a few parameters:
$\Lambda_{QCD}, M_{max}, m_u,m_d,m_s,Q_o $.
Additional features of the Webber 
model include successively reducing the opening angles of parton emission
to account for leading interference effects, and  decaying 
clusters exceeding some
$M_{max}$ as a string. Initially, c and b quarks decayed weakly before
the clusters were formed. Now c quarks are kept in the shower and
form charmed clusters which decay into $D^{\ast}$ and a $\pi$ or K.
While this is not directly suited to predicting charm yields, charm
fragmentation is generated directly via the cluster decay algorithm.
While Cluster Fragmentation is still in the developmental phase, it has shown
itself able to compete with SF and IF.  The purpose of this cursory
treatment of CF is to demonstrate that we may be finally approaching
the point where fragmentation is able to proceed naturally out of our
Monte Carlo models without the imposition of phenomenological kludges.
\subsection{Reciprocity} 

The reciprocity relation is often invoked as a check on
fragmentation functions. It is essentially a boundary condition that
proposes
$$ D^H_q(z)_{z \rightarrow 1} \Rightarrow
f^q_H(x)_{z \equiv {1\over x} \rightarrow 1}$$
What this means is that the fragmentation function of a hadron
containing virtually all of the original quarks momentum should
be equal to the structure function of a hadron where that quark 
posses a momentum fraction that approaches one.      
 This can be understood in that 
as a quark obtains a large momentum fraction, the remaining
quarks in the hadron must compensate for this by giving up their
momentum. As $x \rightarrow 1$, the other valence quarks
must be almost at rest.  If this reaction is reversed in time, one
obtains fragmentation.

If reciprocity holds, the appropriate question is  how the structure
functions behave as $x \rightarrow 1$ ? Structure functions are on a
somewhat better theoretical footing than fragmentation functions, but
the answer is still uncertain. The `standard' textbook
formula \cite{ql}[p. 200]
  is
$$ f^q_H(x)_{x \rightarrow 1} = \left(1 - x\right)^{2n_s-1} \eqno(2)$$
where $n_s$ is the number of spectator valence quarks.  This relation
predicts $(1-x)$ for mesons and $(1-x)^3$ for baryons.  An
alternate \cite{gplp}
form is 
$$ f^q_H(x)_{x \rightarrow 1} = \left(1 - x\right)^{2n_s-1+2\Delta S}
\eqno(3)$$
The $\Delta S$ term is the absolute difference between the initial
hadron spin and the quark spin $({1\over 2})$ . This conflicts with the
above expression in the prediction for mesons, yielding $(1-x)^2$.
As we shall see, almost all the fragmentation functions show a
$(1-z)$ behavior as $z \rightarrow 1$, agreeing with the first
prediction.

\subsection{Odds and Ends}

Finally, we present three items that will be needed
in order to coherently proceed to the next section. 
These are the meaning of z,
the  light-cone variables, and some elementary
relations of fragmentation functions.

Fragmentation functions are parametrized in terms of `z', but what
exactly is z?  It is meant to describe the fraction of energy-momentum
or momentum of the primary quark retained by the hadron in the
fragmentation process. Two definitions that appear in the literature are
$$ z^+ \equiv  {\left(E + p_{\parallel}\right)_{had}\over
\left(E + p\right)_{quark} } $$
and
$$ z_p = \left({p_{had}\over p_{quark}}\right)$$
The first definition is Lorentz invariant for boosts along the
quark direction, hence is more desirable theoretically.
This is true because $L(E \pm p) = f^{\pm}(\beta)(E \pm p)$, where
$f^{\pm}(\beta)$ is a constant that is a function of the boost parameter
$\beta$. These constants trivially cancel in numerator
 and denominator.  
 We adopt the convention that z will refer to the first definition. 

Unfortunately, the
kinematical variables E and p of the quark are unavailable to the
experimentalist. Fragmentation functions are measured in terms of
x, where x also has two definitions 
$$x_E \equiv \left( {E_{had}\over E_{beam} } \right) $$
$$x_p \equiv \left( {p_{had}\over \sqrt{E^2_{beam} - {m}^2_{had}
}}\right)$$ Both $x_E$ and $x_p$ have been used by experimentalists,
although $x_p$ has the advantage of ranging from 0 to 1 for all
experiments. The variables x and z are not equivalent. In general
$z\geq x$ because perturbative QCD gluon radiation and initial state
photon radiation tend to make $E_{quark} \leq E_{beam}$.  
                       
The light cone formalism is a handy tool when working with
fragmentation. 
$$ x^\pm \equiv x^0 \pm x^3 $$
$$ p^\pm \equiv p^0 \pm p^3 $$
$p_{\perp}$ is absorbed in the mass term.
As mentioned earlier these combinations transform trivially under 
Lorentz 
transformations, making frame transformations straightforward. 
They have
the added convenience that
$$ z \equiv  {\left(E + p_{\parallel}\right)_{had}\over
\left(E + p\right)_{quark} } \equiv \left(p^+_{had}\over p^+_{quark}
\right) $$
when $p_\perp$ is neglected. Thus, the light cone variables lend
themselves to a more natural expression of z. One is still faced with
a dilemma as to what the denominator should be.  It has been
suggested \cite{galh}
that  $$ p^+_{quark} \simeq  E_{max} + p_{max} = E_{beam}
+ \sqrt{E^2_{beam} - {\mu}^2}$$
however it is not obvious that this is a physically reasonable
assumption.

Measurements of fragmentation functions often appear as plots of
$s\left( {d\sigma \over dz}\right)$ vs. $(X_E$ or $X_p)$.  
The master equation for fragmentation
in $e^+e^-$ collisions is \cite{qp}
$$ \left( {1\over \sigma_{had}} \right)
{d\sigma \left(e^+e^- \rightarrow HX\right) \over dz}
={ \Sigma_i e_i^2 \left[D^H_i(z) + D^H_{\bar i}(z)\right] \over
\Sigma_i e^2_i }$$
where $i$ sums over the quark flavors participating in the reaction.
The fragmentation functions of both i and \=i are summed over if the
hadron H could be produced in the jets of $q_i$ and $\bar q_i$.
Applying $\sigma_{had} = \sigma_{\mu \mu} \left( 3\Sigma_i e^2_i \right)$
and $ \sigma_{\mu \mu} = \left( {4\over 3}{ {\pi \alpha^2}\over s}\right)$,
this can be recast into                                                  
$$ s \left( {d\sigma \over dz} \right) =
4 \pi \alpha^2 \Sigma_i \left[D^H_i(z) + D^H_{\bar i}(z) \right] $$ 
which explicitly shows the proportionality of $s\left( {d\sigma \over dz}
\right)$ to the fragmentation functions.  For the production of heavy
mesons $H_Q$ = $(Q\bar q)$ the form is slightly different. 
$H_Q$ can only be only be found in the debris of Q since heavy quarks
are not produced in the color field. The first equation becomes 
$$ \left( {1\over \sigma_{had}} \right)
{d\sigma \left(e^+e^- \rightarrow H_Q X\right) \over dz}
={ 3e_Q^2 D^{H_Q}_Q(z)  \over
\Sigma_i e^2_i }$$
\section{Fragmentation Functions}

FF arise out of our basic inability to calculate hadronization
from perturbative QCD. None of them represent a great tribute to
theoretical physics, and actually only one has received
considerable attention from experimentalists.  
We analyze
five different functions; two based on IF, two on SF, and one derived 
from the reciprocity relation. 
An important fact to bear in mind is that these functions describe 
the initial  fragmentation of a primary heavy quark Q into a heavy hadron 
$H_Q$ 
$(Q\bar q)$
containing Q.

The gross qualitative features of
heavy quark fragmentation were predicted by several theorists in the
late 70's.  Bjorken \cite{bjork}
 on 
the basis of simple kinematics,
arrived at
$$ <z_p> \sim 1 - \left( {1 GeV \over {\mu} } \right) $$
where $\mu$ is the quark mass and 1 GeV is a constant chosen for `didactic
convenience.' His argument was that ordinary hadrons were produced with
$\gamma$ less than or the order of that of the primary Q. Since 
$p = \gamma m v$, the heavier objects would take a larger fraction of the
available momentum. 
\subsection{Functions Based on Reciprocity}

One of the first charm fragmentation functions was that of
Kartvelishvili
 \cite{kart}.  
Their efforts were essentially directed toward
computing the structure function of a heavy charmed meson using the
Kuti-Weisskopf \cite{kuti}
model, and connecting that to the fragmentation
function via reciprocity. They arrived at the structure function:
$$ f^c_{H_c}(z) = { {\Gamma(2 + \gamma -\alpha_c -\alpha_q)z^{-\alpha_c}
(1-z)^{\gamma - \alpha_q} } \over {\Gamma(1-\alpha_c) \Gamma(1 + \gamma
- \alpha_q) }  } $$
Where $\gamma$ is a constant that equals $3\over 2$ and $\alpha_c$ is the 
intercept of the Regge trajectory for the c quark. $\alpha_q$
is the intercept of the light quark Regge trajectory, set 
to $1\over 2$ based on  the $\rho$, $\omega$, $A_2$, and $f$.
This expression can be used to calculate the average momentum carried
by the valence quarks 
$$<z_c> =  { {1- \alpha_c}\over {2 + \gamma - \alpha_c -\alpha_q} }$$
Their rational was that since the structure function peaks at high z,
the structure function can be set equal to the fragmentation function
at all z. As can be easily seen from the expression for the structure
function, the fragmentation function becomes 
$$D{^H_Q}(z)=Nz^{-\alpha_c} (1 - z)$$
The parameter $\alpha_c$ was unknown, the initial guess was -3.
In a later and often referenced paper \cite{kart2}
they choose
$13.44 z^{2.2}(1-z)$ without explanation. It certainly follows,
though, from the
logic of their earlier work. Claiming that reciprocity is valid at all
z (or at least from .6 to 1) is certainly dubious and must raise
doubts  as to the credibility of the Kartvelishvili function. 
The Kartvelishvili function
is based on the structure functions of charmed mesons.
Differences between charmed meson and baryon structure functions
would have to be accounted for before it could properly be applied to
charmed baryons.
\subsection{Functions Based on Independent Fragmentation}

The most celebrated heavy quark FF is that of
Peterson \cite{peter2}
 \it{ et al}. \rm 
Its success is based on the simplicity of its form and its flexibility.
The function is based on quantum mechanics and IF. 
Basically a q\=q pair is produced in the color field of a heavy quark Q. 
Q then combines with the appropriate anti-quark, and one quark is left
uncombined to continue the fragmentation process. The rate goes as
${\Delta E}^{-2}$, the difference in energy between the initial heavy
quark state and the final state hadron + q,
$$\Delta E = \sqrt{ m_H^2 + z^2p^2} + \sqrt{m_q^2 + (1-z)^2p^2}
- \sqrt{m_Q^2 +p^2} $$
Approximating $m_H = m_Q$ and after a binomial expansion we obtain
$$ \Delta E \propto 1- {1\over z} - {\epsilon_Q \over 1-z} $$
Squaring, and tacking on a factor of $1 \over z$ for longitudinal phase space 
produces the well known result
$$D{^H_Q}(z) = {N \left(z\left[1-{1 \over z}-{\epsilon_Q\over(1-z)}\right]
^2\right)^{-1}}$$
The parameter $\epsilon_Q$ is proportional to
$m_{q\perp}^2/m_{Q\perp}^2$ , and the
transverse mass is defined as ${m_\perp}^2 \equiv {m^2 + {p_\perp}^2}$.  
As defined, z is the fraction of the momentum of the heavy quark Q
retained by the hadron. The binomial expansion is a delicate issue,
especially in the region $z \rightarrow 0$. 
It is mentioned in the paper that
the expansion is strictly valid in the $p \rightarrow \infty$ limit.
The function peaks at
$$ z_{peak} \simeq 1 - 2\epsilon_Q $$
with width $\simeq \epsilon_Q$.

The function
was not initially derived in terms of the light cone variables, 
but it is pointed out
that light cone variables may be more appropriate at finite energies,
and the above expression may be carried over provided it is cut off below
${p^+}_{min}$.

It is reasonable to assume that a di-quark pair
could have been popped and ${{m_q}_\perp}^2
\rightarrow {{m_d}_\perp}^2   $ in the formula for $\epsilon$. 
Both the Peterson and Andersson functions contain a term
that depends on $m_\perp$.  Often the $p_\perp$ is neglected and 
$m_\perp  \rightarrow  m$. A consistent interpretation needs to be
arrived at.     

The Peterson function goes as $(1-z)^2$ as $z \rightarrow 1$.
This is in disagreement with reciprocity using Eq. $2$.  This 
potentially disquieting fact motivated some  
recent work by 
Collins and Spiller \cite{collin}
 who set out to produce a  fragmentation
function that did not `violate' reciprocity and dimensional counting rules.
They calculate the cross
section $\sigma\left({e^+}{e^-} \rightarrow HX \right)$ which depends on
the fragmentation function $D^H_Q(z)$ . They use Independent
Fragmentation and compute the vertex $Q\rightarrow Hq$. This depends
on the transverse momentum distributions of H and q with respect to
the Q direction, and the longitudinal momentum distributions. The
longitudinal momentum distribution, they claim, happens to be equal
to the function which describes the momentum distributions of the
two valence quarks in a meson.   The final result with the
approximations $ m_H = m_Q \gg <k_T^2 >$
$$D{^H_Q}(z{^+}) \simeq N\left({1-z{^+}\over z{^+}}+ {2-z{^+}\over 1-z{^+}}
{\varepsilon_Q}\right)(1+z{^+}^2){\left(1-{1 \over z{^+}}-{{\varepsilon_Q}
\over(1-z{^+})}\right)}^{-2}$$ 
We have elected to use $z^+$ to indicate functions explicitly derived
in terms of light cone  variables.
The parameter $\varepsilon_Q$ is defined $\equiv \left( <k{^2_T}>
\over m{^2_Q}\right)$. $<{k_T}^2> = (0.45 \ GeV)^2$
`represents the size of the hadron in momentum space'.  A Peterson
like term is apparent, and the two functions are noticeably similar.
This function, in contrast to the Peterson function, 
exhibits a $(1-z^+)$ behavior as $z^+ \rightarrow 1$.
It is uncertain how this function would differ for \D\ and \DF\ 
mesons, since it is unclear if $k_T$ is a constant or depends on 
$u,\ d\
{\rm or}\ s$ quarks. 
This function would not be directly suitable for baryons.
Like the Kartvelishvili function, it is partially
derived from the structure function of mesons.

They have also proposed a kernel for $\rm vector\Rightarrow
pseudoscalar$ decays, to allow calculation of inclusive fragmentation
functions. This is an important problem when attempting to determine
\DZ\ or \DP\ fragmentation, since the feed down from \DS\
must be taken into account.
They calculate the fragmentation function for the decay
$ H^{\ast}_Q \rightarrow H_Q \pi$, z being the fraction of momentum
retained by  $H_Q$
$$ P_Q(z) = {N\over z(1-z)\left( {m^2_{H^\ast} \over <k_T^2>}
-  { {m^2_H + <k_T^2> } \over z} - { {m^2_{\pi} + <k_T^2>} \over 1-z} 
\right)^2 }$$
The inclusive fragmentation is the given by
$$ D^{H_Q}_Q(z) = \delta D^{H_Q}_Q(z) + \left(1-\delta \right)
\int_z^1 { dy \over y} D^{H^{\ast}}_Q(y) P_Q({y\over z}) $$
$0 \leq \delta \leq 1$ is the fraction of direct pseudoscalar production.
The first term represents direct pseudoscalar production and the
second feed down from vector mesons.
\subsection{Functions Based on String Fragmentation}

The function of Andersson \cite{ander2}
\etal\ is  
based on the symmetric Lund 
String Fragmentation model. 
The symmetric
Lund function assures that starting at either the q or \=q end and
iterating will produce the same result. 
The left right symmetry is a non-trivial property of an iteratively
generated cascade. The Andersson function is derived by solving
a set of coupled equations that demand this property. The resulting
form is
$$D{^H_Q}(z{^+})={N\over z{^+}}(1-z{^+})^a exp({-b
\cdot {{m_H}_\perp}^2\over z{^+}} )$$
Where $a$ and $b$ are constants related to the production of q\=q
vertices in spacetime.  The constant $b$ is flavor independent, 
while $a$ may depend on
flavor.  The above expression assumes all $a$'s are equal. The constant
$m_\perp$ is the transverse mass of the hadron produced. It is anticipated
that $ 0\leq a \leq 2$ and $ b\geq 0 $,with the current values
set at $a \sim 1$ and $b \sim \left(1 \over 2.25\right)$. 
The mean is predicted to be
$$ <z^+> \simeq 1 - {  {(a +1)} \over {bM^2} } $$
M is the mass of the meson produced.

The Andersson expression was derived
using massless quarks moving along light cones. It is expected to carry
over to an initial heavy Q\=Q pair, which move along hyperbolas with
light cone asymptotes. A definite difficulty with this function is that
it differentiates light and heavy quark fragmentation only by the mass
of the hadron formed. Although it is derived in the full glory of
the string model, it has to be recognized that symmetry was the
main impetus for this function.  

\begin{figure}[htp!]
\centering
\includegraphics[scale=0.55]{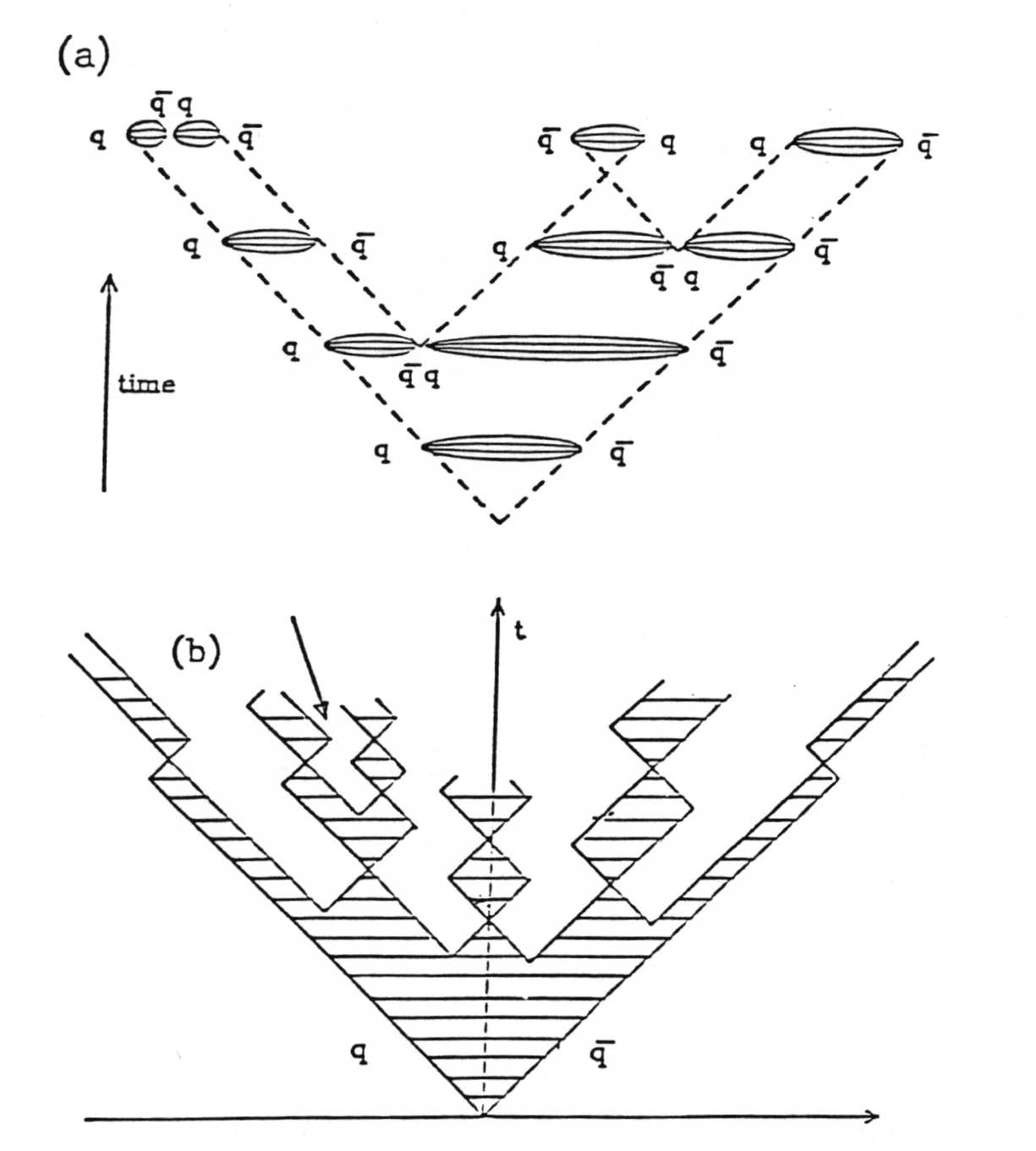}
\caption{Models of string fragmentation.   The top diagram
(a) shows a model where the energy stored in the quark flux tube
becomes so great that additional quark pairs are produced. The
bottom (b) diagram shows a light cone view.}
\label{fig:c2f2}
\end{figure}

An event generated using SF has the following structure (Figure~\ref{fig:c2f2}),
q\=q vertices appear in spacetime in the light cone of the initial
$q_0 \bar q_0$ pair. When quark world lines cross a hadron is formed.
The hadron mass is proportional to the area of the color field between
the quarks (hatched region). We know from Eq. $1$ that massive
objects must move along hyperbolas in spacetime. In order to produce
a hadron of mass m, the secondary q\=q pair must lie on this
hyperbola. The heart of the Lund model is that when a string breaks
to produce a hadron of mass m, vertices of the secondary q\=q will
be uniformly distributed along the production hyperbola. 
In this scheme, quarks fragment into  hadrons with a discrete mass
spectrum.
An
alternative point of view was taken by Bowler \cite{bowl}
   who  choose to break the
string in a more classical way.  The probability $dP$ of a break
occurring at spacetime coordinates $(x + dx,t + dt)$ is
$$ dP = \Pi dx dt $$
$\Pi$ is a constant per unit length of string and unit time.
The function is derived utilizing the
basic equations  of the string model. The primary Q\=Q pair
is massive and the secondary quark pairs are massless. Two sets of 
coordinates are considered in the fragmentation process, $(x_1,t_1)$
the point where secondary q\=q pair is produced, and $(x_m,t_m)$ the
coordinates where the world lines cross and the hadron is formed.
The derivation is straightforward, applying
$$ dP = \Pi dA exp(-\Pi A) $$
the probability of creating the secondary pair within $dA$
where A is the area of the color field in the absolute past of
the point $(x_m,t_m)$ gives
$$D{^H_Q}(z)=\left(B\over z\right)exp(-Bm{^2_Q}
( {m{^2_H}\over m{^2_Q}z} - 1 - ln( {m{^2_H}\over m{^2_Q}z})))$$
B equals $\left(\Pi / 2k^2\right)$. $\Pi$ being the constant
probability the
string will break, and k is the string constant.

The function has recently been modified \cite{galh}
 to include light
cone formalism and be made more suitable for CESR energies. Bowler also
later suggested the addition of the term $ (1-z)^{\beta}$ to account
for the fact that the string is not `straight' due to soft gluon emission.
The modified Bowler function is:
$$D{^H_Q}(z{^+})=(1-z{^+})^{\beta}\left(B\over z{^+}\right)exp(-Bm{^2_Q}
( {m{^2_H}\over m{^2_Q}z{^+}} - 1 - ln( {m{^2_H}\over m{^2_Q}z{^+}})))$$
It is expected that $\beta$ will be close to 1. One criticism of this
approach is that the Bowler function is singular as $m_H \rightarrow
0 $ and would highly favor 0 mass mesons unless a low mass cutoff is
imposed. 

\subsection{ Comparison of the Functions}
\begin{figure}[htp!]
\centering
\includegraphics[scale=0.60]{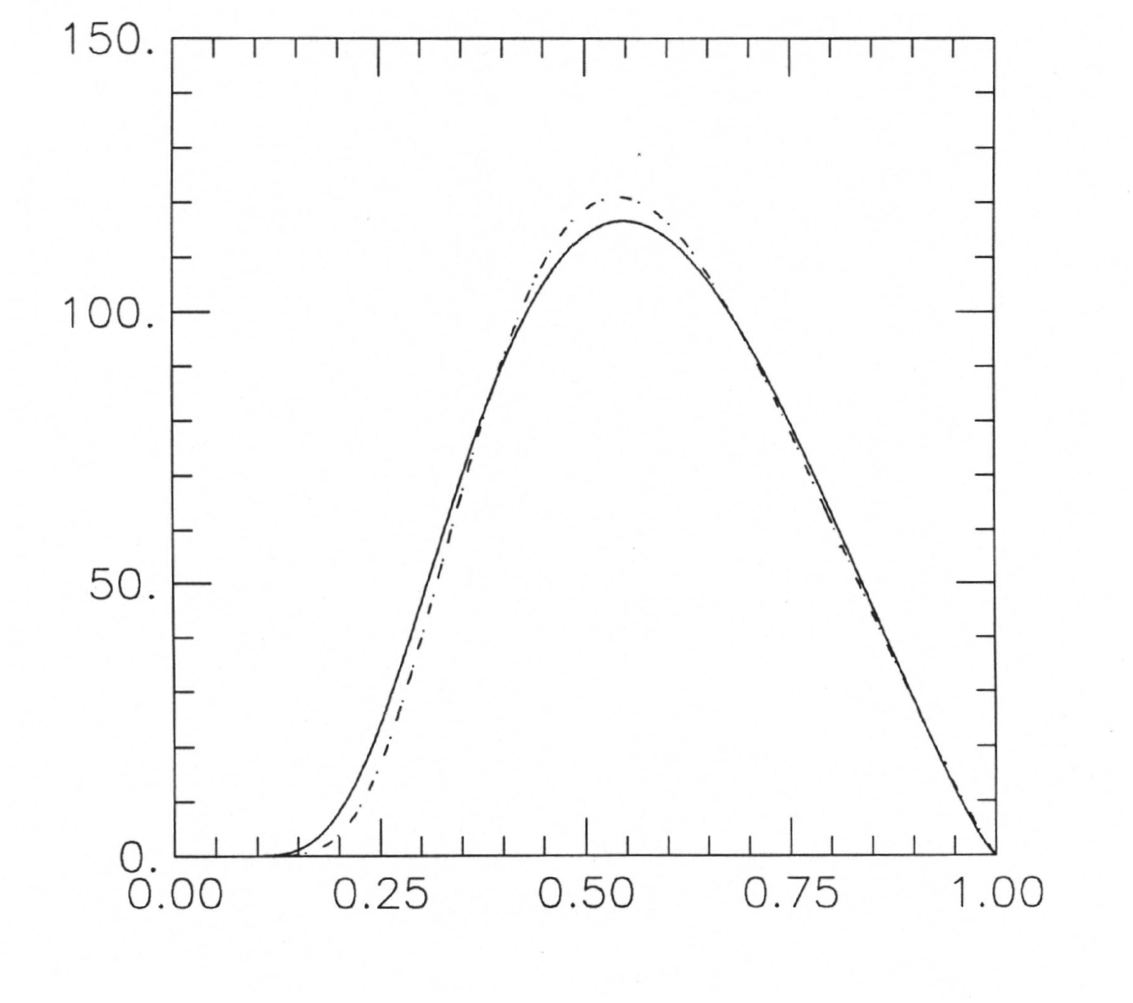}
\caption{String
inspired functions.   Andersson fragmentation function (solid
line),  and Modified Bowler function (dash-dotted line).}
\label{fig:c2f3}
\end{figure}

\begin{figure}[htp!]
\centering
\includegraphics[scale=0.55]{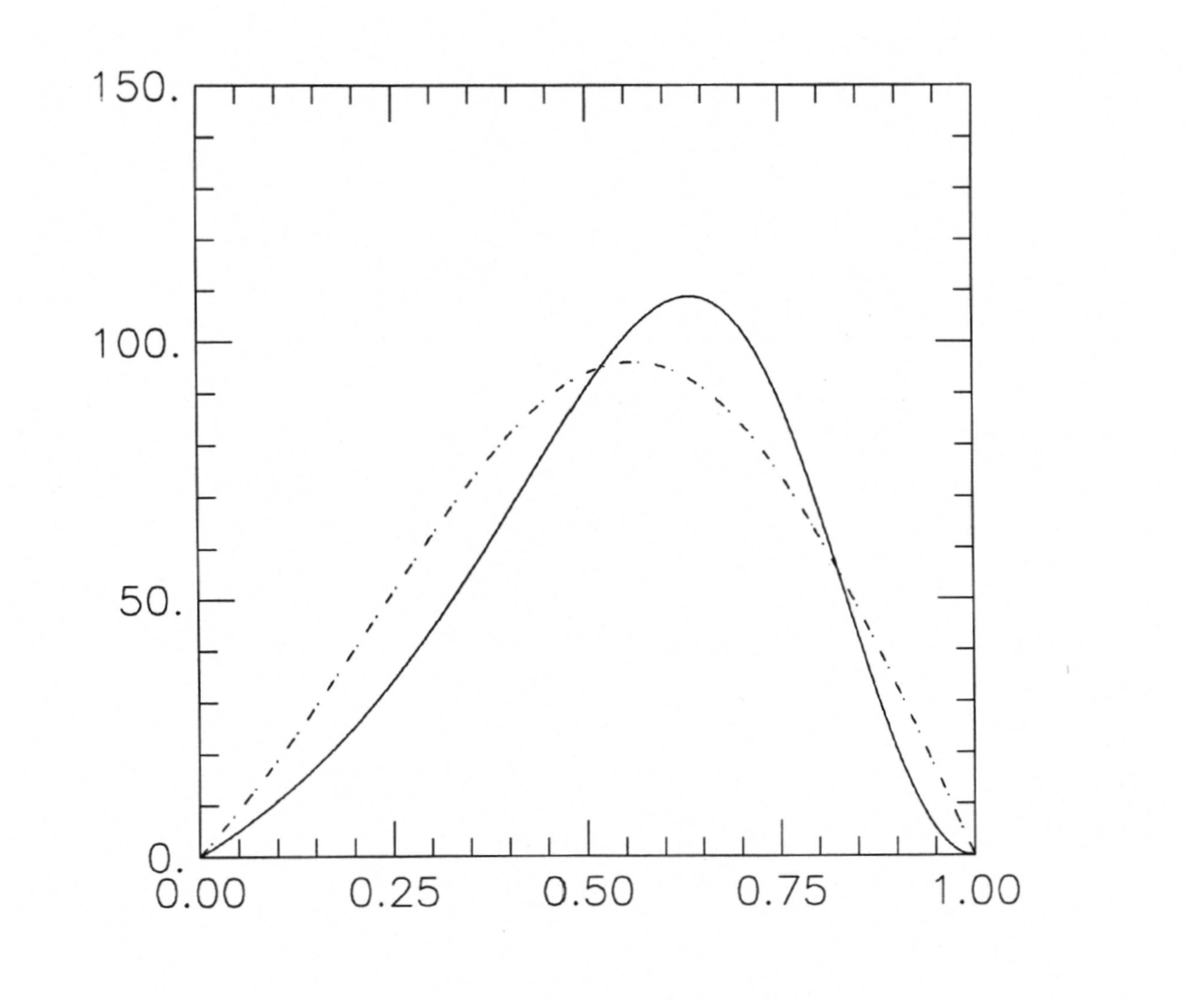}
\caption{Independent fragmentation
inspired functions. Collins  
function (dash-dotted line), and the Peterson  function
(solid line).}
\label{fig:c2f4}
\end{figure}

\begin{table}[ht!]
\centering
\caption{Properties of Fragmentation
Functions}
\begin{tabular}{|c|c|c|c|c|c|}
\hline
Function & Model & Reciprocity & Variable & Baryons &         $ z\rightarrow 1$ 
\\ \hline
Kartvelishvili & - & Yes & CT & No & $(1-z)$ \\ \hline
Peterson & IF & No & CT,LC & Possible & $(1-z)^2$ \\ \hline
Collins & IF & Yes & LC & No & $(1-z)$ \\ \hline
Andersson & SF & No & LC & ? & $ \sim (1-z)$ \\ \hline
Modified Bowler& SF & No & LC & ? & $\sim (1-z)$\\ \hline
\multicolumn{3}{|c|}{ LC = Light Cone} & \multicolumn{3}{c|}{ CT = Cartesian} \\ \hline
\end{tabular}
\label{t:2p1}
\end{table}
Table~\ref{t:2p1} presents the collected features of the various fragmentation
functions.  
The two functions derived from IF are similar,
as are
the two based on SF. There are  differences
between the two models. The most distinct difference is the
exponential term in the SF functions. This causes the fragmentation
function to go to 0 prematurely. SF also appears to be flatter, but
this feature strongly depends on $\epsilon$ for the IF functions.
We show the Andersson and Bowler functions in Figure~\ref{fig:c2f3}, 
, and the
Collins and Peterson forms in Figure~\ref{fig:c2f4}.
Another difference is the $z \rightarrow 1$ behavior, where 
reciprocity can be used as a guide. All the functions, save the
Peterson, lean toward a $(1-z)$ behavior. The Peterson function has
a $(1-z)^2$ as $z \rightarrow 1$.  From an experimental perspective,
the low z region is difficult to measure because of poor
efficiencies of low momentum tracks. For the high z end, a large
amount of statistics would be necessary to accurately determine the
power dependence of the curves. So experimental clarification of
these issues will only come with great effort.  A detailed
comparison of these five functions to CLEO's \DSP\ fragmentation
distribution can be found  elsewhere \cite{frag}.
One of the more serious shortcomings of IF stems from the so-called
string effect
\begin{figure}[htp!]
\centering
\includegraphics[scale=0.7]{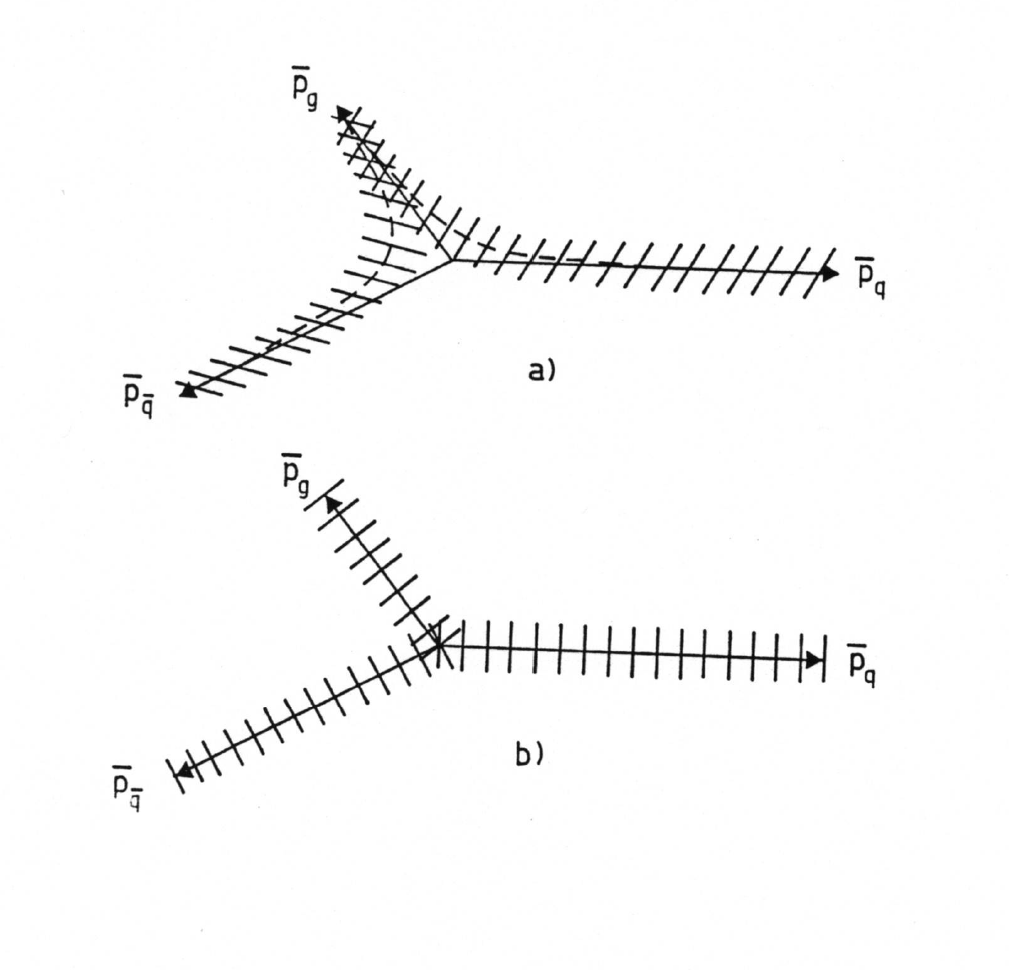}
\caption{A qqg event in SF (a) and IF (b) models.}
\label{fig:c2f5}
\end{figure}
(Figure~\ref{fig:c2f5}).
In IF models all the partons fragment independently,
while in SF models a color ``web'' stretches from the quarks to the
gluons. This acts to increase the density of hadrons in the vicinity of
the gluon direction. This has been observed experimentally \cite{bart}.
 While the
exact cause of this effect  in SF Monte Carlos is unclear, it cannot be
accommodated in IF schemes. This, along with other undesirable features
has all but terminated interest in IF as a viable \EPEM\ fragmentation
model.
\section{Theoretical Approaches to Charm Decay}
Here we briefly recount the evolution of  our understanding of the
  weak decay of charmed mesons. The weak force has supplied the
physics community with a great deal of beauty and bedazzlement, with
the charm sector being no exception. This field also attests to the
vital interplay of experimental and theoretical physics, as
theoretical predictions for charm decays have been consistently
defied by experimental observations. Here we trace our
understanding of charm decays from the failings of the simple
spectator model to more advanced approaches. We also detail
outstanding conflicts with  experimental measurements. Voluminous
amounts of material has previously been written on this subject.
The reader may find of particular use the experimental summaries
of Hitlin \cite{Hitlin}, \
 Schindler \cite{rafe},\
and the
theoretical  treatments of R{\"u}ckl \cite{ruck},
Shifman \cite{shif} 
 and the recent review by Bigi \cite{bigi}
to append  the rather quick distillation
presented here. 
\subsection{The Trivial Spectator Model}
\begin{figure}[htp!]
\centering
\includegraphics[scale=0.6]{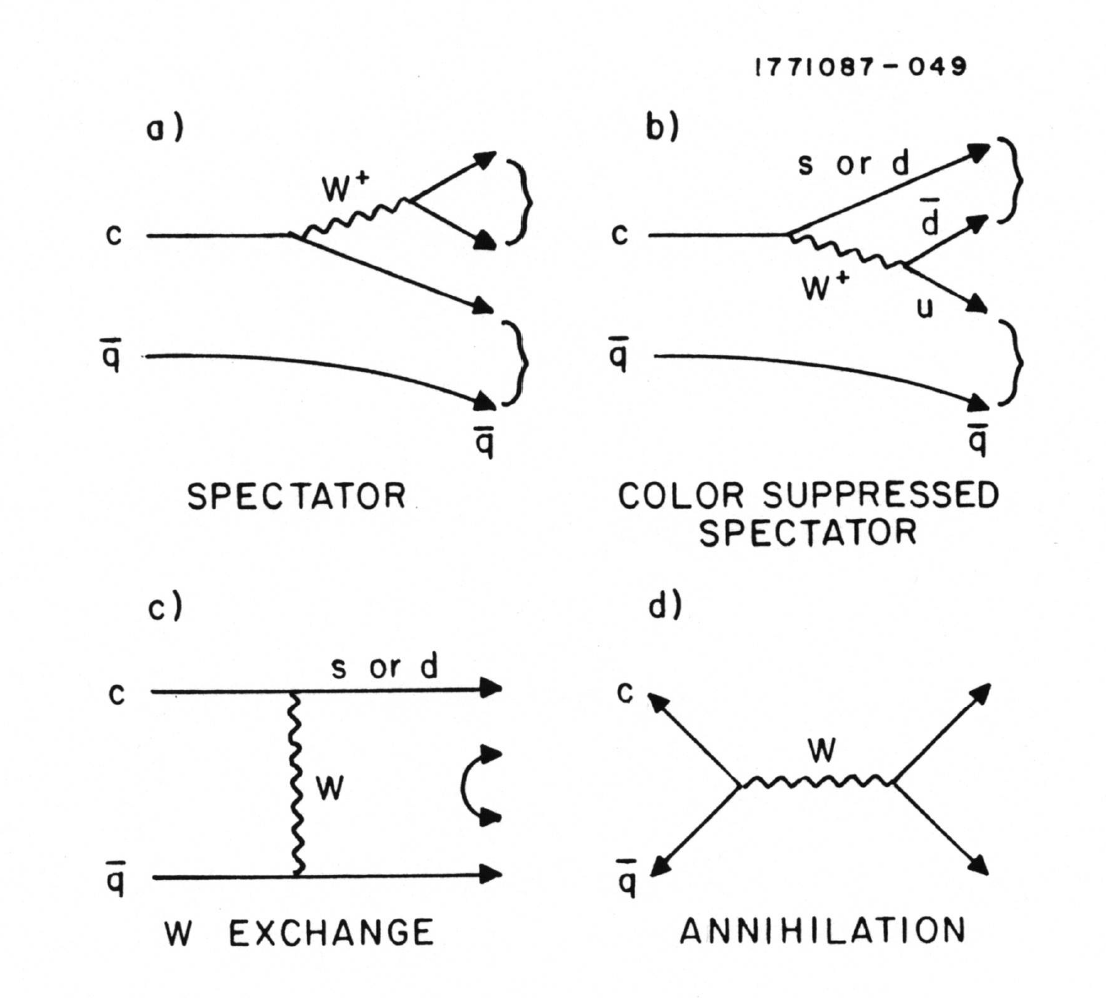}
\caption{Lowest order decay diagrams for charmed mesons.  a) the
spectator diagram, b) color matched or color suppressed spectator
diagram, c) \W\ exchange diagram, d) \W\ annihilation.}
\label{fig:c2f6}
\end{figure}

The lowest level of understanding of charm decays utilizes the
method of quark diagrams. The first order processes are collected in Figure~\ref{fig:c2f6}.

They may be distinctly be separated topologically into two classes;
the ``spectator class" (a - b), and the nonspectator or
``annihilation class"  (c - d). The primary difference is in  the
the location of the \W\ vertices. In spectator processes the \W\
vertex occurs only on the c quark world line, it then hadronizes
either independently (spectator) or in conjunction with the light
quark (color suppressed). In  annihilation processes the \W\
vertex touches both quark lines of the initial meson. The trivial
spectator model makes the following predictions:
\begin{enumerate} 
\item
The dominant process in the weak decay of charmed mesons is the
spectator diagram. This approximately corresponds to the beta decay
of the charmed quark. The rate for charm decay $(\Gamma_0)$ can  
be obtained (to first order) by scaling the muon lifetime $
\tau_{\mu} = {192\pi^3 \over G^2_F m^5_\mu} \approx 2.2 \times
10^{-6}  \rm sec   $
by a factor of $   {1 \over 5} \left ( { m_{\mu} \over m_{c}
} \right)^5 \approx 7 \times 10^{-13} \rm sec$,  to account for the
charmed quark mass and the additional number of decay channels
available to the \W. This form suffers from an exceptional (fifth
power) dependence on the charmed quark mass which is an undefined
quantity. It is generally approximated to be in the range of
$1.5-1.6$ GeV.
 \item
The color suppressed spectator diagram occurs at a rate reduced
by a factor of $\approx$ 10. The is because the \W\ has to hadronize a quark
pair with the right color combination to make color singlet hadrons.
\item
The exchange diagram, which is contributes to \DZ\ decay on the
Cabbibo allowed level is expected to be strongly suppressed by a
small wave function overlap $ | \psi(0) |^2 \propto {f^2_D \over
M_c^2} $ where $f_D \sim .15  $ GeV. Similarly the \W\ annihilation
is expected to suffer helicity suppression (as in pion decay) by
a factor $ {m^2_q \over
M_c^2}$
\item
Taking these factors into account, the \DZ\ and \DP\ are expected to
have the same lifetimes and semi-leptonic branching ratios.
\end{enumerate}

\subsection{ Experimental Controversies}
\begin{table}[ht!]
\centering
\caption{Lifetimes of Charmed Mesons}
\begin{tabular}{|c|c|c|c|}
\hline
Group &  \DP\ $\times 10^{-13} \rm sec$ & \DZ\ $\times 10^{-13} 
\rm sec$ &  \DF\ $\times 10^{-13} \rm sec$ \\ \hline
CLEO & 11.4 \pp\ 1.6 \pp\ 0.7 & 5.0 \pp\ 0.7 \pp\ 0.4 & 4.7 \pp\
2.2 \pp\ 0.5 \\ \hline
World Average & $10.45^{+0.31}_{-0.29}$ & 4.27 \pp\ 0.10 &
$4.31^{+0.36}_{- 0.32}$ \\ \hline
\end{tabular}
\label{t:2p2}
\end{table}
After the experimental picture of charm decays began to unfold,
several failings of the trivial Spectator model were revealed.
\begin{enumerate}
\item
The \DP\ and \DZ\ mesons were found to have very different
decay rates. This has been determined by measuring the lifetimes
of the \DZ\ and \DP, as collected in Table~\ref{t:2p2}.
 We have presented the measurements of CLEO \cite{csor}
 and the
current world average \cite{Hitlin}.
Empirically, we measure $ {
\tau(\DP) \over \tau(\DZ) } = 2.45 \pm\ 0.09$.
 This difference has also been
established by a large discrepancy in the semileptonic branching
ratios of the \DZ\ and \DP. A recent Measurement by the MARK
III \cite{balt}
 has determined $ \rm { B(\DP\ \decays\
\EP X) \over  B(\DZ\ \decays\ \EP  X)} =2.3^{+0.5+0.1}_{-0.4-0.1}$,
which is in good agreement with the measured lifetime difference.
\item
Several color suppressed \D\ decays have been observed to occur with
a healthy rate. Among them \DZ\ \decays\ \KZB\PIZ\ has been found
to occur at a rate of 0.45 \pp\ 0.9 times that of \DZ\ \decays\ \KM\PIP.
CLEO \cite{haas}
and ARGUS \cite{halb1}
 have  also observed a color suppressed decay of the \B\ meson,
B( B \decays\ \PS X) $\sim$ 1.2 \%.
\item
Evidence for annihilation processes appears to exist. The first
was the  observation of \DZ\ \decays\ \PH\KZB.  This mode was first
observed by ARGUS \cite{halb2}
and later by CLEO and
MARK III.  These measurements have confirmed a branching 
fraction of $\simeq 1\%$.
Other clean annihilation class signatures have been sought in \DF\
decay. The E-691 experiment \cite{anjos}
 has placed a stringent limit on the
\DF\ \decays\ \RHZ\PIP\ mode, finding
$ \rm { B(\DF\ \decays\ \RHZ\PIP) \over 
B(\DF\ \decays\  \PH\PIP)} = < 0.08$ at the 90 \% CL. This same
group has, however,  has observed the annihilation decay candidate 
  $ \rm { B(\DF\ \decays\ \PIP\PIP\PIM)_{nonres} \over 
B(\DF\ \decays\  \PH\PIP)} =  0.29 \pm\ 0.07 \pm\ 0.05$.
\end{enumerate}
\subsection{Patching up the Spectator Model using QCD}
In light of these difficulties, reexamination of the spectator
model was in order.
The charged weak current has as its hadronic component:
 $$  J_{-}^\mu = \left(J_{+}^\mu \right)^t = \left(\bar u , \bar s ,
\bar t \right) \gamma^\mu \left ( 1 - \gamma^5 \right)
     \left( \matrix{ V_{ud} & V_{us} & V_{ub} \cr  
 V_{cd} & V_{cs} & V_{cb} \cr
 V_{td} & V_{ts} & V_{tb} \cr} \right)
\left ( \matrix{   d  \cr   s  \cr  t  \cr } 
\right )
$$
Where the $V_{xy}$ are the familiar K-M matrix elements.  The
current-current approximation  $( q^2 << M^2_W )$ of the weak
Hamiltonian leads to the ``Bare" Hamiltonian for hadronic charm
changing interactions of
$$ H^{(0)}_W(\Delta c= -1) = { G_F \over \sqrt{2}}
V_{cs}V_{ud}^{\ast} \big [ (\bar s c)_L (\bar du)_L \big ]$$
where the notation $(\bar a b)_L$ implies the canonical V-A
structure $ \bar a \gamma^{\mu}(1-\gamma_5)b $. 
\begin{figure}[h!]
\centering
\includegraphics[scale=0.55]{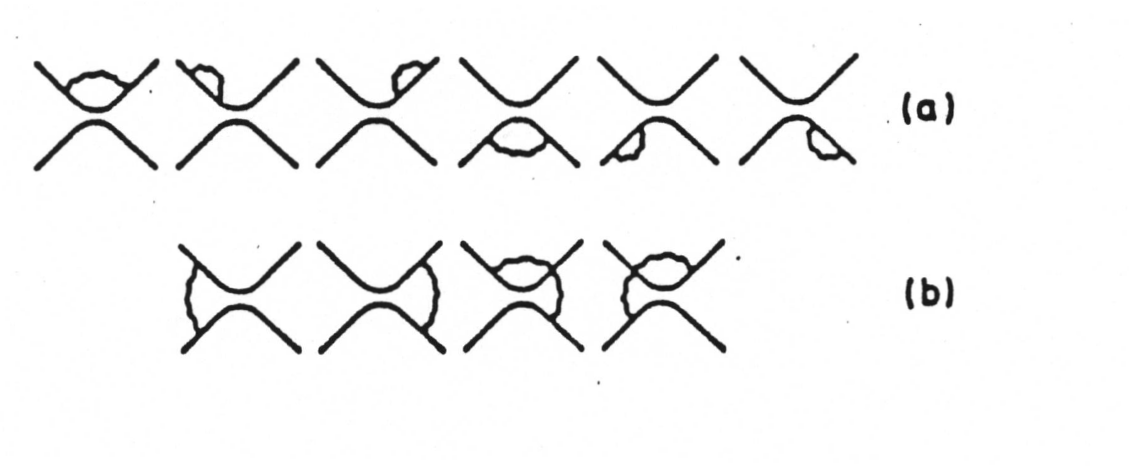}
\caption{One loop gluon corrections at a four quark vertex.}
\label{fig:c2f7}
\end{figure}
\pagebreak
Since a \W\ vertex
can be thought of as a four fermion interaction,
R{\"u}ckl \cite{ruck} first studied the effects of 1 loop gluon
exchange at a \W\ vertex. The diagrams for these processes are shown in Figure~\ref{fig:c2f7}.
The
diagrams of  a) are absorbed in the renormalization of  $G_F$, while
those of b) introduce new effects. We note that none of the diagrams
in b) are possible in semileptonic decay. R{\"u}ckl predicted the
following ramifications of the hard gluon exchanges:
\begin{enumerate}
\item
weak couplings would be renormalized.
\item
distortions to the color structure of the interaction from octet
currents
\item
possibility of new chiral structures, (V+A) for example
\item
possibility of new Lorentz structures (scalar or tensor)
\end{enumerate}
\par
\begin{figure}[t!]
\centering
\includegraphics[scale=0.40]{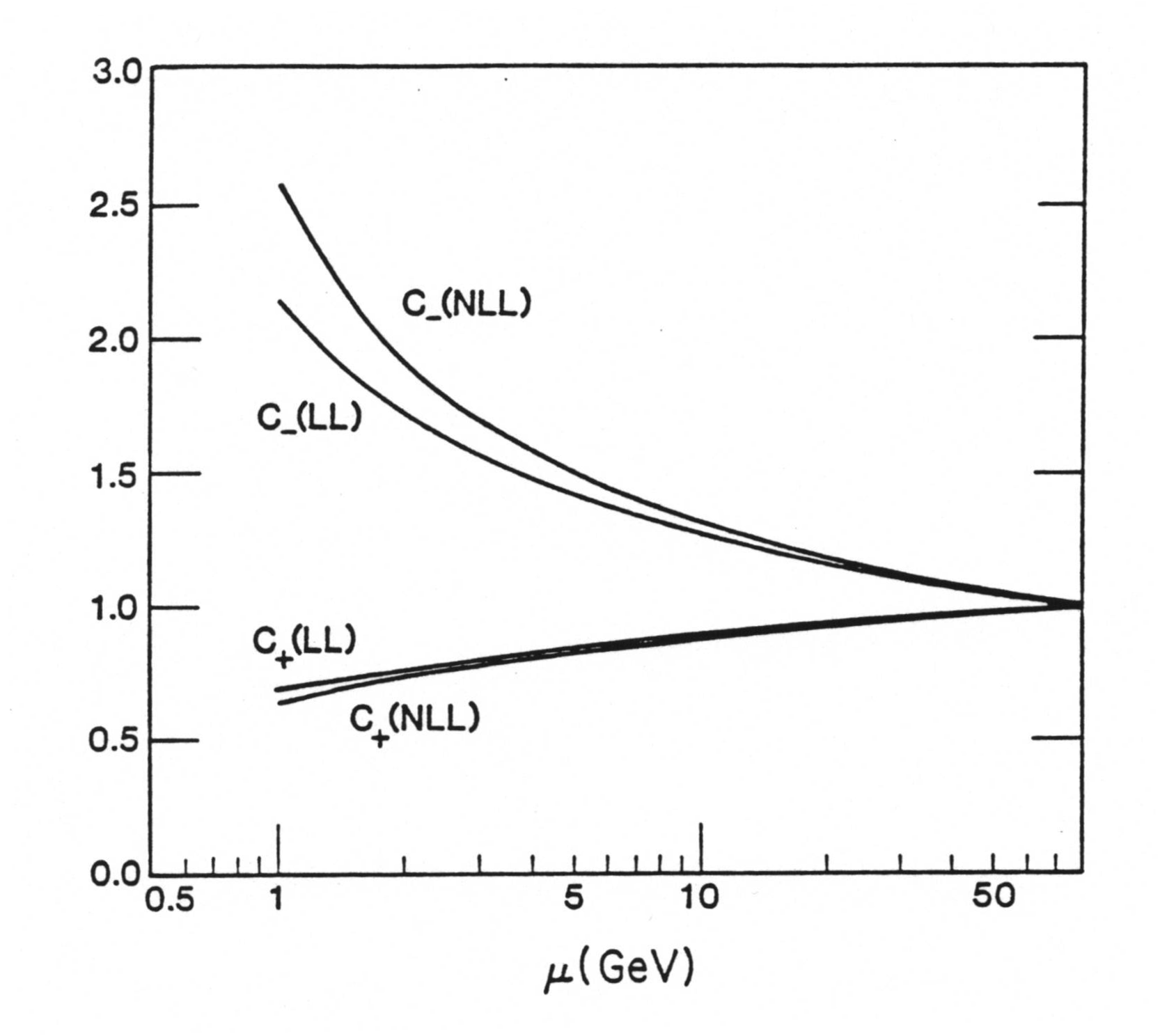}
\caption{Variations of the coefficients $ c_+$ and $c_-$ versus mass scale.}
\label{fig:c2f8}
\end{figure}
\begin{figure}[t!]
\centering
\includegraphics[scale=0.7]{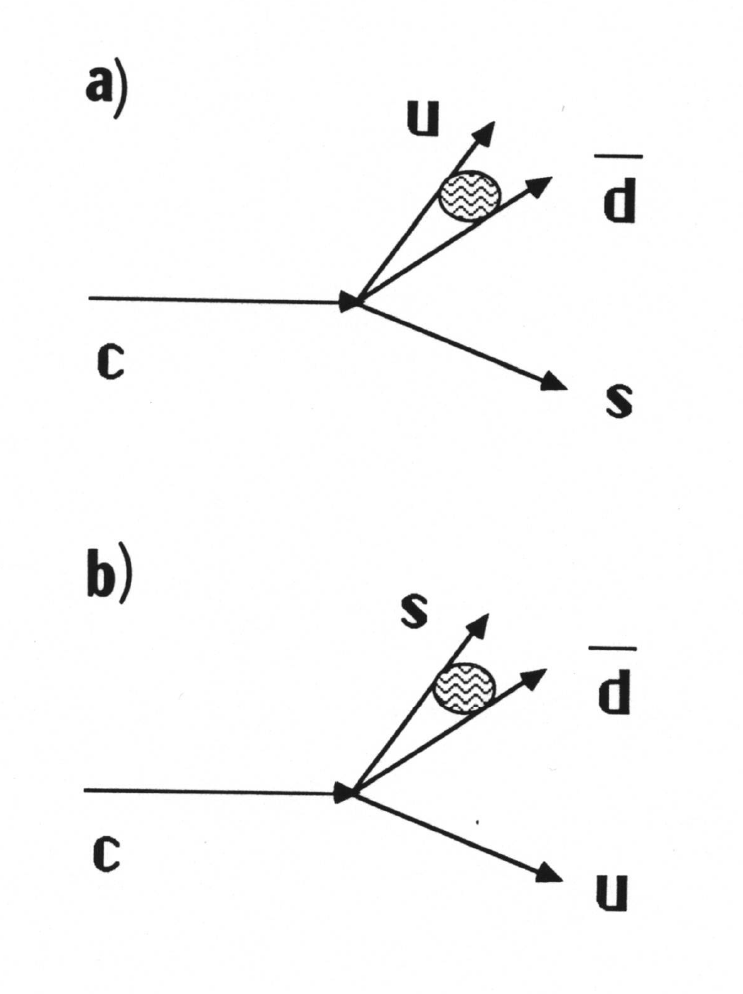}
\caption{The $c_1$ 
(a) and $c_2$ (b) processes.}
\label{fig:c2f9}
\end{figure}

The last two effects should be small, but the first two are not.
He calculated the first order correction to the Hamiltonian to be
$$ H^{(1)} =  { G_F \over \sqrt{2}}V_{cs}V_{ud}^{\ast} {3\alpha_s
\over 8 \pi} \log \left( {M^2_W \over \mu^2} \right) (\bar s
\lambda_a c)_L(\bar d \lambda^a u )_L $$
The effective weak Hamiltonian then becomes
$$ H_{eff}(\Delta c = -1) = { G_F \over \sqrt{2}}
V_{cs}V_{ud}^{\ast}\big [ c_+ O_+ + c_- O_- \big ] $$
where
$$ O_{\pm} = {1 \over 2} \big [ (\bar s c)_L (\bar  ud)_L 
 + (\bar s d)_L (\bar u c)_L \big ]$$
and to lowest order
$$ c_{\pm} =  1 \mp\ { \alpha_s
\over 2 \pi} \log \left( {M^2_W \over \mu^2} \right)$$ 
Where the $c_{\pm}$ coefficients obey the relation $ c_- = { 1
\over \sqrt{c_+}}$ and are approximately 2.0 and 0.7, respectively
at $q^2$ = 1.5 GeV. A plot of  $c_{\pm}$ verses mass scale for
leading log (LL) and next to leading log (NLL) is displayed in Figure~\ref{fig:c2f8}
Making one final transformation allows for a convenient representation
of the Hamiltonian. We define
$$ c_1 = { c_+ + c_- \over 2} $$
$$ c_2 = { c_+ - c_- \over 2} $$
The Hamiltonian then becomes
$$ H_{eff}(\Delta c = -1) = { G_F \over \sqrt{2}}
V_{cs}V_{ud}^{\ast}\big [ c_1 (\bar s c)_L (\bar  ud)_L 
 + c_2(\bar s d)_L (\bar u c)_L
 \big ] 
$$
The resulting  $c_1$ and $c_2$ processes are displayed in Figure~ \ref{fig:c2f9}.
\par 
Several important relationships can be derived based on the 
$c_{\pm}$ coefficients.
The non-leptonic width becomes enhanced by a factor
$$ \Gamma_{NL} =(2c^2_+ + c^2_-)\Gamma_0$$
 where we would naively expect three.  The semi leptonic branching
ratio is also modified
$$ B(c \decays\ eX) =  { 1 \over (2c^2_+ + c^2_- + 2) } \approx 15
\% $$ 
Both of these are clearly pushing things in the correct direction.
An induced neutral current ($c_2$ process)
 has also arisen which has precisely the
form of the color suppressed spectator diagram, and may neatly
account for such processes. In spite of these successes, the
spectator model still does not predict the large difference in the
\DP\ and \DZ\ lifetimes which has been confirmed experimentally, nor
the annihilation type processes.
To this end we must either consider the addition of nonspectator
processes or resort to the alternate approach of final state
interactions. 
\subsection{ Final State Interactions}
We examine final state interactions on the quark level and the
hadron level.
\par
\begin{figure}[t!]
\centering
\includegraphics[scale=0.85]{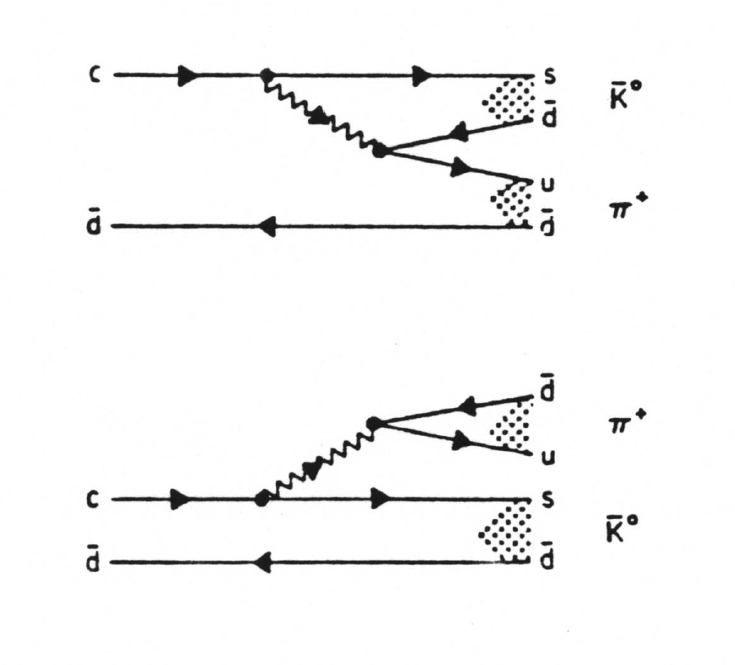}
\caption{
The decay mode \DP\ \decays\ \KZB\PIP.  
 Both color clustered terms contain a $\bar d$ quark.}
\label{fig:c2f10}
\end{figure}

 Several \DP\ decay modes  contain identical quarks in the
final state. An example is the decay mode \DP\ \decays\ \KZB\PIP\
depicted in Figure~\ref{fig:c2f10}.
Since there
are two identical fermions in the final state, the Pauli principle
demands that the wave function be suitably altered. This
induces an additional term to the \DP\ rate 
$$ \Gamma^{pauli} =   (2c^2_+ - c^2_-){ 12 \pi^2 \over M_D^2}f_D^2
\Gamma_0$$ 
Since this term is linked to $f_D^2$ which depends on the
wave function overlap, uncertainty is introduced into the magnitude
of this effect. General considerations predict that the size of the
interference term compared to the spectator term will be about 20
\%. We note that for $c_- >> c_+$ the overall term is negative and
acts to increase the \DP\ lifetime, as expected.
\par
It is also possible that in the case of two body decays, the two
final state hadrons may re-scatter into a different final state
(post hadronization).
This has been proposed \cite{dono}
 as a mechanism by which ordinary
spectator decays may produce more exotic nonspectator final states.
Hadron re-scattering is also required by unitarity.
\subsection{Summary}
An easy lesson to be learned  from the  previous discussion is that
simple quark diagrams are meaningless without fully considering the
role of gluons. Their presence substantially alters the weak
Hamiltonian, and may well act to catalyze the nonspectator diagrams
(by removing helicity suppression in \W\ exchange, for example).
While the spectator model with QCD improvements is a more reliable
theory, it is still incomplete. Both the nonspectator processes
along with pre- and post- hadronization final state interactions seem
to have established a foothold in the picture of charm decay.
Some of the more modern theoretical systems have been successful at
predicting the coarse features of relative \D\ decays, and these
have been limited to the two body case (fortunately a large number
of \D\ decays are two body or quasi-two body). Stumbling blocks
persist however. Implementation of nonspectator processes will
require some effort.  The Bauer-Stech-Wirbel \cite{bsw}
group
has typically neglected annihilation processes, and thus has a
difficult time explaining \PH\KZB. The  1/N expansion method of
Buras-Gerard-Ruckl \cite{buras}
includes such processes, however they are often of
higher order an are then dropped out. It is also unclear whether
final state hadron scattering can ever be integrated into
calculations in a systematic way. Although the $\DZ \leftrightarrow
\DP$ lifetime difference has been known for
almost a decade, we still await a quantified theoretical answer.


\chapter{Apparatus}
 
\begin{figure}[htp!]
\centering
\includegraphics[scale=0.6]{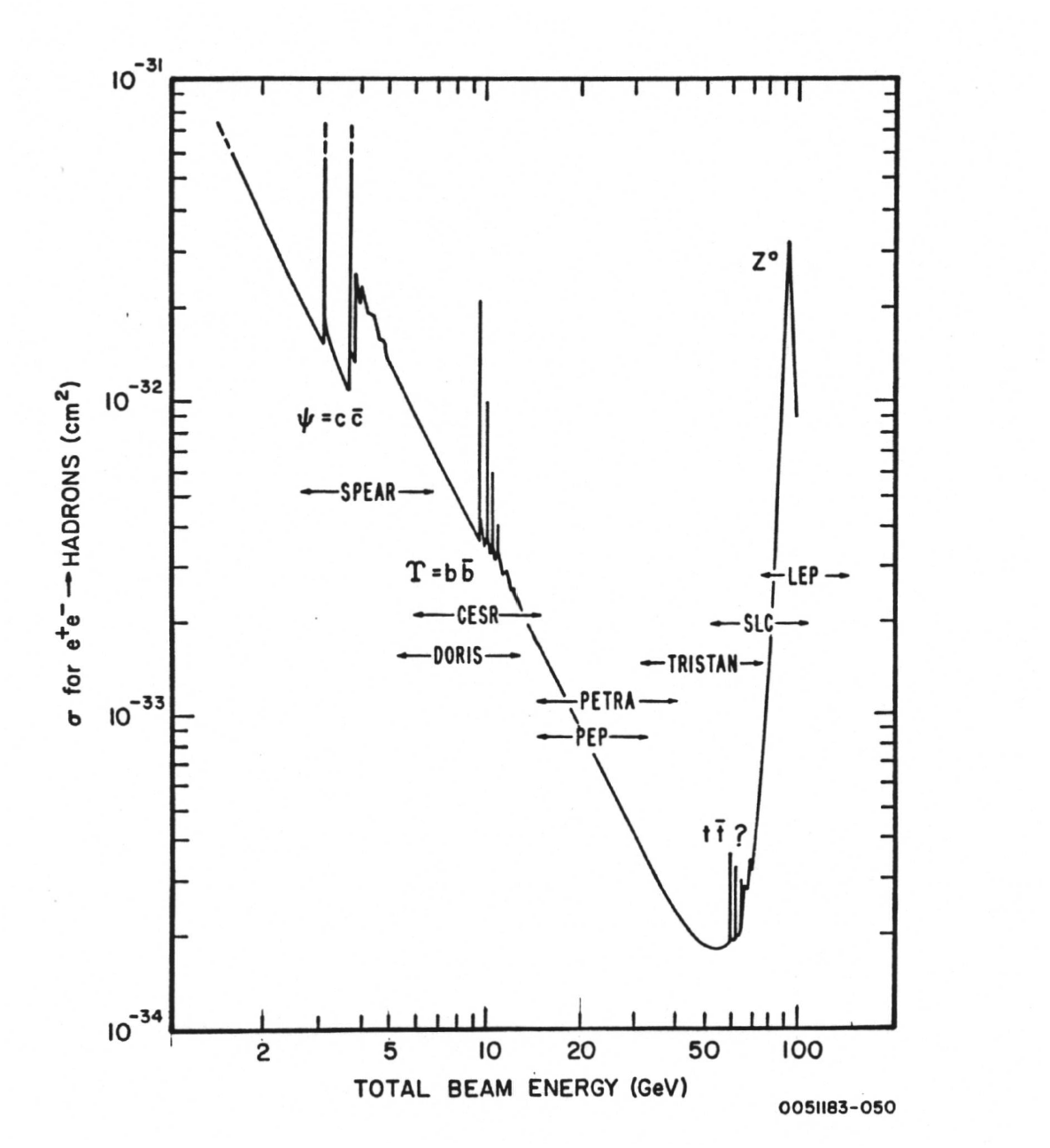}
\caption{Electron-positron cross section as a function of center of mass
energy. The cross section has been measured up to 50 GeV. Above this
energy only the Z resonance is established.}
\label{fig:c3f1}
\end{figure}

In order to perform high energy physics experiments, one needs   to
generate and record high energy interactions. Two general methods
exist for producing interactions, accelerating a particle and firing
in into a target (fixed target), or accelerating two particles
(usually particle-antiparticle) and colliding them.  \EPEM\
colliders are useful as both particles completely annihilate into a
state of pure energy. The  non resonant cross section for production
of a fermion-antifermion pair in an \EPEM\ collision is
approximately: 
$$ \sum_{nf} \ {4 \pi\ \alpha^2 e^2_f \over 3 s }$$
where $nf$ is the number of species of the particular fermion,
$\alpha$ the fine structure constant, and s the center-of-mass
energy squared $\rm (GeV^2)$. In addition the \EPEM\ annihilations,
\EPEM\ scattering (Bhabha) and two \G\ process also occur.
Of  greater importance to the experimentalist are the several
resonant enhancements to the \EPEM\ cross section. These are shown
in Figure~\ref{fig:c3f1}.
\par  
Reactions from proton colliders are extremely rich and dense, since
the protons themselves are composite objects. Because the
interaction rates for proton colliders are substantially larger, and
the physics more difficult to disentangle, they have assumed the
role of discovering phenomena, while \EPEM\ machines excel at
refined measurements. 
\section{CESR} 
\begin{figure}[p!]
\centering
\includegraphics[scale=0.6]{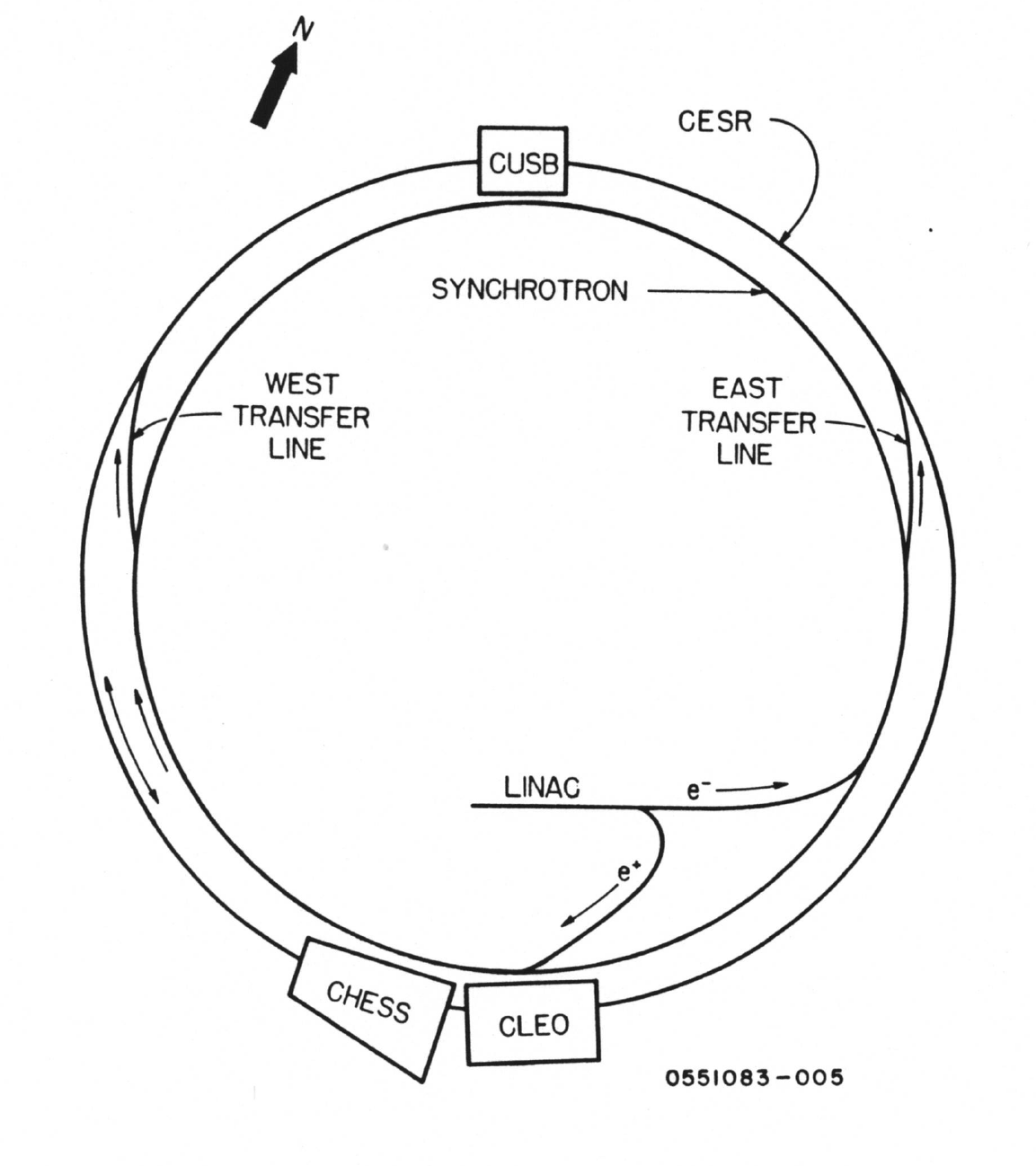}
\caption{The Cornell electron storage ring and
associated components.}
\label{fig:c3f2}
\end{figure}

The data for this research
project was gathered at the Cornell Electron Storage Ring (CESR),
located at Wilson Synchrotron Laboratory on the campus of Cornell
University in Ithaca, New York. It is capable of producing \EPEM\
collisions  at a center-of-mass energy from 7 to 14 GeV (here, and
throughout this thesis $\rm c=1$ unless explicitly stated
otherwise). Electrons and positrons are produced in a linear
accelerator and injected into a synchrotron where they are
accelerated to near collision energy. Finally the counter-rotating
beams are transferred into a storage ring 770 m in circumference
where they are made to collide in two interaction regions    at
multiples of a
fundamental frequency of 390 KhZ. The layout of this facility is
shown in Figure~\ref{fig:c3f2}.
Individual data runs last an average of
1-2 hours. The energy resolution, which is largely a function of the
bending radius of the machine is of order 3 MeV. \par Storage rings
maximize the opportunity for obtaining collisions from a generated
group of electrons and positrons. This is offset by synchrotron
radiation, the energy lost by the beams as they are accelerated so
as to travel in a (near) circular orbit. The power lost per electron
per turn is: $$ P_{\gamma} = {2\over 3} \ {{c r_e E^4} \over {
(mc^2)^3 \rho^2 } }$$ $r_e $ is the classical electron radius, E the
beam energy, m the mass of the particles in the beam, $\rho$ the
radius of curvature. At CESR   energies this amounts to
$\approx 1 \rm \ MeV$.  To maintain a constant beam energy, this
energy must be constantly replenished by boosts through R.F.
cavities.   The R.F. power costs makes a large contribution to the
operating expenses of a storage ring.  The cost of storage ring is
predicted by the Richter relation: $$ \$ = \alpha R + \beta {E^4
\over R} $$  which the first term accounts for magnets, tunnels,
etc., and the second for R.F. expenditures.
 This predicts both the size and the cost of a storage ring increase
as the energy squared.
\par   The critical performance parameter in any
collider facility is the luminosity (L). The number of events
observed $\rm N_{obs}$ is given by: $$\rm N_{obs}  = \sigma B
\epsilon \cdot  \int L  $$ The integrated luminosity enters linearly
 in the rate, along with the cross section, branching ratio and
reconstruction efficiency. Since the only variables available to the
experimenter are the luminosity and efficiency (which is largely
determined by detector design) it is desirable to accumulate as much
luminosity as possible. The instantaneous luminosity, for a
storage ring such as CESR, is governed by: $$ L = {{f N s\xi_v  }
\over {2 r_e \beta^{\ast}}} $$ where $f$ is the frequency  of the
collisions, N the number of particles per beam, $\xi_v$ the vertical
tune shift, and $\beta^{\ast}$ is a parameter which represents how
tightly the two beams can be focused at the interaction point.
During this data taking the CESR group began a program to increase
the luminosity by simultaneously circulating several bunches of
electrons and positrons. This project was successful, although the
strain on the R.F. system caused, at first, the machine to operate
less reliably. The average luminosity during this running period was
roughly $\rm 3 \times 10^{31} \ cm^{-2}sec^{-1}$.
\subsection{Data Sample}

  The data sample was acquired from August 1984 through February
1986 with the CLEO
detector, operating in what shall be referred
to as the 85VD configuration. Table~\ref{t:3p1}
\begin{table}[ht!]
\centering
\caption{85VD Data Sample}
\begin{tabular}{|c|c|c|c|}
\hline
Resonance region & Machine
timing & \# bunches & $\int \rm L$
 $\rm (pb^{-1})$ \\ \hline \US &  3 & 3& 17.0\\ \hline \USSSS & 7 & 7 &10.5\\ \hline
\USSSS & 7 & 3 & 63.3\\ \hline \US & 7 & 3 & 3.4\\ \hline \USSSS & 3 & 3 &
43.8\\ \hline \USSS & 3 & 3 & 33.3 \\ \hline
\end{tabular}
\label{t:3p1}
\end{table}
collects the operating conditions for collection of the data sample.
The  various states of timing and bunches reflect the multibunch
upgrade program. The luminosity reflects the runs to first order
which are considered good,  the data analysis  utilizes   a subset
of the above luminosity. The major impetus for  CLEO physics running
is to study the decays of \B\ mesons which are produced on the
\USSSS\ resonance. To efficiently separate processes connected with
the \USSSS\ from those of the surrounding continuum, the \USSSS\
running time is divided between on resonance and a center-of-mass
energy  approximately  60 MeV below resonance in a $2:1$ ratio.
Also, running on the \US\ is partially motivated to study lepton
faking backgrounds.
\section{The CLEO Detector}
 The results of the
\EPEM\ collisions were recorded by the CLEO \cite{cleonim}
 detector, a large
magnetic spectrometer with excellent charged particle tracking 
capabilities.  Operational since 1979, in the summer of 1984 the
first stages of a major detector upgrade \cite{cleo2}
 were implemented. This
consisted of the installation of a precision vertex detector (VD)
and instrumentation of the central drift chamber (DR1.5) to
simultaneously perform drift time and specific ionization (dE/dx)
measurements in all 17 layers. The CLEO detector so configured shall
be referred to the 85VD configuration.

\begin{figure}[htp!]
\centering
\includegraphics[scale=0.65]{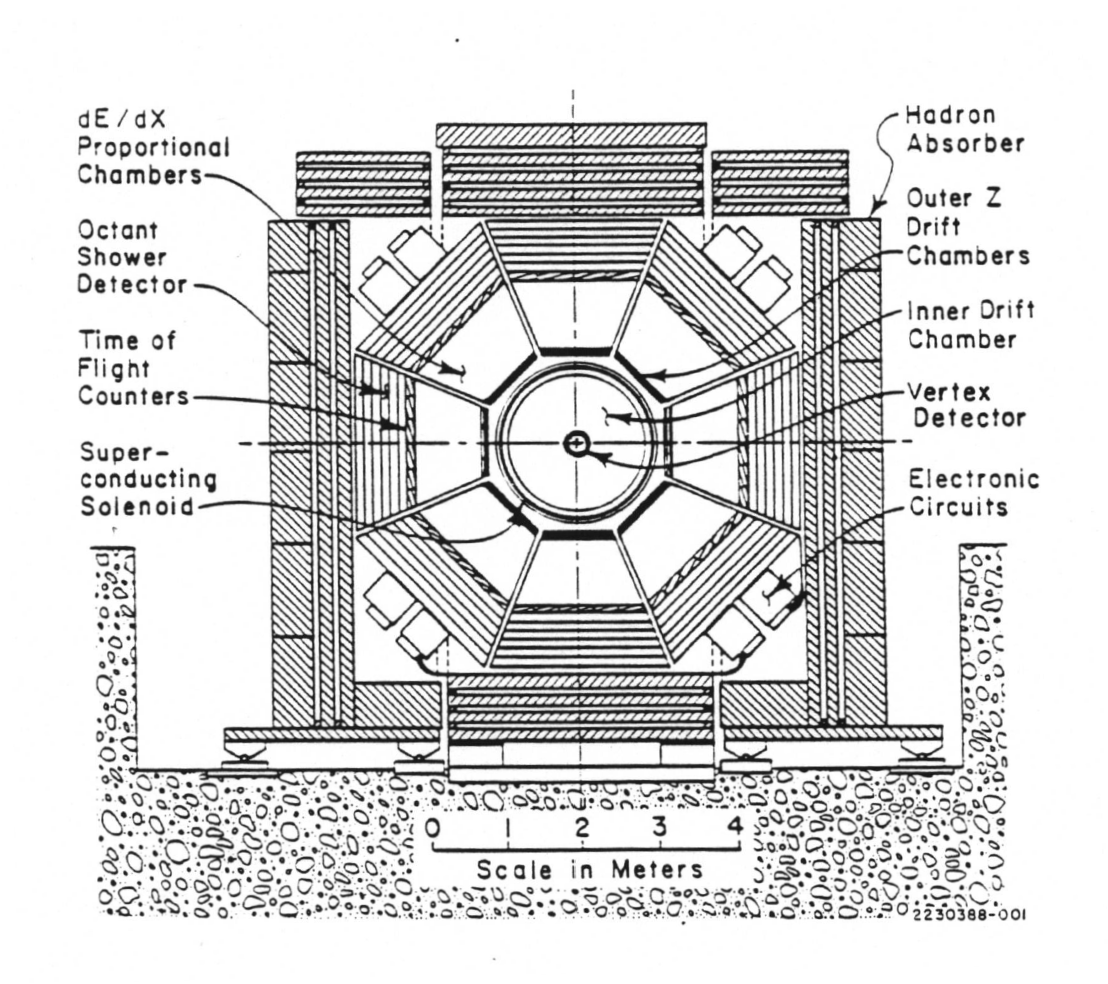}
\caption{Beam's eye view of the CLEO
detector.}
\label{fig:c3f3}
\end{figure}

\begin{figure}[htp!]
\centering
\includegraphics[scale=0.65]{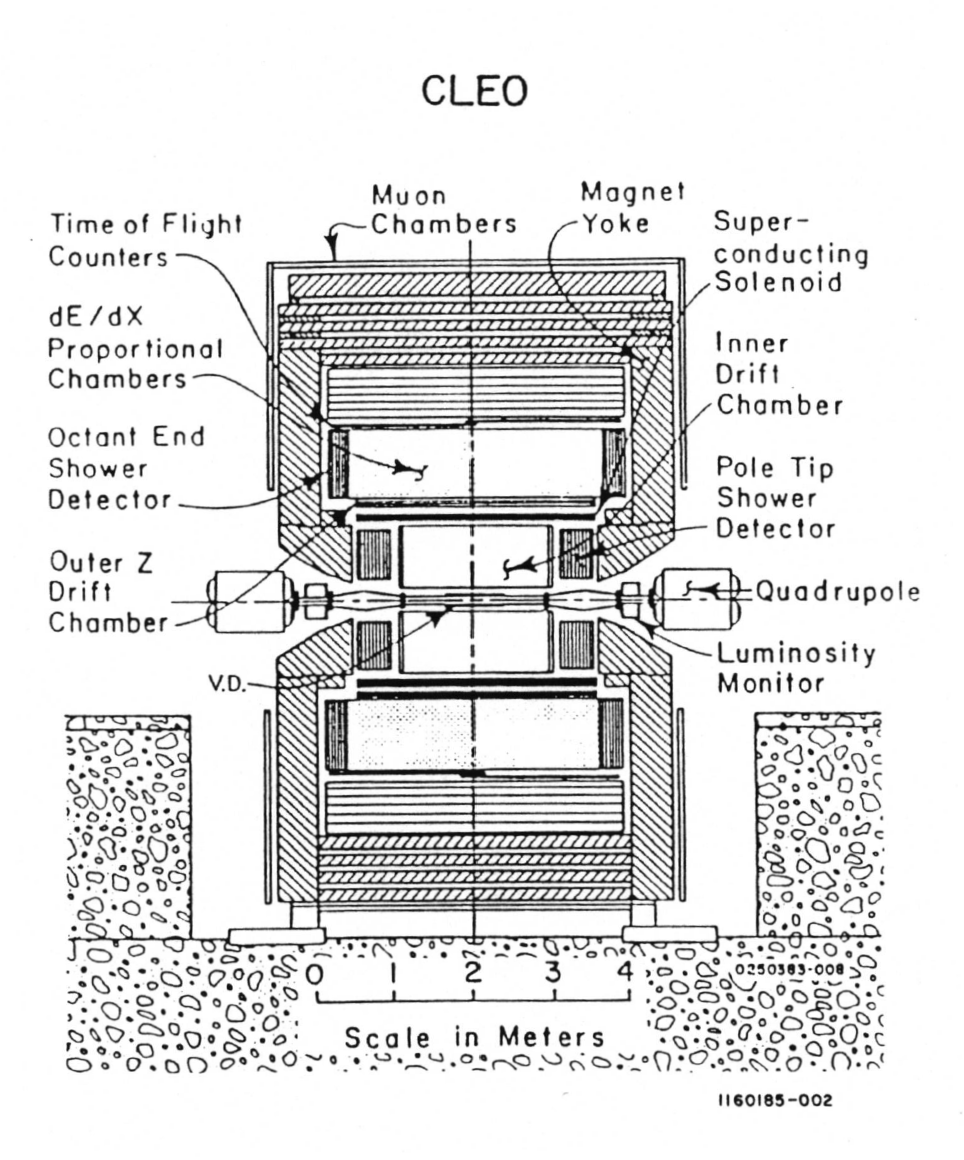}
\caption{Side view of the CLEO detector.}
\label{fig:c3f4}
\end{figure}

The CLEO detector, not unlike a high fidelity stereo system,
consists of several distinct detector elements, each of which
performs a specific function in the  event reconstruction. The
single most important property  that determines the detector
response is whether or not the particle posses electric charge. The
methods of detecting and analyzing charged and neutral particles
differ such that detectors of CLEO's generation were often polarized
to one extreme. 
  In brief, the CLEO consists of an inner detector dedicated to
charged particle tracking and and outer detector dedicated to  the
identification of both  charged  and neutral particles. The natural
boundary between the inner and outer detectors is a 1.0 Tesla
superconducting solenoid  magnet. The inner detector consists of
the two drift chamber devices mentioned above, as well as shower
counters (ES) on either end of the drift chamber face. The outer
detector contains a dedicated dE/dx system (DX) for charged 
particle identification, a time of flight scintillator detector
(TF), also for particle identification, and an electromagnetic
shower detector (RS) for \G's, \PIZ's and electron identification.
These three devices are arranged in eight identical octant sections
around the coil.
Surrounding the entire detector is a steel-drift chamber sandwich
used to identify muons (MU). Several other detector elements exist
from previous detector configurations, they do not effect this
analysis and will not be discusses here. A beams eye view of the
CLEO detector is displayed in Figure~\ref{fig:c3f3}.
 The octant structure is readily apparent, and a side
view is provided in Figure~\ref{fig:c3f4}.
 We now turn to a more
involved discussion of the  detector function.
\subsection{Tracking Fundamentals}
 The inner detector is globally cylindrical in geometry, consisting
of fine wires strung parallel to the beam direction, on circles of
constantly increasing radius. Drift chamber detectors are gaseous
detectors which detect the passage of charged particles through
ionization. The ionized electrons drift toward sense wires held at
  high electrostatic potential.  The extent to which the
sense wires are isolated and the way the geometry of the electric
field is defined determines the intrinsic performance characteristics
of the chamber. As the electrons  are accelerated toward the sense
wire, they begin to acquire enough energy to liberate other
electrons. This develops into a process referred to as avalanche
multiplication, which the electron gain is on order $10^4$. The gas
chosen is typically argon, which has high specific ionization, good
gas gain, and undergoes roughly 30 ionizing collisions/cm at STP.
Since argon is in the same chemical family as neon, the high voltage
conditions could result in breakdown or self sustaining discharge.
This is obviated by the addition of an organic vapor.   Nobel gases
can only be excited by the emission or absorption of photons, while
organics posses a myriad of rotational and vibrational states. This
leads to a substantial amount of energy dissipation is radiation-less
transitions. Organics also tend to increase the drift velocity of
the gas, thereby  decreasing diffusion effects. CLEO operates its
wire chambers in a 50-50 argon-ethane mixture, which has a mean     
drift velocity of $ \rm  50 \ \mu m/nsec ${.} 
 
 The central operating
principle of a drift chamber is drift velocity saturation. As the
electric field is increased, the drift velocity plateaus or
saturates. Insuring a constant drift velocity across the cell allows
for a theoretically simple (in practice the drift-time relationship
often represents the most difficult aspect of drift chamber
calibration) method for reconstructing the  trajectory through the
cell by measuring the time difference between entry into the cell
and a hit on the wire. The accuracy with which trajectories can be
reconstructed is in the range of hundreds of microns, and requires
timing precision on the nanosecond level.
\par
\begin{figure}[htp!]
\centering
\includegraphics[scale=0.55]{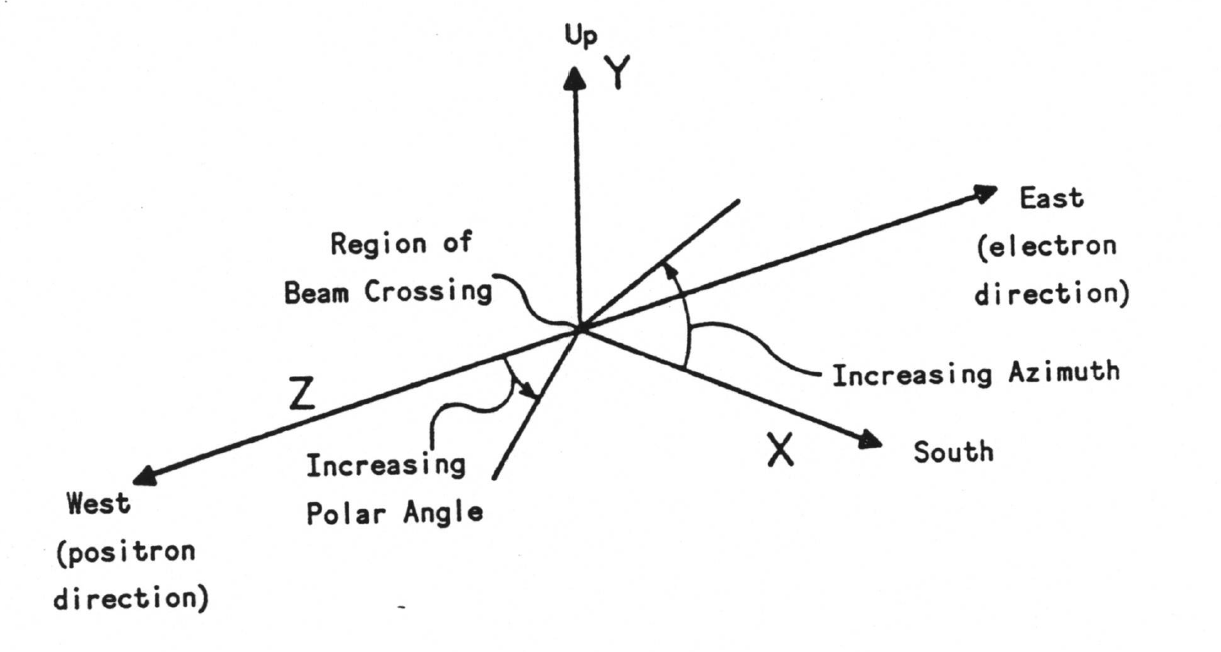}
\caption{CLEO coordinate system.}
\label{fig:c3f5}
\end{figure}

Immersing the drift
chamber in a solenoidal magnetic field (B) allows for the measurement
of two vital quantities for a charged track, the sign of the electric
charge and the momentum. The path of a charged particle moving a a
magnetic field is a helix. The CLEO reference coordinate system is
depicted in Figure~\ref{fig:c3f5}.
CLEO uses a track coordinate system that
consists of three $r-\phi$ parameters;  CUDR is one half times the 
reciprocal of the radius of curvature and is signed, FIDR is
measured from the distance of closest approach to the beam line,
DADR is the signed impact parameter, and two additional parameters
CTDR, the cotangent of the angle between the track and the beam
line (polar angle), and  Z0DR, the z coordinate of the distance of
closest approach. The transverse momentum $  p_{\perp}$ can then be
calculated by: $$  p_{\perp} \rm = 0.015 \bigg| {B \over CUDR}
\bigg | $$
where the dimensions are GeV, Kilogauss, and meters.
 Once the transverse momentum is determined the total
momentum is trivially extracted from  the measurement of the polar
angle.

Drift chamber performance is gauged by the spatial resolution in the
$r-\phi$ plane $\sigma_{r\phi}$ and the momentum resolution
${\delta p}\over p $. The two are   linked by the relation:
 $$ \left({{\delta p_{\perp}}\over p_{\perp} } \right)_{res} =
{p_{\perp} \sigma_{r\phi} \over (0.03){\rm L^2B}} \sqrt{ \rm 750
\over N+5}$$ for a drift chamber with N equally spaced measurements
over a lever arm L in an axial magnetic field B (the units are GeV,
meters, and kilogauss). The other dominant term comes from multiple
scattering: $$ \left({{\delta p_{\perp}}\over p_{\perp} }
\right)_{ms} = \rm { 0.5 \over LB} \sqrt{1.45 {L \over X_0} }$$ 
here $X_0$ is the average
radiation length of the detector in meters. The resolution dominated
term is prominent for high momentum tracks whilst the multiple
scattering limits the resolution at low momentum. The  CLEO VD+DR1.5
tracking system achieves a momentum resolution of $ \left({ {\delta
p}\over p } \right)^2 = ( 0.007p)^2 + (0.006)^2$  ($p$ in GeV/c).
\subsection{Vertex Detector} 
\begin{figure}[htp!]
\centering
\includegraphics[scale=0.6]{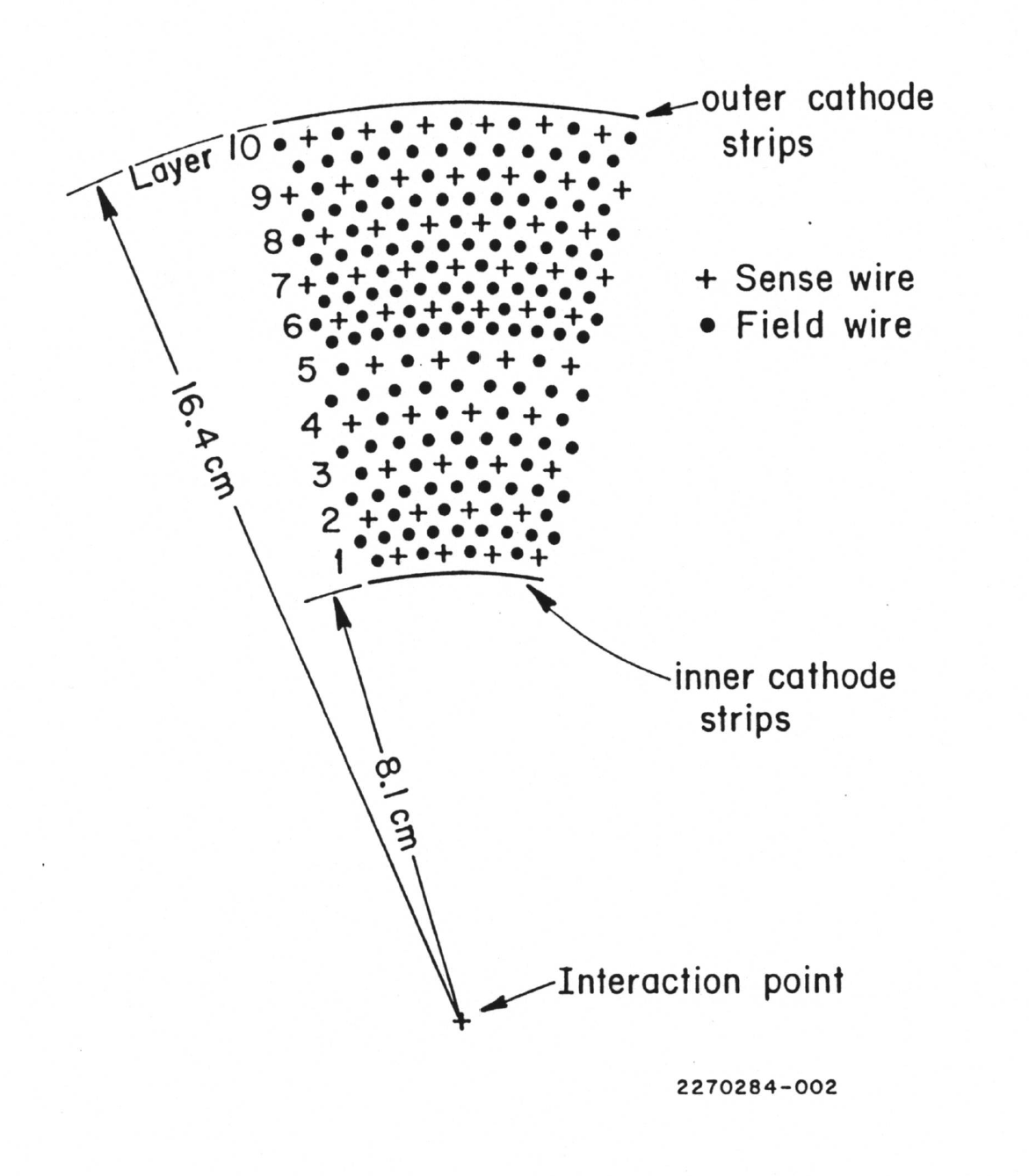}
\caption{Vertex detector cell arrangement.}
\label{fig:c3f6}
\end{figure}

The CLEO vertex detector is a high
precision drift chamber which forms the innermost detector element.
800 cells are divided among 10  axial layers, which range in radius
from 8 to 16 cm (see Figure~\ref{fig:c3f6}).
The inner 5 layers have 64 cells per layer, and the outer 5 layers
each have 96 cells. The cells are hexagonal in shape, and expand in
size so as to subtend a constant angle in $\phi$.  The
$\sigma_{r\phi}$ resolution for the VD \cite{sob}
on average is 100 microns. 
 
The active z length is 70 cm.  Two methods are provided for z
measurements. A resistive sense wire is read out at both ends
allowing a measurement by charge division.
 The resolution for this method is in the realm of 9 mm. More precise
information is obtained from the conducting cathode strips on the
inner and outer support tubes. The strips are  segmented into 8
$\phi$ regions which are further divided into 64 (96) sections on
the inner and outer tubes. The cathode strips z measurements have a
resolution of 750 microns.

The device derives its name from the design purpose to extrapolate
tracks back to their production vertex.  The instrument is separated
from the interaction point by a silver coated beryllium beam pipe
and a carbon inner support tube. The interceding material amounts to
0.1 \% of a  radiation length. The extrapolation error is
given by: $$ \sigma_{ext} = \sqrt{{ {100}^2 + {115 \over \sin\Theta
(p \beta)^2}}} $$ $\Theta$ is the polar angle and $\beta = v/c$. $p$
is in GeV and the result has the units of microns.
\par
 A plethora of benefits accompanied the installation of the VD.
The tracking momentum resolution was greatly improved, and the
momentum range for low momentum track reconstruction  was extended.
Beam wall and beam gas events were efficiently removed, and the
capabilities of the experimental trigger were extended. 

 \subsection{Drift Chamber} 
In contrast to the VD, the design of the drift chamber relies on the
repetition of a single cell geometry. The unit cell consists of a 20
$\rm \mu m$ diameter gold plated tungsten sense wire responsible for
a radial region of 11.3 mm. The cell boundaries  are formed by three
115 micron diameter silver plated beryllium copper field wires
located on either side of the sense wire, and arranged in a straight
line perpendicular to the radius of the cylinder over a region of 10
mm centered about the sense wire. The inner and outer boundaries of
the device are kept at ground, which coupled with the open
rectangular cell geometry leads to certain peculiarities in the
function of the detector which shall be addressed.

\begin{figure}[htp!]
\centering
\includegraphics[scale=0.6]{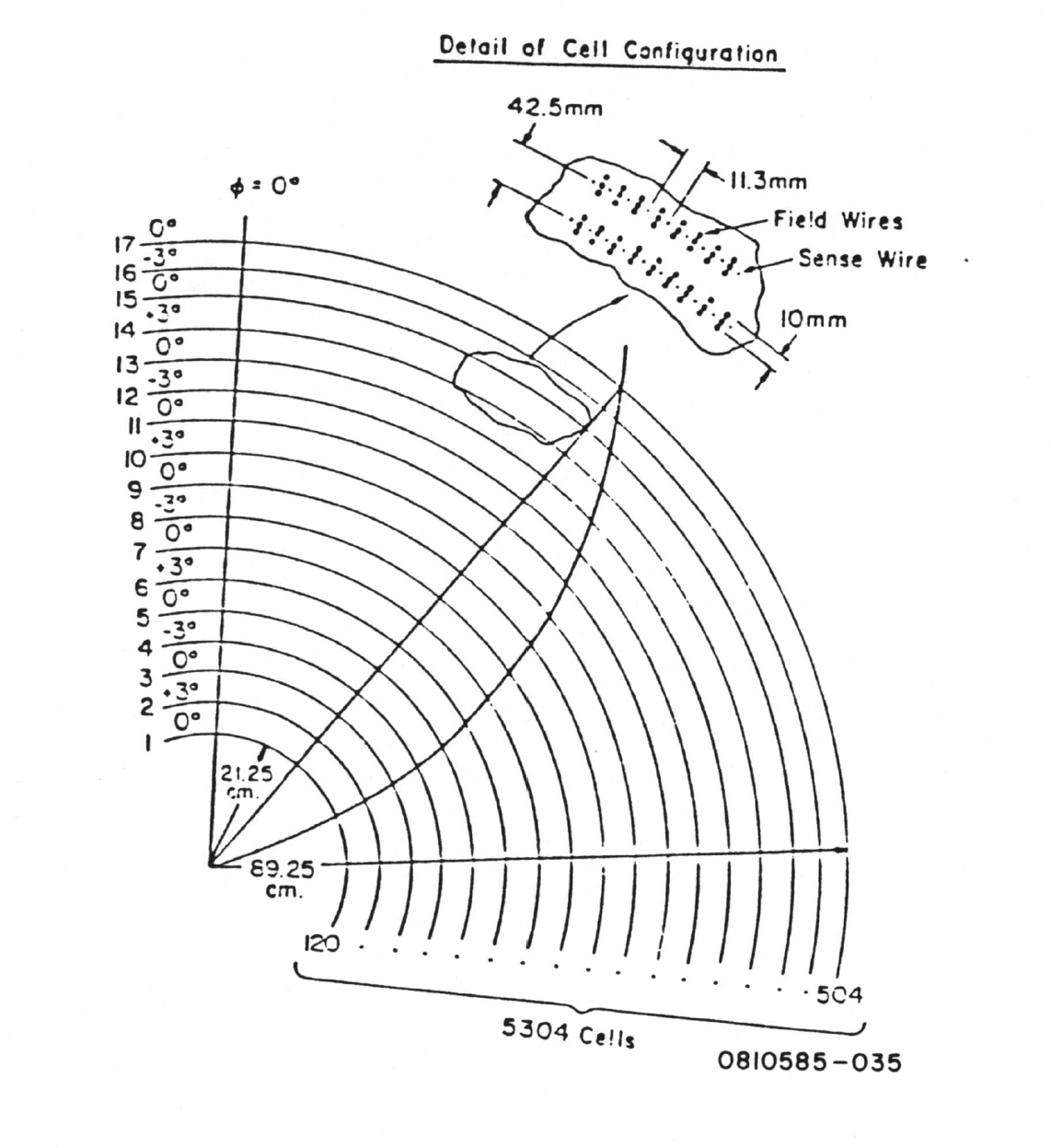}
\caption{The CLEO central drift
chamber.}
\label{fig:c3f7}
\end{figure}

The arrangement of cells is rigidly described by the prescription 
$\rm cells/layer = 24 \times (N+4)$  which are located on nearly
concentric cylinders of radius $\rm R_{N} = 42.49 \times (4+N)$ (mm)
and N ranges from 1 to 17. The lever arm of the drift chamber
ranges from 212.5 to 892.5 mm. A schematic illustration of the
drift chamber can be found in Figure~\ref{fig:c3f7}.
Also distinct from the VD is the method of z measurement.
Starting with the second cylinder, the 8 even numbered layers are
strung in a ``stereo'' mode at alternating  $ + -$ angles of
$2.9^{\circ}$ to the beam axis. The odd numbered layers are all
axial, strung parallel to the beam line. High voltage is distributed
through the field wires, with the sense wires kept at ground for
ease of electronic readout.  During August 1984, the readout
electronics for the drift chamber was replaced, with the new system
allowing both timing and pulse height measurements. With the
addition of the vertex detector, the drift chamber underwent  
its most through calibration, revealing several biases. It is this
mode of operation which shall be recounted here.

\begin{figure}[htp!]
\centering
\includegraphics[scale=0.55]{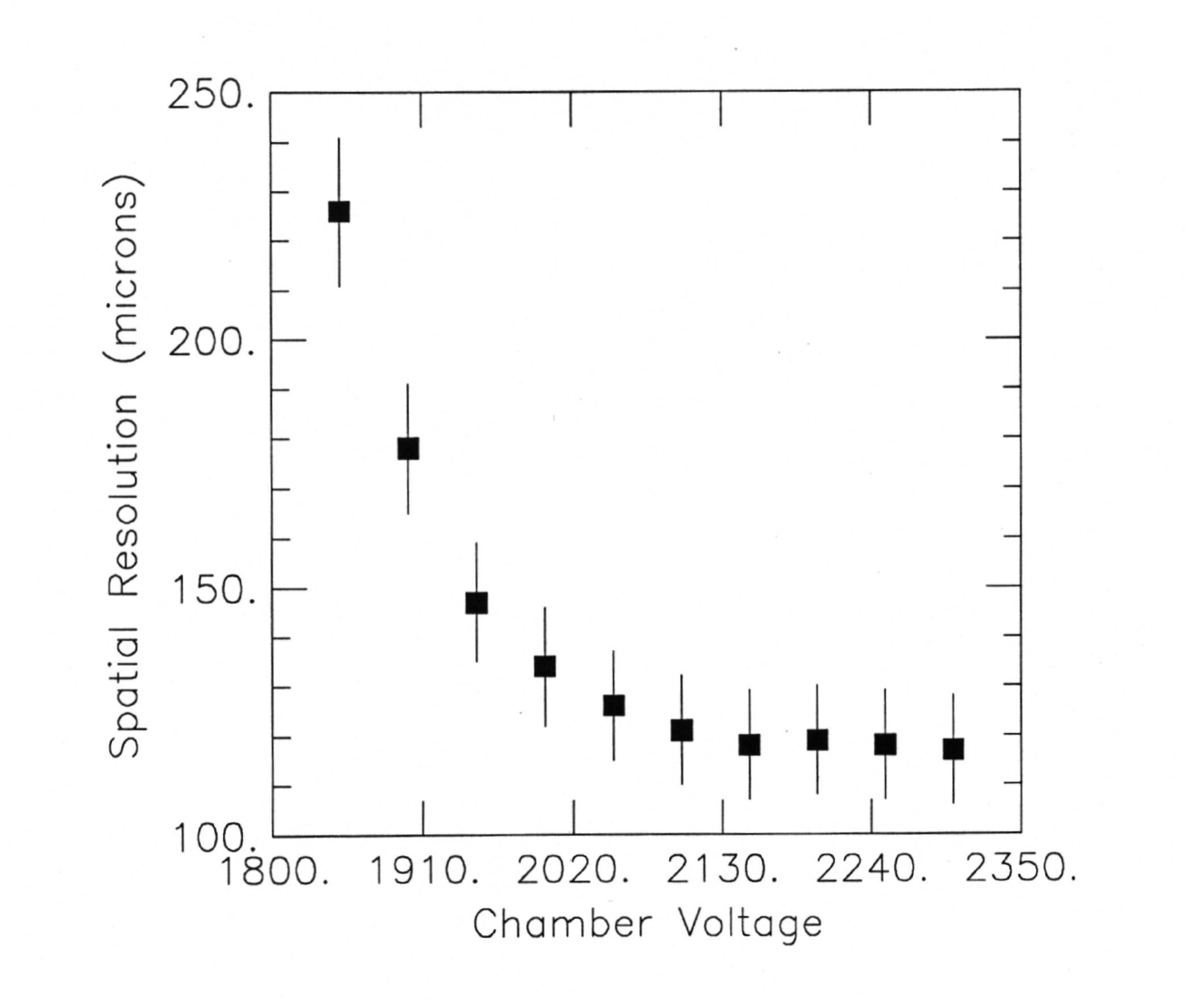}
\caption{Measured test lab spatial resolution $\sigma_{r\phi}$ verses chamber
voltage.}
\label{fig:c3f8}
\end{figure}

The chamber was operated at 2050V,
corresponding to an approximate electric field strength of 3630 
v/cm and gas gain $\approx 2 \times 10^4${.} Performance tests \cite{avery1}
of the new electronics were done by T. Copi\'e and the author
using  a  9 layer, 220 channel $\phi$ slice
 mock-up of the drift chamber using cosmic ray muons. Plots
of the timing and pulse height performance are shown in Figure~\ref{fig:c3f8}
 and
 \begin{figure}[htp!]
 \centering
\includegraphics[scale=0.55]{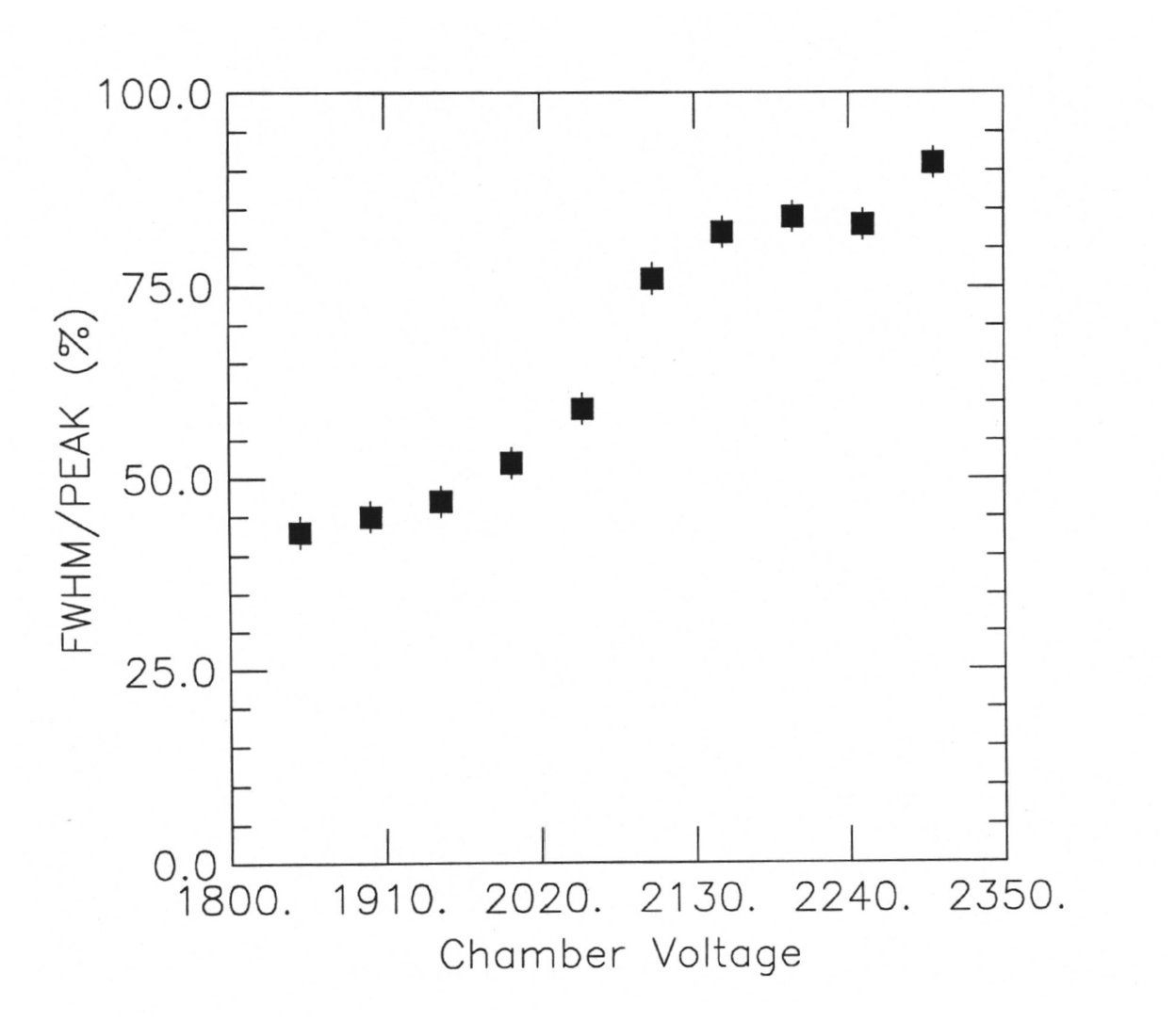}
\caption{Test lab pulse height  resolution $\rm \sigma \over peak$ versus chamber voltage.}
\label{fig:c3f9}
\end{figure}
and Figure~\ref{fig:c3f9}.
 Because of the open cell geometry, a symbiotic relationship
exists among the gains of the cylinders. Shown in Table~\ref{t:3p2}
\begin{table}[ht!]
\centering
\caption{Drift Chamber Cylinder Gain Interdependence}
\begin{tabular}{|c|c|c|}
\hline
Layer 16  $\langle PH \rangle$ & Layer 16 Voltage (V) & layer 17 Voltage (V)\\ \hline
252 & 2050& 2150\\ \hline
148 & 2050 & 0\\ \hline
186 & 2100 & 0\\ \hline
\end{tabular}
\label{t:3p2}
\end{table}
are
the effects on the average measured pulse height in layer 16
subject to variations in the outermost layer 17. Layer 15 was held
at 2050 V. Since the inner and outermost cylinders have one high
voltage layer and ground as their two  radial nearest neighbors,
one can deduce from the above table that they will also suffer from
reduced gain. To compensate for this problem, these two cylinders
were operated at 100V higher than the nominal operating voltage of
the chamber. A more deleterious effect from the outer ground planes
was that the field would ``leak" out of the cell creating
asymmetries in the drift-time relationship on the right and left
sides of the cell.  The measured versus fit drift distance for the
right and left sides of a wire are shown for an outer layer (Figure~\ref{fig:c3f10}) and an inner layer (Figure \ref{fig:c3f11}). Note the asymmetry in the outer layer.
 \begin{figure}[p!]
\centering
\includegraphics[scale=0.65]{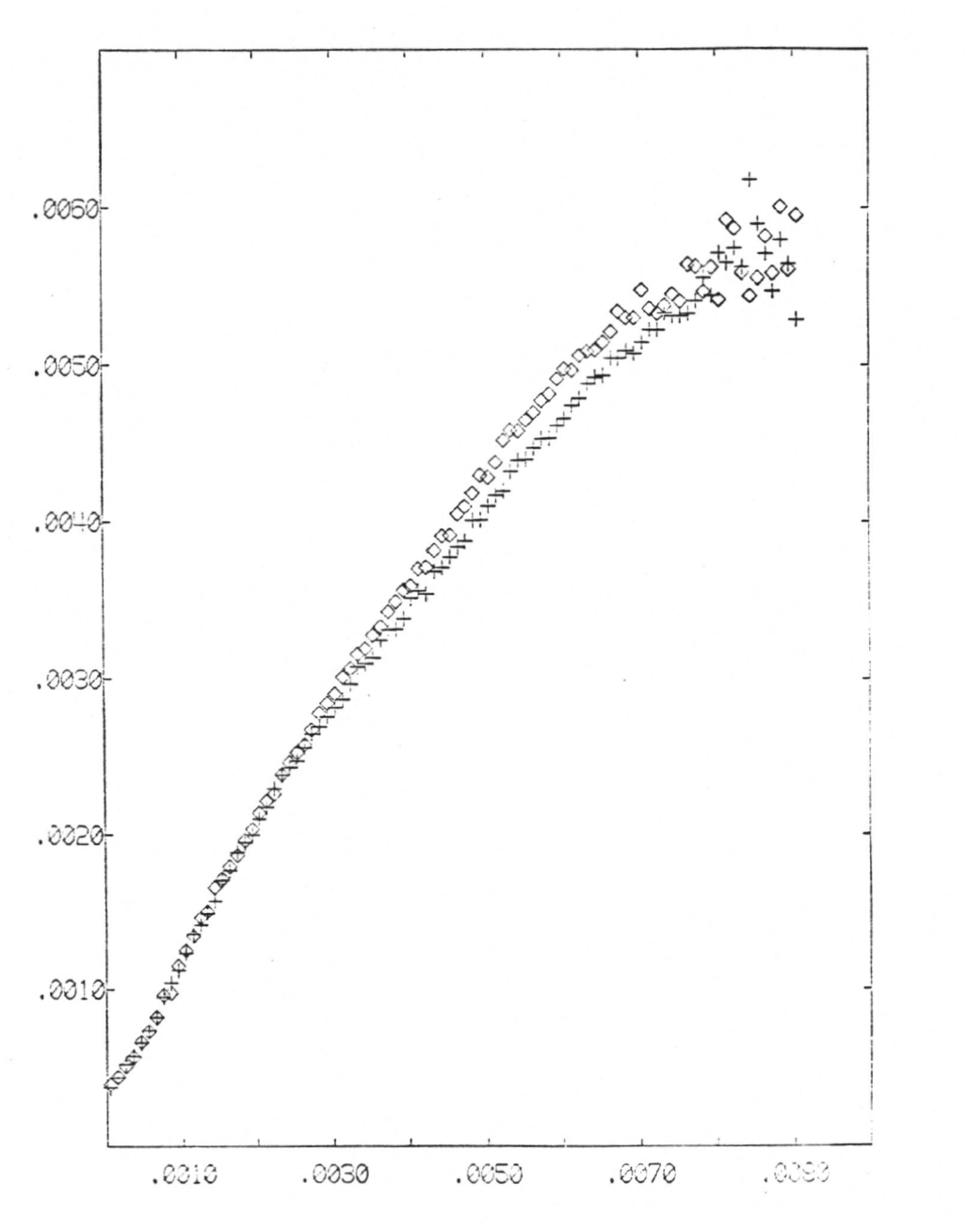}
\caption{Measured vs fit drift distance for the left (cross) and right
(diamonds) side of the cell in layer 17. The units are in meters.}
\label{fig:c3f10}
\end{figure}
\begin{figure}[p!]
\centering
\includegraphics[scale=0.65]{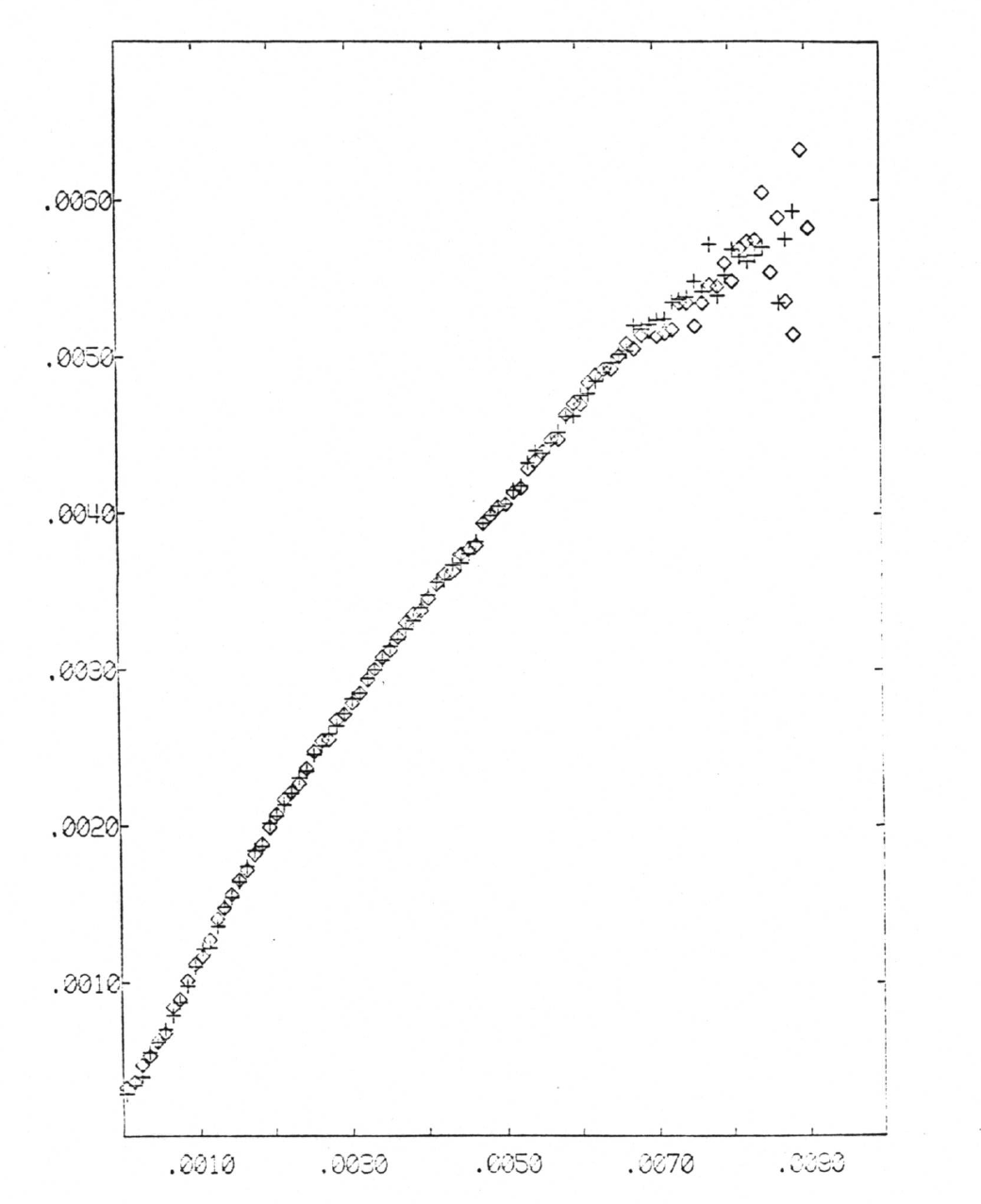}
\caption{
Drift time relations for left and right
sides of a normal layer.}
\label{fig:c3f11}
\end{figure}
An analogous effect was observed in the vertex detector, which has a
similar grounding arrangement. This problem was partially calibrated
away by having separate drift-time relationships with right-left
asymmetries for the inner and outer layers. A plot of the $r-\phi$
residuals from hadronic events is shown in  Figure~\ref{fig:c3f12}.
\begin{figure}[htp!]
\centering
\includegraphics[scale=0.6]{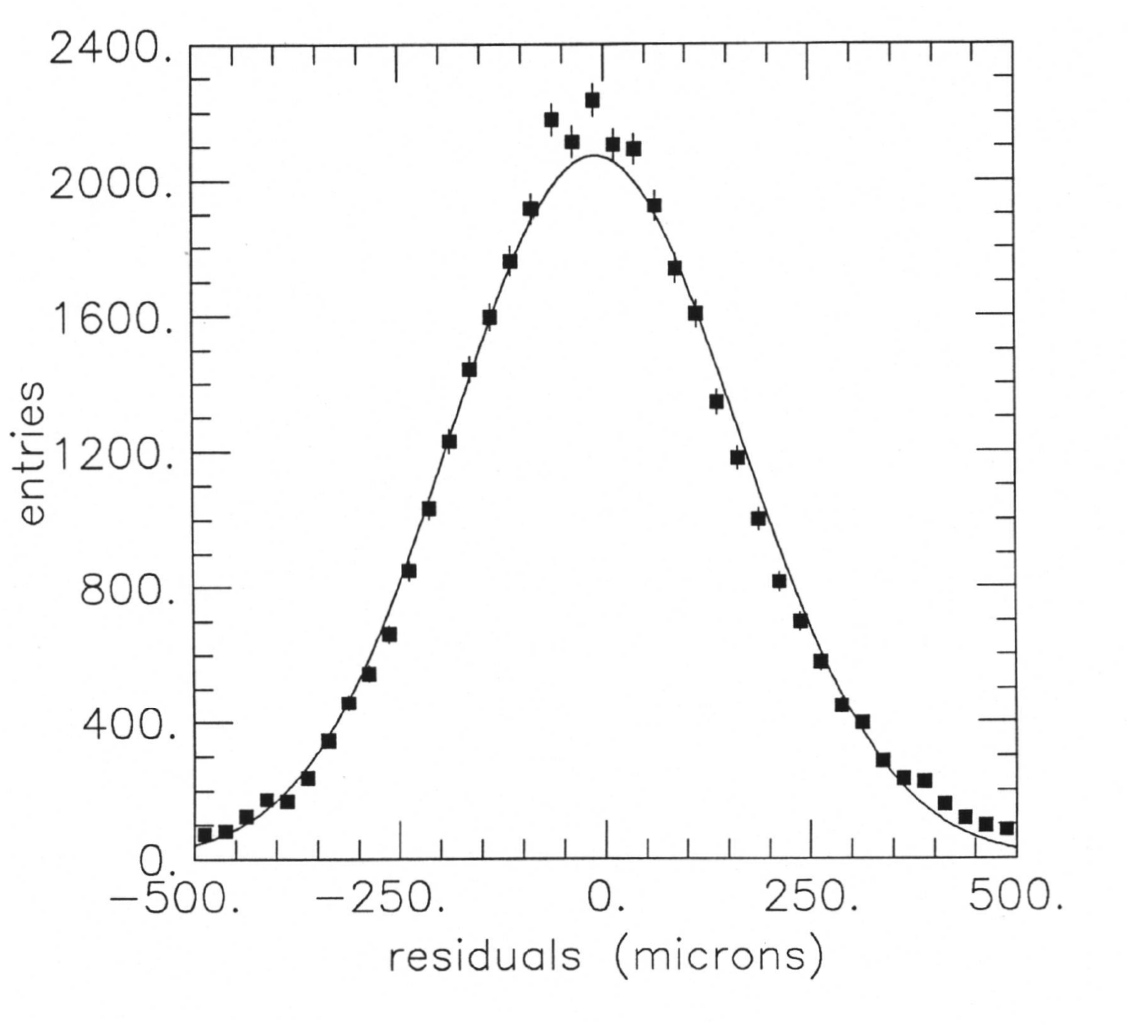}
\caption{Measured -
fit drift distance from all drift chamber layers, from hadronic type
events.
}
\label{fig:c3f12}
\end{figure}
 Fitting
the central peak to a Gaussian yields $\sigma_{r \phi} = 160 \pm 10$
microns.

Since the primary interaction of charged particles and a drift
chamber detector is through ionization, it becomes possible to use
the measured pulse heights to perform particle identification
based on energy loss (dE/dx).
The amount of energy lost for a relativistic charged particle
traversing a medium is predicted from the Bethe-Bloch equation:
$$\rm  {dE \over dx} = { A \over \sin\Theta} \cdot {1 \over
\beta^2} \cdot \Bigl [ \ln(2m_e \beta^2 \gamma^2 / I_0) -\beta^2
\Bigr]$$
 where A is a term related to the medium traversed, $\ 1
\over {\sin\Theta}$ the path length, $\rm I_0$ the ionization
potential, $\beta, \gamma$ the canonical relativistic variables. 
To engineer such measurements a precarious balance must be struck
between preserving the spatial resolution which is improved with
increasing   gas gain, and the dE/dx measurements which
deteriorate with increasing gas gain.
  Timing circuits are equipped with   discriminator circuits which
trigger once a pulse amplitude greater then a specified size is
registered. It is desirable to have the circuit respond as quickly
as possible to a given pulse in order to obtain the the best
resolution. The path of least resistance is to  make the gas
gain as large as possible, providing hefty pulses from the chamber
which are easy to trigger on. This approach was the initial
operational mode of the CLEO drift chamber, where only timing
information was collected. To also make ionization measurements,
the gas gain must be lowered. Since the incoming pulses are smaller
the system must be capable of operating at a reduced threshold. The
major obstacle is that lowering the sensitivity means that noise
hits may penetrate the system, and then real
information can be washed out by the background. 
Noise immunity was enhanced by mounting preamplifiers, 24 channels
to a preamplifier card, directly on to the drift chamber face. The
entire drift chamber was read out one end face. The boosted
signals were transmitted via a symmetric differential transmission
system on a 7.5m flat twisted pair cable, to receivers  on 48
channel data cards, located in standard
CLEO readout crates outside the detector. Each data card channel
contained timing and pulse height circuits which analyzed the same
incoming pulse. The operating sensitivity was 300 nanoamps, and the
resulting dE/dx resolution was in the range of 10-14\%. 
Part of the difficulty in performing dE/dx measurements was that
the open rectangular cell geometry was not ideally suited for
charge collection. The system was engineered and implemented by John
Dobbins and Don Hartill. The calibration and analysis of the dE/dx
information was primarily done by Thierry Copie and Tom Ferguson,
while the timing performance system was the responsibility of Paul
Avery and the author. Although the dE/dx resolution was somewhat
inferior to the outer detector, simply  being able to have
information associated with every track (especially at low momentum)
dramatically improved the particle identification performance of
the detector.
\subsection{ End Cap Shower Counters}
Beyond the drift chamber face on both sides of the detector are
the end cap shower counters. This detector is formed from aluminum
proportional tubes and lead, with 21 layers of the device oriented
in 3 groups oriented at $120^{\circ}$ to each other. The energy
resolution is $\sigma_E/E = 0.39/\sqrt{E}$. The primary function of
the device is to provide a measurement of the luminosity from
Bhabha scattering.
 \subsection{Superconducting Coil}
Since in the initial CLEO conceptual design the particle
identification detectors were located beyond the coil, it was 
necessary to produce the field using as little material as possible,
hence the decision was made  to use a superconducting coil.  The coil
is 3.1 m in length and 2 m in  diameter. The winding is made of
Nb-45\% Ti in a copper matrix. The standard running conditions
mandate a 1 Tesla field which requires a current of 1500 amps.
Measurements of the field have determined uniformity to within
2\%{.} The net material, including the cryostat is 0.7 radiation
lengths. This material ranges out pions with momenta less than 150
MeV, kaons with momenta less than 400 MeV, and protons with momenta
less than 600 MeV.

 \subsection{ dE/dx} Immediately outside the coil is
the dedicated dE/dx system. Each octant contains 124 modules, each of
which contains 117 wires oriented perpendicular to the beam line. The
large number of wires was desirable to to obtain a high statistics
measurement of the average pulse height, which CLEO calculates from
the lowest 50\% of the measurements to avoid the long tail
associated with the Landau distribution. The detector was operated
on a gas mixture of 91\% argon and 9\% methane at a regulated
pressure of 45 psia. The
 resolution, as measured by the peak divided by the standard
deviation of the distribution is 5.8 \% for hadrons. Despite this
fine performance, the inability to distinguish multiple tracks
and the loss of low momentum tracks where hadrons are most easily
separated seriously undermined the usefulness of this device.
\subsection{Time of Flight}
\begin{figure}[htp!]
\centering
\includegraphics[scale=0.6]{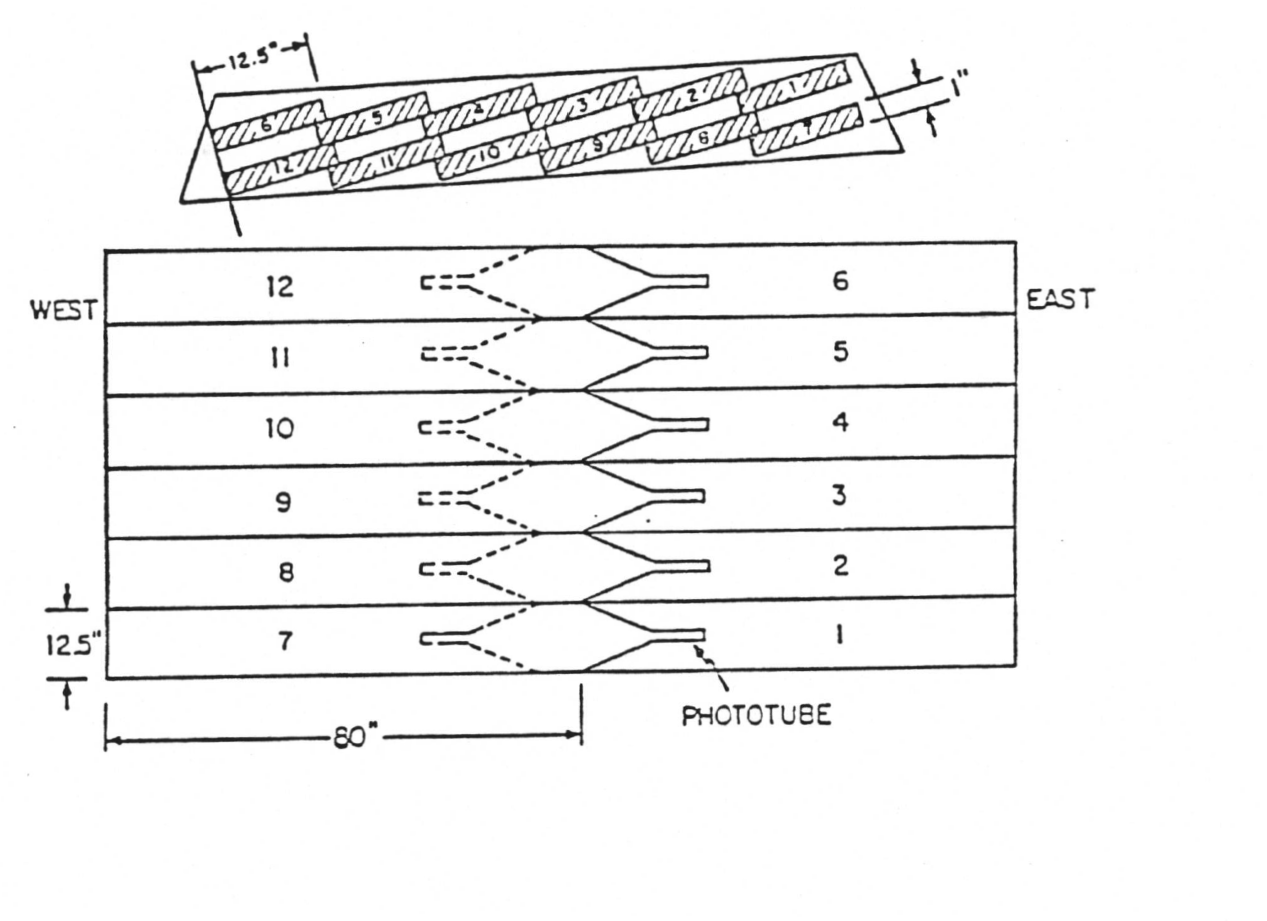}
\caption{ Time of flight detector.}
\label{fig:c3f13}
\end{figure}

 Occupying the next radial position in the octants is the time of
flight system. Each octant contains twelve $2.03 \times 0.312 \times
0.0023$ m scintillation counters. Each scintillator is read out on
one end by an Amperex XP-2020 photomultiplier tube. The TF
geometry per octant in shown in  Figure~\ref{fig:c3f13}.
The counters operate using two discriminators sensitive to pulse
height from the  photomultiplier tube. The time for a given hit is
then determined by extrapolating to zero pulse height and correcting
for travel time in the scintillator medium, as well as measured
pulse height.  The rms resolution of the
detector is 350 psec. 

TF hits are matched to drift chamber
tracks, and the flight time coupled with the tracks momentum is
used to determine the mass of the track. In addition the TF plays a
vital role in the experimental trigger.
\subsection{Octant Shower Counters}
The last device in the octant is the octant shower counter usually
referred to as the RS. It is a twelve radiation length thick lead
proportional tube sandwich, operated on
on a gas mixture of 91\% argon and 9\% methane. It covers a solid
angle of $\Omega/4\pi=0.47$ and the device performance can be
characterized by
$\sigma_E/E = 0.17/\sqrt{E}$ with E in GeV. Neural particle generate
characteristic electromagnetic showers and can thus be identified.
 Charged particles also are detected, though the response is much
weaker, allowing electrons to be distinguished from other charged
particles. The instrument is also used to calibrate the luminosity
using large angle Bhabha events, and serves in the experimental
trigger.
 \subsection{Muon Detectors}
The muon system is the outermost element and surrounds the CLEO
detector in a box-like geometry. Beyond a steel hadron absorber of
6 to 12 hadron interaction lengths are two orthogonal sets of drift
chamber planes with cell widths of 10 cm, operated in a 50-50 argon
ethane gas mixture. The solid angle coverage is $\Omega/4\pi=0.72$.
The hadronic faking for tracks in the 1 GeV range is 1\% while the
muon detection efficiency is $\sim$ 30 \%.
\subsection{Luminosity Monitors}
CLEO uses 3 devices to measure the luminosity, the two shower
counters described above, and a of set two scintillation-lead shower
counter units (LM) to detect small angle bhabha scattering. Each unit
contains 4 sections, and are mounted at   small angles to the beam
trajectory. The device is triggered by hits in   any of
sections
  diametrically opposite from the beam spot. This device is
useful for measuring relative luminosity, since the  
orientation of the two units cannot be measured reliably enough to
match the rapidly varying Bhabha cross section at those angles. The
absolute
measurement of the luminosity comes from the two shower counters,
and is determined with a systematic error of 2\%.
\subsection{Experimental Control} 
\begin{figure}[p!]
\centering
\includegraphics[scale=0.7]{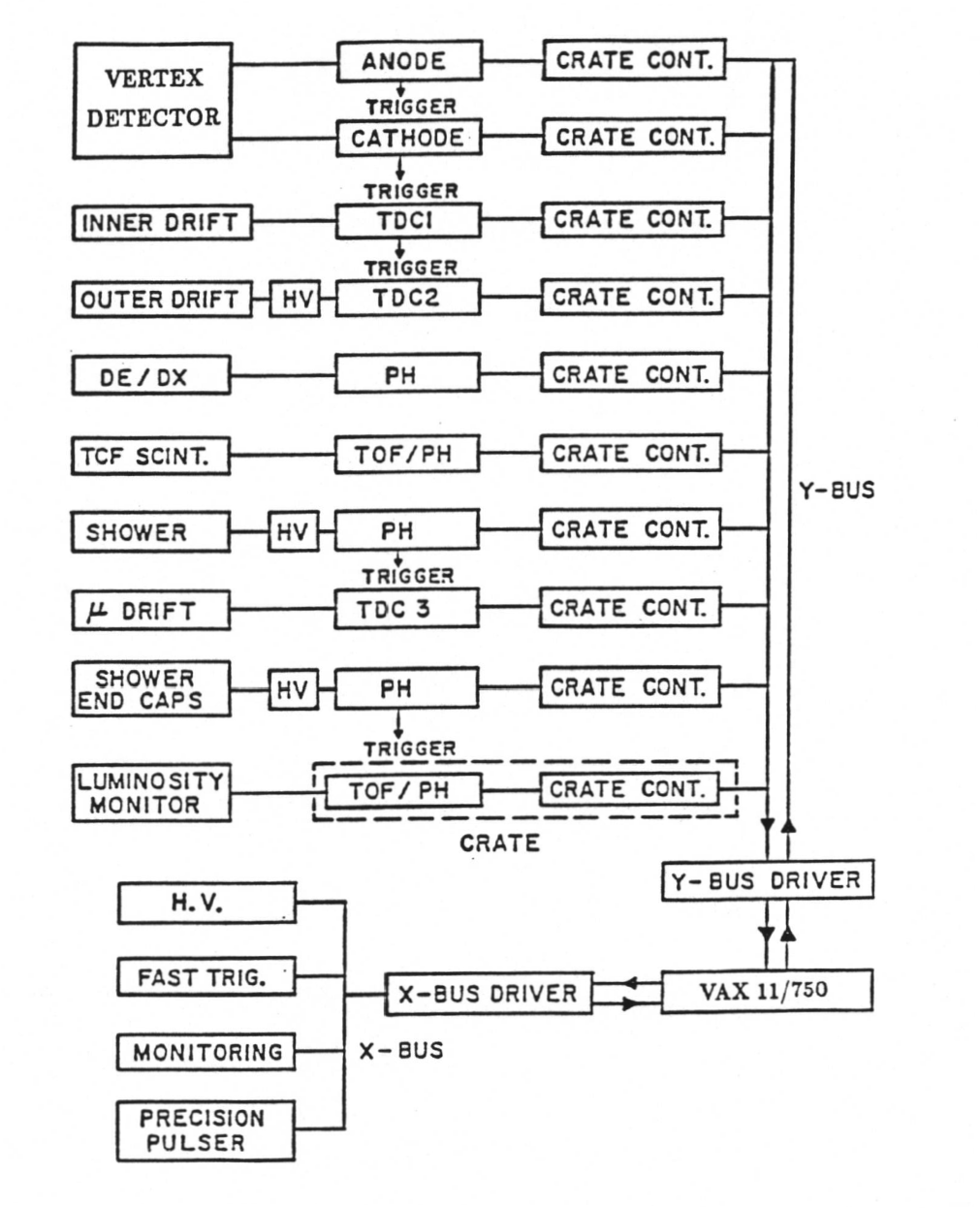}
\caption{CLEO control system.}
\label{fig:c3f14}
\end{figure}

Information from the
individual channels must be sampled,  organized, and stored in a
coherent fashion for each accepted collision. In the CLEO system each
associated detector circuit occurs redundantly in groups of 24-60 on
discrete data cards housed in water cooled ``crates." The crate
supplies power for the electronics, and contains a ``controller"
which directs the readout of the digitizing electronics. Data from
the channels is converted to a voltage and stored on buffered
capacitors which are digitized by the 12 bit ADC of the controller.
All crates are connected to a 16 bit data bus known as the y-bus,
which is interfaced to the crate via the controller. The y-bus was
driven by a VAX-11/750 computer during this run period. The 750
collects the data and writes  it to 6250 bpi magnetic tape.  An 8
bit data line (x-bus), also driven by the 750, is used to control
and monitor detector functions such as high voltage and calibration
pulsing. In all, nearly 80 crates comprise the data acquisition
system. Virtually all of the readout electronics and crate
controller system were designed by members of the CLEO
collaboration. A block diagram of the control system is shown in
Figure~\ref{fig:c3f14}.
\section{Data Selection} In the vast
majority of beam crossings no interesting interaction takes place. A
fast trigger must be able to sample the initial response of the
detector and decide to pursue the event before the next collision
takes place. Accepted events are recorded merely as addresses and
values, these must be converted into a format suitable for physics
analysis. This ``arduous" process is collectively referred to as
compress.
 Detectors must be calibrated and
highly complex algorithms sift through  the data to reconstruct 
the interactions of the decay products with the detector. At this
stage the data is analyzable, however since the trigger
requirements are fairly loose, some of the events at this stage are
still uninteresting. Selective event classification schemes are
activated to filter out the types of events desired for  particular
analysis needs. These facets of the analysis shall be examined here.
\subsection{Experimental Trigger}
The experimental trigger allows for reduction of the  1-3 MHz
crossing rate to a raw data recording rate of 1-2 Hz, with an
average live time fraction of 0.9. The detector elements which play
a pivotal role in the trigger system are the VD, DR1.5, TF and the
RS. To a minor extent MU and ES are also involved. Channels from the
various detector elements are ganged together to provide a course
overview of the detector response. Because of accelerator
development work that transpired concurrently with the data
acquisition program, two trigger modes were used in this data set.
They are equivalent from a physics standpoint, differing primarily
with the  speed in which the decisions had to be made. In 7 (3) bunch
mode the crossings occurred every 360 (854) nsec. 

In the tracking chambers, four wires are OR-ed together to form a 
fast-or bit. 
The patterns of these fast-or bits are correlated in a
track segment processor, which selects acceptable topologies.
In the vertex detector, the inner and outer five layers are each
segmented into blocks, where a block consists of  the fast-or bits of
 five vertically adjacent layers. The five layers in a block are
or-ed, the block being on if four of the fast-or bits are on. A
physical combination of an inner and outer block turns on a VD bit.
In the drift chamber blocks are formed from 3 vertically adjacent
axial layers, containing all the fast-or bits in an approximate
$24^{\circ}$ phi slice for each layer. A DR bit is set if 2 of the
3 layers in a block are hit. The nine axial DR layers contribute to
a total of 3 possible drift chamber bits. The VD and DR bits are
correlated in a loose road in $\phi$ to form track track segments.
A ``medium track" consists of a VD bit and the first two DR bits,
and a long track is defined by all four bits being on.
\par
Two signals are sent from each time of flight octant,  which are
the or's of groups of six scintillators which  share the same z
side of the octant. The TF bit pattern is analyzed for acceptable
patterns, such as TFNADJ (two TF hits in nonadjacent octants) or
BBTF (TF hits in diametrically opposite octant segments).
Trigger information is also examined from the shower detector.
Pulse heights are accumulated with an analog or of all 24 channels
on a data card, these are likewise summed on the crate level and
then on the octant level (two crates). The octant energy threshold
(OCT) was approximately 1 GeV. Due to the analog summation, the
input  to the trigger discriminator was  susceptible to noise. In
particular an odious glitch in the baseline of the discriminators
occurred after analog reset. This partially contributed to the loss
of 2 (1) crossings in 7 (3) bunch mode to allow for settling time.
Analog reset was fired every 19 (16) crossing during this running
period. This effect coupled with other engineering instabilities 
prompted a sweeping renovation of the energy trigger in the fall of
1986. This upgrade was developed by John Dobbins with assistance
from the author. Preamplifiers (gain 25) were placed at the octant
sum junction making the transmitted signals less susceptible to
noise. The boosted signals were differentiated with a time constant
of 560 nsec at the discriminator input to smooth out the $\rm
\overline{AR}$ glitch. The improvement system has proved to be
significantly more robust than its predecessor. Similar information
from the end cap shower detectors (ES) was supplied to the trigger.
\par
The trigger logic analyzes the input from the detectors in a two
tiered approach. The level 1 trigger is  enabled by a hit in the
TF,  OCT or ES lines. The singles rate with beam in the machine is
about 5 KhZ for the ES and OCT lines. Level 1 initiates a search
through all of the possible level 2 topologies, which are much more
restrictive. The  CLEO trigger logic
\begin{table}[ht!]
\centering
\caption{CLEO Trigger Logic}
\begin{tabular}{|c|c|c|c|c|c|}
\hline
Level & option1 & option2 & option3 & LOGIC & prescale \\ \hline
1 & TF & ES & OCT & OR & N/A \\ \hline
2& 1M2L & TFNADJ &  &  AND& N/A \\ \hline
2& 2M&  BBTF &  &  AND& N/ A \\ \hline
2& 1M1L &  1 + TF & 1 + OCT  &  AND& N/A \\ \hline
2& OCTOPP &   &  &  AND& N/A \\ \hline
2&  2ES &   &  &  AND&  1/64 \\ \hline
2&  MU2 & 1 + TF  &  &  AND&  1/64 (16/64) \\ \hline
\end{tabular}
\label{t:3p3}
\end{table}
 implemented during this data taking is organized
 in Table~\ref{t:3p3}.
 The level 2 lines  can be physically
grouped into a hadronic element which fires the tracking devices
and some of the outer detectors (to avoid beam-wall, beam gas
triggers) and a two track trigger for QED processes. Because of
higher rates, some lines are prescaled.  Only 1 of 64  2ES 
triggers in accepted, while the MU2 1 + TF line is accepted 1/64 th
 (16/64) of the time in 7 (3) bunch mode. Each level 2 search
 takes roughly 2.5 $\mu$sec, and a successful $\rm L1\cdot L2$
trigger initiates a digitization and data readout sequence lasting
20 msec.
\subsection{Compress}
\begin{figure}[p!]
\centering
\includegraphics[scale=0.7]{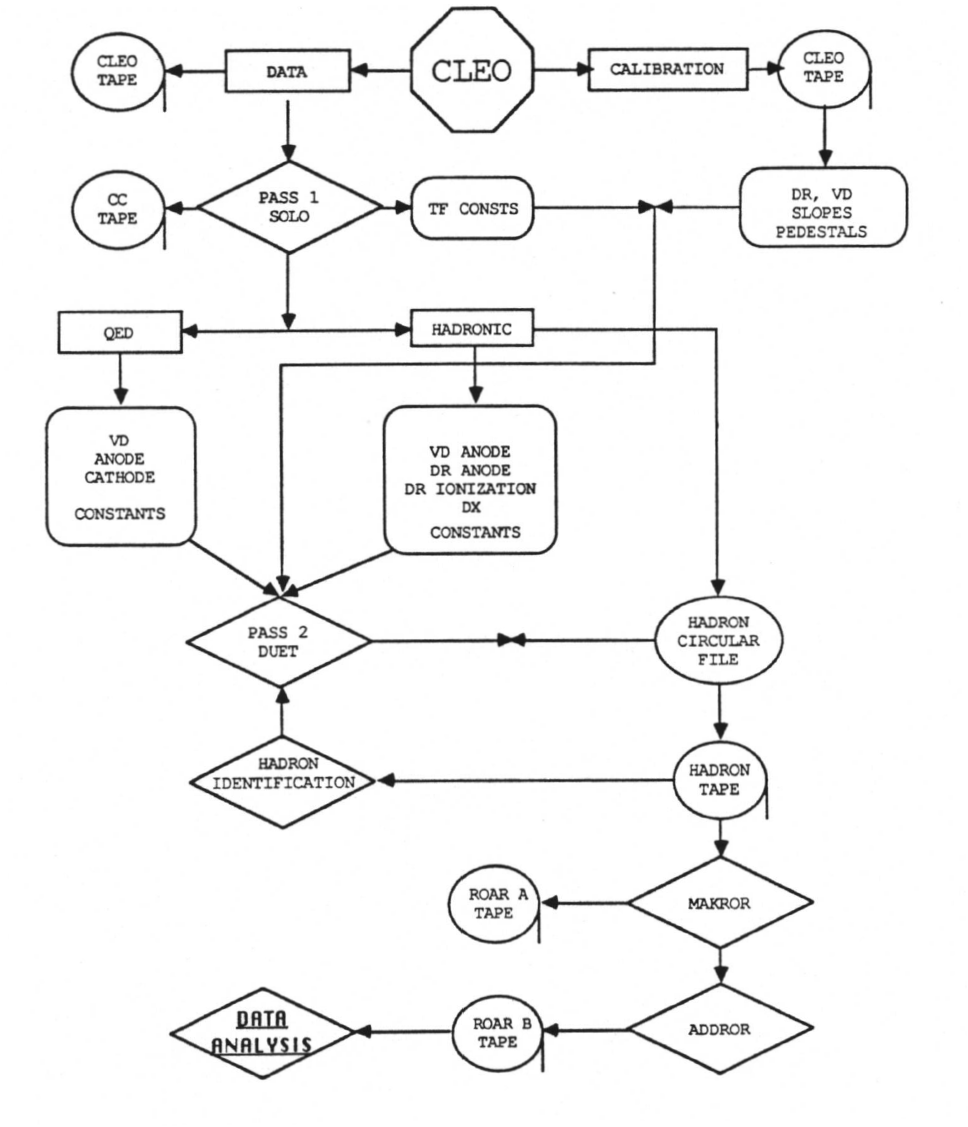}
\caption{CLEO compress data flow for hadronic type events. }
\label{fig:c3f15}
\end{figure}

The process of turning the raw data into analyzed events is a
complex, iterative, feedback system. A diagram representing the
compress system used to analyze this data set is shown in Figure~\ref{fig:c3f15}.
Data runs are taken in two modes, physics running and calibration.
As seen from the diagram, the full calibration of the detector uses
both calibration and physics data. Calibration runs for the drift
chamber were taken every 2 to 3 days, and consisted of pulsing the
chamber to determine the pedestals and gains for the timing channels
and the pedestals for the pulse height channels.

 As the physics data was gathered, it was subjected to a ``first
pass" track finding algorithm (SOLO \cite{solo})
to find seed tracks. Those
events which were judged interesting by the track finder were then
segregated into QED and hadronic event candidates.  For
the drift chamber,  information from hadronic events at this stage
was used to calibrate out global changes (such as machine timing) and
systematic shifts in the electronics or performance of the drift
chamber (non standard operating conditions). After completion of this
stage all the devices have been fully tuned, and the final analysis
algorithms are run. A second, more rigorous, track finding
algorithm (DUET \cite{duet})
uses the seed
tracks as initial conditions and systematically searches the data for
additional tracks. At this stage information from the vertex
detector is applied, and the intervening material between the vertex
detector and drift chamber is compensated for by allowing a ``kink"
in the track. \par
Those fully analyzed events which are classified as hadronic are
compacted into a format called ROAR \cite{avery2}.
This format strips the event down to the minimal amount of
information necessary for analysis, and computes and stores a
variety of invariant mass combinations for quick analysis. The
\USSSS\ data sample of 113.7 $\rm pb^{-1}$ produced 113 2400 foot,
6250 bpi tapes of hadronic events. After being ROARed, the set fit
on to five such tapes and could be analyzed in 1-2 VAX 8600 cpu
hours.

 \subsection{Hadronic Event Selection}
All events used in this analysis have been classified as hadronic
in nature. The primary selection  criteria evolve from charged
particle tracking in the sequence single tracks, event vertex,
and event shape. To eliminate beam gas, beam wall, cosmic ray
showers, and radiative bhabhas,the energy response of the
octant shower detectors is also employed. \par
To begin, the charged multiplicity of the event is examined. A
charged track is considered good if no more than one following
tests is failed:  
\begin{enumerate}
\item
 The z value at the distance of closest approach to the
beam line (DOCA) is less than 50 mm from the origin. 
 \item
The track's DOCA  with respect to the nominal beam spot in the
r-$\phi$ plane is less than 5 mm.
 \item
 The average residual from the 5 parameter helical fit must
be less than 700 microns. 
\item
The track is defined using at least 8 drift chamber layers. 
\end{enumerate}

 The next level of selection is a crude classification of a
hadronic event:
\begin{enumerate}
\item
Tracking.
\begin{enumerate}
\item
At least 3 good charged tracks, which may come from the primary or a
secondary vertex, excluding those consistent with originating from a
converted photon.
\item
At a least one of the tracks must originate from the primary vertex
with a DOCA less than 5 mm.
\end{enumerate}
\item
Vertex.
\begin{enumerate}
\item
The x and y position of the  vertex must be within 20 mm of the
nominal x-y beam position.
\item
The z position of the vertex must be within 50 mm of the origin.
\end{enumerate}
\item
Energy.
\begin{enumerate}
\item
The sum of the energies of the charged and neutral tracks in the
event must exceed 15 \% of the center-of-mass energy.
\end{enumerate}
\end{enumerate}
All events that have survived to this stage are finally subjected
to a stringent, well developed set of cuts known as
morcut \cite{fhm}
\begin{enumerate}
\item
Tracking.
\begin{enumerate}
\item
At least 15 \% of the hits in the drift chamber must be associated
with tracks.
\item
The fraction of bad tracks to good tracks must be less than 1.15
\end{enumerate}
\item
Energy.
\begin{enumerate}

\item
The total energy of the charged and neutral tracks in the event must
exceed 30\% of the center-of-mass energy.
 \item 
The energy measured in the octant shower counters must exceed 250
MeV.  
\end{enumerate}
 \item 
Topology.
\begin{enumerate}

\item
The event must not be consistent with a radiative bhabha.
\item
The event must not be classified as a beam wall collision.
\item
The Fox-Wolfram \cite{foxwo}
parameter $H_1/H_0$ which
measures momentum imbalance must be less than 0.4.
\item
The Fox-Wolfram parameter $H_2/H_0$ which measures event shape
( 0 = spherical, 1 = back to back 2 track event) must be less than
0.98.
\end{enumerate}
\end{enumerate}
The non hadronic background for events which have passed morcut is
only a few percent, while a Monte Carlo simulation of D decays
shows that 98 percent of the events that are successfully
reconstructed pass morcut. 
\chapter{Technics}

After the data has been collected and  processed, additional
``second order" analysis systems need to be developed to fully
exploit the underlying physics. Here we detail   the other
relevant tools essential to this analysis. They are; reconstruction
of secondary vertices, charged particle momentum correction,
hadron identification, and Monte Carlo simulation. Also presented
is a discussion of  particle decay kinematics,  which forms the
basis for a number of analysis procedures used in the following
chapters. To conclude, a presentation of the analysis architecture
used to derive the results of this thesis shall be crystallized.
\section{Reconstruction of   Secondary Vertices}
As discussed in chapter 1, the tremendous diversity in the strength
of the four forces causes a difference in the decay times
associated with each force  of several orders of magnitude.  At
CESR energies, strong and electromagnetic processes evolve at such a
rate that they cannot be distinguished from the primary vertex.
Particles that decay weakly, in contrast, can move from order 100
microns (charm, \T\ decays)  to  tens of millimeters (\KSH, \LA\
decays). At CLEO, these particles are  reconstructed
from their decays into purely charged modes.
  For the first class of decays, the decay length is in the
realm of the extrapolation error of the individual tracks. Evidence
for these vertices comes from the observation of systematic
offsets from zero in distributions of extrapolated decay lengths and
impact parameters. The vertices of the second group are often
clearly visible with decent event display graphics.  Decays of
\KSH's and \LA's have been historically referred to as ``vees"
because their 2 prong decay topology resembles the letter v.

\begin{figure}[htp!]
\centering
\includegraphics[scale=0.56]{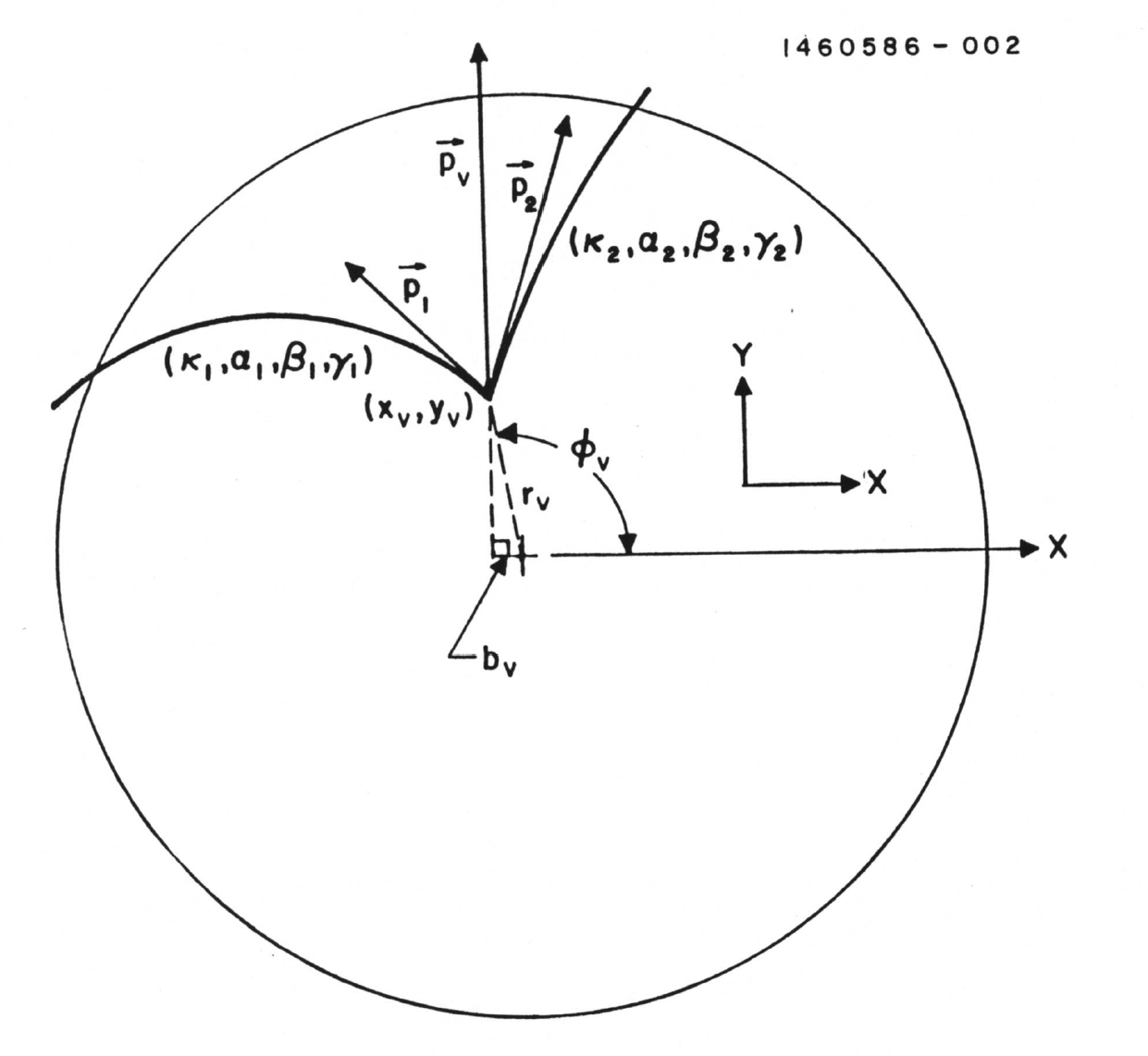}
\caption{The r-$\phi$
intersection of two helical tracks. The intersection occurs at the
point $\rm (x_V,y_V)$ at radius $r_V$. The reconstructed vector has
impact parameter $b_V${.}}
\label{fig:c4f1}
\end{figure}

Since the track finding algorithm performs searches for individual
tracks, an  additional  program is required to isolate vees.
The vee finder examines all  good tracks  that have a full 3-D
reconstruction, with the two fold mission of finding pairs of tracks
which come from a vertex distinct from the common event vertex, and
properly determining the momentum of the original neutral particle.
Vectors of the two daughter particles must be re-evaluated, since
they were determined at the point of closest approach to the drift
chamber origin. The algorithm used in this analysis was developed by
M. Ogg and M. Mestayer. The search is initiated by  
parametrizing the tracks as   circles in the r-$\phi$ plane. This
is a preferred starting point since the r-$\phi$ tracking
resolution is about a factor of five better than that in z.  Vee
candidates are formed from pairs of oppositely charged tracks 
 where the sum of the absolute values of the track  DOCAs   exceeds
1mm.  The two circles are tested for intersection, and only those
candidates with 2 intersection points are considered. If either
solution is consistent with originating from the primary vertex
(z of the vertex less than 3 cm from the origin, and radius of the
vertex in the r-$\phi$ plane  with respect to the average beam
spot $\rm (r_v)$ less than 0.3 cm ) the candidate is rejected. Each
solution is required to have $\rm r_v$ in the range of 0.5 cm to 50
cm, and the reconstructed vee vector must point away from the
origin, as determined by  requiring the normalized  dot product of
the vee vector and the vector drawn from the origin to the secondary
vertex be positive. The selection procedure is completed with
the definition of a vee quality factor:
$$\chi^2_{_V} ={ \left( { {\Delta z}\over{{\sigma}_z}
}\right)}^2
 + { \left( { { b_V}\over{ {\sigma}_b} }\right)}^2 $$ where the 
parameter $\Delta z$ is the z difference of the two candidate
charged tracks at their r-$\phi$ intersection point, and $b_V$
represents the impact parameter of the fit secondary vertex with
respect to the run by run average x-y position of the beam spot. 
The nominal rms errors ${\sigma}_z$ and ${\sigma}_b$ were chosen 
from Monte Carlo studies to be 10 mm and 2 mm, respectively.
Application of a cut on $\chi^2_{_V}$ improves the purity of the
\KSH\ sample by rejecting poorly determined and incorrect  vertices.
Vee candidates are subject to a loose cut of $ \chi^2_{_V} \leq
12.0$ in order to provide a maximal vee base available to the user.
In practice, the author uses $\chi^2_{_V}$ cuts of  2-3 for \KSH's.
Should both solutions pass all cuts, the solution with smaller
$\Delta z$ is selected. The variables involved in the vee finder as
discussed above are visually presented in Figure \ref{fig:c4f1}.
Once the position of the secondary vertex has been determined, the   momenta of the two daughter tracks are
redetermined with respect to the new vertex, and added to form the
vee  three vector. 
\par Two measures of the robustness of the vee
finder are the reconstruction efficiency and FWHM (width) of the vee
invariant mass peak. We examine the performance with respect to the
decay \KSH\ \decays\ \PIP\PIM, which is vital to this analysis.
Details on \LA\ reconstruction can be found
elsewhere \cite{skip}.
\begin{figure}[htp!]
\centering
\includegraphics[scale=0.6]{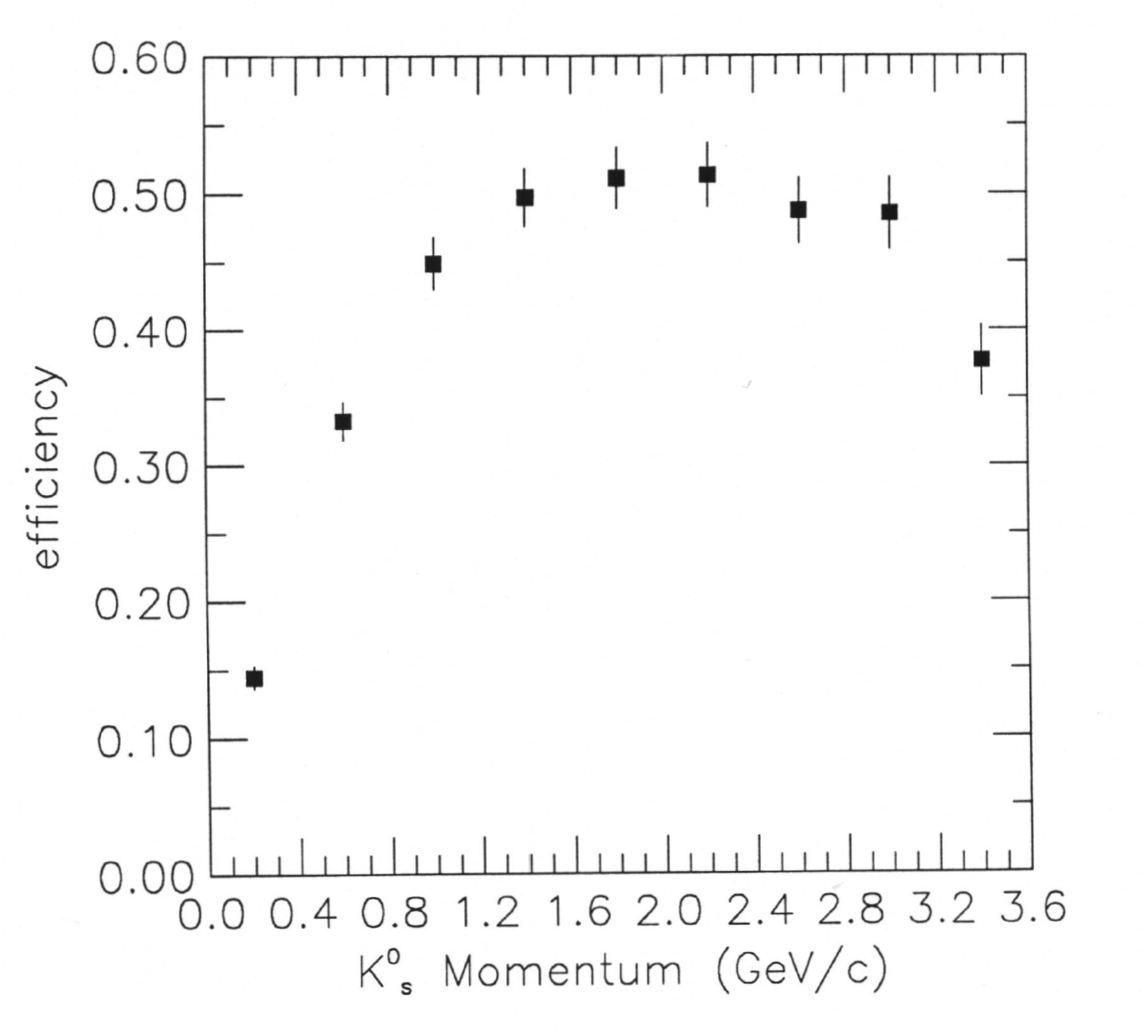}
\caption{Efficiency for reconstructing the
decay \KSH\ \decays\ \PIP\PIM\ in $\rm c \bar c$ events versus \KSH\
momentum.}
\label{fig:c4f2}
\end{figure}
Figure~\ref{fig:c4f2} contains a plot of the efficiency for reconstructing \KSH\
\decays\ \PIP\PIM\ versus \KSH\ momentum, as determined form Monte
Carlo simulation. The efficiency folds in hadronic event selection,
single particle tracking, and secondary vertex finding. The
efficiency is seen to plateau at $\approx 50 \%$ from 1 to 3 GeV.
 The momentum dependence of the mean and width of the \KSH\
observed in the data are shown in Figure~\ref{fig:c4f3}
\begin{figure}[htp!]
\centering
\includegraphics[scale=0.525]{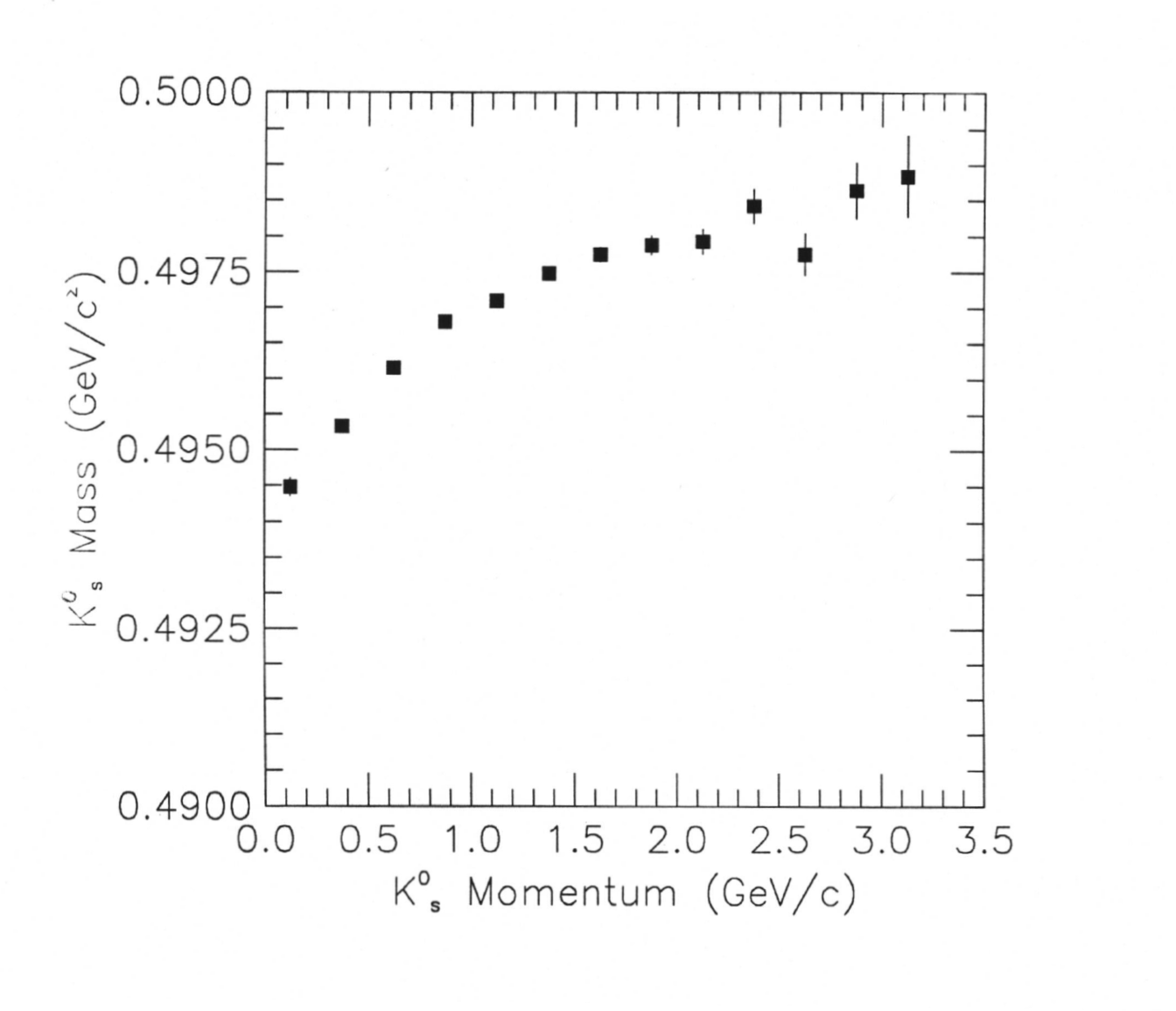}
\caption{Observed mean of the \KSH\
mass peak versus \KSH\ momentum.}
\label{fig:c4f3}
\end{figure}
and  Figure~\ref{fig:c4f4}.
\begin{figure}[htp!]
\centering
\includegraphics[scale=0.55]{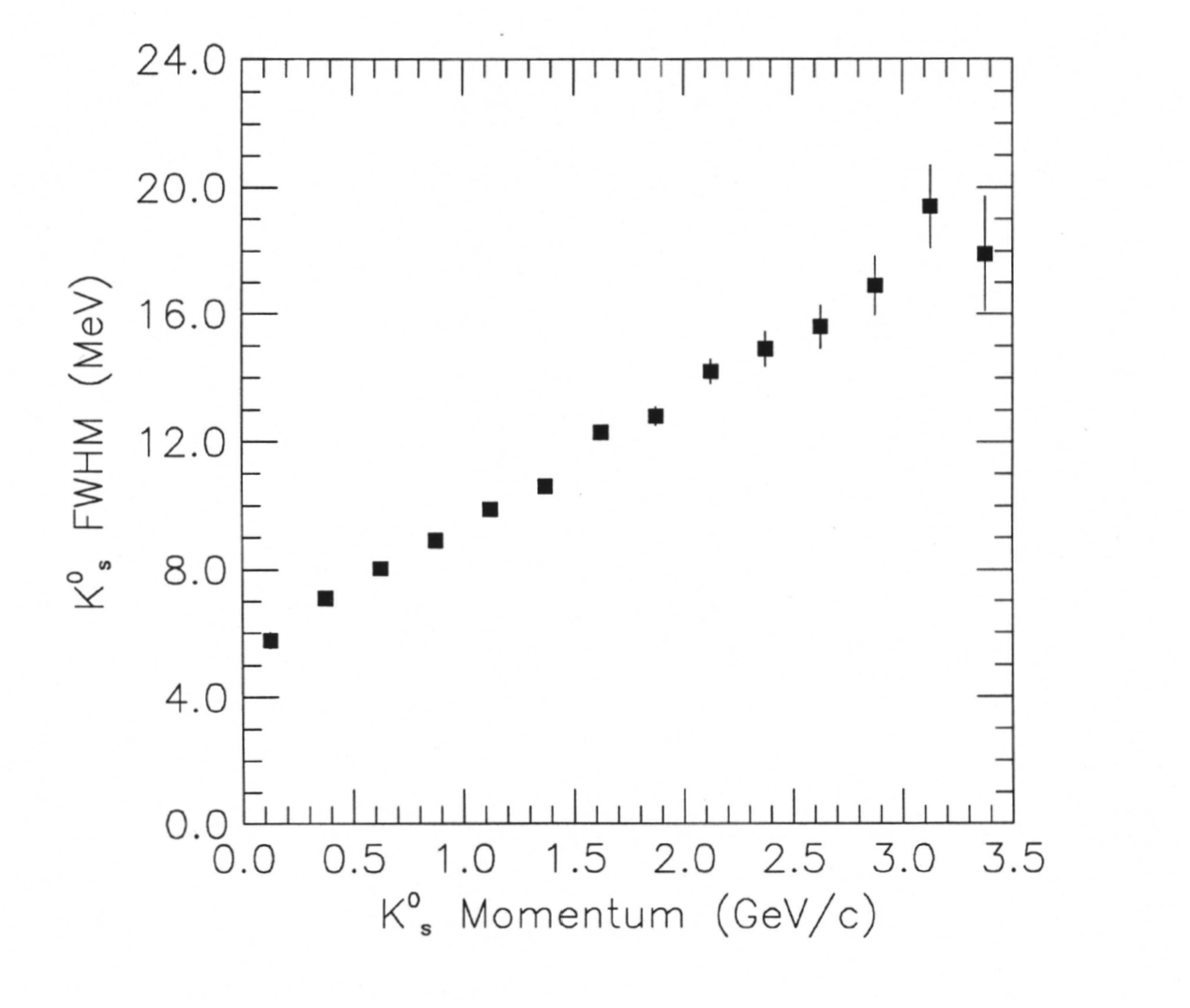}
\caption{Observed FWHM of the \KSH\
mass peak versus \KSH\ momentum.}
\label{fig:c4f4}
\end{figure}
The invariant mass
spectrum  is fit to a Gaussian  signal and
a polynomial background. A
systematic deviation from the known \KSH\ mass versus momentum is
prominent at at low \KSH\ momentum. This is due to energy loss of
the \KSH\ daughters. The correction for this effect will be
presented in the next section. The width  obeys an approximately
linear relationship as a function of momentum. Convolving this with
the observed \KSH\ momentum spectrum yields an average value for the
width of the \KSH\ mass peak of 9 MeV.  The behavior of both the
mean and width are in agreement with Monte Carlo simulation of this
decay in the CLEO detector.

\section{Corrections to Charged Particle
Momenta} 

\begin{figure}[htp!]
\centering
\includegraphics[scale=0.55]{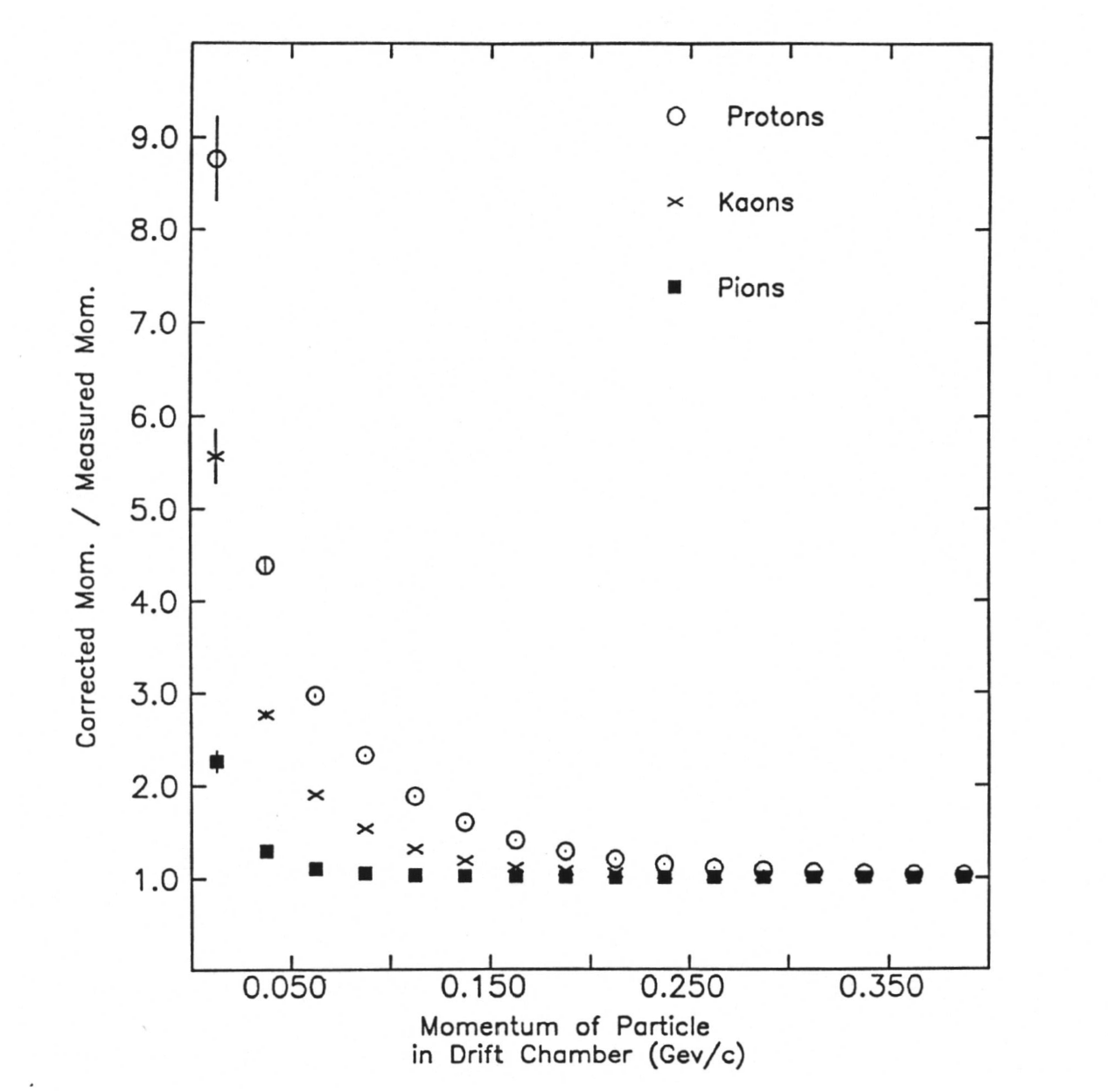}
\caption{ Average energy loss corrections for
tracks as a function of momentum and mass hypothesis. }
\label{fig:c4f5}
\end{figure}

Tracks in the vertex detector appear largely as straight line
segments, making track curvature (hence momentum) measurements the
sole responsibility of the drift chamber. Energy lost in
interactions with material before the drift chamber will manifest
itself in systematically lower invariant masses for reconstructed
particles.
\begin{table}[ht!]
\centering
\caption{Material Preceding the Drift Chamber Detector}
\begin{tabular}{|c|c|c|}
\hline
outer radius  (cm) & Material & dE/dx (MeV)  \\ \hline
7.57 & silver coated beryllium & 0.246 \\ \hline
8.02 & Carbon filament tube & 0.322  \\ \hline
16.44 & VD gas + Carbon filament tube & 0.208   \\ \hline
17.44 & Carbon filament tube & 0.270 \\ \hline
\end{tabular}
\label{t:4p1}
\end{table}
The material preceding the drift chamber is collected in Table~\ref{t:4p1}.
The material
corresponds to the beam pipe, inner and outer VD supports,  and drift
chamber inner tube, respectively. As described earlier, we choose the
Bethe-Bloch form $\rm  {dE \over dx} = { A \over \sin\Theta} \cdot
{1 \over \beta^2} \cdot \Bigl [ \ln(2m_e \beta^2 \gamma^2 / I_0)
-\beta^2 \Bigr]$ to model the expected energy loss. The algorithm
used was developed by P Avery \cite{avery3}.
The average momentum loss $\rm {dp\over dx} =
{1\over \beta}\cdot{dE \over dx}$ is calculated   based  on the
track's momentum and cell entrance angle,  for   several
different mass hypotheses.  The momentum loss is added to the
measured track in such a way that only the magnitude, not the
direction is altered. The average corrections for pions, kaons, and
protons are shown in Figure \ref{fig:c4f5}.
The net
result when calculating the invariant mass of several charged tracks
is to shift the peak of the mass distribution upward 1-2 MeV, with
the FWHM of the distribution largely unaffected. Monte Carlo
simulations mirror the effects of the correction observed in the
data. 
 \par
\begin{figure}[htp!]
\centering
\includegraphics[scale=0.575]{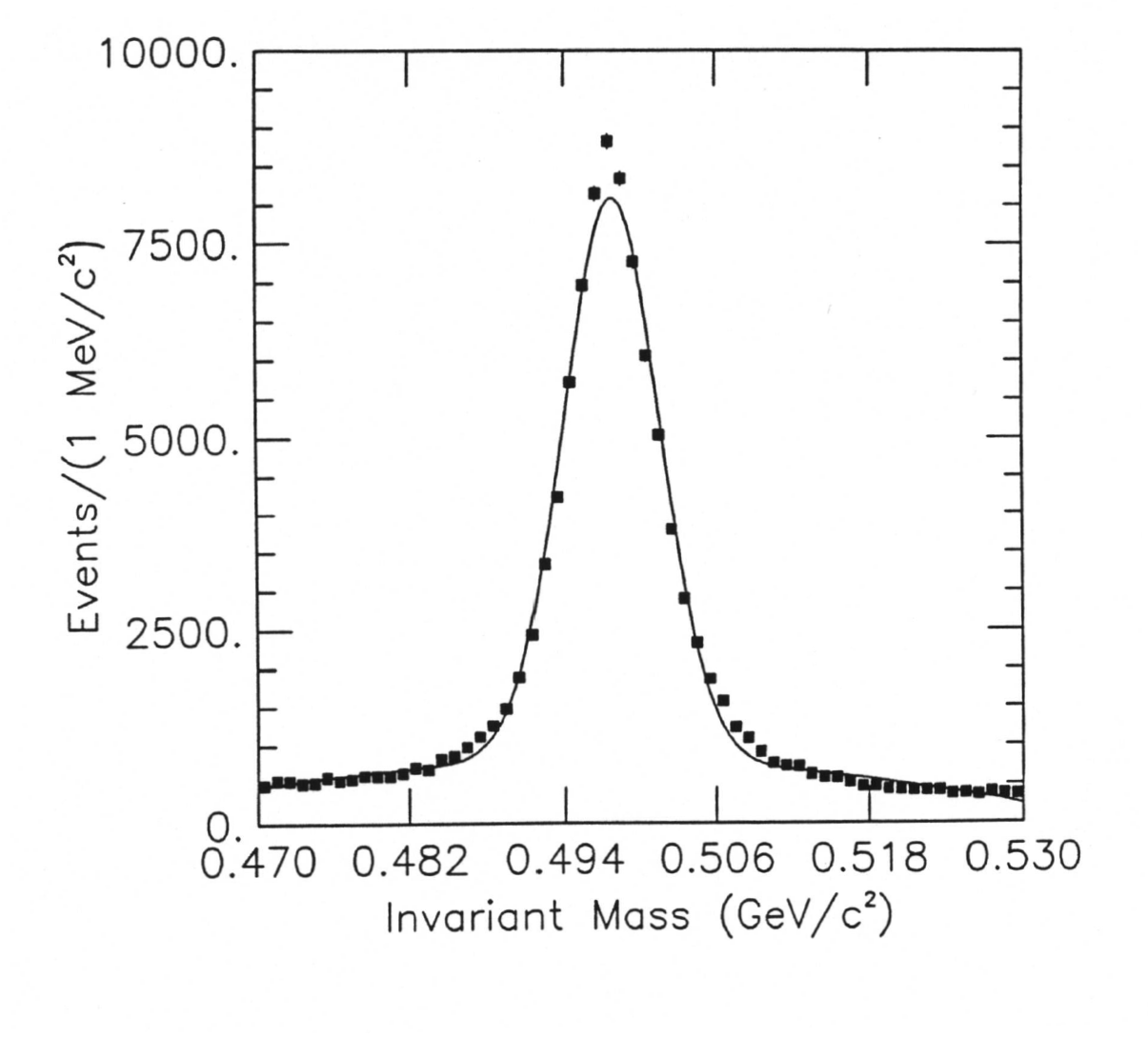}
\caption{Invariant mass of \KSH\ candidates with $ \chi^2_{_V} \leq 3.0$ from data.}
\label{fig:c4f6}
\end{figure}
\begin{figure}[htp!]
\centering
\includegraphics[scale=0.575]{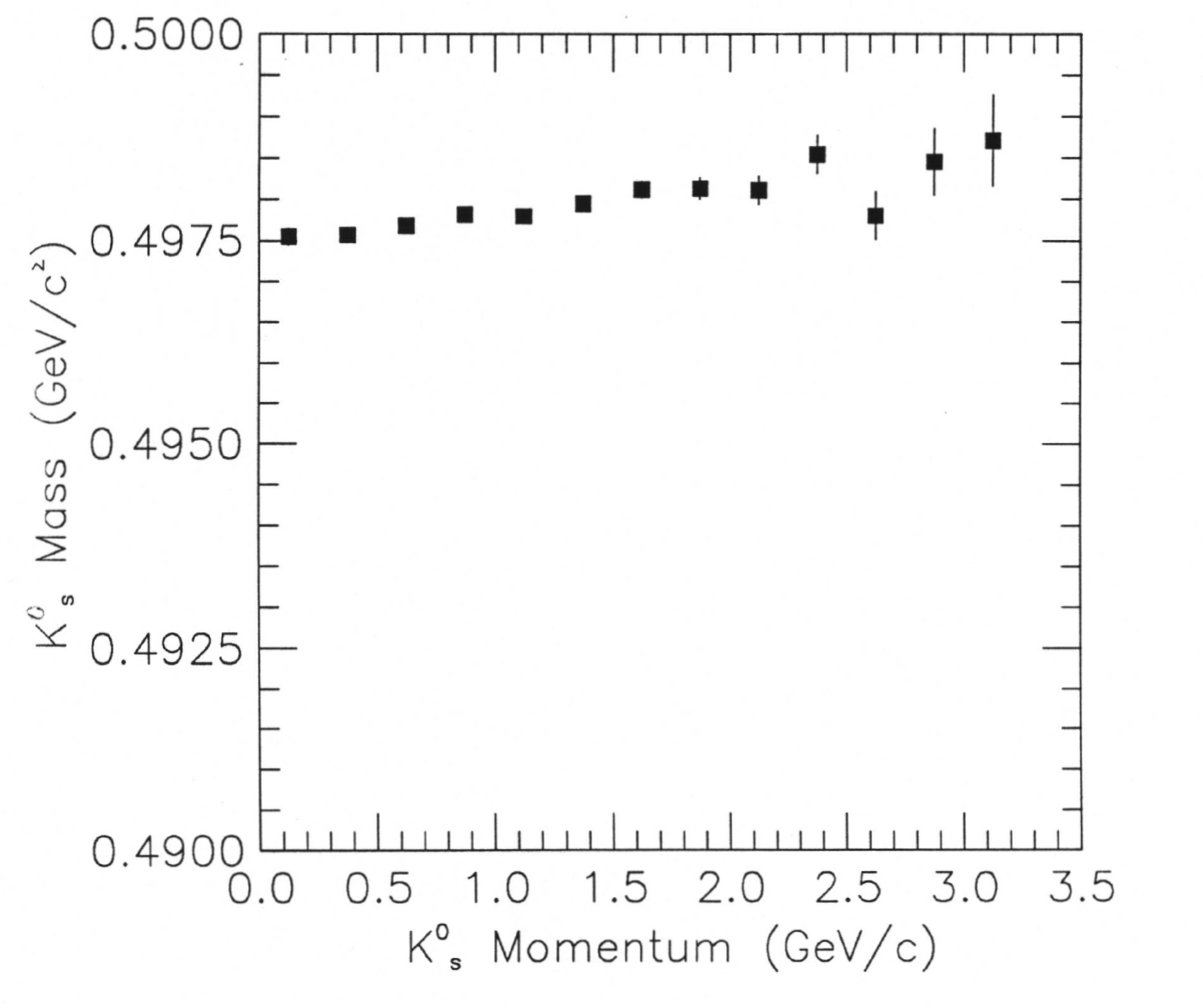}
\caption{The \KSH\ mass as a function of momentum, after
application of the energy loss correction.}
\label{fig:c4f7}
\end{figure}

We can now re-address the properties of \KSH's after energy loss
corrections are applied . The observed \KSH\ mass ( Figure~\ref{fig:c4f6})
 is found to be
$497.8 \pm 0.1$ MeV, in good agreement with the world average of
$497.72 \pm 0.07$ MeV. The \KSH\ mass as a function of momentum
(Figure~\ref{fig:c4f7})
is now centered about the correct mass to within a few tenths of
an MeV over the  entire momentum range.  
\section{Hadron Identification}
 Hadron (\PI, \K, \PR) identification is accomplished by coherently
combining information from the three detector elements with this
capability; DR1.5, dE/dX, and TF. 
Complete details of the hadron identification system can be found
in \cite{skip}.
The hadron identification   algorithm is
unsophisticated in nature yet powerful in performance.
 For each
device, a ``probability" is calculated for each of the three mass
hypothesis $\rm p_{HYP i}$ ( 1 = \PI, 2 = \K, 3 = \PR) defined by:
$$ \rm p_{HYP i} = \exp^{- {1 \over2}\left( { M_{DEV} - E_{HYPi}}
\over {  \sigma_{HYPi} } \right)^{\Large 2}}$$
where $\rm M_{DEV}$ are the measured values (mean of lowest 50\%
pulse height for DR1.5 and  dE/dx, time for the TF), $\rm E_{HYPi},
\sigma_{HYPi}$ are the expected measurement and resolution for each
device, respectively.  The distributions of $\rm M_{DEV}$ for each of the three devices
\begin{figure}[htp!]
\centering
\includegraphics[scale=0.525]{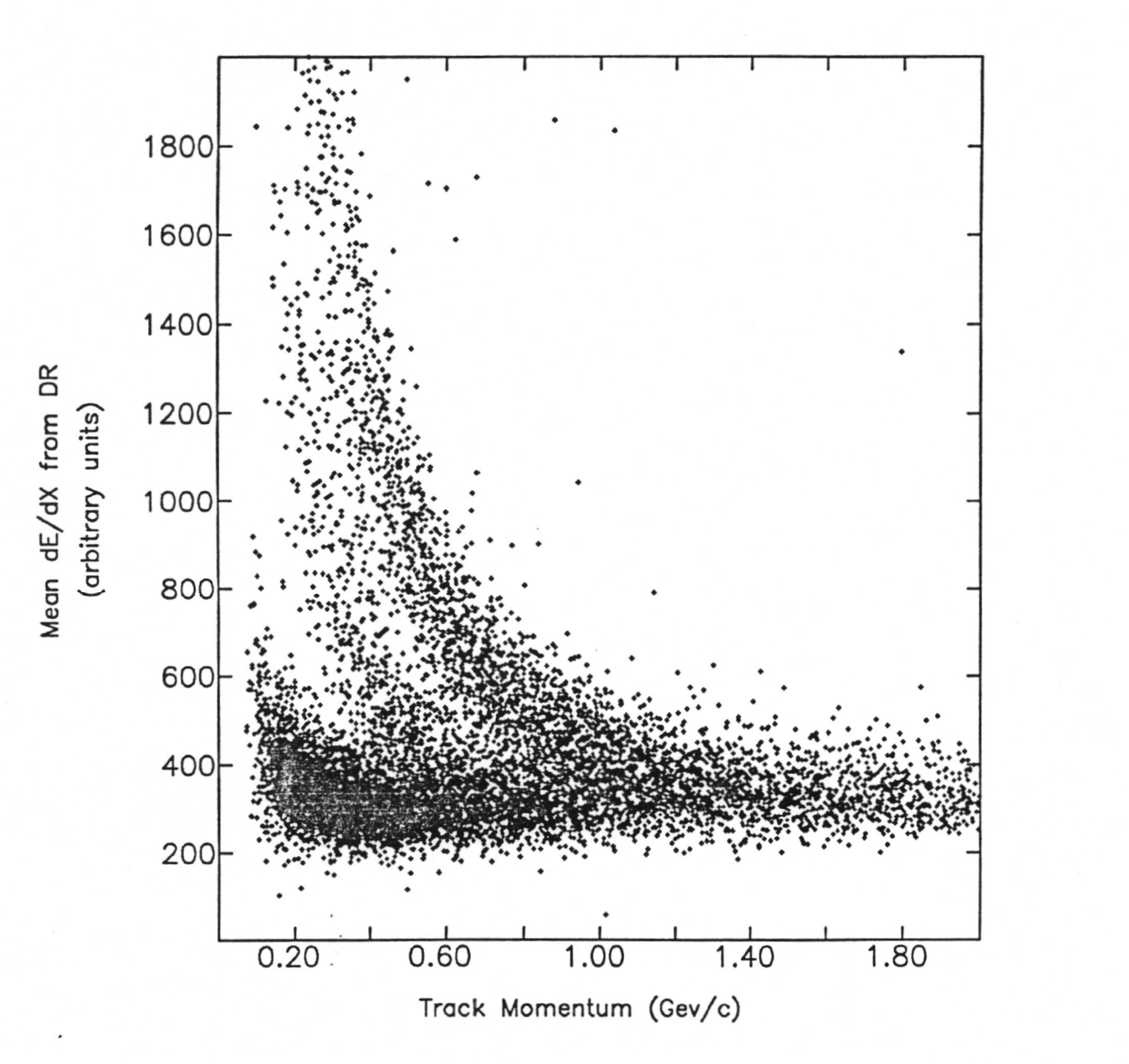}
\caption{Mean of the lowest 50\% of the pulse heights
from the drift chamber detector versus track momentum.}
\label{fig:c4f8}
\end{figure}
\begin{figure}[htp!]
\centering
\includegraphics[scale=0.525]{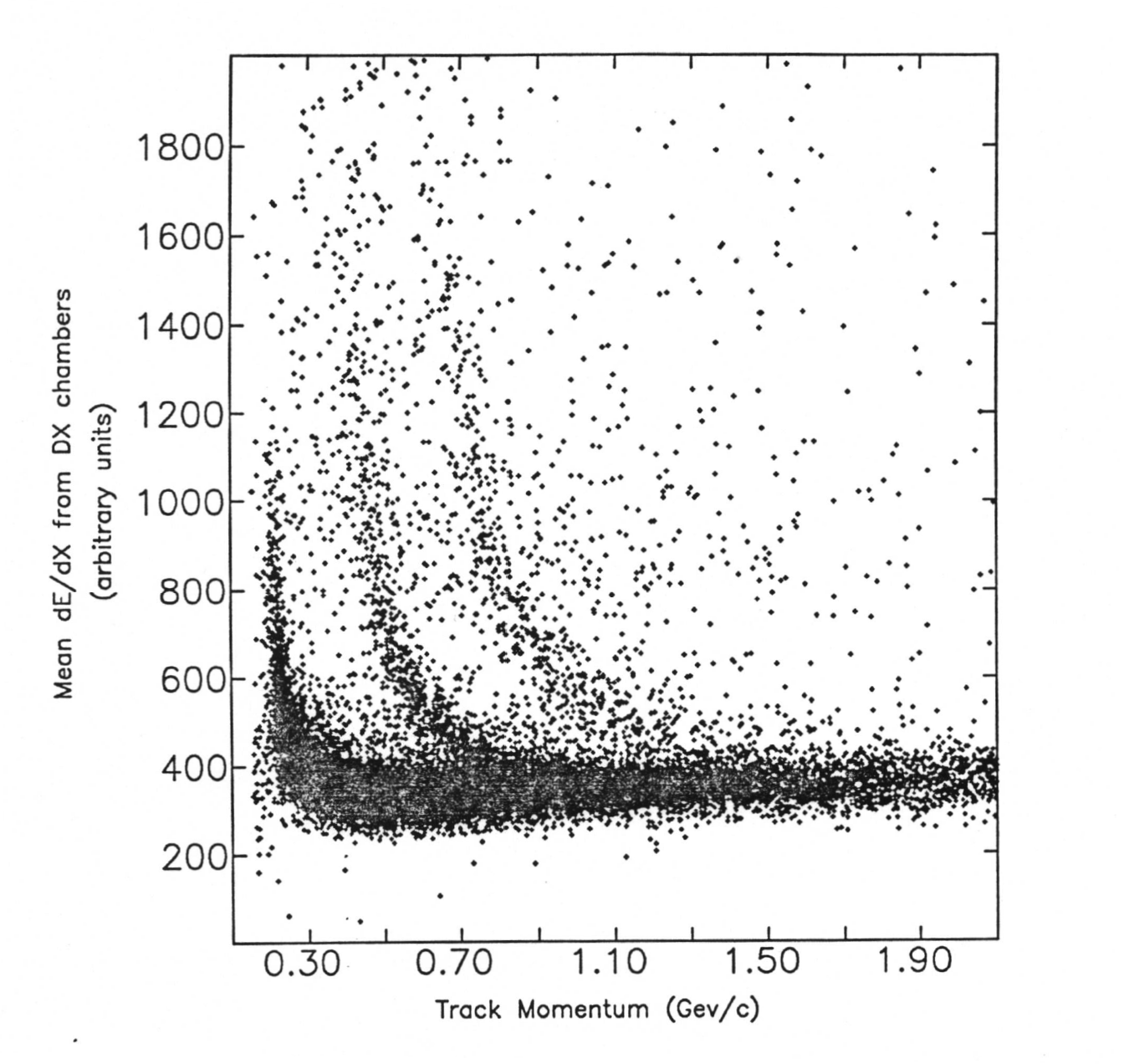}
\caption{Mean of lowest 50\% of the pulse heights from the dE/dx
chamber detector versus track momentum.}
\label{fig:c4f9}
\end{figure}
\begin{figure}[htp!]
\centering
\includegraphics[scale=0.575]{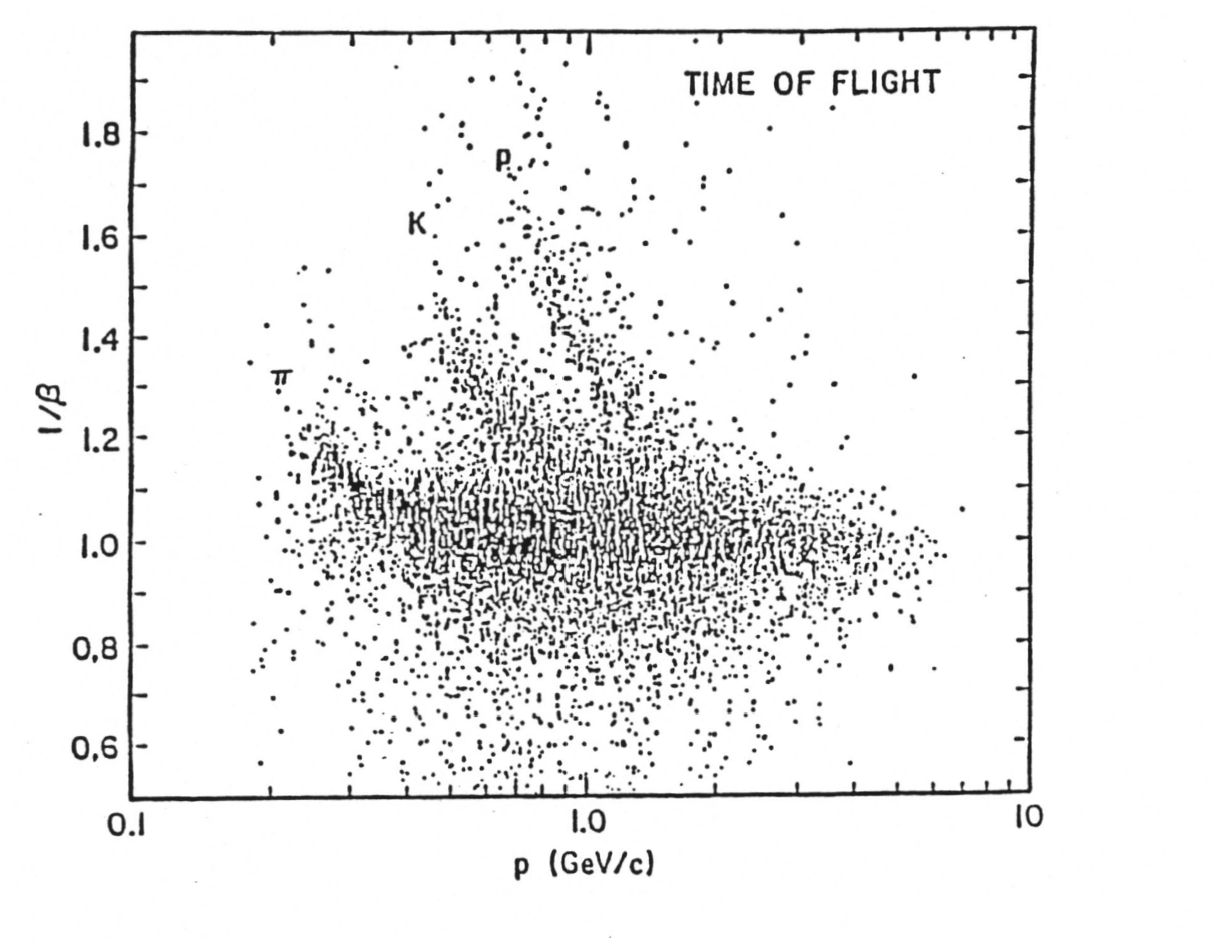}
\caption{Measured velocity in the Time of Flight detector
versus track momentum measured in the drift chamber.}
\label{fig:c4f10}
\end{figure}
(Figures \ref{fig:c4f8} - \ref{fig:c4f10}) all show distinctive \PI, \K, and \PR\ bands.
$\rm E_{HYPi} \ and \ \sigma_{HYPi}$ are the expected detector
responses for the mean and width of three species, as determined from
non-trivial measurements from the data. Since dE/dx and the TF
detectors are separated form the drift chamber by the
superconducting coil, these devices can only be used if  successful
matches are made to drift chamber tracks. The overall hypothesis is
calculated from the product of the  probabilities for each functional
device: $$\rm  P_{HYPi} = \prod p_{HYPi}$$
$\rm P_{HYPi}$ is set to 1 for each hypothesis if there was no
information available. To discern among the species, we define a
normalized weight:
$$\rm W_{HYPj} = { P_{HYPj} \over  {\sum_{i=1}^3 P_{HYPi}} }$$
The normalized weight is set to zero if there was no information
available $ \left( \rm \sum_{i=1}^3   P_{HYPi} =3 \right) $  or the
information available favored none of the three hypotheses $\rm
\left( \sum_{i=1}^3 
  P_{HYPi} \leq 0.001 \right) $
\par
The utility of this approach is pragmatically illustrated for the
case of \DP\ \decays\ \KM\PIP\PIP. The \DP\ candidates have
momentum greater than 2.5 GeV, with the momentum of pions greater
than 0.3 GeV.
\begin{figure}[p!]
\centering
\includegraphics[scale=0.6]{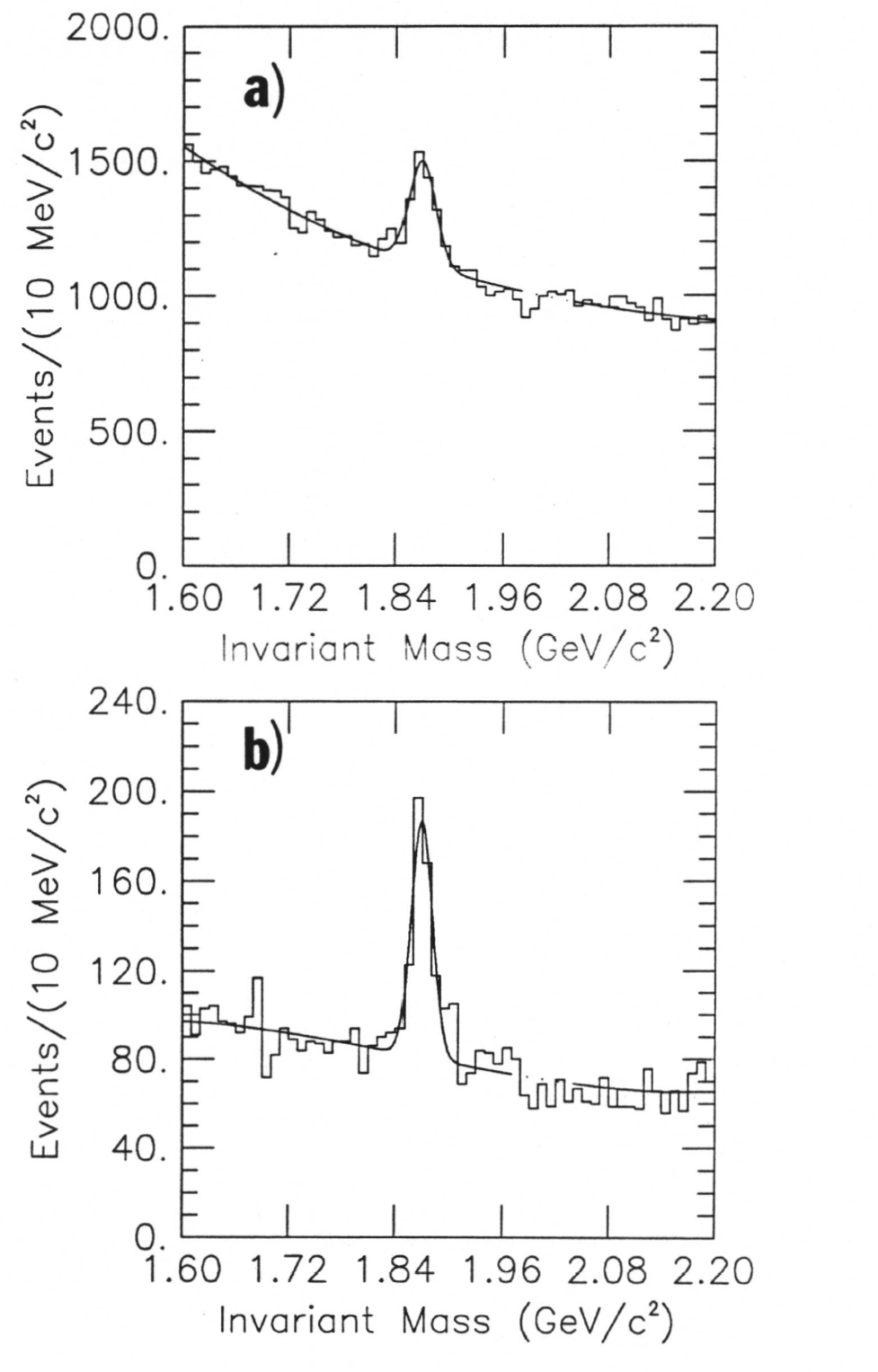}
\caption{
\DP\ \decays\ \KM\PIP\PIP\ candidates with momentum greater than 2.5 GeV. a) $ W_{K} \geq 0.1  $. b) $\rm W_{K} \geq
0.7$ and  $\rm W_{\pi} \geq 0.2$. }
\label{fig:c4f11}
\end{figure}

Figure~\ref{fig:c4f11} a) shows the invariant mass plot where the kaon 
candidate has $\rm W_{\K} \geq 0.1$, while in b) the kaon
candidate is tightly identified $\rm W_{\K} \geq 0.7$ and the
pion candidates are loosely identified$\rm W_{\PI} \geq 0.2$. The
reduction in the background is startling. Obtaining a 
\DP\ \decays\ \KM\PIP\PIP\ mass peak with a signal to noise of $1:1$
was a significant experimental achievement. This has allowed CLEO to
measure the \DP\ lifetime with the world's second largest \DP\
sample.
\section{Monte Carlo Simulation} In order to complete
physics measurements, computer modeling of the detector and
 the 10 GeV \EPEM environment is required. For this analysis, Monte
Carlo methods are used to determine the detector acceptance for
specific \D\ decay modes, along with backgrounds from \DF\ decays. We
simulate the  byproducts of \EPEM\ annihilations using a
modified \cite{art}
version of the LUND \cite{lund}
QQJET generator. For this analysis
all Monte Carlo events were generated using $\rm c\bar c$ jets,
including gluon radiation. The center-of-mass energy was defined to
be 10.56 GeV, approximately the average run energy of this data set.
Effects of initial state radiation of the beams were not
compensated.
  The strategy for studying a particular decay mode was to force
the particle (\DZ, \DP) to decay into the mode in question, while
the anti-particle (\DZB, \DM) was allowed to decay freely. Each $\rm
c\bar c$ event thrown was selected for further analysis only if it
contained the particle under study. Since much of this analysis
involves neutral kaons, when the \D\ meson decayed it was forced
only to decay to a \KZ\ or \KZB, the event being selected only if
the final state \KSH\ \decays\ \PIP\PIM\ was thrown. In this way the
Monte Carlo was free from bias from either an excessive number of
\KSH's or charged pions. All efficiencies are calculated with
respect to the final state \KSH\ \decays\ \PIP\PIM, and
later rescaled for the branching ratio  $\rm B( \KZB\ \decays\ \KSH)
\cdot  B(\KSH\ \decays\ \PIP\PIM)$ \par
\D\ mesons are fragmented according to the Peterson recipe.
Because of the symmetric LUND fragmentation scheme, when two objects
are produced in the fragmentation process each with a mass of order
the beam energy, the first object tends to receive a larger share of
the available energy-momentum. This distorts the fragmentation
distribution. This effect is mitigated by by calculating
efficiencies over a small momentum range and performing a summation.
This method also makes calculation of Monte Carlo parameters
insensitive to the fragmentation model.
\par
The collection of vectors and vertices for each selected event are
then propagated through a detector simulation to mimic CLEO raw
data. Detector response in terms of effective efficiency and
resolution are folded into the process.  
The ``false data" is  passed through the CLEO compress system,
and is subjected to the final tracking algorithm DUET. The data at
this stage is identical to real data, and can be analyzed by 
the same program.
The entire process requires $\simeq$ 1.5 VAX 8600 cpu hours for
about 1000 events, making Monte Carlo generation one of the most
cpu intensive tasks in an analysis project.
\par
\begin{figure}[htp!]
\centering
\includegraphics[scale=0.62]{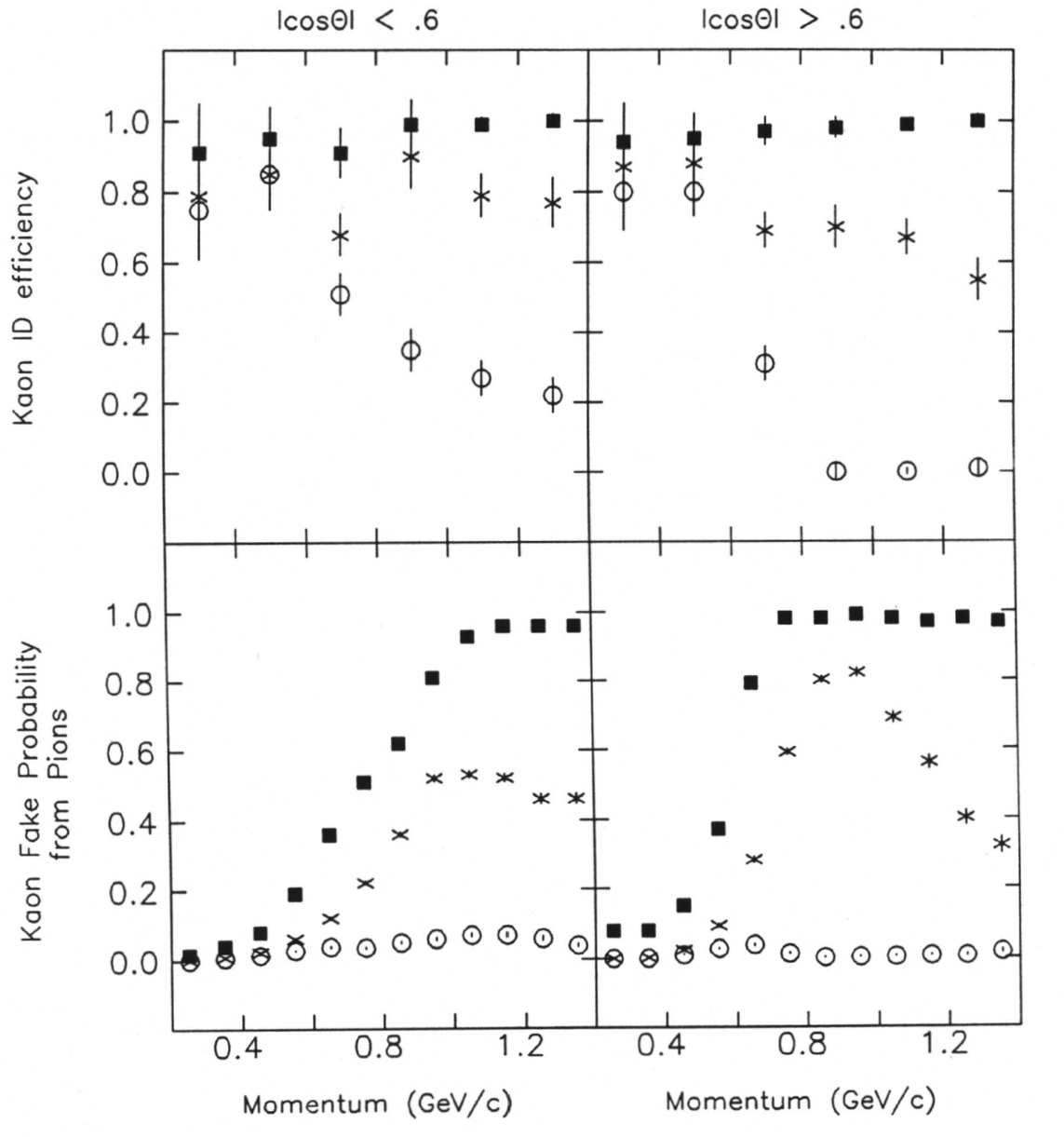}
\caption{Efficiencies for detecting kaons with weights $
W_K \geq 0.1$ (squares),  $ W_K \geq 0.4 $ (crosses), and  $
W_K \geq 0.7$ (circles). Also shown are the efficiencies for
measuring pions with the same kaon weights. For tracks with
$|\cos\Theta| > 0.6$ only information from the drift chamber is available.}
\label{fig:c4f12}
\end{figure}

The detector resolution in most cases dominates the width of an
observed mass peak. Monte Carlo  predicted widths have  been in
respectable agreement with the data, indicating a proper simulation
of the tracking. 
 Information for
neutral particles is approximated to first order based on their
trajectories and shower counter efficiencies. The hadronic event
selection criteria requires 250 MeV deposited in the octant shower
counters. Thorough modeling of this detector would require the cpu
intensive EGS shower simulation, so this constraint was relaxed for
Monte Carlo events.
While charged particles lose energy as they traverse the detector,
response of the devices that measure this energy loss is not
simulated.
  The efficiency for  identifying hadrons is  
measured \cite{skip} directly from the data (Figure \ref{fig:c4f12})
$|\cos\Theta| > 0.6$ only information from the drift chamber
 using ``pure" samples of pions (\KSH\ \decays\
\PIP\PIM), kaons (\PH\ \decays\ \KP\KM {\rm and } \DSP\ \decays\
\DZ\PIP, \DZ\ \decays\ \KM\ \PIP), and protons (\LA\ \decays\
\PR\PI). 
 These efficiencies are injected into the Monte Carlo by
simply throwing a random number between 0 and 1, and comparing it to
the expected efficiency. While this empirical approach assures to
first order proper detector response, the selection of the modes to
produce the ``pure sample" might possibly introduce systematic
effects. 
\section{Two Body Decay Kinematics} Important insight into
the decay mechanism of a parent particle can be gleaned  by
examining the angular distribution of the decay products.
Observation of an angular distribution consistent with an
anticipated polarization (or lack thereof) can help substantiate a
particular decay hypothesis. Also, an anisotropy of the angular
distribution of the background can be used to enhance the signal. A
well quantified formalism \cite{jeffr}
exists for analyzing decays which are  either two body or consist of
a series of sequential two body decays. Termed the helicity
formalism, the following relation can be derived for the angular
distribution of a two body decay in the  rest frame frame of the
decaying particle: $$ \rm { d \sigma \over d\Omega_f }(\theta_f,
\phi_f) = \sum_{\lambda_1,\lambda_2} {\Biggl | {\left( { 2J + 1
\over 4\pi } \right)}^{1\over 2} D^{J \ \ast}_{M \lambda}
\left(\phi_f, \theta_f, \phi_f \right) A_{\lambda_1,\lambda_2}
\Biggr |}^{\lower 1.6ex \hbox{2}} $$ 
Here $\theta_f, \phi_f$ are the
angles of the two body decay axis with respect to the spin
quantization (z) axis of the parent. J and M are the spin and z
projection of the parent particle. ${\lambda_1,\lambda_2}$ are the
helicities of the daughters, as defined by $\left( \lambda_i = \vec
{s_i} \cdot \hat{p_i} \right)$, and $\lambda$ is the overall helicity
$\lambda_1-\lambda_2$,
 which is subject to the constraint $\rm J \geq | \lambda |$. The
 $D^{j }_{m' m}$ functions have the definition:
$$ D^{j }_{m' m} \left( \alpha, \beta, \gamma \right)=
e^{-i\alpha m'} d^j_{m'  m}(\beta)e^{-i\gamma m}$$ 
where $d^j_{m'  m}$
are the ``D" functions \cite{rpp}.
The ``D"
functions are connected to the spherical harmonics through the
relation:
$$ d^{l }_{m 0} \propto Y^l_m(\theta,\phi)e^{-im\phi}$$
The factor $A_{\lambda_1,\lambda_2}$ is related to the decay
matrix element, but does not contain any angular information. We can
neglect the overall normalization (and in most instances the $\phi$
dependence) to get at the the general character of the decay: $$\rm
{ d \sigma \over d\Omega_f }(\theta_f, \phi_f)  \propto
\sum_{\lambda_1,\lambda_2} | d^J_{m \lambda} |^2$$

The first case to consider is the frequently encountered decay of a
pseudoscalar  decaying into two pseudoscalars (P \decays\ PP).
Since everything in this decay is in a state of zero angular
momentum, J=M=$\lambda$=0 and
$$\rm { d \sigma \over d\Omega_{PP} }(\theta_{PP}, \phi_{PP})  
\propto | Y^0_0|^2 = 1$$
hence the decay axis (PP) is isotropically distributed.
\begin{figure}[p!]
\centering
\includegraphics[scale=0.6]{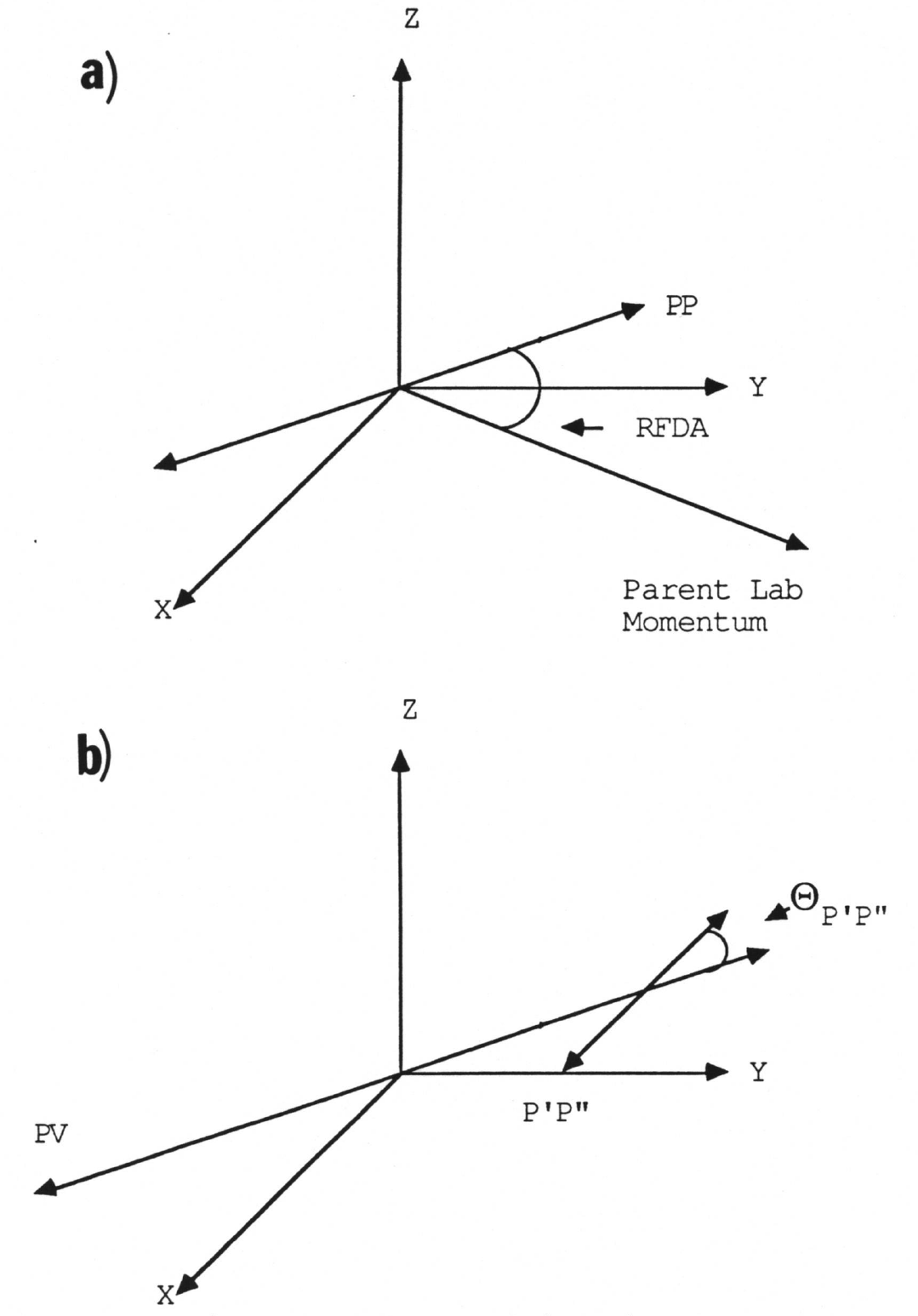}
\caption{
a) Rest frame decay angle (RFDA) for a P \decays\ PP
decay. b) polarization angle $\cos\Theta_{P'P''}$ for P \decays \ PV,
V\decays\ PP.}
\label{fig:c4f13}
\end{figure}
 Since there exists no particular
quantization axis, we choose as the reference direction to be that
of the momentum  of the parent particle in the laboratory frame. We
define the normalized dot product of the decay axis and the
reference vector to be the rest frame decay angle (RFDA).

The second angle of interest results from a polarization in the
decay chain pseudoscalar \decays\ pseudoscalar, vector (P \decays\
PV), where the vector particle subsequently   decays into two
pseudoscalars (V \decays\ $\rm  P'P''$). Because the parent has J=0,
the value for the difference of the helicity state of the daughters
$\lambda_1-\lambda_2 = \lambda \equiv 0$. This polarizes the
vector particle  into the helicity zero state.   The PV decay
axis is also isotropically distributed.
 $$\rm { d \sigma \over d\Omega_{PV} }(\theta_{PV}, \phi_{PV})  
\propto | Y^0_0|^2 = 1$$ 
We now examine the
decay of the vector particle in its rest frame. The quantization
direction has already been defined as the PV decay axis.
 The two vector daughters $\rm  P'P''$  are then boosted into the
vector rest frame. The vector particle has $J=1, M=0$, and both
daughter particles have helicity zero, by definition. The angular
distribution of the $\rm  P'P''$ decay axis  in the V rest frame
with respect to the PV decay axis (Figure \ref{fig:c4f13}) is
$$\rm { d \sigma \over d\Omega_{P'P''} }(\theta_{P'P''},
\phi_{P'P''})   \propto | Y^1_0|^2 = \cos^2\theta _{P'P''}$$
\subsection{Satellite Mass Peaks}
The ramifications of this polarization are profound. In the
laboratory frame, the two vector daughters will have diametrically
opposite values for their total momentum, with one being produced
almost at rest. The invariant mass formed from 2 of the 3 final
state particles, neglecting the particle produced with very small
momentum, will differ in mass from the parent mass largely by the
mass of the neglected daughter. Such a distribution has a distinctly
non Gaussian shape, and is deemed a ``satellite'' mass peak.
\begin{figure}[htp!]
\centering
\includegraphics[scale=0.65]{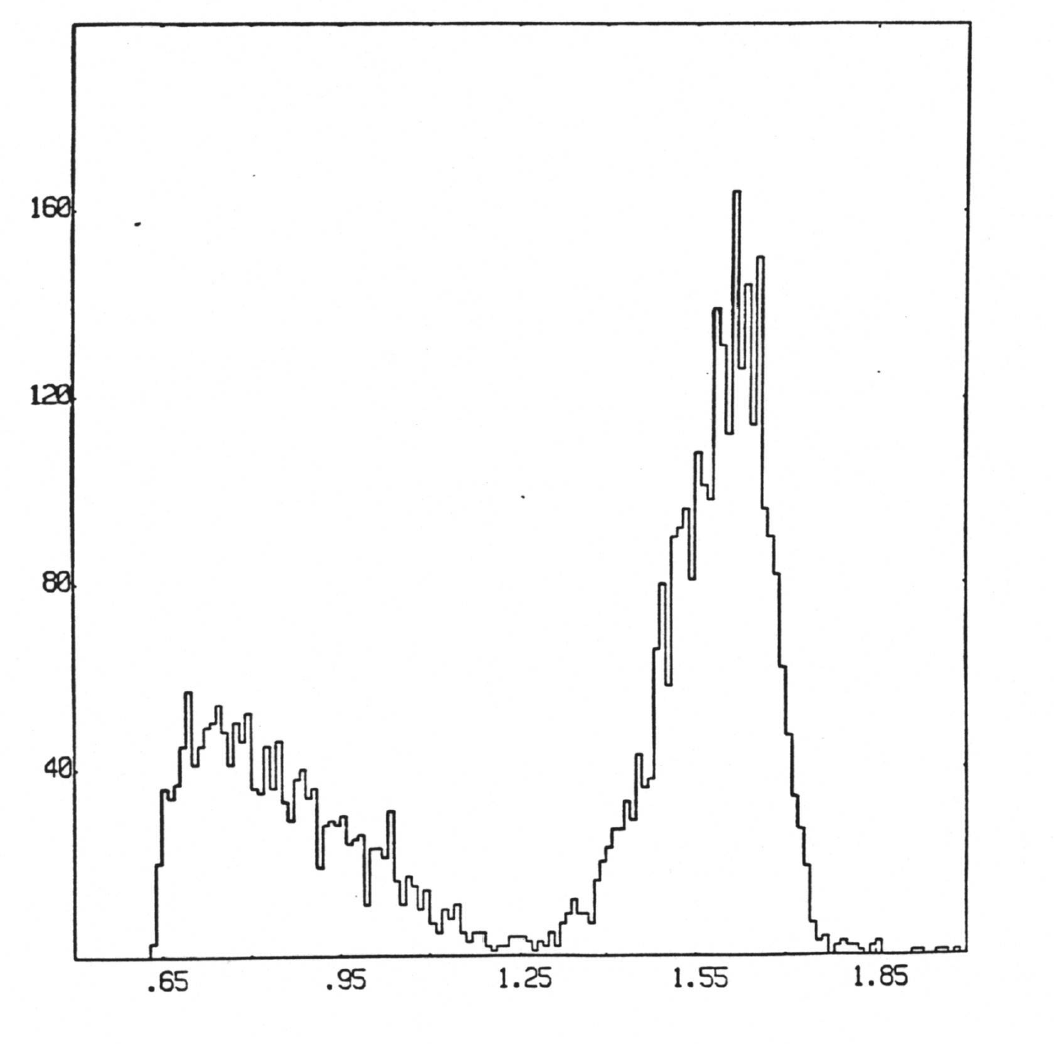}
\caption{Monte Carlo simulation of the decay \DZ\ \decays\
\KM\RHP, \RHP\ \decays \PIP\PIZ. The Invariant mass spectrum is
formed from the \KM\ and \PIP.
}
\label{fig:c4f14}
\end{figure}
Figure~\ref{fig:c4f14}
shows a Monte Carlo simulation of
the decay \DZ\ \decays\ \KM\RHP, \RHP\ \decays\ \PIP\PIZ, where the
invariant mass is formed from the \KM\ and \PIP. The two lobed
structure corresponds to the cases when the \PIP\ is the slow and
fast particle from the \RHP\ polarization. Because the low mass lobe
is much broader and in a region of extremely high background, it is
not visible in the data. 
Information can be extracted from satellite peaks, provided a
proper form can be determined to fit the peak. Extensive study by
the author has  demonstrated that a form termed the Bifurcated
Gaussian provides very agreeable fits to satellite peaks. This
curve is composed of two Gaussian peaks that share a mean, but have
different areas and widths on either side of the  mean. Continuity
of the curve is guaranteed by forcing the constraint $ \rm { A_1
\over F_1} = { A_2 \over F_2}$.
\begin{figure}[htp!]
\centering
\includegraphics[scale=0.560]{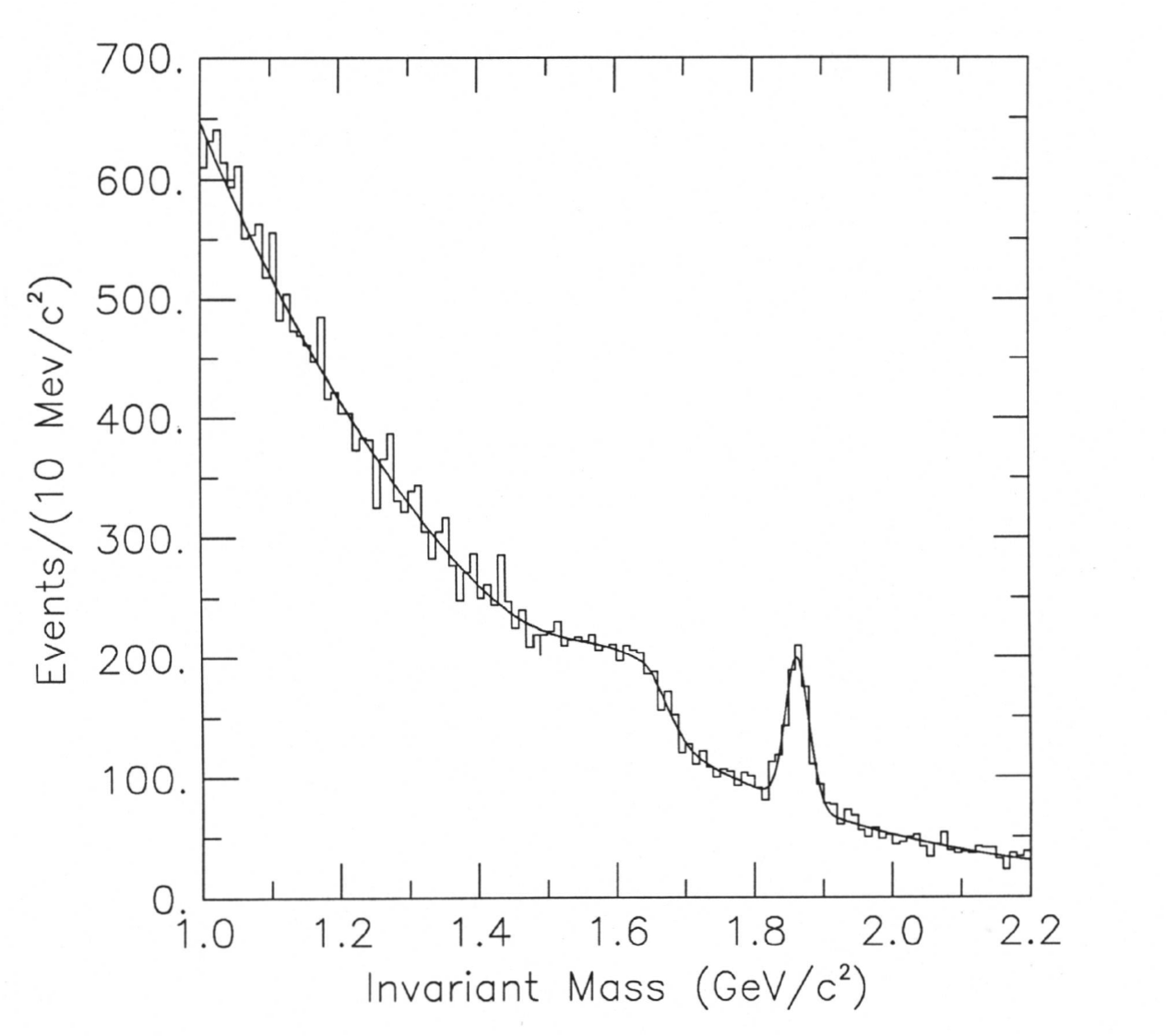}
\caption{Invariant mass spectrum for the decay \DZ\  \decays\
\KM\PIP. A satellite peak is observed in the 1.4 - 1.6 GeV region.
}
\label{fig:c4f15}
\end{figure}
Figure~\ref{fig:c4f15}  
shows the invariant mass distribution for  of \DZ\
\decays\ \KM\PIP. A satellite mass peak is clearly evident in the
1.4-1.6 GeV region. The \DZ\ peak is fit to a Gaussian peak, while
the the satellite peak is fit to a Bifurcated Gaussian. Fitting a
satellite peak to a Gaussian shape underestimates the area by
10-15\%.
  \section{Analysis Architecture}
 Having collected all the techniques and machinery employed in this
project, let us examine how they are used in concert to derive
results. Since most of this analysis involves the reconstruction of
exclusive decays using  charged particle tracking and the 
invariant mass technique, the first step required is to prepare a
combinatoric driver. This code generates suitable combinations of
charged tracks and secondary vertices specific for each decay mode.
The most difficult task in using the invariant mass technique is to
isolate a signal. This is accomplished by developing a set of
physically motivated cuts to simultaneously maximize the observed
signal and  provide the highest possible rejection of spurious
combinations which form the background.  For this analysis, the
 cut development was accomplished by writing out a binary word for
each event containing all the relevant information into a disk file,
which ranged in length from 2.5 - 100 K. This facilitated a 
decrease in cut development by a large factor, since after passing
through the data once to create the file, the entire set could be
reanalyzed in a matter of minutes. More importantly, this made the
cut development largely an interactive process.

\begin{figure}[htp!]
\centering
\includegraphics[scale=0.425]{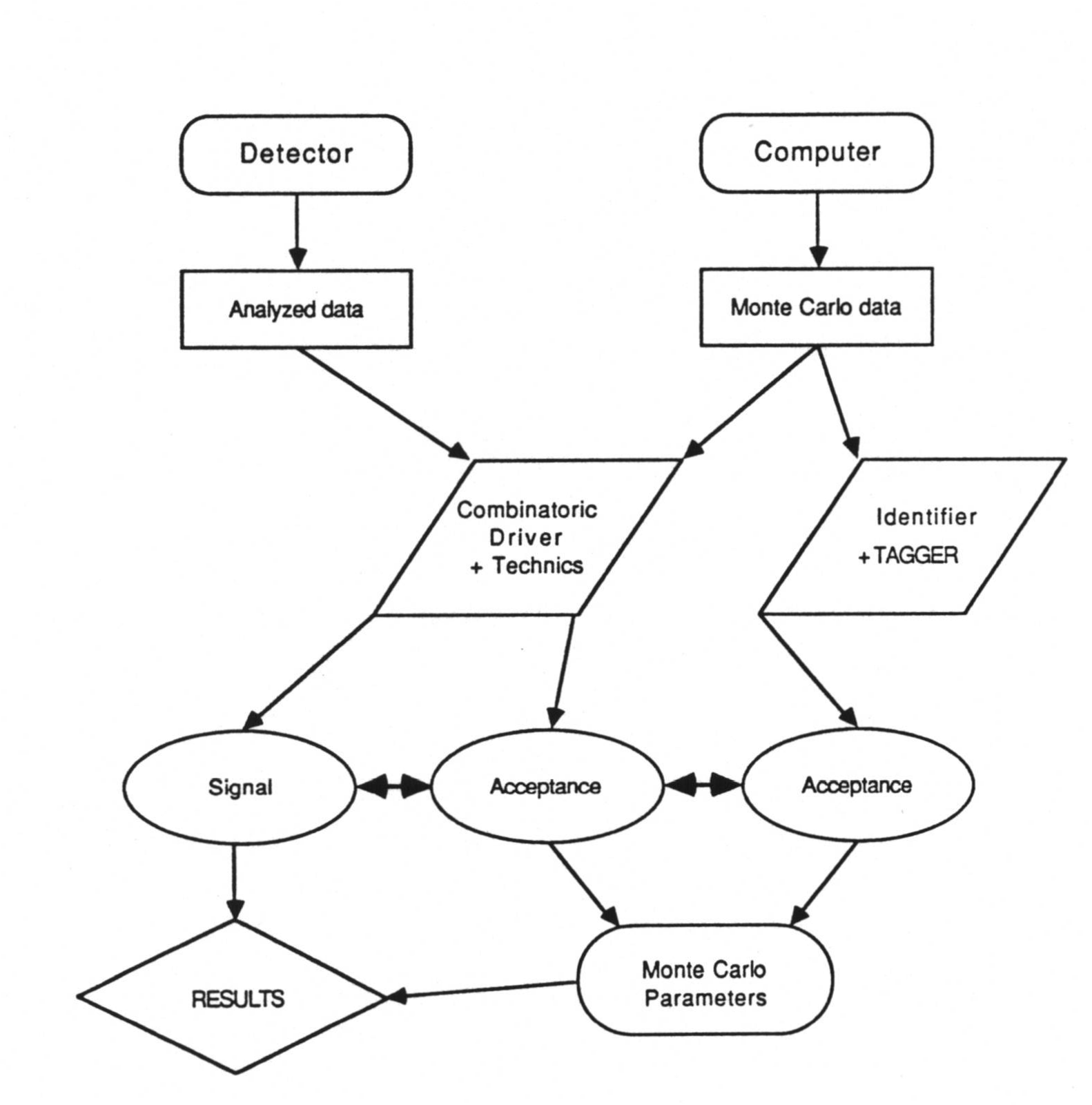}
\caption{Data analysis flow diagram.
}
\label{fig:c4f16}
\end{figure}

After a signal has been isolated, the detector acceptance must be
determined from Monte Carlo. Two methods are used to  extract the
Monte Carlo parameters. This redundancy assures that these
parameters are properly determined, as well as demonstrating the
correctness of the combinatoric driver. In the first method the
Monte Carlo events are subjected to the full data analysis system.
Since the Monte Carlo was generated  with \DZ's and \DP's decaying
into the desired final state, decays of the antiparticles into
analogous final states are eliminated. The second  (TAGGER) method
involves using additional stored information on the Monte Carlo data
which
 ``tags'' the original Monte Carlo tracks to the tracks 
eventually found  in the simulated event by the track finder. 
The decay products in the event are traced down to particles
observable in the detector, these trajectories are compared to
the found tracks, and the Monte Carlo trajectory associated with
the most hits on the track is ``assigned" to the track. For each
event the initial \DP\ or \DZ\   Monte Carlo track is found, and its
generated momentum is calculated.  The decay daughters are found,
and  tested for tagged matches to drift chamber tracks. If all
daughter tracks are matched,  all standard analysis parameters for
that  group of daughter tracks are calculated.
The Monte Carlo signals are fit to a Gaussian signal and a
polynomial background, and the Monte Carlo parameters  are
calculated from a weighted average of the two methods. The agreement
\begin{table}[ht! ]
\centering
\caption{Comparison of Monte Carlo Efficiencies for
Reconstructing the Decay Mode \DP\ \decays\ \KSH\PIP}
\begin{tabular}{|c|c|c|}
\hline
x range &  Method 1 &  Method 2 (TAGGER) \\ \hline
$ 0 \leq {\rm x} < 0.375 $   &$0.381 \pm\ 0.009 $& $0.373 \pm
0.009$\\ \hline
 $ 0.375 \leq {\rm x} < 0.51 $&$0.346 \pm\ 0.009 $& $0.347 \pm
0.009$\\ \hline
$  0.51 \leq {\rm x} < 0.625 $&$0.358 \pm\ 0.009 $& $0.352 \pm
0.009$\\ \hline 
$ 0.625 \leq {\rm x} < 0.750 $&$0.371 \pm\ 0.009 $& $0.365 \pm
0.009$\\ \hline 
$ 0.750 \leq {\rm x} < 0.875 $&$0.363 \pm\ 0.010 $& $0.357 \pm
0.010$\\ \hline 
$ 0.875 \leq {\rm x} \leq 1.0 $&$0.369 \pm\ 0.020 $& $0.365 \pm
0.020$\\ \hline  
\end{tabular}
\label{t:4p2}
\end{table}
between the two methods is excellent, as evidenced from the
example listed in Table \ref{t:4p2}
 which
compares  the two methods in calculating the parameters for the
decay \DP\ \decays\ \KSH\PIP. Although we believe the extracted
parameters to be   correct for a given Monte Carlo,
 we add an additional 5\% systematic error for uncertainties in
the generator and the simulation of the detector.
When determining fragmentation distributions, because of limited
statistics, when performing fits to the data we choose to constrain
the mean and the FWHM of the Gaussian signal to be within $\PM$ 3
standard deviations of the Monte Carlo parameters. This is a more
physical approach  than an exactly constrained fit (with or without
smoothing), as it can adjust for systematic errors in the Monte
Carlo simulation. The analysis structure is summarized in Figure~\ref{fig:c4f16}.

\chapter{ Production and Decay of Charged  \D\  Mesons}
 The \DP\ meson is the spin 0  charged ground state of the 
charmed,nonstrange meson. Its mass is measured to be $1.8693 \pm\ 0.0006$ GeV.
Initially, the properties of \D\ mesons were studied in \EPEM\
experiments running on the \PSPP\ resonance. This was an optimal
running location since the \PSPP\ decays into a \D\DB\ pair
produced almost at rest. Experiments  performed at this energy were
able to accumulate large samples of \D\ mesons, and successfully
reconstruct a vast number of \D\ decay modes. Not accessible to these
experiments was the opportunity to study   the charm hadronization
process or to measure the lifetimes of the \D\ mesons.
\par
In \EPEM\ annihilations at higher center of mass energies, charm
production accounts for $ 4 \over 10$ ($4 \over 11$ beyond the
$\Upsilon$ region) of the total hadronic cross section, allowing
for a fairly copious production of charmed particles. This is offset
by a superabundant  combinatorial background, which impedes the
isolation of charm decay signals. For the \DZ\ and \DSP\ mesons, this
problem is obviated by exploiting the well known mass difference
of the two states using the decay chain  \DSP\ \decays\ \DZ\PIP, \DZ\
\decays\ X. Using this technique, \EPEM\ experiments in the energy
range of  $10-50$ GeV have been able to study the properties of
these two mesons. No such kinematical artifice exists for \DP\
mesons, as the mass difference between the \DSZ\ and the \DP\
forbids charged pion transitions. Additional difficulties
complicate the reconstruction of \DP\ hadronic  decays. The \DP\
has a large branching ratio into semileptonic modes   $( \rm B  (
\DP\ \decays\ \EP X   ) = 17.0 \pm 1.9 \pm 0.7 \ \%   )$
\cite{dmc}.
and the mass splitting of the \D\ and \DS\
mesons allows cascades from the \DS's to \DP's to occur only one
third of the time. Since  \DS\ mesons are favored in the charm
hadronization process $\rm     \left( {\EPEM\ \decays\ \DS\ \over
\EPEM\  \decays\ \D\ + \DS\ } \right) \simeq 0.7 \pm 0.2 $,   this
manifests itself in an inclusive \DZ\ cross section which is about a
factor of two larger than the \DP. Thus, the lack of a
 convenient reconstruction method and low production rates have
inhibited experimental study of \DP\ mesons.
\par
The excellent charged particle tracking system coupled with  
hadron identification capabilities make the CLEO experiment a
highly competitive facility for studying charmed particles,
including \DP\ mesons. The sample of reconstructed \DP\ mesons used
in this analysis, on a mode by mode basis, is larger that that of the
MARK III experiment.  The MARK III group has been a leading
source of information on \D\ decays, and  have measured the
the  greatest number of \D\ meson decay modes. Here we will detail
the measurement of three Cabibbo allowed, hadronic decays of the \DP.
They are \DP\ \decays\ \KZB\PIP, \DP\ \decays\ \KZB\PIP\PIP\PIM, and
\DP\ \decays\ \KM\PIP\PIP. For comparison and use in future
chapters,
we also present a measurement of \DZ\ \decays\ \KZB\PIP\PIM.
 Signal isolation techniques and
corrections for physics backgrounds will be discussed. Fragmentation
distributions and relative branching ratios will be compared, and
total cross sections for \EPEM\ continuum production  will be estimated. 
Information of the CLEO
measurement of the \DP\ lifetime can be found elsewhere \cite{haasp}.
 The
measurements of the \DP\ \decays\ \KZB\ modes are the first to be done at 
an \EPEM\
experiment at an energy greater than the \PSPP, and the measurements
of the fragmentation distributions of these decay modes are the
first such measurements.
\section{Preliminaries}
\begin{table}[ht!]
\centering
\caption{MARK III  \D\  Meson  Branching  Fractions}
\begin{tabular}{|c|c|}
\hline
Decay Mode & Branching Fraction (\% )\\ \hline
\DP\ \decays \KZB\PIP\ \hfill & 3.2\pp\ 0.5 \pp\ 0.2\\ \hline
\DP\ \decays \KZB\PIP\PIP\PIM\  \hfill & 6.6 \pp\ 1.5 \pp\ 0.5\\ \hline
\DP\ \decays\ \KM\PIP\PIP\ \hfill & 9.1 \pp\ 1.3 \pp\ 0.4\\ \hline
\DZ\ \decays\ \KZB\PIP\PIM\ \hfill & 6.4 \pp\ 0.5 \pp\ 1.0\\ \hline
\DZ\ \decays\ \KP\KM  \hfill & 0.51 \pp\ 0.09 \pp\ 0.07\\ \hline
\DZ\ \decays\ \KM\PIP  \hfill &  4.2  \pp\ 0.4 \pp\ 0.4 \\ \hline
\end{tabular}
\label{t:5p1}
\end{table}
 All events under consideration here have passed the
hadronic selection criteria. The invariant mass for \D\ candidates
is formed only from ``good tracks" which  have been corrected for
energy loss in the material preceeding the drift chamber detector.
For calculation of  continuum cross sections, we used only data
taken in the region of the \USSSS\ (77.7  \pb\ on resonance, 36 \pb\
below resonance). The 33 \pb\ of \USSS\ data is also used for
specific measurements, we choose not to use this data to calculate
cross sections because of the high combinatorial  backgrounds  in
this region and the possibility of contamination form the process
\USSS\ \decays\  $c\bar c${.}
\par
 Since \DP\ mesons can also be produced in the
 decay of \B\ mesons, this  \DP\ production mechanism
must be excluded. A Monte Carlo study of 11.5 K \B\BB\ decays where
there was at least 1 \DP\ in the event found no \B\ \decays\ \DP X
decays where the \DP\ had a momentum greater than 2.5 GeV. We
prudently select  momentum cutoff for \DP\ candidates of $\rm  p
\geq 2.52 GeV \ (\rm x = {p\over p_{max}} \geq 0.51)$ 
 to use the on resonance data.
To calculate the differential cross section we elect to use the 
kinematical variable $ \rm x = {  p\over p_{max}} = {p \over 
\sqrt{E^2_{beam} - m^2_{had}} }$. While the approximate ``light-cone" variable
$ x^+ = \rm \ {{\kern -1.1em(E+P)}\over (E+P)_{max}}  $ is described as
being the most suitable for comparing fragmentation distributions at
different energies, this is mitigated by radiative effects. 
This variable also suffers  from  systematic distortions at low 
$x^+$.
The x
variable ranges from 0 to 1 for all experiments and is much more
useful in fitting and  visualizing the data.

Finally, for  comparison  of the \DP\ meson decay rates, and  to
 convert those measurements into cross sections we collect the most
recent MARK III \D\ meson branching ratios \cite{adler1}
in Table \ref{t:5p1} 
and the measurements
\begin{table}[ht!]
\centering
\caption{MARK III Resonant Substructure of Three Body Decays}
\begin{tabular}{|c|c|c|}
\hline
 Decay Mode & Fraction & Branching Fraction (\% )\\ \hline
 \multicolumn{3}{|c|}{\DP\ \decays\ \KM\PIP\PIP\ } \\ \hline
\KSZB \PIP \hfill & 13 \pp\ 1 \pp\ 7 & 1.8 \pp\ 0.2 \pp\ 1.0 \\ \hline
non-resonant \hfill & 79 \pp\ 7 \pp\ 15 &  7.2 \pp\ 0.6 \pp\ 1.8
\\ \hline
\multicolumn{3}{|c|}{\DZ\ \decays\ \KZB\PIP\PIM\ } \\ \hline
\KZB\RHZ\ \hfill & 12 \pp\ 1 \pp\ 7 & 0.8 \pp\ 0.1\pp\ 0.5 \\ \hline
\KSM\PIP\ \hfill & 56 \pp\ 4 \pp\ 5 & 5.3 \pp\ 0.4 \pp\ 1.0 \\ \hline
non-resonant \hfill & 33 \pp\ 5 \pp\ 10 &
 2.2 \pp\ 0.3 \pp\ 0.7 \\ \hline
\end{tabular}
\label{t:5p2}
\end{table}  
of resonant substructure of three body \D\ meson
decays \cite{adler2}
in Table \ref{t:5p2}.
\section{ \DP\ \decays\ \KZB\PIP} 
\subsection{Signal Isolation}
\begin{figure}[htp!]
\centering
\includegraphics[scale=0.6]{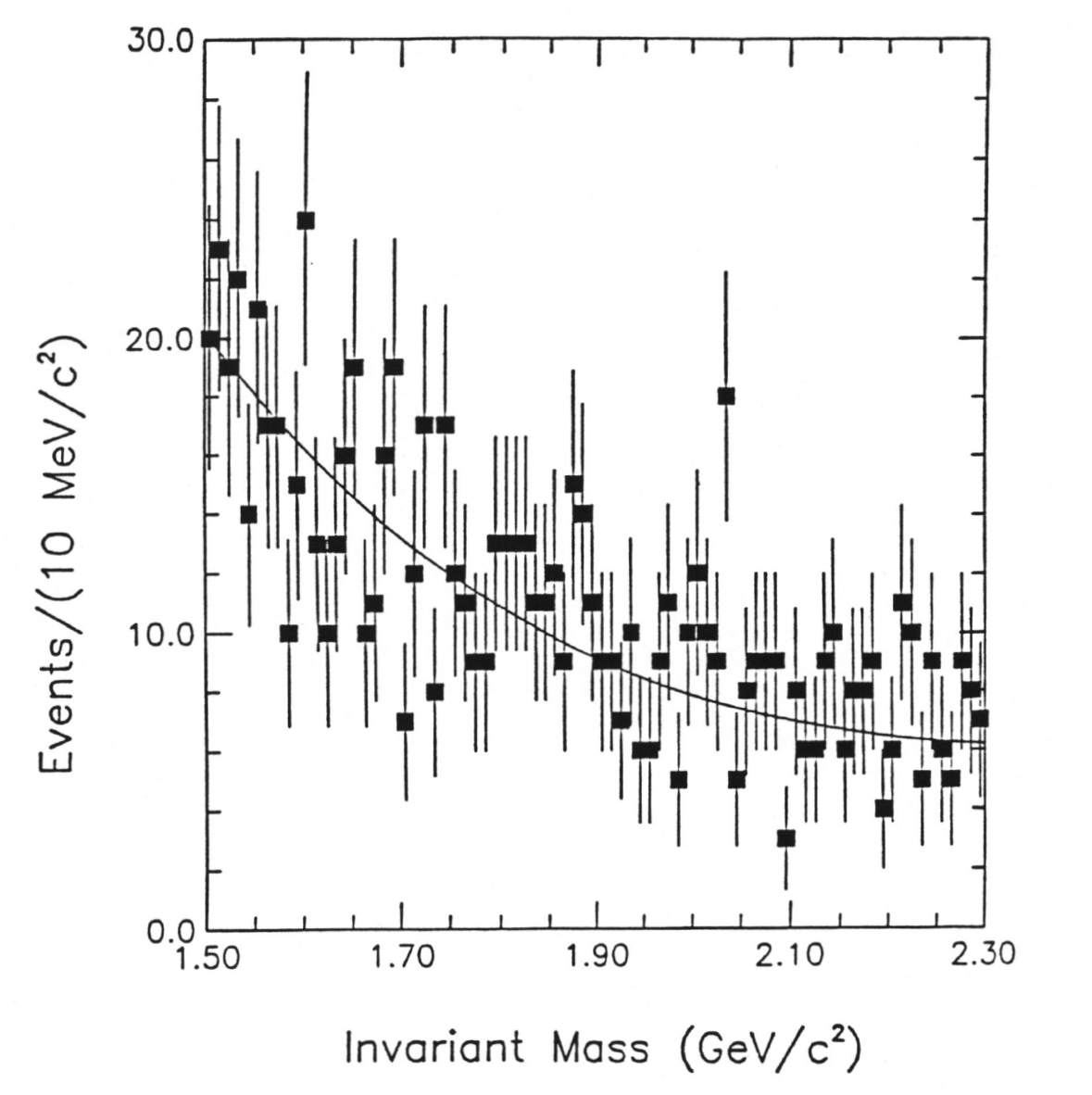}
\caption{ \DP\ \decays\ \KSH\PIP\ candidates
where the \KSH\ mass was selected from a side band region centered at
400 MeV.}
\label{fig:c5f1}
\end{figure}

The first approach used in isolating a \DP\ signal is to use decay
modes which contain a secondary vertex, in this circumstance a \KSH.
Since  approximately 10 \% of the events contain a \KSH\ candidate,
this dramatically reduces the combinatorial background. We observe
the \KSH\ through its decay into two charged pions. To enhance the
\KSH\ signal we  require that the secondary vertex quality factor,
\cv, be less than 2.0, and that the invariant mass of the \KSH\
formed from its two daughter tracks be within 30 MeV of the nominal
\KSH\ mass.  The effect of the \cv\ cut was tested by calculating
the efficiency corrected number of \DP\ candidates for several
different values of \cv.  The numbers were found to be completely
consistent. To test the loose mass cut, we formed \DP\ candidates
where the \KSH\ candidates were selected from a similar mass band
centered at 400 MeV (Figure \ref{fig:c5f1})
, no enhancement is evident.
 Since this is a two body decay, restrictive cuts are
placed on the on the momentum of the track not associated with
the \KSH. We require that that the single pion track have a
momentum greater than 0.4 GeV. This track was also required not to
be consistent with originating from a secondary vertex.
\par
\begin{figure}[htp!]
\centering
\includegraphics[scale=0.6]{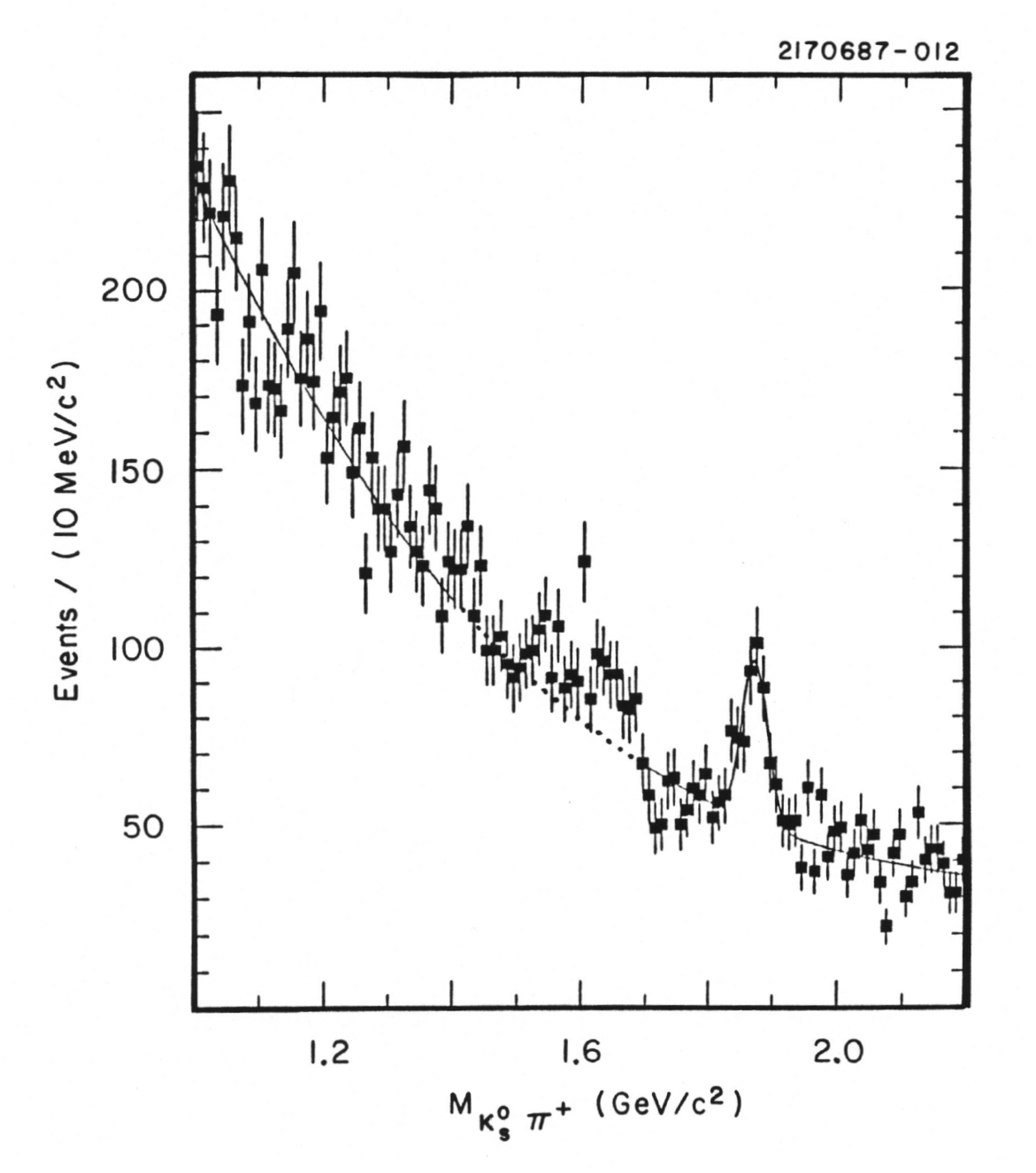}
\caption{Mass spectrum for \DP\ \decays\ \KSH\PIP\ candidates with $\rm x
 \geq 0.51${.}}
\label{fig:c5f2}
\end{figure}

The \DP\ candidate four momentum is  calculated by adding the four
momenta of the \KSH\ and \PIP\ candidates, where the  \KSH\ four
momentum was calculated from the three-momentum of the \KSH\ as
determined from the secondary vertex finding algorithm, and defining
the mass to be that of the \KSH. A plot of the invariant mass of
\DP\ candidates with $\rm x \geq 0.51$ is shown in Figure \ref{fig:c5f2}.

We fit the mass spectrum to
 a Gaussian signal plus a third order polynomial background.
 The
signal was found at a mass of $ 1871.4 \pm 3.1 $ MeV/$c^2$ and a 
FWHM of
$48.3 \pm 8.0$ MeV/$c^2$ which are consistent with a Monte Carlo 
simulation of
this decay in the CLEO detector. 
\subsection{Background}
In addition to the combinatorial background, which is a smoothly
varying function of \DP\ candidate momentum, certain physics
processes cause anomalous enhancements to the combinatorial
background. The final state \KSH\PIP\ is also a subset of the decays
\DP\ \decays\ \KZB\RHP, \DZ\ \decays\ \KSM\PIP. Because of the
polarization of the vector particle, combining the
pseudoscalar daughter of the \D\ with  only one of the daughters 
of the vector particle
will produce a satellite peak.
Evidence for a satellite  peak appears in the region
of 1.4  to  1.7 GeV/$c^2$ and was excluded from the fit.
\par
A particularly difficult situation arises when the
enhancement occurs in the signal region. This is a known problem
for the \DP, which is plagued by reflections from the \DF.  A \DF\
final state which is identical to a \DP\ final state with the
exception of one pion in the \DP\ state replaced with a kaon in the
\DF\ final state can produce an enhancement in the \DP\ region if
the \DF\ kaon is misconstrued as a pion. Possibilities also exist
for a similar confusion of \PR\ from a \LC\ decay, however the
reflection from this decay does not significantly overlap with the
\DP\ region.
\par
\begin{figure}[htp!]
\centering
\includegraphics[scale=0.60]{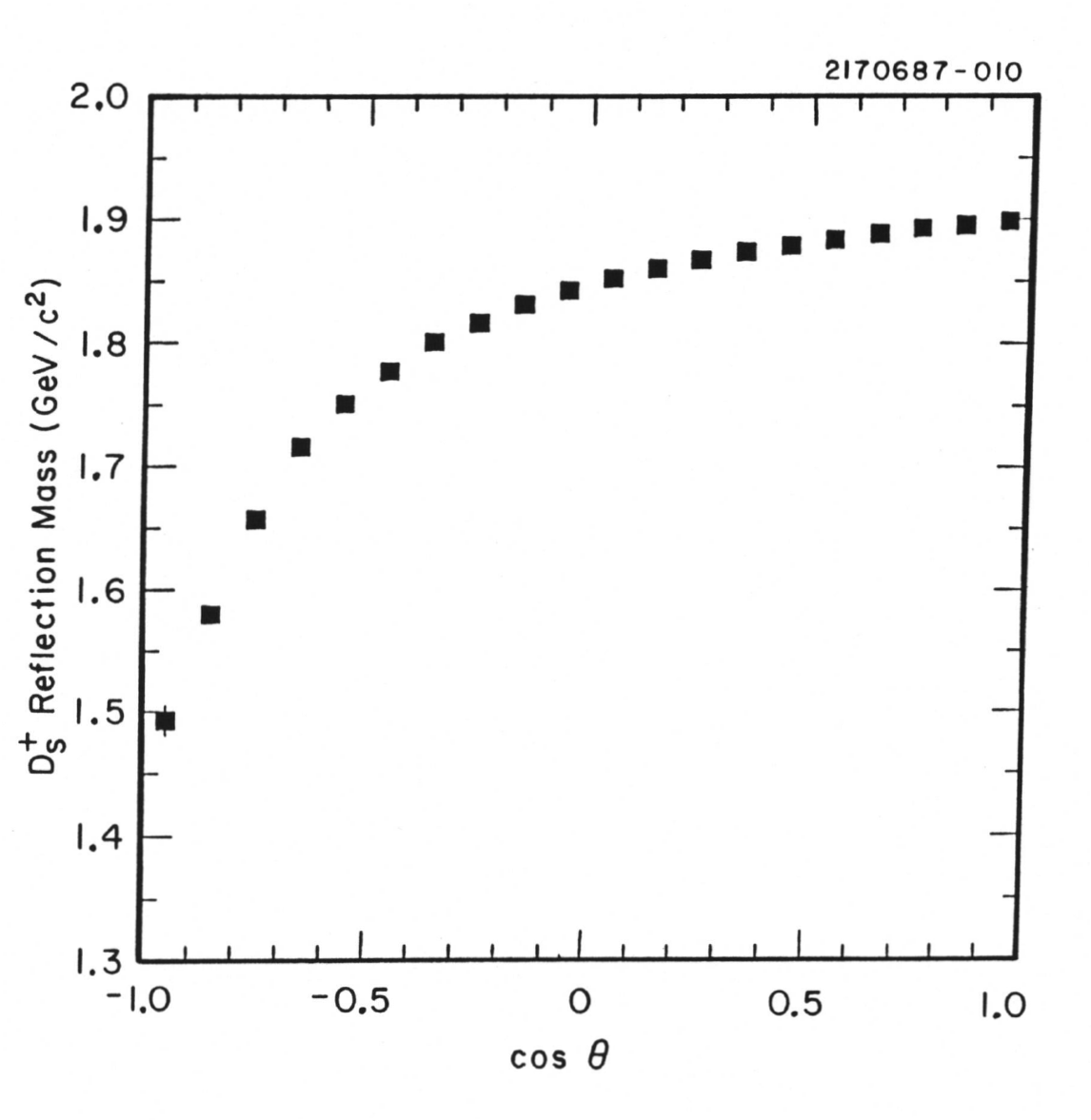}
\caption{Reflected mass
of \DF\ \decays\  \KSH\KP\ candidates with momentum  $\geq$ 2.5 GeV
versus the   RFDA (see text).}
\label{fig:c5f3}
\end{figure}

 The author has developed a technique for quantifying
$\DP \Leftrightarrow \DF$ reflections  for two body decays.
While examining the properties of the invariant mass \DF\ \decays\ 
\KSH\KP\ when the \KP\ was misidentified as a \PIP, it was noticed
that  at high momentum $(\rm p \geq\ 2.5 \ GeV)$
 the reflected mass had a strong dependence on
the  Rest Frame Decay Angle (RFDA - section 4.5), which is defined
here as 
  the cosine of the angle $\theta$ of the `pion' in the `\DP'
center-of-mass frame with respect to the `\DP' direction in the
laboratory frame.  A Monte Carlo 
prediction for this dependence is shown in Figure \ref{fig:c5f3}.
In a certain  sector of RFDA  $(
\rm RFDA \leq -0.2) $ the \DF\  reflection does not contaminate the
\DP\ region. Thus it becomes possible to decompose the \DF\ \decays\ 
 \KSH\KP\ reflection into parts which do (contaminated region) and
do not (pure region) populate the \DP\ region.
\begin{figure}[p!]
\centering
\includegraphics[scale=0.65]{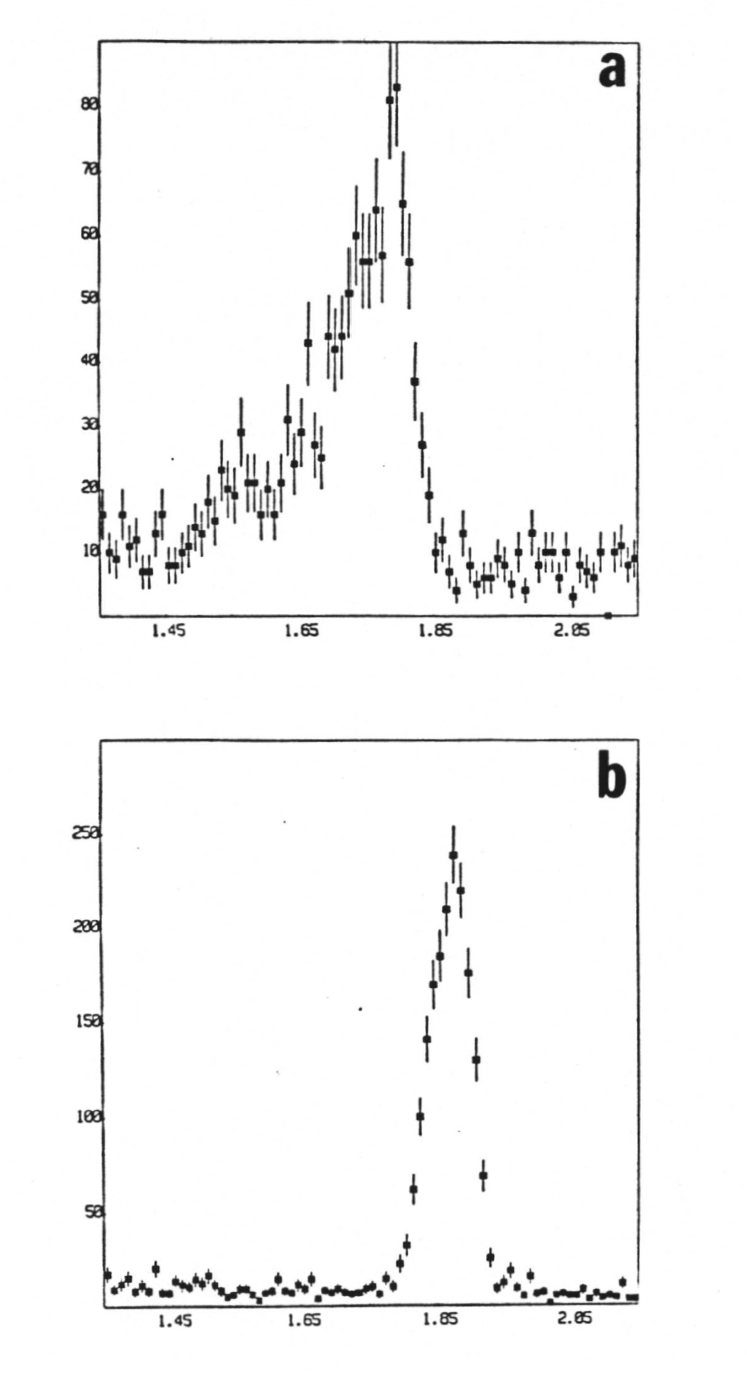}
\caption{Monte
Carlo Simulation of the decay \DF\ \decays\  \KSH\KP\ where the
 invariant mass has been calculated calling the \KP\ a \PIP. The
invariant mass is plotted for two regions of RFDA; a) RFDA  
$ < -0.2$ (pure region) and b) RFDA $\geq -0.2$ (contaminated
region). Only the events in b) significantly overlap the \DP\
region.}
\label{fig:c5f4}
\end{figure}
Figure \ref{fig:c5f4}
 illustrates a Monte Carlo simulation of the distribution of
reflected \DF\ mass when segregated by RFDA. In addition, the broad
peak  observed in the  contaminated region would change the width of
the observed \DP\ peak in that region depending on the \DP\ : \DF\
ratio.
\begin{table}[ht!]
\centering
\caption{Monte Carlo Predicted 
Contamination Ratios, $p \geq 2.5$ GeV }
\begin{tabular}{|c|c|c|c|}
\hline
$ \DP\ :  \DF\ $ &P = $\rm
{N_{contaminated}\over N_{pure}}$&pure FWHM (MeV)&cont FWHM (MeV)\\ \hline
Pure $\rm D^+$ & 1.59 $\pm$ 0.05 &40 $\pm$ 1 & 42 $\pm$ 1\\ \hline
4 : 1 & 2.01 $\pm$ 0.06 &39 $\pm$ 1 & 47 $\pm$ 1\\ \hline
2 : 1 & 2.47 $\pm$ 0.08& 39 $\pm$ 1 & 52 $\pm$ 1\\ \hline
\end{tabular}
\label{t:5p3}
\end{table}
Table \ref{t:5p3}
contains a Monte Carlo
study of the properties of the signal at the \DP\ mass for various
ratios of \DP\ : \DF\ production. The presence of a reflection from
the decay  \DF\ \decays\  \KSH\KP\ appears in the broadening of the
\DP\ signal width in the contaminated region and an increase in the
 ratio of events in the contaminated to pure region.
To perform this measurement in the data we include the 33 \pb\ of
\USSS\ data.
\begin{figure}[p!]
\centering
\includegraphics[scale=0.65]{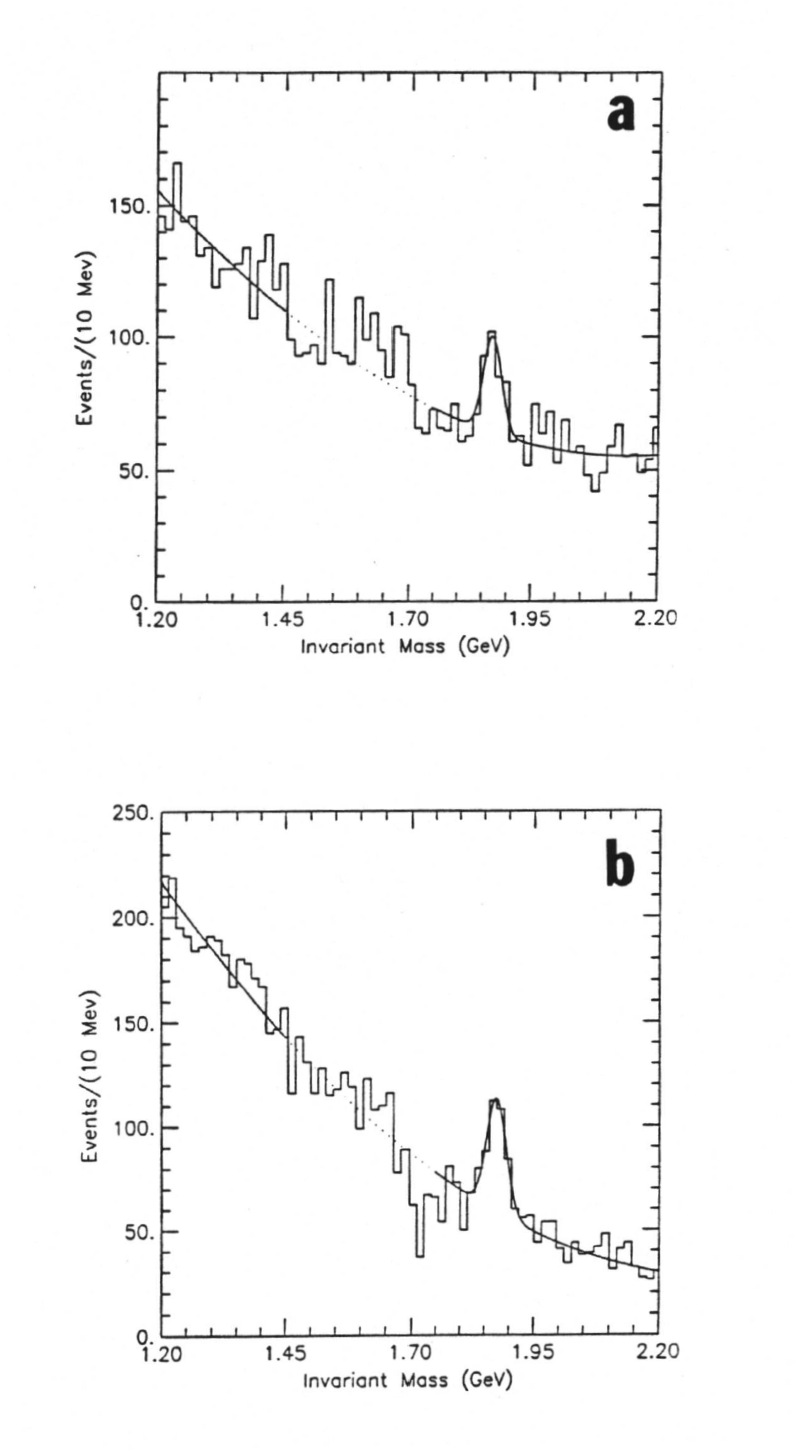}
\caption{ The \DP\ \decays\ \KSH\PIP split into
two regions on the basis of RFDA. a) Pure region (RFDA $< -0.2$)
b) contaminated region (RFDA $\geq -0.2$). }
\label{fig:c5f5}
\end{figure}
Figure \ref{fig:c5f5}
shows the \KSH\PIP\ signal decomposed into pure and contaminated
regions. We fit the  distributions to the  same  form   as the other
\KSH\PIP\ signals, however we make no constraints on the mean  and
width of the signal since these properties could be altered by
\DF\ reflections.
 The efficiency corrected ratio of events found in the  contaminated
region to those found in the  pure region is 1.6 \pp\ 0.4. The large
error, which is due to the  combinatorial background,  unfortunately
prevents a quantitative statement being made about the amount of
\DF\ contamination. The width of the invariant mass peak is found
to be 43.7 \pp\ 9.4 ( 47.4 \pp\  8.0 ) in the pure (contaminated)
regions, and are also consistent. While we are prevented from
making a strong statistical statement, it would seem unlikely that
\DF\ \decays\ \KSH\KP\ contaminates \DP\ \decays\ \KSH\PIP\ at more
than a $\approx 20 \%$ level.
 \par
From a theoretical perspective \DF\ \decays\ \KZB\KP, which
proceeds through non spectator diagrams,
is not anticipated to be large. Kamal \cite{kamal}
 predicts B(\DF\ \decays\ \KZB\KP)/B(\DP\
\decays\ \KZB\PIP) = 0.04 to 0.08. Approximating that there are
nearly 3 times as many \DP's as \DF's, it is expected that the ratio 
$\sigma(e^+e^{-}\rightarrow  \DF\ )$B(\DF\ \decays\ \KZB\KP\ )
/ $\sigma(e^+e^{-}\rightarrow  \DP\ )$B(\DP\ \decays\ \KZB\PIP\ )
 would be only a few percent. 
CLEO has searched \cite{pla}
 for the decay \DF\ \decays \KSH\KM\  and determined the upper limit
the upper limit B(\DF\ \decays\ \KZB\KP\ )/ B(\DF\ \decays\ \PH\PIP\ )
 $<$ 0.55.  Only one group has seen a signal for this decay
mode.
MARK III  \cite{toki} finds the
ratio of the above two \DF\ branching ratios to be $0.88 \pm
0.50${.} This number, however, has continued to enjoy  preliminary
status. 
  The amount 
of contamination from \DF\ reflection for this decay mode can be
described by
$$ C = \left[ \sigma_{\DF}
 \cdot {\rm B}( \DF\ \decays\ \KZB\KP) \right] K_R  $$
where $K_R$ is a kinematical factor relating how much of the
\DF\ signal actually reflects into the \DP\ region and is 
approximately 0.5 for this case. Based on  CLEO's
measurement \cite{frag}
 of  
${\sigma_{\DF} \cdot  {\rm B}(\DF \decays\ \PH\PIP) =  5.8 \pm  1.0} $ $(x
\geq 0.5)$ we estimate a systematic error of 1.3 \pb\ for potential
contamination of the \DP\ \decays\ \KSH\PIP\ cross section from
\DF\ \decays\ \KZB\KP reflections.
\subsection{ Fragmentation Distribution}
\begin{table}[ht!]
\centering
\caption{Fragmentation
Distribution for \DP\ \decays\ \KZB\PIP }
\begin{tabular}{|c|c|c|c|}
\hline
x range & $\rm N_{obs}$ & efficiency & ${\rm {{d\sigma} / {dx}}
}\cdot B$ (pb) \\ \hline
0.000 -  0.375 & 43 $\pm$ 32 &  0.367 $\pm$ 0.020  &25 $\pm$ 18 \\ \hline
0.375 -  0.510& 17 $\pm$ 12 &0.347 $\pm$ 0.019    &30 $\pm$ 21 \\ \hline
0.510 - 0.625 & 73 $\pm$ 17 &0.355 $\pm$ 0.019    &46 $\pm$ 11 \\ \hline
0.625 - 0.750 & 73 $\pm$ 14 &0.368 $\pm$ 0.019    &41 $\pm$ 9 \\ \hline
0.750 - 0.825 & 48 $\pm$ 10 &0.360 $\pm$ 0.019    &27 $\pm$ 6 \\ \hline
0.825 - 1.000 & 14 $\pm$ 5   &0.367 $\pm$ 0.024 &8 $\pm$ 2 \\ \hline   
\end{tabular}
\label{t:5p4}
\end{table}

The results of the fits for the differential cross sections for the
\KZB\PIP\ mode are collected in Table \ref{t:5p4}.
When quoting fragmentation
distributions for modes which contain a \KSH, we adopt the
convention that $\rm N_{obs}$ is number observed in the \KSH\ mode,
where the error is statistical only. The quoted efficiency
($\epsilon_{r}$)   is the reconstruction efficiency which accounts
for
 geometrical
acceptance, tracking efficiency, and  cuts for this decay mode
 where the \KZB\ has decayed 
via the chain \KZB\ \decays\ \KSH, \KSH\ \decays\ \PIP\PIM. The
error on the efficiency includes both statistical and systematic
effects combined in quadrature. 
 The value ${\rm
{{d\sigma} / {dx}} }\cdot B$ is defined as $ (2.91)N_{\rm
obs}/{\epsilon_{r} {\Delta x} L}$. Here  L is the 
integrated luminosity, $\Delta x$ is the width of the bin, and the
 factor
of 2.91 accounts for $B$( \KZB\ \decays\ \KSH)$ \cdot $
$B$(\KSH\ \decays\ \PIP\PIM). Differential cross sections will
always be referenced to the original state of a \KZB.
 Because of the greatly reduced
luminosity and immense combinatorial background, the measurements
below $\rm x  = .51$ have disproportionately large errors. The most
statistically significant information about the production cross
section comes from a summation of the points above x = .51.
Performing this summation yields $ \rm \sigma_{\DP} \cdot B(\DP\
\decays\ \KZB\PIP) =  14.8 \pm 1.7 \pm 0.6 \pm 1.3{}$ pb $(\rm x
\geq 0.51)$, where the errors are statistical, systematic (fitting
procedure and Monte Carlo simulation), and an estimate for \DF\
reflection contamination. 
\section{ \DP\ \decays\  \KZB\PIP\PIP\PIM}
\begin{figure}[htp!]
\centering
\includegraphics[scale=0.6]{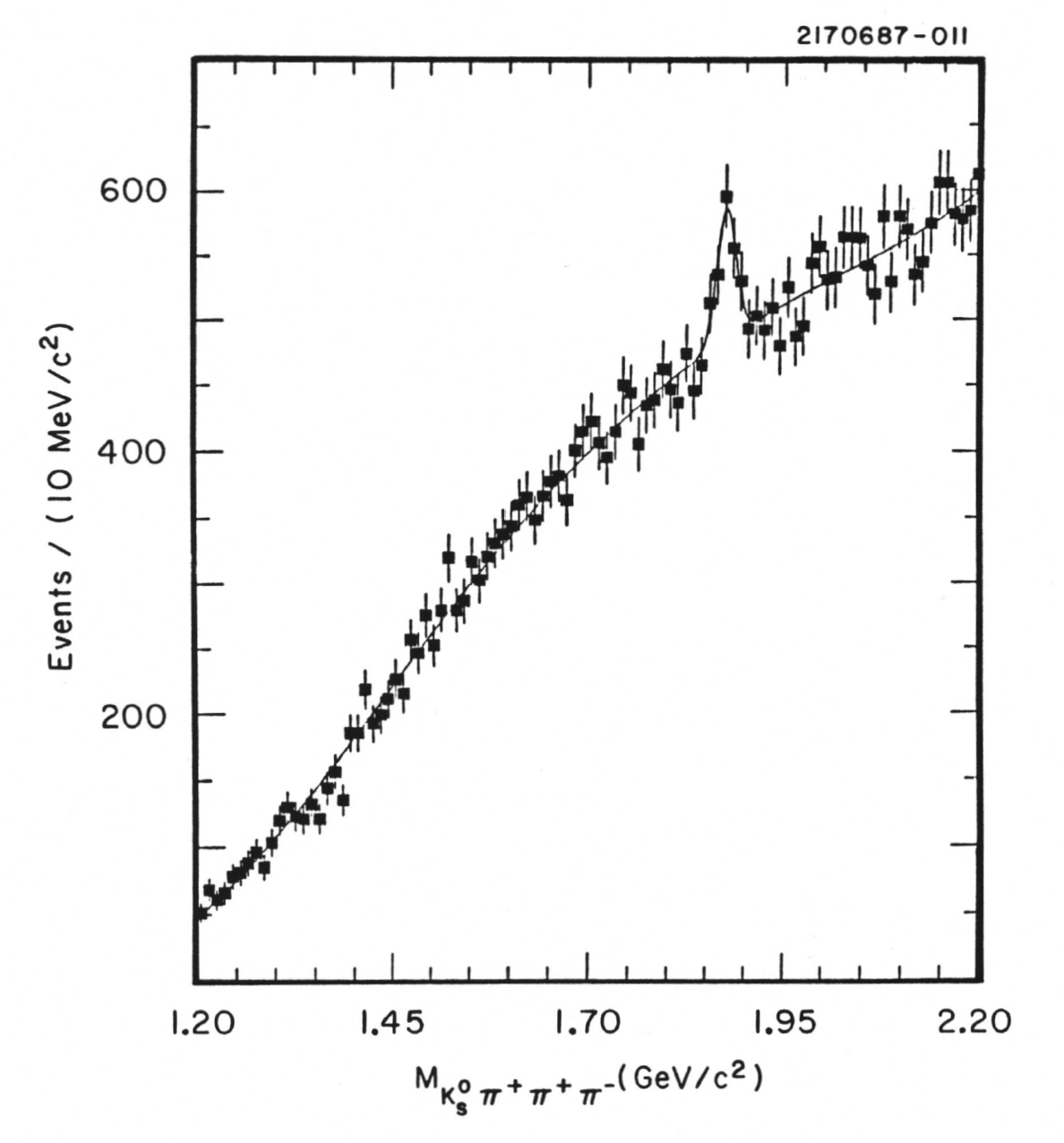}
\caption{ Mass spectrum for \DP\ \decays\
\KSH\PIP\PIP\PIM\ candidates having x $\geq\ 0.51${.}}
\label{fig:c5f6}
\end{figure}

The approach for isolating this decay is quite similar to \KSH\PIP.
 We again rely on the \cv\ cut, though it is relaxed to 2.5, since
in this case the \KSH\ is slower and more difficult to reconstruct.
 All tracks not associated with the \KSH\   were required
to have a momentum greater than 200 MeV.   The mass spectrum for
\DP\ candidates with x greater then 0.51, displayed in Figure \ref{fig:c5f6},
 was fit to a
Gaussian signal plus a fourth order polynomial background.  A signal
from the  decay \DSP\ \decays\ \DZ\PIP, \DZ\ \decays\ \KSH\PIP\PIM\
appears in the mass spectrum in the high x bins 
and is excluded from the fit. The
signal was found at a mass of $ 1876.0 \pm 2.6 $ MeV/$c^2$ and a FWHM of
$26.3.0 \pm 6.3$ MeV/$c^2$.
The FWHM is consistent with Monte Carlo simulation; however the mass is
2.5 standard deviations away from the expected mass.   This is caused by 
an upward fluctuation in the mass spectrum in the
region $ 0.51 \leq x < 0.625${.}  
Fitting the mass spectrum for x
greater than  0.625 yields a \DP\ mass of $1869.9 \pm 2.7 $ MeV/$c^2$,
as expected.  Because this is a four body decay, \DF\ reflections
are extremely broad and do not significantly overlap with the \DP\
region.
\subsection{Fragmentation Distribution}
\begin{table}[ht!]
\centering
\caption{Fragmentation Distribution  for 
\DP\ \decays\ \KZB\PIP\PIP\PIM }
\begin{tabular}{|c|c|c|c|}
\hline
 x range & $\rm N_{observed}$ & efficiency & ${\rm {{d\sigma} / 
{dx}}}\cdot B$ (pb) \\ \hline
0.510 - 0.625 & 99 $\pm$ 33 &0.183 $\pm$ 0.010 &120 $\pm$ 41 \\ \hline
0.625 - 0.750 & 98 $\pm$ 28 &0.206 $\pm$ 0.011 &98 $\pm$ 29 \\ \hline
0.750 - 0.825 & 44 $\pm$ 15 &0.236 $\pm$ 0.013  &38 $\pm$ 13 \\ \hline
0.825 - 1.000 & 28 $\pm$ 8   &0.275 $\pm$ 0.014  &21 $\pm$ 6
\\ \hline   
\end{tabular}
\label{t:5p5}
\end{table}
The fragmentation distribution for the \KSH\PIP\PIP\PIM\ mode
is collected in Table \ref{t:5p5},
where the definitions are the same
as the previous section. $\rm \sigma_{\DP} \cdot 
B(\DP\ \decays\ \KZB\PIP\PIP\PIM) = 33.4 \pm
6.0 \pm 1.4$
  pb,   $(x\geq 0.51)$. The errors are statistical and systematic,
respectively.
\section{ \DP\ \decays\ \KM\PIP\PIP}
The isolation of \KM\PIP\PIP\ requires a completely different
approach. Here we rely primarily on the hadron identification 
system to produce a discernible signal. An additional advantage of
this decay is that the kaon is produced with the opposite sign of
the two pions. This reduces the combinatorial background along
with providing signal validation and reflection rejection through
the use of right and wrong sign combinations.
\par
\begin{figure}[htp!]
\centering
\includegraphics[scale=0.55]{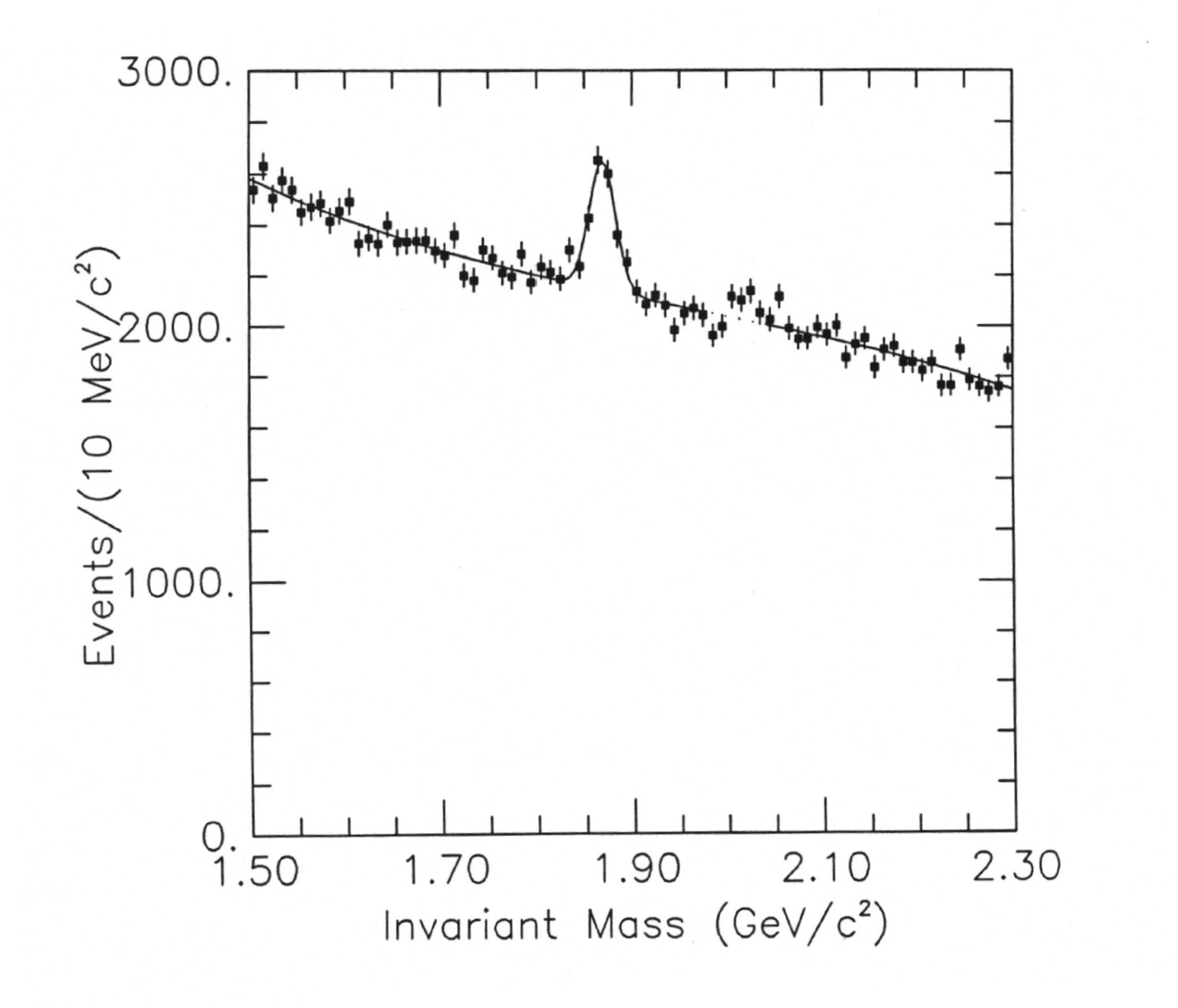}
\caption{\DP\ \decays\ \KM\PIP\PIP\ candidates with $\rm x \geq\ 0.51$. The
\KM\ has been observed with $\rm W_K \geq\ 0.1$.}
\label{fig:c5f7}
\end{figure}

To enhance the signal we require that the kaon candidate have a
weight $\rm W_{\K} \geq 0.1$. In addition we require that the kaon
candidate posses at least 200 MeV of momentum. The motivation for
this cut is simply that at low momentum it is difficult to produce
a pure sample of kaons to measure the efficiency. Since a low
momentum a particle's energy loss is maximal, saturation effects in
the electronics may produce unusual responses. Since these effects
may not be properly reflected in the measured  efficiency, it is
best to avoid this particular group. We also make the usual cuts on
the momenta of the pions (200 MeV). To further reduce the
combinatorial background we require that all tracks have their
DOCA's less than 8 mm.
The invariant mass spectrum for  \DP\ \decays\ \KM\PIP\PIP\
candidates with $\rm x \geq\ 0.51$ is presented in Figure \ref{fig:c5f7}.
The mean of the
\DP\ peak is measured to be $1.8693 \pm\ .0013$ GeV, which is in
excellent agreement with the world average of $1.8693 \pm\ 
0.0006$ GeV.
 \subsection{Background}
\begin{figure}[htp!]
\centering
\includegraphics[scale=0.5]{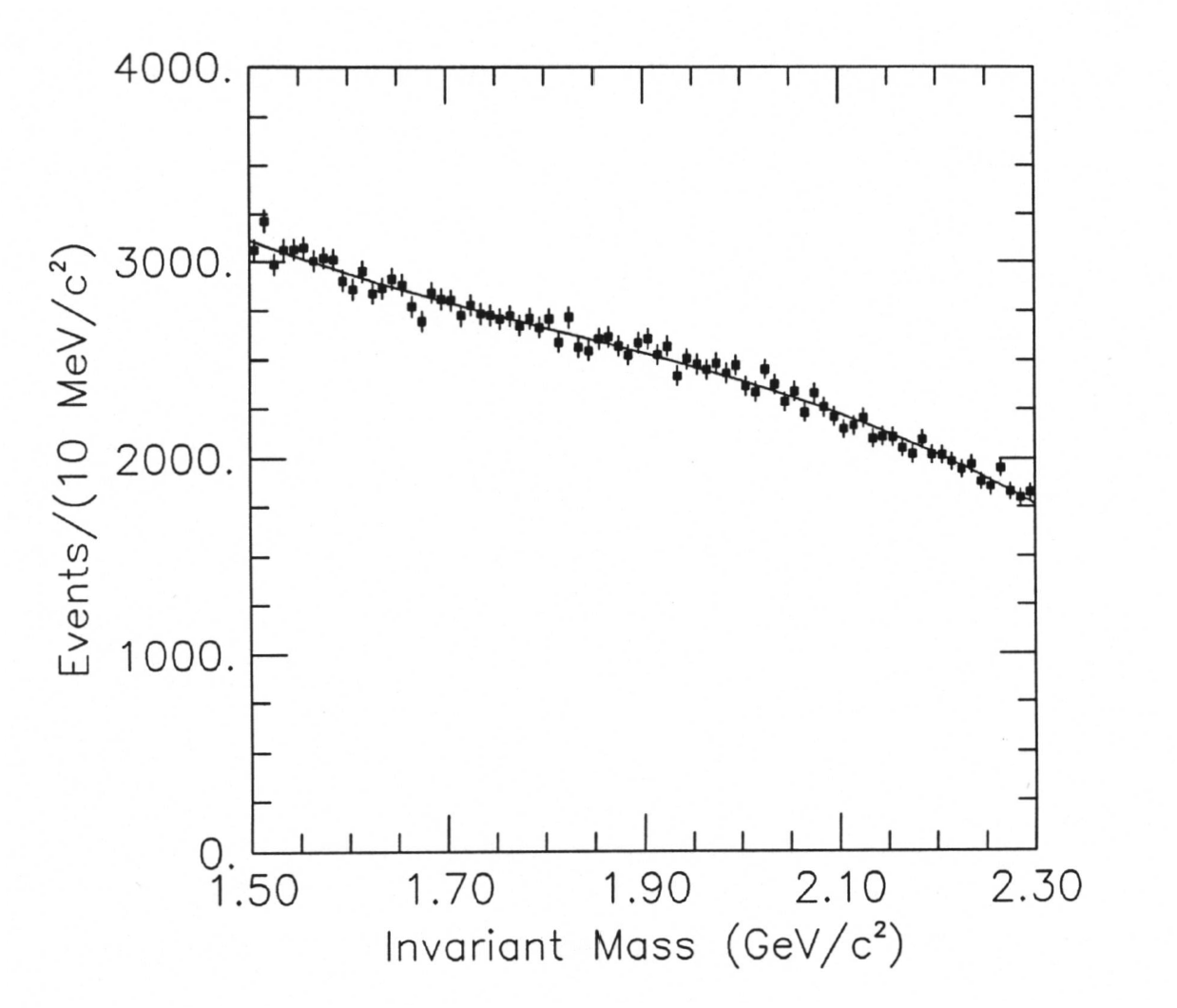}
\caption{ \DP\ candidates formed from
combinations of the type \KP\PIP\PIM. No enhancement is observable.}
\label{fig:c5f8}
\end{figure}

This decay mode can be contaminated by the \DF\ final state
\KP\KM\PIP.  If  either (but only one) of the kaons  are
misidentified
  a reflection may occur. The misidentification \KM\PIP\PIP\ could
count as a valid \DP\ candidate but \KP\PIM\PIP\ would not. We are
therefore able look for the effects of \DF\ reflections by examining
wrong sign combinations. Figure \ref{fig:c5f8}
contains a plot of wrong sign combinations. No enhancements to the
background are observed. We also note that because this is a three
body decay the reflected peak is much broader and peaks well below
the \DP\ region. For this case $K_R \approx 0.1$ as opposed to  the
\KSH\PIP mode where $K_R \sim  0.5$.  Enhancement to the background
also occurs from the decay chain \DS\ \decays\ \DZ\PIP, \DZ\ \decays
\KM\PIP, which shares the same final state. This is simply removed 
by
excluding the \DS\ region from the fit.
 \subsection{Fragmentation Distribution}
\begin{table}[ht!]
\centering
\caption{Fragmentation Distribution
for \DP\ \decays\ \KM\PIP\PIP}
\begin{tabular}{|c|c|c|c|}
\hline
x range & $\rm N_{obs}$ & efficiency & ${\rm {{d\sigma} / {dx}}
}\cdot B$ (pb) \\ \hline
0.000 -  0.375 &  -201  $\pm$  149 &  0.405  $\pm$ 0.046  & -36 
$\pm$
 30 \\ \hline 
0.375 -  0.510&  221 $\pm$  67 &0.421 $\pm$ 0.047    & 107 $\pm$
39 \\ \hline 
0.510 - 0.625 &  702 $\pm$  96 &0.416 $\pm$ 0.047    & 129 $\pm$
 23 \\ \hline 
0.625 - 0.750 &  609 $\pm$  68 &0.461 $\pm$ 0.052    & 93 $\pm$
 15  \\ \hline 
0.750 - 0.825 &  384 $\pm$  49 &0.475 $\pm$ 0.052    & 57 $\pm$
 10 \\ \hline 
0.825 - 1.000 &  119 $\pm$  24   &0.493 $\pm$ 0.059 & 17 $\pm$  4 
\\ \hline    
\end{tabular}
\label{t:5p6}
\end{table}
The Monte Carlo simulation of this decay consisted of the nonresonant
\KM\PIP\PIP\   and \KSZ\PIP\  \decays\ \KM\PIP\PIP\ 
in a ratio of $6.1 : 1$ in
accord with Table 5.2. Since the kaon efficiencies are not
determined from Monte Carlo simulation, we include an  additional
term which is combined
 in
quadrature with the error on the Monte Carlo efficiency and the
error from the fitting procedure to estimate the overall systematic
error. The summary of the differential cross section for the
\KM\PIP\PIP\ mode is contained in Table \ref{t:5p6}.
 The data are fit to a Gaussian signal and a third order polynomial.
Performing a summation  of  the points with $\rm x \geq 0.51$
 We find a differential cross section  $\rm \sigma_{\DP} \cdot 
B(\DP\ \decays\ \KM\PIP\PIP ) = 35.6 \pm
 2.7 \pm  2.4$
  pb,   $(\rm x \geq\ 0.51)$.

\section{\DZ\ \decays\ \KZB\PIP\PIM}
\begin{table}[ht!]
\centering
\caption{Fragmentation Distribution
for \DZ\ \decays\ \KZB\PIP\PIM}
\begin{tabular}{|c|c|c|c|}
\hline
  x range & $\rm N_{obs}$ & efficiency & ${\rm {{d\sigma} / {dx}}
}\cdot B$ (pb) \\ \hline
0.000 -  0.375 &  -50  $\pm$  44 &  0.175  $\pm$ 0.010  & -61 $\pm$
  62 \\ \hline 
0.375 -  0.510&   53 $\pm$  21 &0.221 $\pm$ 0.012    & 143 $\pm$
 63 \\ \hline 
0.510 - 0.625 &   239 $\pm$  32 &0.249 $\pm$ 0.014    &  213 $\pm$
 33 \\ \hline 
0.625 - 0.750 &   212 $\pm$  25 &0.275 $\pm$ 0.016    &  159 $\pm$
  21 \\ \hline 
0.750 - 0.825 &  134 $\pm$   17 &0.284 $\pm$ 0.017    &  96 $\pm$
  14 \\ \hline 
0.825 - 1.000 &   39 $\pm$   8  &0.286 $\pm$ 0.025 &  28 $\pm$  7.0
\\ \hline  
\end{tabular}
\label{t:5p7}
\end{table}
For comparison with the \DP\ measurements and for use in the
following chapter we document the properties of the \DZ\ \decays\ 
\KZB\PIP\PIM. To detect this signal we make identical cuts on the
\KSH\ and associated pions that we use in reconstructing \DP\
\decays\ \KZB\PIP\PIP\PIM. 
Our simulation of this decay included \KSM\PIP, \KZB\RHZ, and
non resonant \KZB\PIP\PIM in the ratio as prescribed by Table 5.2.
This decay mode is  free from reflections
of any kind.
\begin{figure}[htp!]
\centering
\includegraphics[scale=0.62]{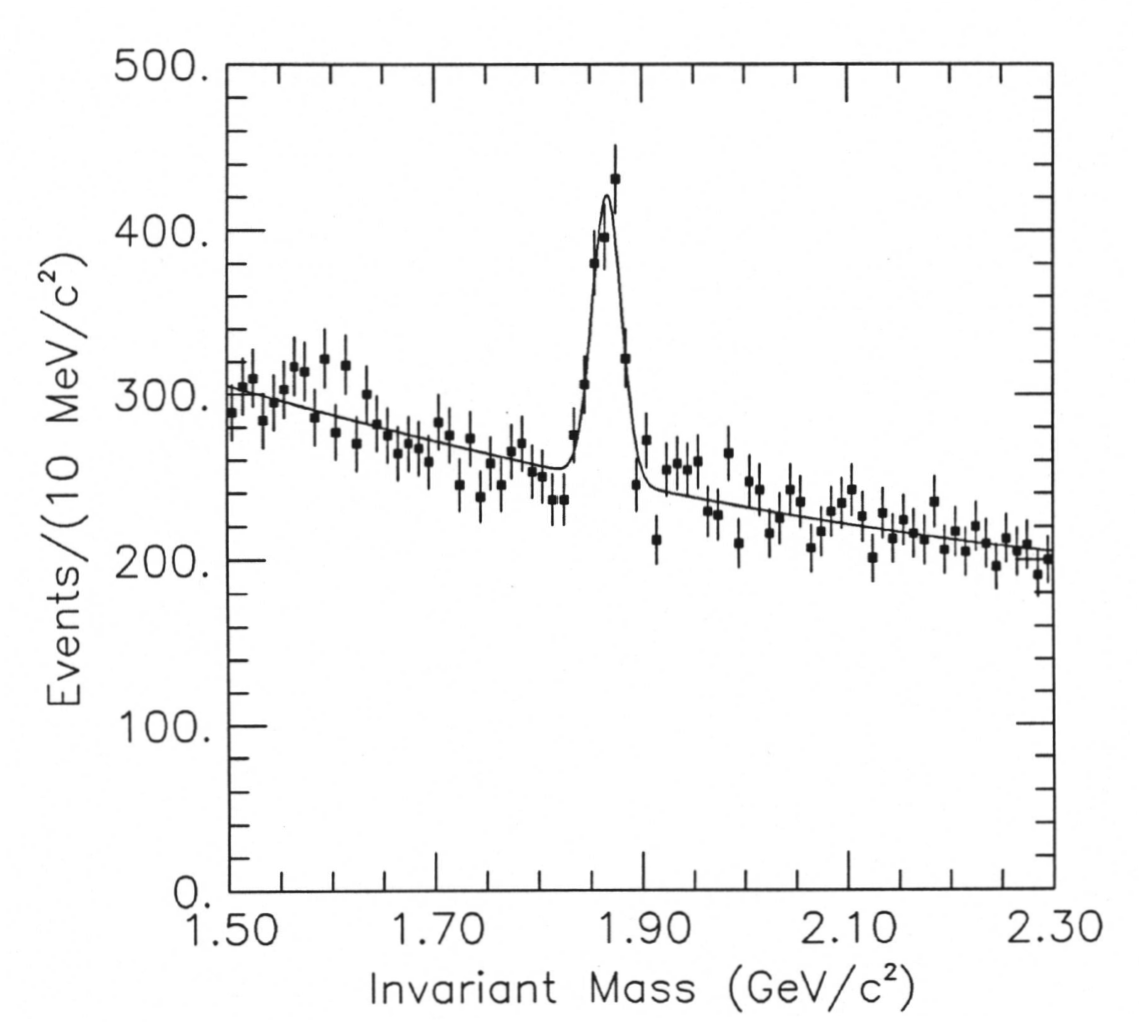}
\caption{Mass spectrum for
\DZ\ \decays\ \KSH\PIP\PIM  candidates having $\rm x \geq .51${.}}
\label{fig:c5f9}
\end{figure}
Figure \ref{fig:c5f9}
displays the invariant mass for \KSH\PIP\PIM candidates. The curve
is   a second order polynomial with a Gaussian signal. Table \ref{t:5p7}
collects the measurements of the
differential cross section for this mode. 
  Performing the canonical summation, we find
 $\rm \sigma_{\DZ} \cdot 
B(\DZ\ \decays\ \KZB\PIP\PIM ) =  59.7 \pm
 4.4 \pm  2.3$
  pb,   $(\rm x \geq\ 0.51)$.
   
\section{ \DP\ Relative Production Rates}
\begin{table}[ht!]
\centering
\caption{ \DP\ Cross Section Measurements}
\begin{tabular}{|c|c|c|}
\hline
 Mode & Experiment & $\rm \sigma  \cdot B $ \\ \hline
\DP\ \decays\ \KZB\PIP\ & LGW & 0.14 \pp\ 0.05 (nb) \\ 
 &  MARK II & 0.14 \pp\ 0.03  (nb) \\ 
  & MARK III  & 0.135 \pp\ 0.012 \pp\ 0.010  (nb)\\ 
 &  CLEO & 14.8 \pp\  1.7 \pp\  0.6 \pp\ 1.3 (pb) \\ \hline
\DP\ \decays\ \KZB\PIP\PIP\PIM\ &   
   MARK II &  0.51 \pp\ 0.18  (nb) \\ 
  & MARK III  & 0.305 \pp\ 0.031 \pp\ 0.030  (nb)\\ 
 &  CLEO & 33.4 \pp\  6.0 \pp\  1.4  (pb) \\ \hline
\DP\ \decays\ \KM\PIP\PIP\ (inclusive) & LGW & 0.36 \pp\ 0.06 (nb)
\\ 
 &  MARK II & 0.38 \pp\ 0.05  (nb) \\ 
  & MARK III  & 0.388 \pp\ 0.013 \pp\ 0.029 (nb)\\ 
 &  CLEO & 35.6 \pp\  2.7 \pp\  2.4  (pb) \\ \hline

\end{tabular}
\label{t:5p8}
\end{table}
\begin{table}[ht!]
\centering
\caption{ B( \DP\ \decays\ \KZB\PIP\PIP\PIM\ )
  / {B(\DP\ \decays\ \KZB\PIP )} }
\begin{tabular}{|c|c|}
\hline
Experiment & Ratio \\ \hline
MARK II &3.6 \pp\ 1.5\\ \hline
MARK III &2.3 \pp\ 0.3\\ \hline
MARK III $ ( 2 \ast T) $&2.1 \pp\ 0.6 \\ \hline
 CLEO &2.3 \pp\ 0.5  \\ \hline
\end{tabular}
\label{t:5p9}
\end{table}
\begin{table}[ht!]
\centering
\caption{ B( \DP\ \decays\  \KM\PIP\PIP\ )
  / {B(\DP\ \decays\  \KZB\PIP)} }
\begin{tabular}{|c|c|}
\hline
Experiment & Ratio \\ \hline
LGW & 2.6 \pp\ 1.0\\ \hline
MARK II &2.7 \pp\ 0.7 \\ \hline
MARK III &2.9 \pp\ 0.4 \\ \hline
MARK III $ ( 2 \ast T) $&2.8 \pp\ 0.6 \\ \hline
 CLEO & 2.4  \pp\ 0.4   \\ \hline
\end{tabular}
\label{t:5p10}
\end{table}
The  production measurements of \DP\ mesons as observed in \EPEM\
annihilations are arranged in Table \ref{t:5p8} \cite{Hitlin}
 The CLEO cross sections are only partial $( x \geq\
0.51)$, they represent the  most statistically significant
information available, and are perfectly acceptable for calculating
relative rates. We have evaluated the relative rates from the
various groups for $\rm \left( { B(\DP\ \decays\ \KZB\PIP\PIP\PIM\ )
\over B(\DP\ \decays\ \KZB\PIP ) } \right)$ (Table \ref{t:5p9})
and $\rm \left( { B(\DP\ \decays\ 
\KM\PIP\PIP ) \over  B(\DP\
\decays\ \KZB\PIP ) } \right)$    (Table \ref{t:5p10}).
 For the MARK III group, we also compare the branching ratios
derived from the global fit double tag method  $( 2 \ast T)$. This
analysis determines \cite{sblp}
$$\rm \left( { B(\DP\ \decays\ \KZB\PIP\PIP\PIM\ )
\over B(\DP\ \decays\ \KZB\PIP ) } \right) = 2.3 \pm\ 0.5$$
and
$$\rm \left( { B(\DP\ \decays\ 
\KM\PIP\PIP ) \over  B(\DP\
\decays\ \KZB\PIP ) } \right) = 2.4 \pm\ 0.4$$
We find these results to agree
  well with previous measurements.
\section{Analysis of Fragmentation Distributions}
\begin{table}[ht!]
\centering
\caption{  Results of Fitting to the Peterson Form}
\begin{tabular}{|c|c|c|}
\hline
Mode & $ \epsilon_Q $& $\chi^2$/d.o.f.\\ \hline
\DP\ \decays\ \KM\PIP\PIP & $0.21^{+6}_{-5} $& 10.7/4 \\ \hline
\DP\ \decays\ \KZB\PIP  & $0.17^{+7}_{-5}$ & 3.7/4 \\ \hline
\DZ\ \decays\ \KZB\PIP\PIM& $0.22^{+6}_{-5}$ & 9.0/4 \\ \hline
\end{tabular}
\label{t:5p11}
\end{table}
\begin{table}[ht!]
\centering
\caption{Results of Fitting to the  Andersson Form}
\begin{tabular}{|c|c|c|c|}
\hline
Mode &  A & B & $\chi^2$/d.o.f.\\ \hline
\DP\ \decays\ \KM\PIP\PIP &  1.3 \pp\ 0.15 & 0.39 \pp\ 0.05 & 2.4/3
\\ \hline \DP\ \decays\ \KZB\PIP  &  1.2 \pp\ 0.26 & 0.47 \pp\ 0.03 &  1.4/3
\\ \hline \DZ\ \decays\ \KZB\PIP\PIM &  1.3 \pp\ 0.17 & 0.42 \pp\ 0.04  &
9.0/4 \\ \hline 
\end{tabular}
\label{t:5p12}
\end{table}

The precise shape of the fragmentation distribution reveals
distinctive features of the hadronization process. Charmed hadrons
make an excellent laboratory to study such phenomenon, since they
can be abundantly isolated in a variety of forms, and at different
center of mass energies.
 Experimentally,
measurements of fragmentation distribution distributions are
difficult to execute. This is largely because  fragmentation models
distinguish themselves most at low momentum, which for reasons of
acceptance, feed down, and or lack of data tends not to be measured
well. None the less qualitative  still exist trends exist.

We choose to compare two  fragmentation functions to our
fragmentation data, one from each prevalent philosophy. We elect not
to use models which explicitly include meson structure functions in
their derivation (Kartvelishvili and Collins).
 The Peterson function $$ \rm D{^H_Q}(z) = {N
\left(z\left[1-{1 \over z}-{\epsilon_Q\over(1-z)} \right]
^2\right)^{-1}}$$ of independent fragmentation is used as is  the 
Andersson form
$$\rm D{^H_Q}(z{^+})={N\over z{^+}}(1-z{^+})^a
exp({-b{{m_H}_\perp}^2\over z{^+}} )$$ representing string
fragmentation. These functions happen to also be the most common in
the literature.
 We note that the Andersson function is derived in
terms of the light cone variable, analysis in terms of the x
variable may not be optimal, but is suitable here for our  mainly
didactic purposes.
The values of A and B are only for comparison with similar
fragmentation
distributions binned in x, and should not be compared with theoretical
expectations. When fitting to the Andersson form, we fix the parameter
${m_H}_\perp$ to be the known hadron mass.
 The results for the fits are exhibited in
Table \ref{t:5p11}  and Table \ref{t:5p12}. Fits of both forms to the 
\KM\PIP\PIP\ and \KZB\PIP\PIM\  modes are shown in 
\begin{figure}[htp!]
\centering
\includegraphics[scale=0.6]{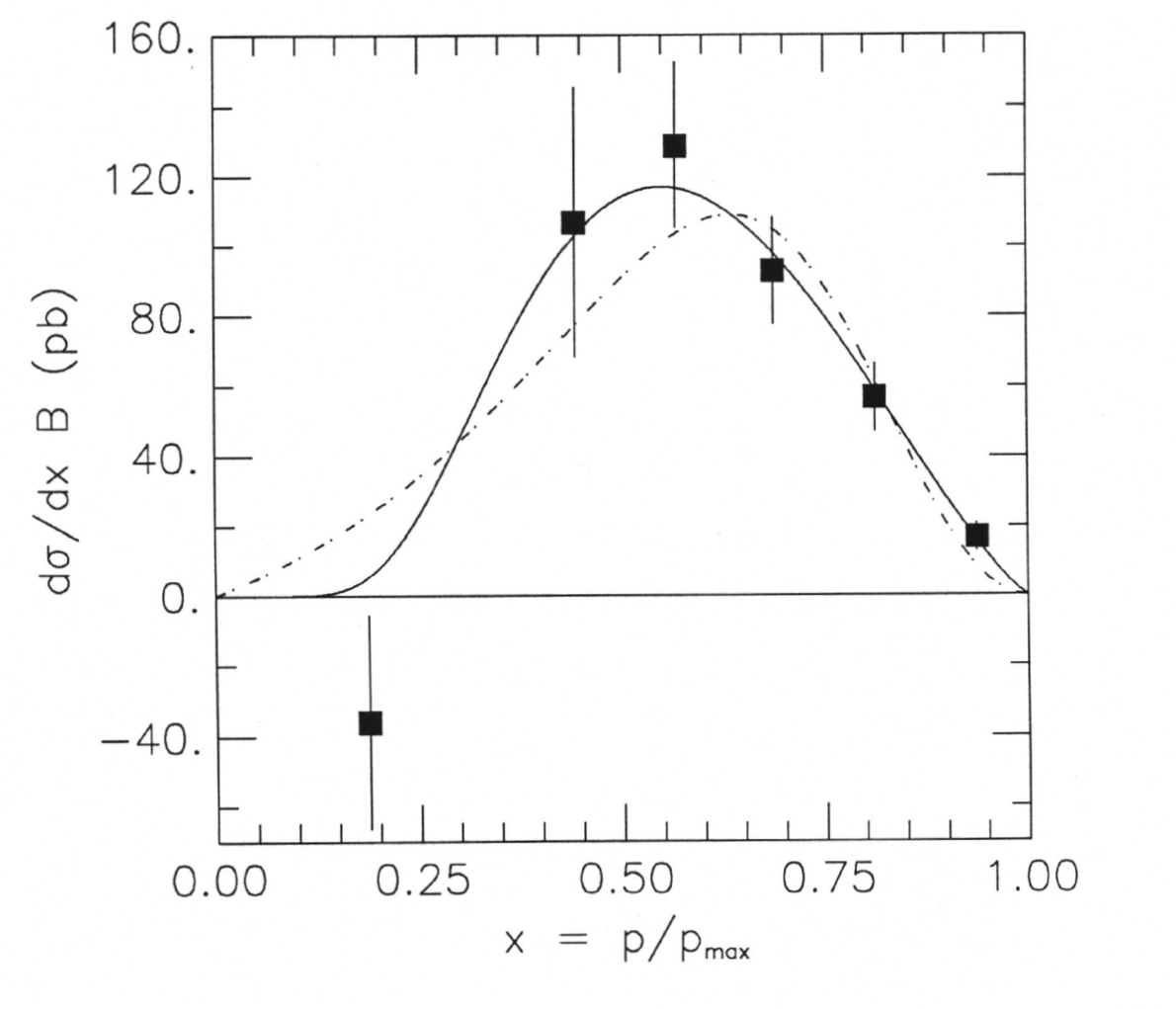}
\caption{Fits to the Andersson form (solid line) and Peterson form (dashed
line) for \DP\ \decays\ \KM\PIP\PIP .}
\label{fig:c5f10}
\end{figure}
\begin{figure}[htp!]
\centering
\includegraphics[scale=0.59]{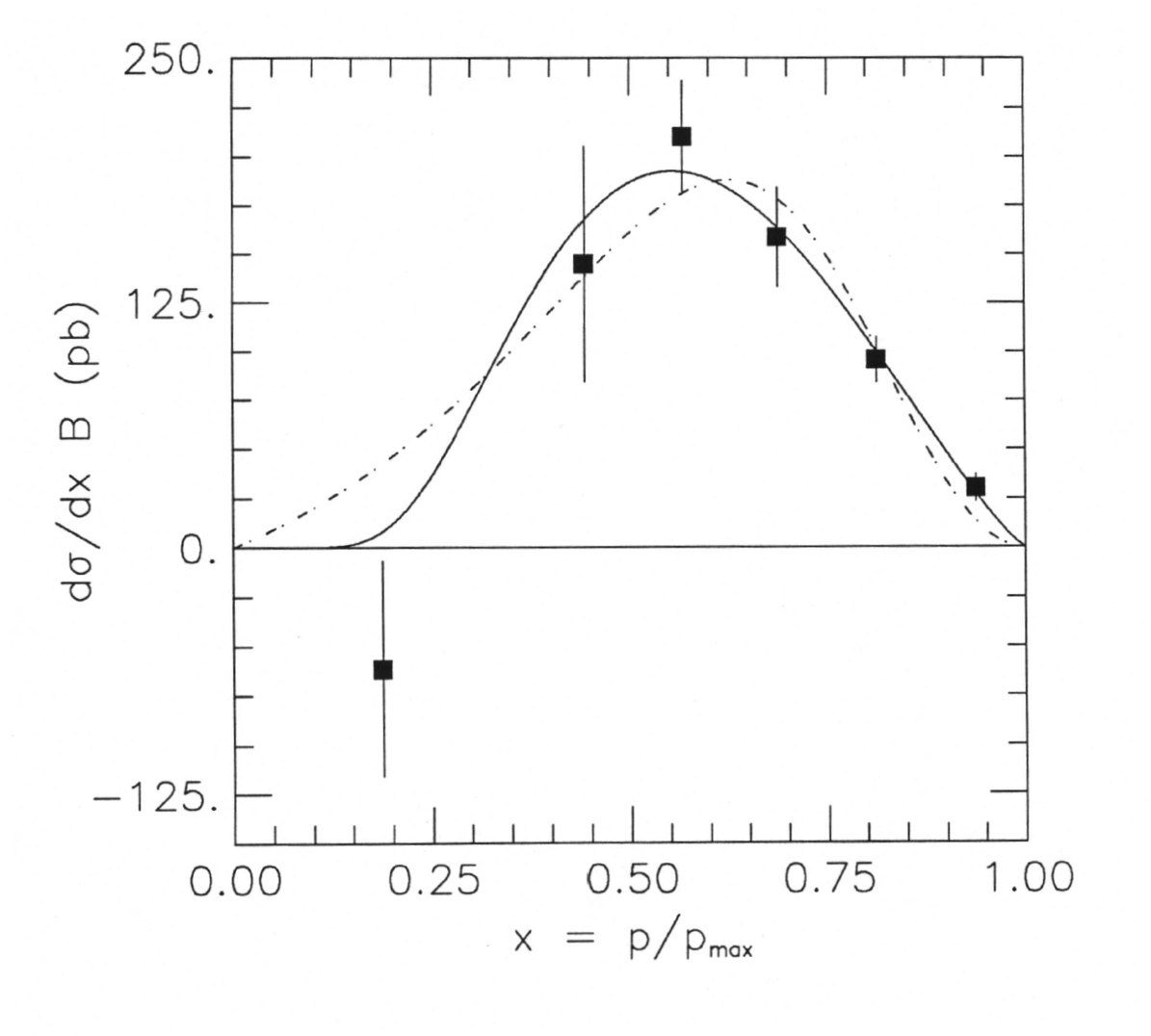}
\caption{Fits to the Andersson form (solid line) and Peterson form (dashed
line) for \DZ\ \decays\ \KSH\PIP\PIM . }
\label{fig:c5f11}
\end{figure}
Figures \ref{fig:c5f10} and \ref{fig:c5f11}.
We have elected not to analyze the fragmentation distribution of the
\KZB\PIP\PIP\PIM\ mode because of the much weaker statistical
significance. For comparison Figure \ref{fig:c5f12}
\begin{figure}[htp!]
\centering
\includegraphics[scale=0.565]{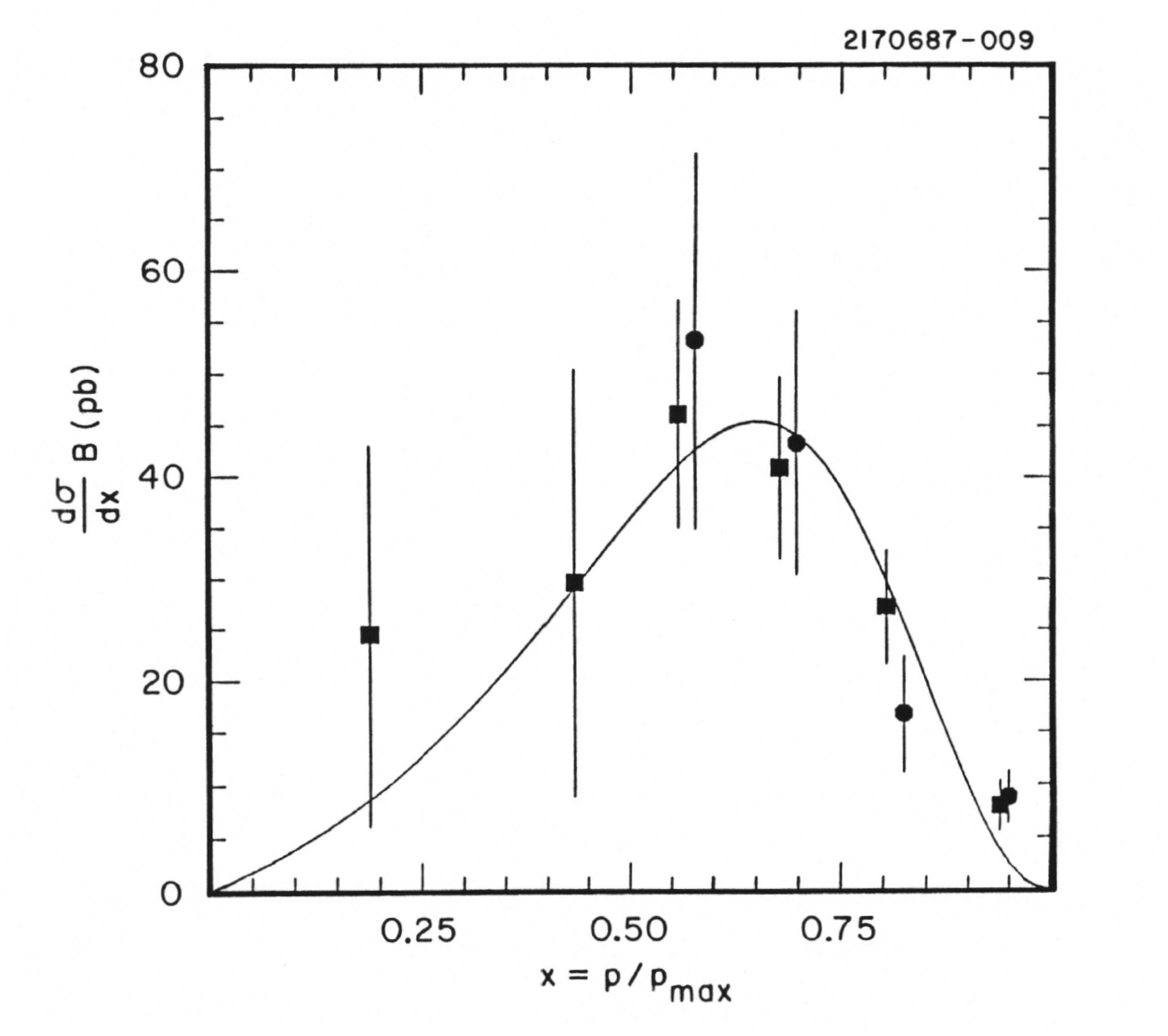}
\caption{The Fragmentation
distributions for \DP\ \decays\ \KSH\PIP\ (solid squares) and  \DP\
\decays\ \KSH\PIP\PIP\PIM\ (solid
 circles). The two modes have been scaled by their
relative  production ratios.}
\label{fig:c5f12}
\end{figure} 
 displays the
fragmentation distributions for the two \DP\ \decays\
\KZB\ X modes, scaled by their relative production
rates. Collectively  we note that the shapes of the fragmentation
distributions are consistent, as endorsed  by the
similarity of the parameters from the various fits.
\par
In interpreting the information from the experimental
observed fragmentation distributions, the historical  trend has been
to compare the values of the Peterson $\epsilon_Q$. This is an
unfortunate tradition which shall be continued here. The
main advantage of this approach is that $\epsilon_Q$ has an easily
interpretable physical meaning, namely the parameter 
$\epsilon_Q$ is proportional to
$m_{q\perp}^2/m_{Q\perp}^2$, where the
transverse mass is defined as ${m_\perp}^2 \equiv 
{m^2 + {p_\perp}^2}$. The heavier the object  is that combines with
the charmed quark, the larger epsilon will be. Also, radiative
effects will produce softer distributions which are again reflected
in larger values of epsilon. CLEO has made a precise \cite{frag}
measurement of the \DSP\ fragmentation distribution and found
$\epsilon_Q = 0.16 \pm\ 0.02$,  and has also
performed a fit to the fragmentation  distribution of several
 \LC\ modes \cite{malp}
and determined $\epsilon_Q = 0.30 \pm\
0.10$. The results of this analysis fit agreeably into this scheme.
The \D\ mesons are softer than the \DS\ mesons, as they should be
since \D's are produced in the cascade of \DS's, and the
fragmentation distribution of charmed baryons is softer than charmed
mesons. 
In conclusion, we note that mapping experimentally determined parameters
of fragmentation functions to theoretical predictions for those parameters
is an extremely perilous task. Data gathered at different energies by 
different experiments must be reconciled for choice of scaling
variable
and more importantly QED and QCD radiative effects. Work by Bethke \cite{bethke}
has attempted such a
reconciliation, and determined $<z_c> = 0.71 \pm\ 0.14$. This
corresponds to a Peterson $\approx\ \epsilon = 0.04$, which is more in
line with naive expectations. Work has also been done by Galik \cite{galik}
to evolve fragmentation
distributions to different energies (A detailed analysis of the CLEO \DS\ 
fragmentation
distribution, with comparison to similar measurements at different
energies, including radiative effects, can be found
elsewhere \cite{frag}).
 
\subsection{Estimates of Total Cross Sections}
\begin{table}[ht!]
\centering
\caption{  Extrapolated Total
Cross Sections (nb)}
\begin{tabular}{|c|c|}
\hline
Decay Mode &  $\sigma_{Total} $ \\ \hline
\DP\ \decays\ \KM\PIP\PIP & 0.63 \pp\ 0.06 \pp\  0.10 \pp\ 0.09 \\ \hline
\DP\ \decays\ \KZB\PIP  &  0.74 \pp\ 0.11 \pp\  0.12 \pp\ 0.11\\ \hline
\DP\ \decays\ \KZB\PIP\PIP\PIM  &  0.81 \pp\ 0.15 \pp\  0.20 \pp\
0.12 \\ \hline 
\DZ\ \decays\ \KZB\PIP\PIM &  1.49  \pp\ 0.13 \pp\  0.26 \pp\
0.22  
\\ \hline
\end{tabular}
\label{t:5p13}
\end{table}
 Because of the large errors in the fragmentation distribution at
low x, a measurement of the total cross section by summing the 
differential cross section will lose much of its statistical
significance. Alternatively, we can attempt such a measurement by
extrapolating from the part which is well measured. Both the
Peterson and Andersson  distributions predict about 63 \% of the
fragmentation distribution resides above $\rm x = 0.51$. We estimate
the systematic error for the extrapolation procedure by varying the
 observed value of $\epsilon_Q$ by $\pm\  1 \ \sigma$ and
determining the extrapolation factor. This, coupled with uncertainty
in theoretical models and the inability to measure the end points of
the fragmentation distribution well leads to an error of $ \simeq
15\%$. We base our extrapolation on the value of epsilon determined
from the \KZB\PIP\PIM\ and \KM\PIP\PIP\ modes, which are determined
with the most precision. Normalizing out the branching ratios (Table
5.1), we find the results presented in Table \ref{t:5p13}.
The first error is the statistical and
systematic combined in quadrature, the second is uncertainty in the
branching ratio, and the third term is due to the extrapolation
error.  CLEO has recently measured \cite{frag} the total
cross section for \DZ\ \decays\ \KM\PIP\ to be $ \sigma_{\DZ } =1.24
\pm\ .21$ (nb) which is consistent with this measurement of  $
\sigma_{\DZ } = 1.49
\pm\ .36$ (nb). We also compare a previous  CLEO \cite{bort}
 measurement for \DP\
\decays\ \KM\PIP\PIP\ of  $ \sigma_{\DP } = 0.52 \pm\ 0.11 $ (nb)
agrees with our weighted average of the three decay modes  $
\sigma_{\DP } =  0.68 \pm\ 0.13 $. The systematic error on the
previous measurement may have been underestimated.
\chapter{Penultimate}
In this chapter we  examine several unique aspects  of the
production and decay of charmed mesons. We shall detail a search
for a rare \DZ\ decay which may provide evidence for the role of
hadronic final state interactions in charm decay. We measure the
probability that a charmed meson will emerge from the \EPEM\
hadronization process in a state of non-zero angular momentum, and
analyze the experimentally important transition rate for a charged
\DS\ meson to decay  into a \DZ\ and a \PIP.
\section{\DZ\ \decays\ \KZ\KZB}
\subsection{Motivation}
\begin{figure}[p!]
\centering
\includegraphics[scale=0.8]{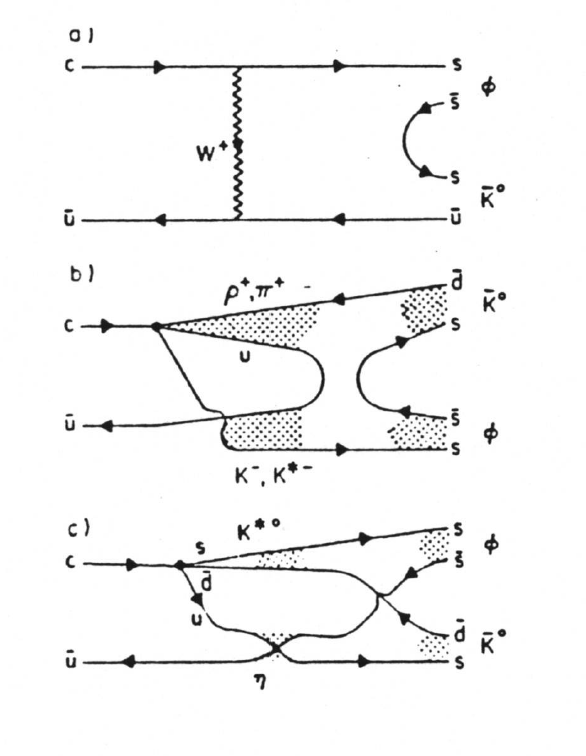}
\caption{\DZ\  \decays\ \PH\KZB\ through a) \W\ exchange and b)-c)
final state inter\\ -actions.}
\label{fig:c6f1}
\end{figure}

Decays of charmed hadrons have provided valuable insight into
the dynamics of the weak interaction. The nonleptonic sector is
believed to be the source of the radically different lifetimes and
semileptonic branching ratios of the charged and neutral \D\ mesons.
The mechanisms which produce these effects are not fully quantified.
A possible component of the solution is to shorten the \DZ\ lifetime
by including a  class of nonspectator decays known as \W\ exchange,
which are accessible to \DZ's  at the Cabibbo allowed level.  The
contribution of these processes was initially anticipated to be
small based on helicity arguments.  The experimental
observations \cite{halb2,bebek}
of the decay \DZ\ \decays\ \PH\KZB\
with a large branching fraction $(\sim 1\%)$ was initially considered
evidence for \W\ exchange. An alternate theory \cite{dono}
 proposed final state
re-scattering of hadrons, not \W\ exchange, as the origin of this
mode.  The  quark diagrams  leading to \DZ\ \decays\ \PH\KZB\ 
through \W\ exchange and final state interactions are can be found in Figure \ref{fig:c6f1}
 The extent to which either of these
theories successfully explains \DZ\ \decays\ \PH\KZB\ remains to be
resolved.

\par   The decay \DZ\ \decays\ \KZ\KZB\ is uniquely suited to 
study \cite{pham}
the effect of final state interactions. At the quark level 
this decay proceeds through two classes of nonspectator diagrams
\begin{figure}[p!]
\centering
\includegraphics[scale=0.675]{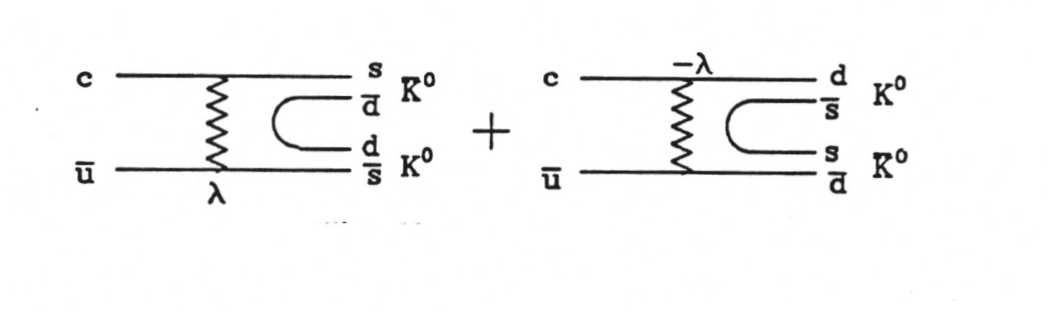}
\caption{Quark diagrams contributing to the decay  \DZ\ \decays\
\KZ\KZB. a) \W\ exchange, b) sideways ``penguin."}
\label{fig:c6f2}
\end{figure}
(Figure \ref{fig:c6f2}).
The unique
feature of this  decay is that both quarks present in the initial
state are absent in the final state, which in this case is composed
of an $\rm  s \bar  s$ and a  $\rm  d \bar  d$ pair. There are two
paths to the final state for each diagram, each of which contains one
Cabbibo suppressed \W\ vertex. In the limit of exact $\rm SU(3)_f$
symmetry, a  $\rm  s \bar  s$ can be popped from the vacuum on
equal footing with a $\rm  d \bar  d$ pair. We  could then 
factorize the term 
 $V_{ud}^{}V_{cd}^{\ast} + V_{us}^{}V_{cs}^{\ast}$, 
 which then 
multiplies the matrix element for each of the two diagrams.
This  expression can be recognized as the product of the first two  
terms of the first two rows of the K-M Matrix. Unitarity of the K-M
matrix demands $$V_{ud}^{}V_{cd}^{\ast} + V_{us}^{}V_{cs}^{\ast}  + 
V_{ub}^{}V_{cb}^{\ast} \equiv 0 $$  
In the limit of vanishing b \decays\ u  coupling, 
$V_{ud}^{}V_{cd}^{\ast} + V_{us}^{}V_{cs}^{\ast} \approx 0$, and
the amplitude for this decay vanishes.
\par 
A calculation by Pham \cite{pham}
 based on the
re-scattering of the modes \DZ\ \decays\ \KP\KM, and \DZ\ \decays\
\PIP\PIM, predicts B(\DZ\ \decays\ \KZ\KZB) $\simeq$ $1\over 2$
B(\DZ\ \decays\ \KP\KM) $\simeq 0.3 \% $.  A branching ratio of 1\%
or greater  for this decay could not be produced by final state
interactions and would represent a violation of the standard model.
If a substantial branching ratio were found for this mode, this
would confirm the role of hadronic final state interactions.
Conversely, if this decay was strictly ruled out, the case for
nonspectator processes would be strengthened. Since the general
theoretical framework for describing charm decays does not
fully implement either of these two processes, a better
experimental understanding of this decay mode would lend
valuable direction to charm decay theorists.
\subsection{Signal Isolation}
 Experimentally this mode can be cleanly observed in \KSH\KSH, given
good \KSH\ mass resolution and reconstruction efficiency.  
In the previous chapter we have demonstrated the robustness of the
CLEO detector in reconstructing \D\ meson  decay modes which contain
a \KSH.  For quick comparison, we note that using the \USSSS\ data
set, we have reconstructed $\sim 600$ high momentum \DZ\ \decays\
\KSH\PIP\PIM\ decays. \DZ\ \decays\ \KSH\KSH\ has a predicted 
branching ratio about twenty times smaller, and tacking on another
factor of 3 to get $\rm B (\KZB\ \decays\ \KSH)  \cdot  
 B(\KSH\ \decays\ \PIP\PIM)$ optimistically reduces our sample 
expectations to  10. Since we cannot hope to make an absolute
measurement of this decay rate with these statistics, we choose
to study the properties of this decay through normalization  to a
well known decay mode. The decay mode which is most similar to
\KSH\KSH\ is clearly \DZ\ \decays\ \KSH\PIP\PIM. They are both final
states containing four charged pions, and differ only by one
secondary vertex. Normalization to this decay mode provides maximal
cancellation of systematic errors.
\par
 To detect such a small signal we need to reduce the background to a 
minimum while maintaining reconstruction efficiency.
 We
restrict our sample to those events in which the candidate \DZ\ has
a momentum greater than 2.5 GeV.   
Despite the loss of the low momentum \D's, this cut has consistently
produced invariant mass peaks with the highest signal to noise in all
observed exclusive charm decays. Since we are performing a
normalization and not a cross section measurement, we are free to
take advantage of the \DZ's which are copiously produced in \B\
decay. Unfortunately, \DZ's produced at low momentum have \KSH\
daughters in the momentum range where the \KSH\ reconstruction
efficiency begins to fall off, producing as much as a 20 \% loss in
efficiency. Coupled with the rising background at low momentum, this
makes searching for this decay mode with low momentum \DZ's
unprofitable. \par
The data  for this analysis includes 113.7 \pb\ taken on the \USSSS\
and 33 \pb\ gathered on the \USSS.
The event selection procedure and track corrections follow those
outlined in section 5.1. Candidate \DZ's are then formed
from two \KSH\ candidates. To further improve the purity of the
sample we also require that the number of track pairs consistent
with secondary vertices in an event be less than five,
and that the
$\chi^2_{V}   \leq 3.0$  per \KSH\ candidate.
Based on the measured \KSH\ FWHM as a function of momentum, we demand
that each \KSH\ be within 2.5 standard deviations of the expected
\KSH\ mass. 
\subsection{Background}
\begin{figure}[htp!]
\centering
\includegraphics[scale=0.62]{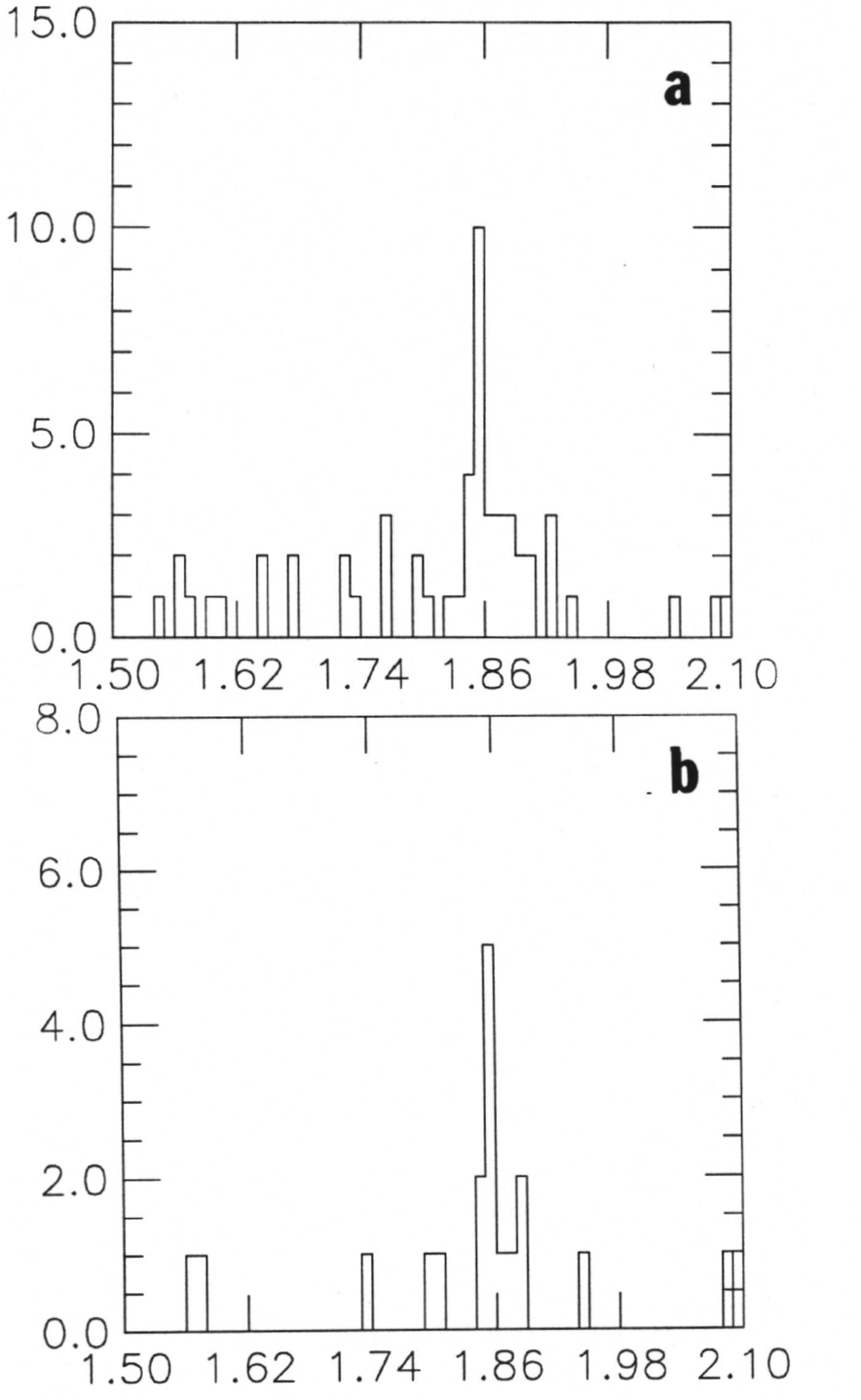}
\caption{Monte Carlo
simulation of the decay  \DZ\ \decays\ \KSM\PIP, \KSM\ \decays\
\KSH\PIM\ which was passed
through the \KSH\KSH\ analysis program. One \KSH\ candidate was
the correct \DZ\ daughter, while the other contained at least one
other  \DZ\ daughter.
a) no cuts, b) with \KSH\
mass and \cv\ cuts.}
\label{fig:c6f3}
\end{figure}

 We have analyzed the background from the decay \DZ\
\decays\ \KZB\PIP\PIM. We simulated this mode using the  
resonant substructure measurement outlined in Table 5.2. 
 For each  Monte Carlo event  we  tagged the four  final
state \DZ\ daughters  for this decay with matches to drift chamber tracks.
We then ran the event through our \KSH\KSH\ driver program, and
analyzed \KSH\KSH\ pairs only if one of the \KSH's was the true
\DZ\ daughter and the vee finder has
accidentally found one of the two \KSH\ tracks  to be
one of the other  \DZ\ daughter pions. For the \KSM\PIP\ mode,
a plot of these events in shown in Figure \ref{fig:c6f3}.

\begin{figure}[p!]
\centering
\includegraphics[scale=0.625]{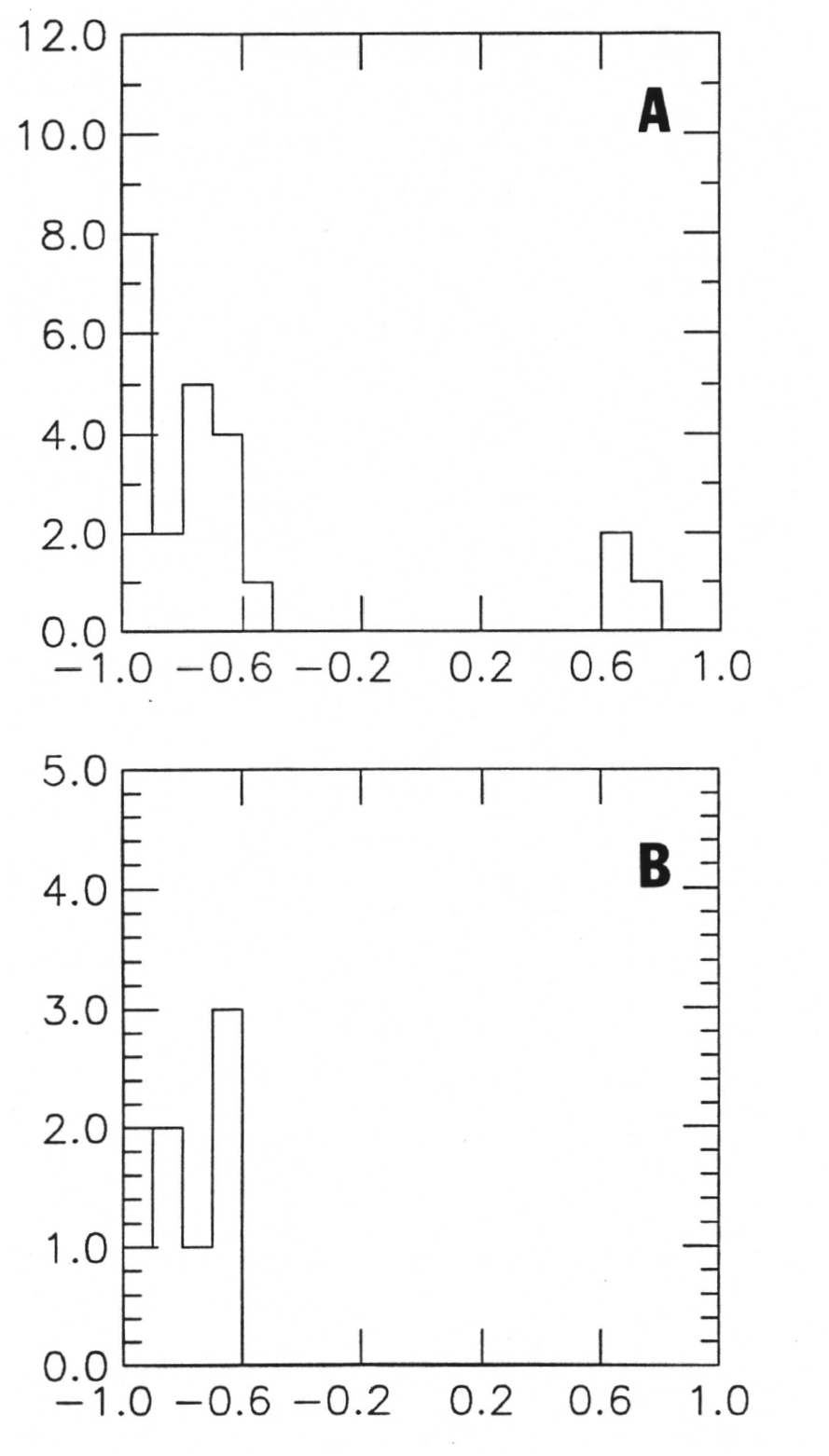}
\caption{$\cos\Theta_{P'P''}$ distributions A) without and B) with a \DZ\
mass cut.}
\label{fig:c6f4}
\end{figure}

An anomalous enhancement occurs at the \DZ\ region. No such
enhancement was found in either the nonresonant \KZB\PIP\PIM\ or
\KZB\RHZ\ modes. To test that this was not a statistical fluctuation,
we analyzed the properties of this distribution using the fact that
the \KSM\ is polarized in the helicity zero state (section 4.2).
The \KSM\ daughters are polarized with a   $\cos^2\Theta_{P'P''}$
distribution in the \KSM\ rest frame with respect to the \KSM\
direction in the \DZ\ rest frame. We calculate   this  angle 
 from the
Monte Carlo for the \PIM\ daughter, and histogram the quantity
for each  ``fake" \KSH\KSH\ candidate. We show this distribution
for all events and for events in the \DZ\ region in Figure \ref{fig:c6f4}.
 The events in the \DZ\ are clustered where 
 $\cos\Theta_{P'P''} \leq\ 0.7$. this would correspond to the case
where the \KSM\ daughter \PIM\ ends up almost at rest in the \DZ\
rest frame, and can end up being boosted in the same direction as
the \DZ\ \PIP\ daughter. We note that while this effect seems to be
caused by the unique kinematics of this particular decay, the
efficiency for such processes is exceedingly small. The efficiency
for a \DZ\ \decays\ \KSM\PIP\ to cause a \KSH\KSH\ candidate (where
both \KSH\ candidates have been classified as good) with a mass in
the \DZ\ region is of order 0.0006, while for a true \DZ\ \decays\
 \KSH\KSH\ decay the Monte Carlo
efficiency is 0.194. We note, however that for a \DZ\ 
\decays\ \KZ\KZB\ branching ratio of 0.1\%, and a \DZ\ \decays\
\KSM\PIP\ rate of 5.6\%, after scaling down the branching ratios
to reach the four pion final state, we find the branching ratio
times efficiency to be $2 \times 10^{-3}$ for \KZ\KZB\ and
$7 \times 10^{-4}$ for \KSM\PIP, differing by only a factor of two! To avoid
any such confusion, we  calculate the invariant mass of of each
\KSH\ candidate with the other two tracks which form the second
\KSH,
and require that the mass not be consistent with a \KSM. This
eliminates this anomalous enhancement while only slightly reducing
the \KSH\KSH\ reconstruction efficiency.  We also note that this
faking caused by \KSM\PIP\ does not seem to result from gross track
mis-measurement errors, as the reconstructed \DZ\ has a momentum on
average of 99\% of the Monte Carlo generated momentum.
\subsection{ Calculation of Upper Limit}
\begin{figure}[htp!]
\centering
\includegraphics[scale=0.62]{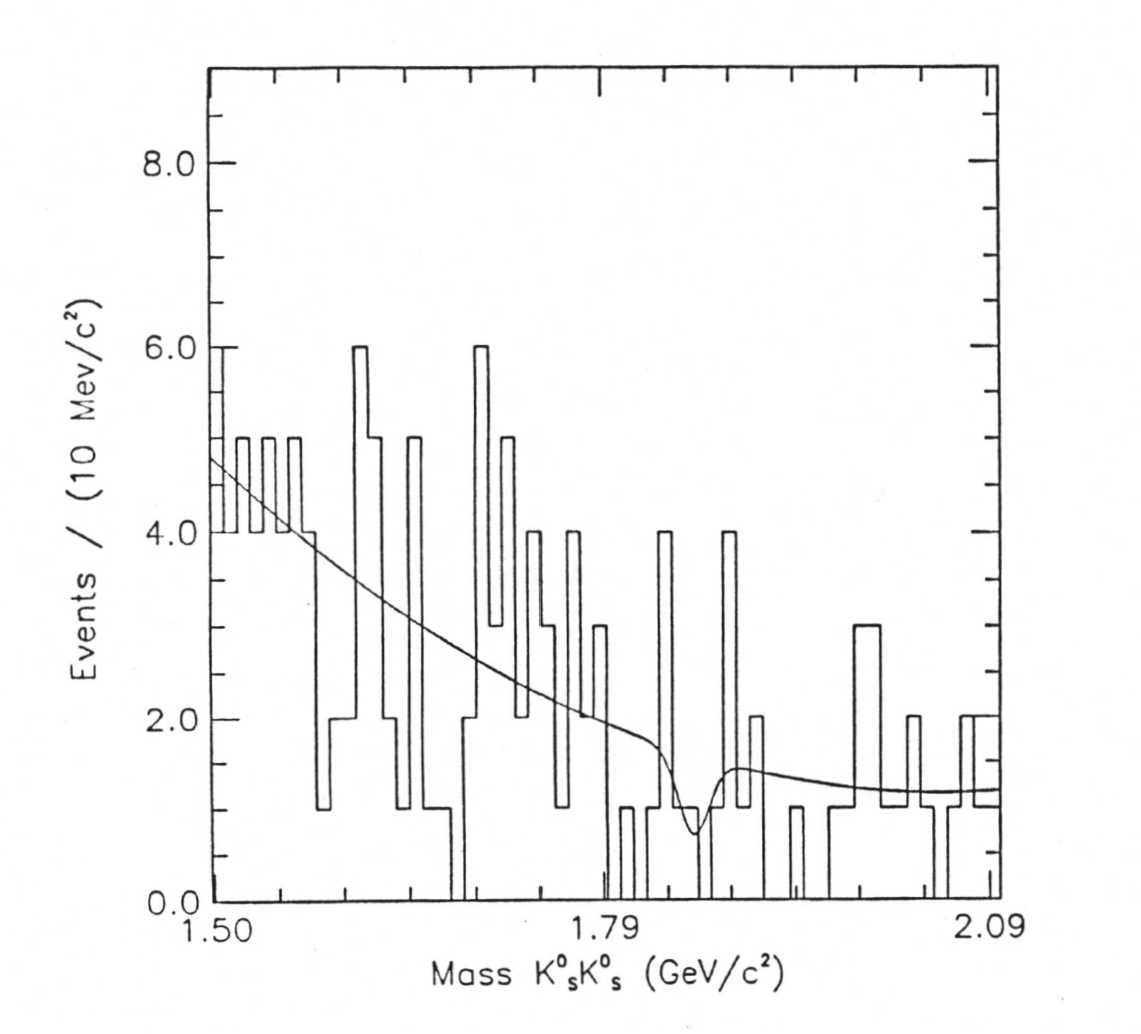}
\caption{Invariant mass spectrum for \DZ\
\decays\ \KSH\KSH. The \DZ\ candidate is required to have a
momentum in excess of 2.5 GeV.}
\label{fig:c6f5}
\end{figure}

 The   mass spectrum for \KSH\KSH\ candidates that have passed all
cuts is displayed in Figure \ref{fig:c6f5}.  
 We   use a Monte Carlo procedure to
determine the expected properties of the \DZ\ \decays\ \KSH\KSH\
signal, where the \KSH\ is allowed to decay into the \PIP\PIM\
mode.  Subject to the cuts described above, we find the mean 1.8645
$\pm$ .0002 GeV, FWHM .029 $\pm$ .003 GeV and an overall
reconstruction efficiency  of $\epsilon_{_{\KSH\KSH}}=.170\pm.007$.
We  fit the  mass spectrum (Fig 3.) using  a Maximum Likelihood 
method to a polynomial background and a Gaussian signal representing
the \DZ\ \decays\ \KSH\KSH\ decay. We observe a  signal $N_{\rm
obs}\KSH\KSH $ of $-2.7^{+2.7}_{-1.9}$   events at the \DZ\ mass,
where the errors are statistical  only. The stability of the
observed signal area and errors has been analyzed subject to
variations in \DZ\ mass, FWHM, and selection of background function.
The results from the various fits are consistent, and we
conservatively estimate a systematic error of 1.0 event.
\begin{figure}[htp!]
\centering
\includegraphics[scale=0.65]{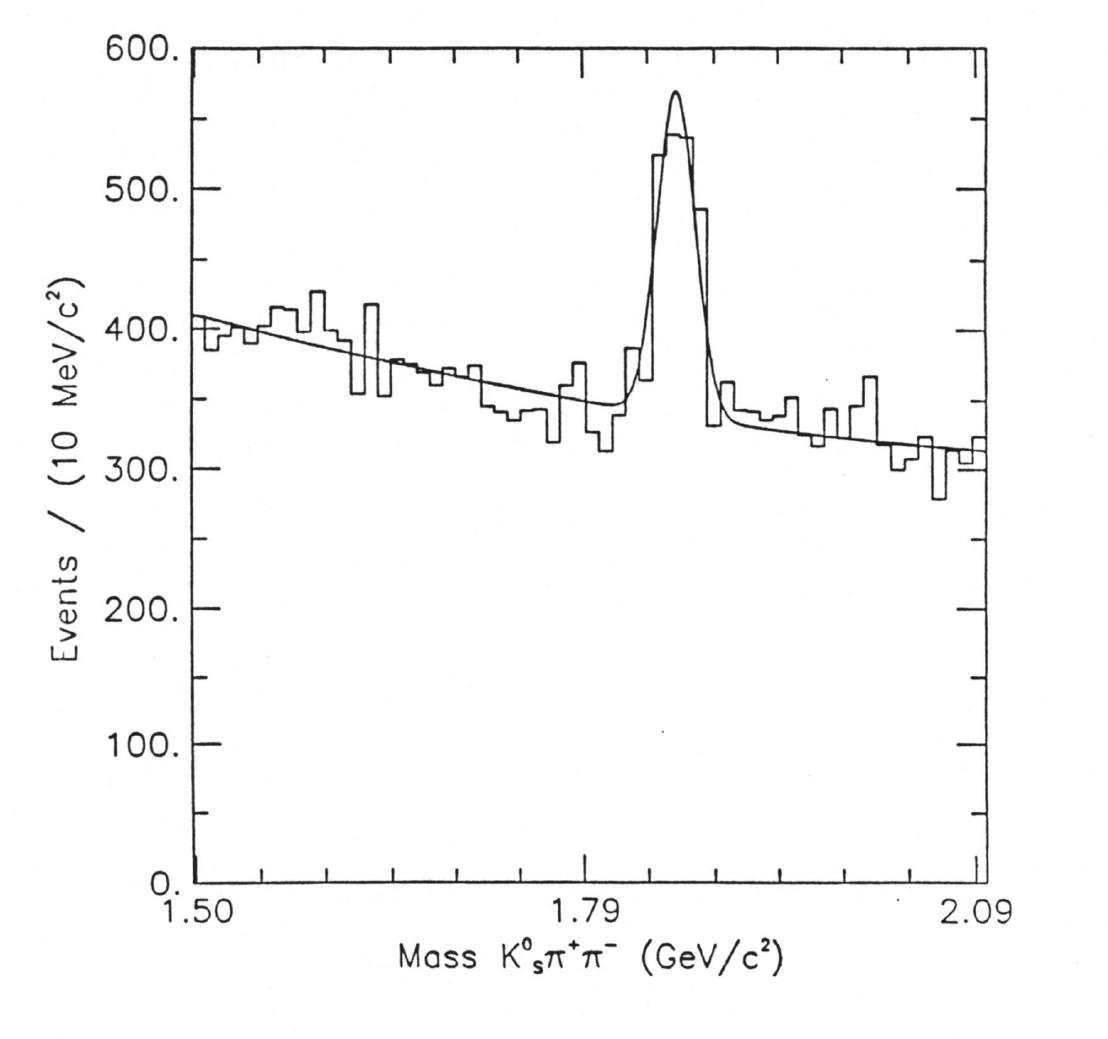}
\caption{Invariant mass spectrum for \DZ\ \decays\ \KSH\PIP\PIM\
which is used to normalize the \KSH\KSH\ signal.}
\label{fig:c6f6}
\end{figure}
Figure \ref{fig:c6f6}
displays the \DZ\ \decays\ \KZB\PIP\PIM\ signal which is used to
normalize the \KSH\KSH\ signal. In reconstructing this decay, we
subject the \KSH\ to the same selection cuts as applied to the
\KSH\KSH\ mode, and we require that the two \DZ\ daughter pions have
a momentum greater than 200 MeV. The signal is fit to a Gaussian signal
consistent with Monte Carlo simulation of this decay and a
polynomial background. For this decay mode we find
$N_{\rm obs}\KSH\PIP\PIM = 811 \pm 64$, and
${\epsilon_{_{\KSH\PIP\PIM}}} = .260 \pm .011$ 
 \par 
 The ratio of the branching fractions is obtained from the following
prescription, $$ { {\rm B} (\DZ\ \decays\ \KZ\KZB) \over
  {\rm B}(\DZ\ \decays\ \KZB\PIP\PIM) } = 2.91
 \left({N_{\rm obs} \KSH\KSH \over
{\epsilon_{_{\KSH\KSH}}}}\right)\cdot
 \left({\epsilon_{_{\KSH\PIP\PIM}}\over{N_{\rm
obs}\KSH\PIP\PIM}}\right)$$ 
which yields $-0.015 \pm 0.016$.  The factor 2.91
accounts for B(\KZ\ \decays\ \KSH)$\cdot$B(\KSH\ \decays\ \PIP\PIM). 
This ratio is converted to an upper limit on B(\DZ\ \decays\
\KZ\KZB)  utilizing the most recent \cite{adler1} Mark III value 
$\rm B(\DZ\ \decays\ \KZB\PIP\PIM) = 6.4 \pm 1.1 \% ${.}  From this
we find a 90\% confidence level upper limit of B(\DZ\ \decays\
\KZ\KZB\ ) $< .12\%$. The consistency of the normalization procedure 
has been checked  by normalizing  to the decay mode \DZ\ \decays\
\KM\PIP. This result is subject to systematic uncertainties in both
\KSH\ reconstruction and \KPM\ identification efficiency. The upper
limit determined in this fashion gives B(\DZ\ \decays\ \KZ\KZB\ ) $<
.17\%$.
\par 
Two other measurements of this decay have been publicized.
 The Mark III \cite{adler1} collaboration  has determined  the upper
limit B(\DZ\ \decays\ \KZ\KZB\ ) $< .46\%$.  
A signal for this decay has been claimed by the E-400 \cite{cuma}
 experiment.
Using the \DS-\DZ mass difference trick, they have observed a
signal of 8.9 \pp\ 2.7 events. They elect to normalize to the
decay mode \DZ\ \decays\ \KP\KM, which would seem to maximize the
prospects for systematic errors. They determine
$ { {\rm B} (\DZ\ \decays\ \KZ\KZB) \over
   {\rm B}(\DZ\ \decays\ \KP\KM ) } = 0.4 \pm 0.3$.
 Normalizing away the denominator using the Mark III branching
fraction $\rm B(\DZ\ \decays\ \KP\KM ) = .51 \pm .11$, E-400 finds
   B(\DZ\ \decays\ \KZ\KZB) = .20 \pp\ .16. This is consistent
with both the   upper limit determined from this analysis and 0. It
should also be noted that the E-400 group has  not  described  any
attempt to study the background to their signal, or examined the
validity of the signal by exploring the \KSH\ side bands. The also
have not normalized their \DZ\ \decays\ \KP\KM\  to any other \DZ\
decay mode.
 \par
In conclusion, more information is required to understand the
decay \DZ\ \decays\ \KZ\KZB. CLEO is currently performing an
analysis of this decay using a substantially larger $(\times 2)$
 data
set taken with a new 52 layer drift chamber.
 CLEO has observed 5 \DZ\
decay modes with high statistical significance, and fails to observe a
signal in this mode. CLEO has also determined that in their tracking
system that a non negligible background for this decay mode can be
produced from the decay \DZ\ \decays\ \KSM\PIP, \KSM\ \decays\
\KSH\PIM.
 The E-400 group has observed a statistically weak effect at
the \DZ\ mass in the \KSH\KSH\ final state. The have not
demonstrated the robustness of their normalization procedure, nor
unambiguously attributed the signal to \DZ\ \decays\ \KZ\KZB.
\section{ B(\DS\ \decays\ \DZ\PIP) and   Spinless \D\ Meson Production.}
\begin{table}[ht!]
\centering
\caption{  Extrapolated Total Cross Sections for Vector \D\ Mesons}
\begin{tabular}{|c|c|}
\hline
Decay Chain & $\sigma_{Total}$  (nb) \\ \hline
\DSP\ \decays\ \DZ\PIP\ & 0.77 \pp\  0.14 \\
\DZ\ \decays\ \KM\PIP\   & \\
\DZ\ \decays\ \KM\PIP\PIP\PIM & \\ \hline
\DSZ\ \decays\ \DZ\PIZ\ & 0.74 \pp\  0.18\\
\DSZ\ \decays\ \DZ\G & \\
\DZ\ \decays\ \KM\PIP\ & \\ \hline
\end{tabular}
\label{t:6p1}
\end{table}
In addition to extensive study of pseudoscalar charmed mesons, CLEO
has made precise measurements of the charged and neutral vector D
mesons. CLEO's  measurements \cite{frag} the total
cross sections for the \DSP\ and \DSZ\ mesons are summarized in
Table \ref{t:6p1}.
It was first noticed by the author that CLEO's unique situation
of possessing a complete set of charged and neutral vector and
pseudoscalar cross sections would allow for novel investigation of
the hadronization process. Let us define $\sigma_{P}$ as the cross
section for direct production of pseudoscalar charmed meson  and
$\sigma_{V}$ as the cross
section for production of  a vector charmed meson. We assume belief
in isospin, making $\sigma_{P}$ and $\sigma_{V}$ the same for
charged and neutral particles. Let us also define $\alpha$ as the
branching ratio B(\DSP\ \decays\ \DZ\PIP ) and $\beta$ as the
branching ratio B(\DSP\ \decays\ \DP X). Where we have the
constraint $ \alpha + \beta = 1$. The  total cross sections of \DZ\
and \DP\ mesons are governed by 
$$ \sigma(\DZ) =   \sigma_{P} + \sigma_{V} +\alpha\sigma_{V}$$
$$ \sigma(\DP) =   \sigma_{P} + \beta\sigma_{V}$$
after applying the constraint this becomes
$$ \sigma(\DZ) =   \sigma_{P} +  (1 + \alpha)\sigma_{V}$$
$$ \sigma(\DP) =   \sigma_{P} +  (1 - \alpha)\sigma_{V}$$
by adding and subtracting these two equations, we decompose the
equations into one which contains $\sigma_{P}$ and one which
contains $\alpha$. Solving these two equations for the aforementioned
quantities we find the relations
$$ \left( {\sigma(\DZ) + \sigma(\DP)  \over 2} \right) - \sigma_{V}
= \sigma_{P}$$
$$ \left( {\sigma(\DZ) - \sigma(\DP)  \over 2\sigma_{V}} \right)   
= \alpha$$
We note that in the derivation of the above
equations we have excluded contribution from the ${\D}^{\ast\ast
0}(2420)$. This is a candidate for the spin 2 charmed meson, its spin
and parity have not been established. Its relative production
rate to the \DS\ is on order \cite{bort2}
 $12 \pm
5 \%$.  This state has only been observed in cascades to the \DS.
Provided this state does not have a substantial decay rate directly
into the ground states, these relations are unaffected.
Using the results derived in the previous Chapter 
 $ \sigma_{\DP } = 
0.68 \pm\ 0.13 $,
$\sigma_{\DZ } = 1.49
\pm\ .36$ (nb), and defining $\sigma_{V}  =  \sigma_{\DSZ } = 
0.74   \pm\ 0.18$
we derive
$$ \sigma_{P} = 0.35 \pm\ 0.26 \rm (nb)$$
$$\rm  \alpha =   B(\DSP\ \decays\ \DZ\PIP ) = 0.53 \pm\ 0.29 $$
To evaluate our measurement of  B(\DSP\ \decays\ \DZ\PIP ), we
first make the simple comparison of the \DS\ to \DSZ\ cross section
where both particles cascade into a \DZ, which then decays into
the \KM\PIP\ final state. This yields
$ B(\DSP\ \decays\ \DZ\PIP ) = 0.55 \pm\ 0.07 \pm\ 0.11$.
Other measurements of this number include those of  MARK I \cite{gold}
 (0.60 \pp\
0.15),
 MARK II \cite{coles}
(0.44 \pp\
0.7), and
 MARK III \cite{hitlp}
(0.55 \pp\ 0.02 \pp\ 0.06).
 The MARK II number does not
assume conservation of isospin. We again find favorable agreement
among the measurements.  
Lastly, using our measurement of the direct \D\ cross section
to the amount of charmed particles which are observed in the vector
state. We evaluate the quantity
$  \sigma_{V} \over \sigma_{P} +  \sigma_{V}$ 
Using our derived value for 
$ \sigma_{P} = 0.35 \pm\ 0.26 \rm (nb)$ and a weighted average of
the \DS\ cross sections  for $\sigma_{V}$ we derive
$$ { \sigma_{V} \over \sigma_{P} +  \sigma_{V}} = 0.68 \pm\ 0.18$$ 
Traditionally the measurement was limited by the knowledge of  
B(\DSP\ \decays\ \DZ\PIP ), which for some time suffered a 20 \%
uncertainty. This method bypasses the need for that number.
Theoretical predictions for this ratio are  not well established,
however we note that simple spin statistics predicts 0.75.
In retrospect, we note that the errors on the numbers derived here make
them somewhat less competitive with previous measurements. We feel the
originality of the method warrants its presentation, which may find
application in the future. We also note that the means of these
measurements are in fine agreement with expectations, attributing to
the
consistency of several distinct detector functions.
\chapter{Summary}
 We have made a broad study of the properties of charmed mesons
produced in \EPEM\ annihilations.
The hadronization properties of charmed quarks were studied in two
ways. The fragmentation distributions of \DZ\ and \DP\ mesons were
compared using string inspired and independent fragmentation
models. The Peterson
function  provided consistently poorer fits to the data. This
was in part dominated by the highest x bin which the function
consistently undershot. It is noted that the Peterson function goes
to 0 at x = 1 with a much higher rate $(1-x)^2$ than the other
models $\sim(1-x)$. Radiative effects may soften the spectrum most
severely at high x, however these effects have not been examined  
(this is in part due to the fact that there is significant feed down
from the vector states in pseudoscalar fragmentation distributions,
making vector mesons the laboratory of choice to study these effects).
The
Peterson function also predicts a much larger distribution at low x
than do the string functions, which sharply cutoff at $\approx$ x =
0.3. This cutoff is more intuitively appealing, as catastrophic 
processes would be required in the hadronization process to produce
mesons with very low x.  We find the fragmentation distributions of
the \DP\ and \DZ\ mesons to be quite similar. The \D\ fragmentation
distribution is softer than the \DS\ and harder than the \LC , as
expected. It
was found that about 70 \% of the charmed mesons are produced with
at least 1 unit of angular momentum. This is further supported by
the large difference in the inclusive \DZ\ and \DP\ inclusive cross
sections.

The relative branching ratios were measured for three different \DP\
decay modes. They were found to be in good agreement with previous
measurements done at SPEAR. No specific theoretical information can
be gleaned form these relative rates. This is partly due to the
fact that in \DP\ decays there are often interfering diagrams 
(Figure \ref{fig:c2f8})
 so rates cannot be purely determined. This is not the case
in \DZ\ decays, where relative rates for two body decays often
provide  useful information. The second difficulty is  that in
general theoretical predictions tend to be limited to two body and
quasi two body decays, and no such predictions exist for three and
four body nonresonant decays.
We have also searched for nonspectator decays of the \DF\ as
a contamination of the decay mode \DP\ \decays\ \KZB\PIP. A novel 
technique was developed for this purpose, which would detect
signs of \DF\ reflections in two ways. Due to limited statistics,
we were unable to make a definitive statement, although no real sign for
this reflection was observed

We also searched for the special decay mode \DZ\ \decays\ \KZ\KZB,
which has been touted as evidence for hadronic final state
interactions in charm decays. The E-400 group has measured a
candidate signal for this decay and found B(\DZ\ \decays\ \KZ\KZB)
= 0.20 \pp\ 0.16 \%. This analysis is unable to confirm the E-400
measurement, and places the 90 \% confidence interval upper limit of
B(\DZ\ \decays\ \KZ\KZB) $<$ 0.12 \%. In addition, we have determined
that there is a non negligible background to this mode from the
decay \DZ\ \decays\ \KSM\PIP.

The theoretical understanding of charm decay has advanced greatly in
recent years. This was fueled, in part, by experiments performed at
\EPEM\ energies above the \PSPP, and fixed target   experiments.
The ARGUS group at DORIS was the first to observe the decay \DZ\
\decays\ \PH\KZB, and the tensor meson candidate
$\D^{\ast\ast 0}(2240)$. CLEO was the first experiment to perform a
high statistics study of the \DZ, \DP, and \DF\ lifetimes. The E-691
group has precisely determined the lifetimes of these three mesons,
along with measuring several rare decay modes. One difficulty of
these experiments has been the reliance on states reconstructed
purely from charged tracks, neglecting final states with single or
multiple neutral particles (this partially contributed to the
undoing of the MARK III double tag  method). The CLEO II detector,
with exceptional charged and neutral particle reconstruction
capabilities should make an excellent tool for studying charmed
decays and spectroscopy. Further advances in charm decay theory
will also require a better understanding of the charmed, strange,
meson \DF, which has thus far only been observed in five decay
modes. In conclusion, much has been learned, and much is to be
learned about the charm sector. The study of charmed particles has
 deepened our knowledge of elementary particle physics. A
fully quantified theory of charm decays will represent a great
triumph for theorists and experimentalists alike.

\bibliography{productionanddecay}

\end{document}